\documentclass[onecolumn,authoryear]{els-mrw} 

\usepackage{amsmath,amssymb,amsfonts,amsthm,makeidx,graphicx}
\usepackage{txfonts}
\usepackage{helvet}
\usepackage{anyfontsize}
\usepackage[colorlinks=true,citecolor=blue,filecolor=blue,linkcolor=black,urlcolor=blue,pdftex]{hyperref}

\renewcommand{\vec}[1]{\boldsymbol{#1}}

\def\mnras{\rm{MNRAS}}
\def\aj{\rm{AJ}}
\def\prl{\rm{Phys.~Rev.~Lett.}}

\def\prd{\rm{Phys.~Rev.~D}}
\def\aap{\rm{A\&A}}

\def\physrep{\rm{Phys.~Rep.}}
\def\apj{\rm{ApJ}}                 
\def\apjl{\rm{ApJ}}                
\def\apjs{\rm{ApJS}}               
\def\jcap{\rm{JCAP}}
\def\pasj{\rm{PASJ}}

\usepackage{color}
\usepackage{enumitem}
\def\beq{\begin{eqnarray}}
\def\eeq{\end{eqnarray}}

\begin{document}

\chapter{Power Spectrum, Bispectrum, 2- and 3-Point Correlation Function, and Beyond}\label{chap1}

\author[1]{Zachary Slepian}%
\author[2]{Farshad Kamalinejad}%
\author[1]{Alessandro Greco}%

\address[1]{\orgname{University of Florida},
\orgdiv{Department of Astronomy}, \orgaddress{211 Bryant Space Science Center, Gainesville, FL 32611, USA}}
\address[2]{\orgname{University of Florida}, \orgdiv{Department of Physics}, \orgaddress{2001 Museum Road, Gainesville, FL 32611, USA}}

\articletag{}

\maketitle

\begin{glossary}[Glossary]
\term{Cosmological Principle} That our location is not special, nor our viewpoint---the Universe is presumed to be statistically the same elsewhere as it is here (homogeneity) and the same as we look in each direction (isotropy).\\
\term{Homogeneity} Invariance under 3D translations.\\
\term{Isotropy} Invariance under 3D rotations.\\
\term{Parity} Transformation by spatial inversion $(x,y,z) \to -(x,y,z)$.\\
\term{Inflation} Hypothesized period of exponential expansion of space in the very early Universe.\\
\term{Baryon Acoustic Oscillations (BAO)} Pressure density waves with pressure provided by photons, fluid approximation justified by Thomson scattering with electrons, from very early Universe to decoupling at $z \sim 1, 020$.\\
\term{Gaussian Random Field (GRF)} Field such that, in Fourier space, at given $\vec{k}$, the real and imaginary parts are each drawn from a Gaussian with mean zero and variance given by the power spectrum at that $k$; thus the complex phase is uniformly distributed\\
\term{Primordial Non-Gaussianity (PNG)} Deviations of the density field, stemming from inflation, away from being a Gaussian Random Field\\
\term{Large-Scale Structure (LSS)} Distribution of matter throughout the Universe, often measured as traced by galaxies or quasars.\\
\term{Redshift Space Distortions (RSD)} Apparent breaking of isotropy in a 3D map of galaxy positions derived from converting redshifts to distances, due to peculiar velocities.\\
\term{Spectroscopic Survey} Map of large-scale structure (usually galaxies, can include quasars) made by taking spectra.\\
\term{Galaxy biasing} Idea that galaxies do not trace the large-scale distribution of matter with perfect fidelity, but also are sensitive to its square, its tidal tensor, etc.\\
\term{Perturbation Theory (PT)} Systematic scheme for constructing higher-order, gravitationally-induced corrections to the density and velocity perturbations in a given fluid (usually the matter, which is approximated as such), built recursively out of lower order perturbations (ultimately reducible to linear perturbations).\\
\term{Fourier Transform (FT)} Representation of a function in terms of a sum (or integral) over sines and cosines (or complex exponentials; or, in 3D, plane waves) of many different frequencies.\\
\term{Line of Sight (LOS)} Direction vector to a galaxy or galaxy pair or triplet; relevant for redshift-space distortions.\\
\term{Feldman-Kaiser-Peacock (FKP) Weight} Optimal inverse-variance weight for measuring a power spectrum.\\
\term{Landy-Szalay} Typically-used estimator for the 2PCF.\\
\term{Szapudi-Szalay} Typically-used estimator for the 3PCF.\\
\term{Power spectrum} Correlation of two Fourier-space density fields; loses all complex phase information; associated geometry is a line (one wave-vector). \\
\term{Bispectrum} Correlation of three Fourier-space density fields; sensitive to phase information; associated geometry is a triangle of wave-vectors.\\
\term{Trispectrum} Correlation of four Fourier-space density fields; sensitive to parity; associated geometry is a tetrahedron of wave-vectors.\\
\term{Polyspectrum} Term for generalization of power spectrum to more than four density fields.\\
\term{2PCF} Correlation of two position-space density fields (excess clustering beyond spatially random), as a function of separation.\\
\term{3PCF} Correlation of three position-space density fields (beyond spatially random), as a function of the triangle capturing their separations.\\
\term{4PCF} Correlation of four position-space density fields (beyond spatially random), as a function of the tetrahedron capturing their separations.\\
\term{Shot noise} Variance due to the discrete nature of galaxies; proportional to number density to an inverse power.\\
\term{Covariance matrix} A matrix describing the correlated error-bars between the $i^{\rm th}$ and $j^{\rm th}$ elements of a given observable, for instance the 2PCF in bins $r_i$ and $r_j$. Its diagonal gives the variance, and it is always symmetric and positive definite.\\
\term{Correlation matrix} The covariance matrix but with each element with index $ij$ divided by the geometric mean of the diagonal elements with indices $ii$ and $jj$.\\
\term{Precision matrix} Inverse of the covariance matrix; larger values for an element correspond to measuring the corresponding element of the observable with higher precision.\\
\term{Fisher matrix} Matrix that measures the sensitivity of a given element of an observable to a given parameter (such as a cosmological parameter), and accounts for errors on each element by including weighting by the inverse covariance. Used in Fisher forecasting; its inverse gives the error-bars achievable on each parameter in the forecast.
\end{glossary}

\begin{glossary}[Nomenclature]
\begin{tabular}{@{}lp{34pc}@{}}
LSS & Large-Scale Structure\\
NPCF & N-Point Correlation Function\\
$P / B / T$ & Power Spectrum / Bispectrum / Trispectrum\\
$\xi / \zeta / \zeta^{(4)}$   & 2PCF / 3PCF / 4PCF\\
$P_{\rm pri}$ & Primordial (density) power spectrum\\
$n_{\rm s}$ & Scalar spectral tilt\\
$A$ & Amplitude of the primordial power spectrum\\
$T_{\rm m}$ & Matter transfer function\\
$P_{\rm lin}$ & Linear power spectrum\\
$b_1 / b_2 / b_{\rm tidal}$  & Linear / Quadratic  / Tidal tensor bias\\
$a / z $ & Scale factor / Redshift\\
$ D / f$ &  Linear growth rate / Logarithmic growth rate\\
$\beta$ & Ratio of logarithmic growth rate to linear bias\\ 
EFT & Effective Field Theory (of Large-Scale Structure; also written EFTofLSS)\\
$c_{\rm ctr}$ & EFT counter-term coefficient\\
$n$ & Number density\\
$V$ & Volume of a survey\\
$\delta_{\rm g} / \delta_{\rm m} / \delta_{\rm lin} $ & Galaxy / Matter / Linear density fluctuation\\
$\tilde{\delta} $ & Fourier-space density fluctuation\\
$\delta^{(n)}$ & $n^{th}$-order density fluctuation\\
$\delta_{\rm D}^{[3]}$ & Dirac delta function\\
GRF & Gaussian Random Field\\
BAO & Baryon Acoustic Oscillations\\
RSD & Redshift-Space Distortions\\
PNG & Primordial Non-Gaussianity\\
FT / FFT & Fourier Transform / Fast Fourier Transform\\
$\hat{\xi}$ & Estimator for $\xi$\\
$D / R$ & Data  / Random particle as used to compute an NPCF\\
$N$ & Number of galaxies\\
$N_{\rm g}$ & Number of grid points for an FFT\\
$R_{\rm max} / V_{\rm max}$ & Maximum length-scale out to which a correlation is measured / Volume of a sphere with radius that length-scale\\
SDSS & Sloan Digital Sky Survey\\
BOSS & Baryon Oscillation Spectroscopic Survey (part of SDSS)\\
eBOSS & extended BOSS\\
DESI & Dark Energy Spectroscopic Instrument\\
LRG & Luminous Red Galaxy\\
sBf / $j_{\ell}$& spherical Bessel function\\
FFTLog & Algorithm to convert multiplicative-argument transforms (\textit{e.g.} against $j_{\ell} (k r)$) to convolutions\\
Hankel transform & Integration against a spherical Bessel function\\
$\mathcal{L}_{\ell}$ & Legendre polynomial\\

\end{tabular}
\end{glossary}

\begin{abstract}[Abstract]
N-Point Correlation Functions, usually with N = 2, 3, and their Fourier-space analogs power spectrum and bispectrum, are major tools used in cosmology to capture the clustering of large-scale structure. We outline how the clustering these functions capture emerges, explain that inflation produces a 2PCF or power spectrum but that subsequent evolution eventually produces a 3PCF or bispectrum, and beyond (and that inflation may do so as well at some level). Furthermore, in principle the Universe also has a 4PCF or trispectrum, and even clustering beyond. For each of these tools, we discuss the motivation, the practical details of how they are estimated, the current algorithms used to compute them, the theory behind them, and recent applications to data. Throughout, we focus on positioning the reader to find and apply these algorithms with some understanding, linking to public code for each algorithm to the fullest extent possible. 
\end{abstract}

{\bf Keywords}: Cosmology---Large-Scale Structure---Methods---Correlation Functions---Power Spectrum---Bispectrum---N-Point Functions---Data Analysis

\begin{BoxTypeA}[chap1:box1]{Key points}
\begin{itemize}
    \item \textbf{Clustering of large-scale structure} is (modulo redshift-space distortions) translation and rotation-invariant, and thus well captured by either correlation functions (of pairs, triplets, quadruplets, or beyond) or polyspectra (power spectrum, bispectrum, trispectrum, and beyond). They average over the translational and rotational degrees of freedom.
    \item \textbf{Redshift-space distortions} induce an apparent dependence of clustering on the line of sight, and can be folded in to the statistics above if the angle of the pair separation (or wave-vector) to the line of sight, or the analogous quantities for the 3PCF or bispectrum, are included.
    \item \textbf{Many public codes} exist to enable one to easily measure these statistics on large data-sets.
    \item \textbf{A wealth of information} on the Universe's contents, such as the matter density, dark energy, and neutrinos, as well as its laws (General Relativity, or any-beyond-GR theory) and its beginning (hypothesized to be inflation) is encoded in the clustering statistics, and they are a major pillar undergirding our current cosmological paradigm. 
    \item \textbf{Numerous current and upcoming} 3D spectroscopic surveys offer rich and massive data-sets to which to apply these statistics and unlock this information. 
\end{itemize}
\end{BoxTypeA}

\section{Overview}
Here we outline major tools used in large-scale structure cosmology---the 2- and 3-Point Correlation Functions (2PCF and 3PCF), their Fourier-space analogs the power spectrum and bispectrum, and briefly, N-Point Correlation Functions (and their Fourier-space analogs polyspectra) beyond $N=3$. 

This article is structured as follows. We begin by summarizing the origins of spatial clustering in the Universe (\S\ref{sec:origin})---after all, how can one appreciate the tools we use to quantify spatial clustering, without reference to the reasons it exists? We next turn to the 2PCF (\S\ref{sec:2pcf}) and power spectrum (\S\ref{sec:power_spec}), and then the 3PCF (\S\ref{sec:3pcf}) and bispectrum (\S\ref{sec:bispec}). We close with brief discussion of N-Point Correlation Functions and polyspectra (\S\ref{sec:beyond}).

Throughout, we will focus primarily on the \textit{current practice} rather than history, though we will certainly outline history where it underlies current practice. For a purely historical but relatively comprehensive (to its publication date) perspective, \citet{peebles_2001} is a rather complete review.
\\
\\
\noindent{\bf Notation}\\
In this work (and frequently in the literature), the 2PCF is $\xi$, the power spectrum $P$, the 3PCF $\zeta$, the bispectrum $B$, and we will use $\zeta^{(4)}$ for the 4PCF and $T$ for its Fourier-space analog the trispectrum. A density fluctuation in position space will be $\delta$, in Fourier space $\tilde{\delta}$, and the density fluctuation is $\delta(\vec{x}) = \rho(\vec{x})/\bar{\rho} - 1$ with $\rho$ the density and $\bar{\rho}$ the mean density of the Universe. We encourage skimming the {\bf Nomenclature} table to become familiarized with the rest of our notation.

\section{Origin of Spatial Clustering in Large-Scale Structure}
\label{sec:origin}
By way of motivation, we first outline the origins of spatial clustering in large-scale structure. Many objects relevant for cosmology, ranging from stars, to clouds of neutral hydrogen gas, all the way up to galaxies and quasars, are not randomly distributed in space, but rather are spatially clustered. For galaxies and quasars, the primary reason for this clustering is as an artifact of {\bf i)} the initial conditions of the Universe, hypothesized to stem from inflation, and {\bf ii)} the first 380,000 years of the Universe's evolution, when it is a hot, dense, ionized plasma of photons, electrons, and protons. Acoustic, \textit{i.e.} pressure-density, waves (Baryon Acoustic Oscillations, BAO) propagate through it. We now expand upon {\bf i)} and {\bf ii)}.

\subsection{Inflationary perturbations predict a slope for the power spectrum, and that pair statistics are sufficient}
\label{subsec:inflation}
Inflation is a rich subject; we provide only the barest pr\'ecis here, focused on what is needed to motivate the use of the 2PCF and power spectrum. We also skip over inflation's theoretical motivation here, instead pointing to some of the earliest papers: \citet{brout_1978, brout_1979, Starobinsky:1982ee}. For a very readable review on inflation, we suggest \citet{guth_2007}, and for a recent whitepaper reviewing theory and outlining future observational targets and surveys, \citet{achu_2022}. Inflation is a hypothesized period of accelerated expansion of the Universe in roughly the first $10^{-35}$ seconds after the Big Bang; the energy scales involved are beyond any probed by terrestrial accelerators, so much of the physics is less than certain. However, the basic outline is as follows: some source of energy, the inflaton, acted like a nearly constant energy density, leading to exponential expansion of the Universe. 

The inflaton is often taken to be a scalar field, $\phi$, and its potential, $V$, is taken to dominate over the kinetic energy term in the Lagrangian. This small kinetic energy in turn means small $d\phi /dt$, meaning that the change in $V$ over time will also be small---in short, $V$ is roughly constant. This is what leads to exponential expansion, and the condition that $d\phi/dt$ is small is known as slow-roll (a second slow-roll condition essentially states that $d^2 \phi/dt^2$ is also small, guaranteeing that the first condition endures over some amount of time). 

The rapid expansion of the Universe in turn is sufficient to stretch quantum mechanical fluctuations onto cosmologically relevant scales---and this is what is of interest for us here. Exponential expansion is scale-free, $d \ln a /dt = {\rm constant}$, with $a$ the scale factor, and typically it is framed that Fourier modes (denoted by $k$) corresponding to scales larger than the causal horizon at a given time are frozen and cannot evolve. Once a mode with a given $k$ ``enters the horizon'', \textit{i.e.} the horizon reaches order $1/k$, that mode begins to grow. The scale-free growth of the horizon thus leads to scale-free growth of Fourier modes---\textit{i.e.} equal power in the fluctuations in each logarithmic interval. This leads to what are termed ``scale-invariant'' initial density perturbations---producing a matter density power spectrum that scales very close to $P(k) \; d \ln k = {\rm constant}$, so that $P(k) \propto k$.

In other words, inflation already makes a prediction for a non-trivial power spectrum. In practice, we term this the ``primordial'' power spectrum and we parametrize the power law with $n_{\rm s}$, the scalar spectral tilt, which for perfectly scale-invariant initial conditions would be unity (the deviations from unity are of order the first slow-roll parameter, which we recall is given by $d \phi/dt$ which is small). A perfectly scale-invariant primordial power spectrum is also often called ``Harrison-Zeldovich''. Finding a primordial power spectrum with $n_{\rm s}$ near unity is a significant piece of evidence for inflation, and in fact a significant motivation to measure the power spectrum. In detail it turns out observationally that $n_{\rm s} \approx 0.9611$ \citep{planck_final}!

The second major prediction of inflation is that the density fluctuations it produces are, in Fourier space, uniformly random in phase, \textit{i.e.} $\tilde{\delta} (\vec{k}) = A (\vec{k}) \exp [ i \phi]$ with $\phi$ uniform in $[0, 2\pi)$. This means that the phase of the field at the end of inflation contains no information. This fact motivates how the power spectrum is constructed---by pairing the Fourier-space field and its complex conjugate, it removes the phase entirely, but this is lossless. 

Mathematically, inflation\footnote{This holds for ``standard'' inflationary models.} is said to produce a ``Gaussian Random Field'', or GRF---where the phase is uniform random, as above, and the amplitudes of the real and the imaginary parts of $\tilde{\delta}(\vec{k})$ are each drawn from a Gaussian with variance $P(k)$ (and mean zero, by construction of $\tilde{\delta}$). Importantly, the power spectrum (or 2PCF) fully characterizes such a field's properties. This is a very strong motivation for the measurement of these statistics in cosmology. In particular, for a GRF, all the odd higher-order statistics vanish (\textit{e.g.} the 3PCF or bispectrum), while the even ones (\textit{e.g.} 4PCF or trispectrum) can be reduced to products of 2PCFs or power spectra by Wick's Theorem, more correctly known in this context as Isserlis' Theorem \citep{isserlis}; for an interesting recent proof, see \citet{iso_isserlis}. 

The importance here cannot be over-stated: for a GRF, if we know the 2PCF or power spectrum, we know all that can be known. For a weakly non-Gaussian field, as it turns out our Universe is on cosmological scales, the 2PCF captures \textit{most} of what can be known.

In detail, we should acknowledge that the seeding of perturbations to the density of matter (as well as of photons and neutrinos) at the end of inflation occurs in an epoch known as ``reheating'' (\textit{e.g.} \citealt{reheating_1994}), where the inflaton field has rolled down its potential and now begins to oscillate, so that $d \phi/dt$ is appreciable, and the kinetic energy thereby liberated goes into particle production. The details of reheating remain an open area of theoretical work, but for the purposes of this review, they are not central. An excellent pedagogical review of inflation including treatment of reheating is \citet{baumann_tasi}.

We also note that the initial conditions for the relative density fluctuation in each species are generally taken to be adiabatic, \textit{i.e.} if one compressed a given volume of space-time slightly, one would have $\delta_{\gamma} = \delta_{\nu} =  (4/3) \delta_{\rm m}$, where the $4/3$ comes from the fact the photon and neutrino energy density grows as $r^{-4}$ with a compression while that of matter grows as $r^{-3}$, and the inflationary perturbations are, fundamentally, curvature perturbations, \textit{i.e.} alterations to the local space-time, affecting all species within that region. 

However, significant exploration has been made of other types of initial conditions, such as isocurvature perturbations (where different species' density fluctuations balance each other so as to produce no overall space-time curvature, \textit{e.g.} \citealt{bartolo_isocurv}); there has as yet been no observational evidence for these latter. Furthermore, it is still not well understood why there is any excess of matter over anti-matter at all, and the Universe is not filled with the radiation that would remain after the equal amounts theoretically expected annihilate. This problem is known as baryogenesis \citep{trodden, riotto}, originally pointed out by \citet{sakh_1967}; the corresponding problem regarding neutrinos is called leptogenesis \citep{lepto}. 

\subsection{BAO Era and Linear Evolution}
\label{subsec:bao}
BAO were first hypothesized by \citet{sakh_1966}, with related works by \citet{sunyaev_1970, peebles_1970}. CMB anisotropy observations by Cosmic Background Explorer (COBE) provided the first observational confirmation \citep{cobe}; BAO lead to the characteristic, degree angular scale of the CMB temperature hot and cold spots, seen most clearly in the successor satellites to COBE, Wilkinson Microwave Anisotropy Probe (WMAP) \citep{wmap_9} and \textit{Planck} \citep{planck_final}. The BAO's imprint in late-time large scale structure was first detected using the 2PCF \citep{Eisenstein_05} of Sloan Digital Sky Survey Data Release 6 (SDSS DR6), and the power spectrum of the 2-degree Field Galaxy Redshift Survey (2dFGRS) \citep{cole_2005}.

For BAO (treated rigorously though mathematically in \citealt{hs_96}), photons provide the dominant pressure, and the electrons providing scattering (Thomson scattering) which in turn isotropizes the photons' momenta and allows the system to be treated as a fluid.\footnote{The full set of equations, without invoking the fluid approximation, is given by conservation of phase-space density (Liouville's Theorem), and called the Boltzmann hierarchy; it is well-presented in \citet{ma_bertschinger}.} These sound waves propagate until the Universe has expanded and cooled enough to undergo recombination $(z$$\sim$$1,100)$ and then decoupling $(z$$\sim$$1,020)$, when the photons no longer can Thomson scatter the electrons, as these latter are all locked up in atoms. At decoupling, the sound waves must halt. There is then a sharp bump in velocity at the sound horizon (distance each sound wave has propagated from its starting point up to this time). This velocity feature serves as the spatial initial condition for the growth of density perturbations after, an effect called ``velocity overshoot'' \citep{press_vishniac}, and results in a small excess of galaxies on a spherical shell one sound horizon (also called the BAO scale, about 100 Mpc/$h$ comoving) away from any initial perturbation. A simple, analytic picture of BAO in position space is given using a two-fluid approximation (baryons+photons, and CDM) in \citet{SE_BAO_2016} (see also \citealt{bashinsky} and \citealt{esw_07}, with a related two-fluid treatment of CMB anisotropies in \citealt{seljak_1994}).

Measuring the BAO scale at different redshifts by looking at spatial clustering of galaxies on slices in $z$ can be used a standard ruler. Since the BAO scale is imprinted far before dark energy begins to drive the Universe's expansion, the stretch on each redshift slice can be calibrated to the pristine, unstretched BAO scale, providing a cosmic expansion history. Since, in detail, the expansion rate depends on the dark energy equation of state $w$, this cosmic expansion history constrains the equation of state, giving us hints as to dark energy's possible microphysical nature. For an observationally-oriented review of dark energy, see \citet{weinberg}; for a more theoretical review, \citet{copeland}. We show the 2PCF of Dark Energy Spectroscopic Instrument (DESI) data \citep{desi_dr1}, focusing in on the BAO feature, in Fig. \ref{fig:desi_2pcf}.

Importantly for the present article, the evolution during the BAO era does not alter the Fourier-space phase of the density field. Hence, the field remains a GRF. Indeed, for a substantial amount of cosmic time after decoupling, from $z$$\sim$$1,020$ all the way down to $z$$\sim$$2-3$, the evolution of the matter density is ``linear''---each Fourier mode evolves independently of the others, according to a linear differential equation. In detail, of course, the governing equations are non-linear (fluid) equations. However, since the density perturbations are small, all but the leading-order, linear terms are negligible in this regime. It is only at late times ($z \lesssim 2$) and on relatively small scales that the density perturbations become large enough that the non-linear terms in the fluid equations must be retained. We return to this very shortly.

\subsection{Beyond Pairs and Beyond Gaussianity}
\label{subsec:beyond_pairs}
There are three important motivations for going beyond pair statistics, which we now briefly present. 

\subsubsection{Primordial Non-Gaussianity (PNG)}
\label{subsub:png}
One further complication to the picture above is that, generically, some level of deviation from a GRF is expected in inflation. If inflation is driven by a single field, this deviation is of order unity or larger in suitable units; if inflation is driven by multiple fields, the deviation in the same units is expected to be much less than unity. This effect is called Primordial Non-Gaussianity, or PNG, and it gives a strong reason to measure statistics beyond pairs, since deviations from a GRF mean that pair statistics alone no longer exhaust the information \citep{bartolo}. PNG is often parametrized via $f_{\rm NL}$, but we should emphasize the form of PNG this parametrization captures is just one of a more general set of beyond-GRF scenarios possible, all of which fall under the umbrella term ``PNG.''

\subsubsection{Non-Linear Evolution}
\label{subsub:nonlin}
While at early times $(z \gtrsim 2)$, the density fluctuations on scales of cosmological interest are quite small, and thus terms of order $\delta_{\rm lin}^2$ can be neglected in solving the gravitational evolution, at low redshift ``non-linear'' evolution occurs and we need to retain such terms. Non-linear evolution can be treated using Perturbation Theory (PT; for a standard review, \citealt{Bernardeau_2002}), where higher-order solutions of the fluid equations for the density $(\delta)$ and velocity divergence $(\theta)$ are developed in terms of lower-order solutions. 

This recursive structure, in particular the $\delta \theta$ coupling in the continuity equation and the $\theta \theta$ coupling the Euler (momentum) equation, with $\theta$ the velocity divergence, leads naturally to interactions of density and velocity fluctuations at different wave-vectors. This ``mode-coupling'' is expressed with celebrated kernels, first developed in the 1940s by Lifshitz (reprinted in \citealt{lifshitz_repub}) and again independently in \citet{goroff, bert_jain}.\footnote{For the evolution of relativistic perturbations to second order, see \citet{matarrese_1998}.} These kernels are then integrated against lower-order solutions to produce higher-order ones.

For our purposes, it suffices to understand that these higher-order solutions, very roughly, stem from squaring or raising to a higher power Gaussian density and velocity fluctuations (at the lowest, linear, order, the velocity field has the same statistics as the density field since they are related by the linearized continuity equation). Hence, the higher-order solutions cannot be a GRF! Schematically, we may write the matter density field as
\begin{align}
\tilde{\delta}_{\rm m} (\vec{k}) = \tilde{\delta}_{\rm lin} (\vec{k}) + \tilde{\delta}^{(2)} (\vec{k}) + \cdots, 
\end{align}
where $\tilde{\delta}^{(2)} (\vec{k}) = \int d^3\vec{q}_1 \, d^3\vec{q}_2\; \tilde{\delta}_{\rm lin}(\vec{q}_1)\; \tilde{\delta}_{\rm lin}(\vec{q}_2)\; F_{2}^{(\rm s)}(\vec{q}_1, \vec{q}_2)\;\delta_{\rm D}^{[3]} (\vec{q}_1 + \vec{q}_2 - \vec{k})$ is an integral over linear fields (subscript ``lin'') at all $\vec{q}_1$ and $\vec{q}_2$ that add up to $\vec{k}$. $F_{2}^{({\rm s})}$ is the second-order density kernel, and superscript ${\rm s}$ means it is symmetrized over its arguments. 

Mode-coupling thus, even at the level of the matter field (forget for a moment halos \footnote{Gravitationally bound structures composed of dark matter particles, halos form through hierarchical growth, where smaller halos merge to create larger ones.} or galaxies!), generates higher-order correlations.

\subsubsection{Galaxy-Matter Connection}
\label{subsub:bias}
 A third source of non-Gaussianity is galaxy biasing---the idea that galaxies do not trace the matter field with perfect fidelity. One can view the galaxy bias expansion as essentially a Taylor series for some true function known to the Universe (but not us) that maps matter to galaxies; schematically we have
\begin{align}
\delta_{\rm g}(\vec{x}) = b_1 \delta_{\rm m}(\vec{x}) + b_2 \left[ \delta_{\rm m}^2(\vec{x}) - \left< \delta_{\rm m}^2(\vec{x}) \right>  \right] + \cdots
\label{Eq:bias}
\end{align}
where subscript ``g'' indicates ``galaxy'' and $\cdots$ here captures a myriad of terms, such as dependence on the local tidal tensor, on the cube of the density field, on baryon-dark matter relative velocity bias, on velocities, etc. $b_1$ and $b_2$ are the ``coefficients'' in our Taylor series---unknown constants to be measured from data, and presumably stemming from the detailed physics of galaxy formation. We note that in the $b_2$ term, the mean square is subtracted off to retain the condition that $\left< \delta_{\rm g} \right> = 0$, and angle brackets mean average over all space. 

Full discussion of biasing merits its own substantial review article and we recommend \citet{desjacques} for a recent treatment. Here we simply comment that the Press-Schechter formalism \citep{press_schechter} predicts $b_1$ as a function of the galaxy mass one is observing, and that $b_2$ (and also tidal tensor bias) can be obtained from $b_1$ if one assumes the biasing is local to a galaxy's initial position (``local Lagrangian biasing'') \citep{baldauf}; there is some support for this from simulations \citep{chan}.\footnote{For instance, $b_{\rm tidal} = -(2/7)[b_1 - 1]$ in this framework.} 

Overall, here, it is sufficient to notice that the non-linear nature of the bias expansion would convert even a GRF $\delta_{\rm m}$ on the right-hand side in Eq. (\ref{Eq:bias}) into a non-GRF $\delta_{\rm g}$. (As we have noted, the matter already weakly deviates from a GRF, too.) Thus, galaxy biasing also gives rise to higher-order statistics.

Now, suppose one points out that galaxy biasing is a theoretical construct, and one would prefer simply to simulate galaxy formation \textit{ab initio} from a primordial fluctuation field? A noble goal if achievable; nonetheless, it would certainly still lead to a non-GRF galaxy density, and indeed this has been found in a variety of simulations. In fact, it turns out that on cosmological scales, a relatively compact bias model often is able to well-capture the results of simulations of galaxy formation.

Finally, we note that non-linearity in the matter usually becomes relevant only at relatively low redshifts, while galaxy biasing is generically active at any redshift one is forming (or observing) galaxies. PNG would also be observable at any redshift if it is present at all---it is, by definition, imprinted on the density field during inflation.

\section{2-Point Correlation Function}
\label{sec:2pcf}
\subsection{Motivation and Definition}
The 2PCF measures excess spatial clustering of pairs of objects, such as galaxies, over and above those which a spatially random distribution would have; it is typically denoted $\xi$. It generally measures the excess clustering as a function of pair separation, $r$; this separation is usually binned, into spherical shells, with the most common choice being uniform bins in $r$ (corresponding to uniform spherical shells), but logarithmic bins are sometimes also used.

\subsection{Estimators}
Several estimators have been historically used for the 2PCF. A common feature is that they utilize random particles to correct for spurious correlations induced by the survey geometry (RA, Dec, and $z$ range, as well as holes for masked bright stars), as well as to convert the number count field (galaxies are discrete) into a density fluctuation field.  Here we denote a random particle $R$, and a data particle $D$. The most naive estimator (indicated with a hat) is, about a point $\vec{x}$, 
\begin{align}
\hat{\xi}_{\rm naive} (\vec{r}; \vec{x}) = \frac{D(\vec{x})  D(\vec{x} + \vec{r})} {R(\vec{x})  R(\vec{x} + \vec{r})} - 1.
\end{align}
This estimator would then be averaged over all choices for $\vec{x}$, so that $\xi(\vec{r}) = V^{-1} \int d^3\vec{x}\; \hat{\xi}_{\rm naive} (\vec{r}; \vec{x})$, with $V$ the survey volume. It can also be averaged over directions $\hat{r}$, though it need not be; we return to this issue later.

This estimator already raises an important question---what should be done if the denominator is zero? Since the random points are thrown randomly (usually with some over-sampling relative to the number density of the data), we cannot guarantee there will be one at each point where there is a data point. Furthermore, if there is a missing patch of the survey, as is often the case due to for instance a masked bright star, or the Galactic plane, while the numerator will be zero, the denominator will be too---how will this limit be taken correctly?

For this reason, the estimators are interpreted as: evaluate the estimator but optimally weighted by inverse noise, which in this case, assuming that the random contribution dominates, is just the randoms' pair counts. Hence, we compute in practice
\begin{align}
\left< \xi_{\rm naive, opt} (\vec{r}) \right> = \frac{\int  d^3\vec{x}\; \frac{D(\vec{x})  D(\vec{x} + \vec{r})} {R(\vec{x})  R(\vec{x} + \vec{r})} \times R(\vec{x})  R(\vec{x} + \vec{r})}{ \int  d^3\vec{x}\;  R(\vec{x})  R(\vec{x} + \vec{r})} - 1,
\label{eqn:naive}
\end{align}
where the overall denominator now is interpreted simply as a normalization of the optimal weights, and the angle brackets denote averaging over all positions $\vec{x}$ within the survey. No division by survey volume is required because this is dealt with by the bottom integral. 

The averaging over $\vec{x}$ is taken to be lossless by invoking the Cosmological Principle's assumption of homogeneity, \textit{i.e.} that the laws of physics, and the Universe, are largely the same over here as over there. On small scales, this is evidently not true, but over the typical volumes associated with modern-day galaxy surveys (cubic Gpc), current observations support it, \textit{e.g.} \citealt{laurent, goncalves}. That said, there are subtle effects that can appear to break the assumed homogeneity---in particular, if one is exploring a fat enough slice in redshift, the galaxy population may actually evolve over the interval (known as ``bias evolution''), breaking the translation invariance (\textit{e.g.} \citealt{fry_bias_evol}).

Estimators other than the naive one above were explored in the early-to-mid 1990s, and \citet{landy_szalay} found that the variance is lower, and there is also less sensitivity to careful normalization of the randoms (this is usually called the integral constraint) if instead one has
\begin{align}
\hat{\zeta}_{\rm LS} (\vec{r}, \vec{x}) = \frac{[D(\vec{x}) - R(\vec{x})]   [D(\vec{x} + \vec{r}) - R(\vec{x} + \vec{r})] } { R(\vec{x}) R(\vec{x} + \vec{r})},
\end{align}
where in practice this estimator should be interpreted in the same was as described for the ``naive'' one, \textit{i.e.} \`a la Eq. (\ref{eqn:naive}). As a shorthand this is often written $(D - R)^2/(RR)$ where it is understood that squaring or multiplication actually means evaluating at two possibly different points. \citet{kerscher_2000} carefully compares different 2PCF estimators.

We now describe direction-averaging and binning. Direction-averaging (\textit{i.e.} integration over $\hat{r}$) is taken to be lossless by invoking the second piece of the Cosmological Principle, isotropy. Specifically, it assumes isotropy about each position $\vec{x}$; an observer sitting at any point in the survey should see the same galaxy clustering in all directions around them. It is implemented as averaging over directions of the pair separation, \textit{i.e.} $(4\pi)^{-1} \int d\Omega_r\, \xi(\vec{r})$, and mathematically, can be done before or after translation-averaging, though in practice is typically done before. 

The 2PCF is usually binned in pair separation as well, as
\begin{align}
\xi( R )  \equiv \int r^2 dr \; \xi(r) \Phi(r; R)
\end{align}
where the direction-averaging has already been performed (hence the dependence on $r$ alone in the integrand) and $\Phi(r; R)$ is a binning function normalized to have integral unity, and non-zero only from $R - \Delta r/2$ to $R + \Delta r/2$, with $R$ the bin center and $\Delta r$ the bin width. \citet{xu_2012} presents the 2PCF definitions, including binning, and discussion of the covariance matrix, cleanly; \citet{Eisenstein_05} \S3 also contains worthwhile discussion of 2PCF estimation on SDSS data. 

We now very briefly explain the idea of a covariance matrix. This is simply the matrix describing the correlated error-bars of an observable; for the 2PCF, its diagonal is the variance on the measured 2PCF in each bin, and its off-diagonals the covariance between the measured 2PCF on two different bins. By construction, a covariance matrix is symmetric, and hence by the Spectral Theorem it is always diagonalizable. A related quantity that is often useful for visualization is the correlation matrix, defined as $C^{\rm corr}_{ij} \equiv C_{ij}/\sqrt{C_{ii} C_{jj}}$, where $C_{ij}$ is the covariance matrix. Notably, the correlation matrix is by construction always unity on the diagonal; thus it loses information about the diagonal, so often correlation matrix plots are accompanied by a 1D plot of the diagonal elements of the covariance. The covariance of a given observable can be obtained either from a theory template (often, a GRF is assumed, \textit{e.g.} as in \citealt{xu_2012} for the 2PCF and \citealt{se_3pt_alg} for the 3PCF) or from mock datasets generated from simulations. 

We here pause to define the covariance: it is simply a matrix ${\bf C}$ with elements
\begin{align}
    C_{ij} \equiv \left<\xi_i \xi_j\right> - \bigg< \xi_i \bigg> \left<\xi_j\right>,
\end{align}
where here the subscripts $i$ and $j$ may be taken to refer to the $i^{th}$ and $j^{th}$ radial bins of the 2PCF, $\xi$, but the definition is in fact general to any observable with elements indexed in this way. For instance, the 3PCF, which depends on three parameters, could be mapped to a 1D ``vector'' and its covariance also written as above. Angle brackets represent an imagined averaging over realizations of the universe.

Often the covariance is estimated by measuring the desired observable on many different mock (simulated) galaxy catalogs, as 
\begin{align}
    C_{ij} = \frac{1}{N_{\rm mocks} - 1}\sum_{n = 1}^{N_{\rm mocks}} \xi_i^{(n)} \xi_j^{(n)} - \bar{\xi}_i \,\bar{\xi}_j,
\end{align}
where $N_{\rm mocks}$ is the number of mocks, the superscript in parentheses, $n$, indexes the mock; we note that the product is formed over two possibly different elements of the ``vector'' of observables, $i$ and $j$, but at the \textit{same} mock, $n$. Bar denotes the average over the set of all the mocks given. The ``averaging'' given by dividing by $N_{\rm mocks} - 1$ has the minus one because the means (barred quantities) are also estimated from the mocks, effectively taking away one independent piece of information. Hence, the average should be taken as $1 / (N_{\rm mocks} - 1)$.

For detailed discussion of issues that arise when a covariance is estimated from a set of $N_{\rm mocks}$ mocks, see \citet{hartlap, sellentin, percival_covar}. The first, \citet{hartlap} presents a rough factor that can be used to de-bias the estimated inverse covariance, the ``Hartlap factor'', roughly $(N_{\rm mocks} - {\rm d.o.f.})/N_{\rm mocks} < 1$, with ${\rm d.o.f.}$ the number of degrees of freedom.\footnote{The Hartlap factor decreases the inverse covariance, also known as the precision matrix---hence, it is simply capturing the fact that using a noisy covariance from a finite number of mocks lowers one's precision, as might be intuitively expected.} The second, \citet{sellentin}, explains that the correct distribution to compare a model's goodness of fit statistic to in this case is not a $\chi^2$ distribution, but rather a $T^2$ distribution, which captures the fact that the covariance used to evaluate $\chi^2$ itself had noise. The third, \citet{percival_covar}, has useful connections to observational LSS. These issues are also nicely summarized in \citet{gouyou}.

\begin{figure}[ht]
    \centering
    {\includegraphics[width=0.8\textwidth]{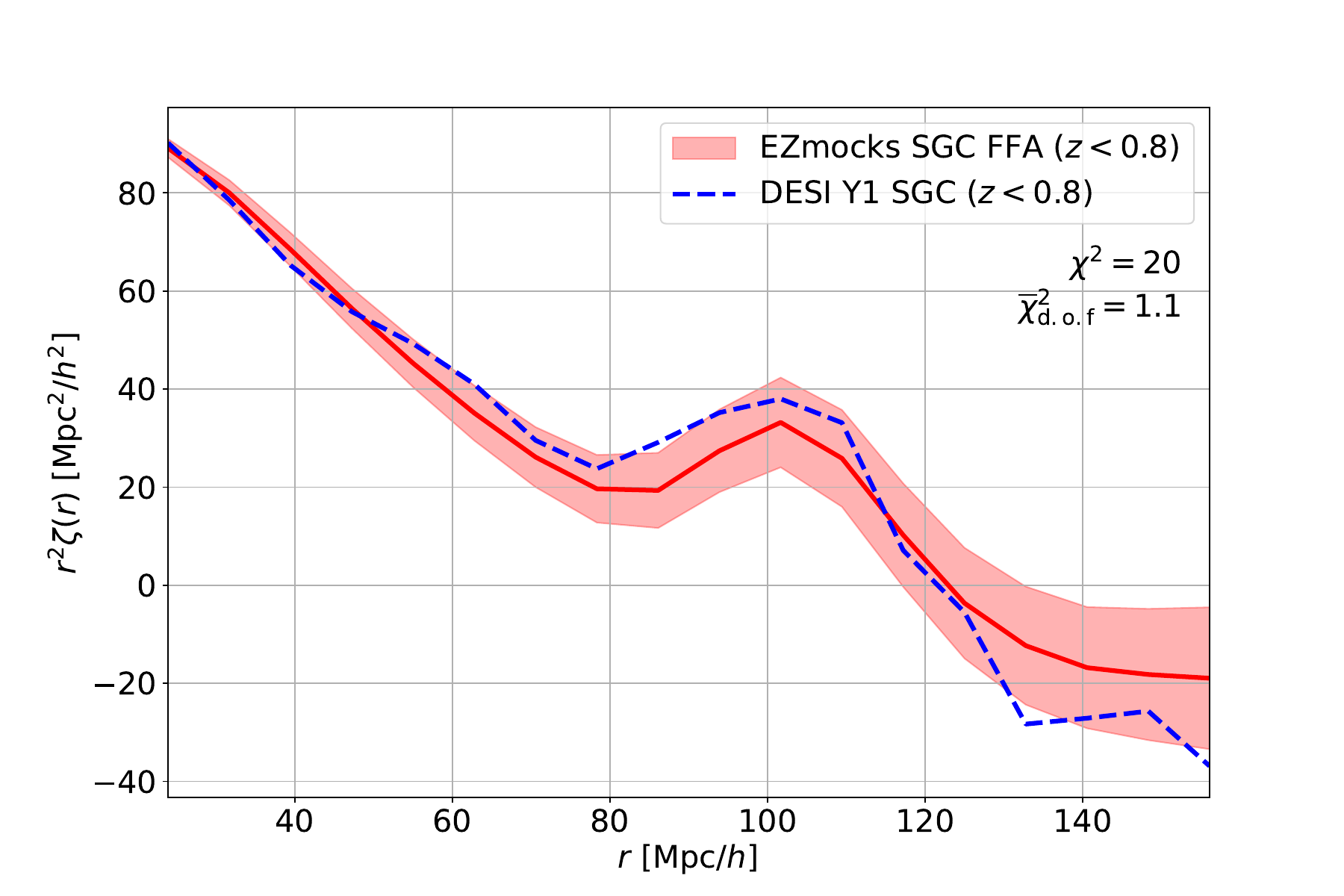} }
    \caption{Here we show the 2PCF of Dark Energy Spectroscopic Instrument (DESI) Year 1 (Y1) Luminous Red Galaxies (LRGs), South Galactic Cap (SGC), from $z = 0.4 - 0.8$, in dashed blue, and the results of averaging 1,000 EZmock \citep{ezmock} simulated catalogs in solid red, with a lighter red region indicating the 1$\sigma$ errors appropriate for the data-set's volume. The plot is weighted by $r^2$ to take out the $1/r^2$ fall-off of the 2PCF. We see that data and mocks are nicely consistent with each other; the errors at large scales are also correlated so the deviation around $140$ Mpc/$h$ should not be over-interpreted. We also display the $\chi^2$ and $\chi^2$ per degree of freedom ($\bar{\chi}^2_{\rm d.o.f}$) of comparing the data to the mean of the mocks, using a covariance matrix estimated from the mocks. The mocks have been run through Fast Fiber Assign (FFA), an approximate algorithm meant to mirror how fiber assignment is in practice performed in the survey \citep{bianchi_fiber_2024} (each galaxy must receive an optical fiber to bring its light to the spectrograph). We draw attention to the prominent bump around 100 Mpc/$h$; this is the BAO feature used as a cosmic standard ruler, as outlined in \S\ref{subsec:bao}. It corresponds to the sound horizon at decoupling.}
    \label{fig:desi_2pcf}
\end{figure}

\subsection{Line of Sight Dependence}
\label{sec:los}
\subsubsection{Breaking of Isotropy}
In reality, the situation is not so simple as regards isotropy. We infer galaxy positions in 3D by converting redshifts to distances while ignoring galaxies' peculiar velocities, but, since galaxies do have them, this approach leads to a map of positions that is distorted. In particular, only the component of a galaxy's velocity that is along our line of sight to it contributes to the distortion, and this means that for a pair of galaxies, our observed clustering will depend on $\vec{v}_i \cdot \hat{n}_i$, $i = 1, 2$ with $\hat{n}_i$ the line of sight to each.

The details of this kind of distortion are treated in for instance \citet{hamilton_1998}, but suffice it to say here that they break both homogeneity and isotropy---we, the observer conducting the survey, serve as a special origin of coordinates, and the RSD have spherical symmetry about \textit{us}. 

However, just as in spherical coordinates, the volume element $r^2 dr \sin \theta d\theta d\phi$ has explicit dependence on two of the absolute coordinates, while in Cartesian coordinates $dx dy dz$ does not, choosing a special origin breaks the translation symmetry. This idea is related to the fact that there is no simple analog spherical coordinates analog of the  Cartesian-basis Shift Theorem for Fourier Transforms. Furthermore, an observer placed on our map but not at the origin will also not see a spherically symmetric clustering pattern around them; one can imagine on the side of them farther from the privileged observer, us, they will see radial vectors diverging more than they do on the side closer to us. 

There is enormous literature on the treatment of redshift-space distortions (RSD), both specific to the effect outlined here (known as ``wide-angle RSD'') and also focused on efficient measurement and modeling of velocities. Here, we will restrict our attention to the most common methods to deal with this in the 2PCF and power spectrum.

\subsubsection{Most Naive Approach: ``Flat-Sky'' Approximation}
The most naive approach to RSD is what is called the ``flat-sky'' approximation---basically, the the whole survey volume is far enough away from us, the actual observer, that it covers a small angle on the sky and we can take a single line of sight to the entire volume; we denote this line of sight $\hat{n}$.
\\
\\
\noindent{\bf Multipoles}
It is common to expand the 2PCF or power spectrum in multipole moments, \textit{i.e.} capturing the dependence on $\hat{r} \cdot \hat {n}$ using Legendre polynomials $\mathcal{L}_{\ell}$:
\begin{align}
\xi( \vec{r} ) = \sum_{\ell} \xi_{\ell}(r) \mathcal{L}_{\ell}(\hat{r} \cdot \hat{n}).
\end{align}
$\xi_{\ell}$ are known as the 2PCF multipoles, and have Fourier Transform (FT) analog $P_{\ell}$, the power spectrum multipoles; in detail these are related by a 1D Hankel transform, \textit{i.e.} integration against $(2 \pi^2)^{-2} \,i^{\ell}\, j_{\ell} (k r) k^2 dk$ to go from Fourier space to position space, and against $4\pi\, i^{\ell}\, j_{\ell}(k r) r^2 dr$ for the reverse. $j_{\ell}$ is the spherical Bessel function (sBf) of order $\ell$. This useful relation can be shown using the definition of one as the 3D FT or inverse FT of the other, and then the plane wave expansion into spherical harmonics, the spherical harmonic addition theorem to expand the Legendres,\footnote{\url{https://mathworld.wolfram.com/SphericalHarmonicAdditionTheorem.html}} and orthogonality.\footnote{\url{https://mathworld.wolfram.com/SphericalHarmonic.html}} The factor of $i^{\ell}$ is important since it implies odd multipoles will be imaginary. We will see that they are not produced in the most naive treatment.
\\
\\
\noindent{\bf Wedges}
A second common means of characterizing the line of sight dependence of the 2PCF or power spectrum is wedges, where one bins the full statistic in $\mu \equiv \hat{r} \cdot \hat{n}$ (if drawn, this would look like wedges of a pie). One has
\begin{align}
\xi(r, \bar{\mu}) \equiv \frac{1}{\mu_+ - \mu_-} \int_{\mu_-}^{\mu_+} d\mu \; \xi(\vec{r}) 
\end{align}
where it is taken that the only dependence of the integrand on $\hat{r}$ is as $\hat{r} \cdot \hat{n}$, and $\bar{\mu} \equiv (\mu_+ + \mu_-)/2$ denotes the bin center. 

How to choose between wedges and multipoles? Often, the choice made is to not choose at all, and use both---for instance as ultimately done in SDSS Baryon Oscillation Spectroscopic Survey (BOSS). Multipoles are arguably more theoretically motivated, since the simplest theory prediction, stemming from assuming a single line of sight to the survey and a linear relationship between peculiar velocity and density fluctuations, is the celebrated Kaiser formula.
\\
\\
\noindent{\bf Kaiser Formula}
The \citet{kaiser_1987} formula is
\begin{align}
\delta_{\rm s} (\vec{x}) = (1 + f \mu^2) \delta(\vec{x}),
\label{eqn:kaiser}
\end{align}
where subscript ``s'' conventionally denotes redshift space and $f = d \ln D/ d\ln a$, $D$ the linear growth factor, and $a$ the scale factor. Here $\mu \equiv \hat{x} \cdot \hat{n}$, with $\hat{n}$ the line of sight. 

As can be seen by inspection, correlating two redshift-space densities will lead to powers up to $\mu^4$, and hence compact support of the multipole series, to $\ell_{\rm max} = 4$. Wedges can be reconstructed from multipoles and vice versa as long as one has at least as many of one as one hopes to obtain of the other; \textit{i.e.} to reconstruct eight wedges, one would need at least eight multipoles \citep{hand_2017}. 

\subsection{Choice of Line of Sight}
\label{subsec:los_choice}
As noted above, the most naive approach is to take a single line of sight to the entire galaxy survey; this is appropriate if the angular extent of the survey (in radians) is small compared to unity; this can be computed as $\theta\;({\rm rad}) \simeq L/d$, where $L$ is the transverse extent of the survey and $d$ is the distance to it. This approximation is usually not good enough for the current generation of surveys, which cover a significant angle on the sky.

An improved choice for line of sight is to use a single member of each pair or triplet of objects to define the line of sight. This is called either ``local'' flat sky or ``Yamamoto'', after the work to first propose it.  To assess whether this choice is suitable, one can again require $\theta \ll 1$, but now defining $\theta\;({\rm rad}) = s/d$, where $s$ is the typical pair separation (or, for a triplet, the typical triangle side length) and $d$ the distance to the survey.

A slightly better choice for line of sight is to use either the midpoint of the pair separation, or the bisector of the pair opening angle (as defined from the observer). It turns out that these methods agree with the single-pair-member line of sight approach at $\mathcal{O}(\theta)$ and differ only beginning at $\mathcal{O}(\theta^2)$ \citep{se_wa}. These methods can also be evaluated using FTs, though via a somewhat complicated set of terms \citep{se_wa, phil_slep_yama}. 

In detail, the observed clustering of pairs or triplets will depend on the true line of sight to \textit{each} of the objects. Appropriate mathematical bases for this have been explored, \textit{e.g.} in \citet{szapudi_wa} (tripolar spherical harmonics; the full 2PCF is given in \citealt{papai}) or with what is known as spherical Fourier-Bessel analysis, which decomposes the clustering around the observer in spherical harmonics for the direction dependence and sBfs for the dependence on distance along the line of sight (for early work see \citealt{heavens_taylor,fisher}; for recent work see \citealt{benabou}; for the analog on a finite range in redshift, \citealt{samushia_sfb}). Generally, this area of research is called ``spherical RSD'' and some of the key works go back to the mid-1990s \citep{hamilton_culhane, tegmark_bromley, fisher, taylor}.

\subsection{Algorithms}
\subsubsection{Direction-Averaged 2PCF}
The most obvious algorithm to compute the 2PCF is simply to count pairs; this scales as $N^2$, with $N$ the number of objects. In more detail, typically one measures the 2PCF only out to some maximum pair separation, $R_{\rm max}$; then, the scaling is as $N (n V_{\rm max})$ where $n$ is the number density of objects and $V_{\rm max} = (4\pi/3)  R_{\rm max}^3$. This is simply the number of pairs of objects that need to be counted. 

The 2PCF can be mathematically rewritten as a set of $N_{\rm bins}$ convolutions of kernels that are spherical shells at the set of $N_{\rm bins}$ bin centers, and where the convolution corresponds to evaluating the contribution to the 2PCF at that bin around a point $\vec{x}$. One then can sum over the convolution to average over $\vec{x}$ (the physical reason being the assumed translation invariance). This is cleanly presented in \citet{se_3pcf_ft}; the idea of using FTs for this goes back fairly far in the history of the field.

This approach is efficiently implemented with an FT, which accelerates it to $N_{\rm g} \log N_{\rm g}$, where $N_{\rm g}$ is the number of (3D) grid points used for the FT. The drawback here is that the FT requires gridding the data to be efficient, and this can lead to loss of precision and artifacts (\textit{e.g.} discussed in \citealt{Jing_05, se_3pcf_ft}) if not handled carefully.

\subsubsection{Multipoles and Wedges}
Multipoles using the local flat-sky approximation can be efficiently computed by splitting the Legendre polynomial into spherical harmonics using the spherical harmonic addition theorem \citep{se_3pcf_ft}; this approach allows one to avoid explicitly looking at pairs of objects to construct the line of sight, and thus use FT methods to evaluate. We note that choosing a single pair member to define the line of sight (\textit{i.e.} local flat sky, or ``Yamamoto'') is equivalent to using the separation midpoint or angle bisector at $\mathcal{O}(\theta)$, where $\theta$ is the opening angle the pair subtends at the observer; the difference appears only at $\mathcal{O}(\theta^2)$ which is generally quite small save for the largest physical scales or very nearby surveys \citep{se_wa}. 

The most efficient way to obtain  wedges is simply to compute the multipoles to relatively high $\ell_{\rm max}$, \textit{e.g.} $\simeq 16$, and then use this series to obtain the full $\mu$ dependence of the 2PCF and thence integrate to get the wedges. Wedges have proven useful in removing systematics that are localized in angle to the line of sight; for instance, SDSS BOSS or Dark Energy Spectroscopic Instrument (DESI) fiber collisions are purely in the 2D plane of the sky $(\mu = 0)$ (for BOSS, \textit{e.g.} \citealt{hahn_fiber}, for DESI, \textit{e.g.} \citealt{burden, pinol, hand}).

\subsection{Theory}
The theoretical modeling of the 2PCF is mainly done by taking the inverse FT of power spectrum models, and so we reserve discussion of theory primarily for the power spectrum section. There are two important exceptions to this point. First, certain Effective Field Theory (EFT) terms in the power spectrum would have divergent inverse FTs, and so cannot be directly propagated to the 2PCF. Second, often in observational analyses of the 2PCF, power law terms in separation (such as $r^{-2}$) are marginalized over to deal with issues such as large-scale modulation by imaging inhomogeneity or failure of the integral constraint. These terms are often different from those used to achieve the analogous goal in the power spectrum. Modeling of the 2PCF is well reviewed in a recent analysis of BOSS data \citep{chen_boss}; see also \citet{white_wang}.


\subsection{Recent Applications}
SDSS Baryon Oscillation Spectroscopic Survey (BOSS) and extended BOSS (eBOSS) are two recent data-sets for which 2PCF measurements have been reported. \citet{alam_2017_boss_2ps} presents the 2PCF of BOSS galaxies, and \citet{bautista} and \citet{hou_eboss} do so for respectively eBOSS galaxies and quasars. The overall cosmological implications of SDSS are outlined in \citet{alam_all_sdss}. DESI Year 1 data has also reported 2PCF measurements \citep{desi_y1_2ps}. Since the 2PCF is the workhorse statistic of cosmological LSS analyses, nearly every redshift survey ever done has measured it, and our outline here is necessarily thus deeply incomplete; however the references above are good starting points.

\section{Power Spectrum}
\label{sec:power_spec}
\subsection{Motivation and Definition}
Due to the cosmological assumption of homogeneity, clustering statistics are taken to be translation-invariant. In detail, this assumption is not perfectly correct because, as discussed earlier in \S\ref{sec:los}, RSD pick out a special origin point, us, the observer. However, these ``wide-angle'' effects are often neglected as relevant only on quite large scales. Hence, taking them aside, we see the desirability of using a translation-invariant basis to understand clustering. Fourier modes are the eigenstates of momentum, which is the conserved quantity associated with the continuous symmetry of translation, fundamentally coming from Noether's theorem. Translations of a Fourier mode simply give it a phase (Bloch's theorem in quantum mechanics is an example of this). Since the phase of the density field after inflation is random, motivating its being averaged over, phase information will disappear when we do so, and any phases picked up by translations will also.

With these qualitative comments in mind, the power spectrum is defined as
\begin{align}
    (2\pi)^3 \delta_{\rm D}^{[3]} (\vec{k} + \vec{k}') P(\vec{k}) = \left<\tilde{\delta}(\vec{k}) \tilde{\delta}(\vec{k}')  \right>. 
\end{align}
The angle brackets mean averaging over realizations of the density field (\textit{i.e.} if we made many draws from the underlying distribution that we presume is associated with inflation), and by the ergodic hypothesis this is, in practice, done by averaging over a large volume of the Universe, usually simply the survey volume of one's galaxy sample. $\delta^{[3]}_{\rm D}$ is a 3D Dirac delta function.

Now, if we apply the Delta function, $\vec{k}' = -\vec{k}$; this is equivalent to taking the complex conjugate of the second density, as we now show. We have the definition of $\tilde{\delta}$ at an arbitrary momentum $\vec{q}$ as
\begin{align}
    \tilde{\delta}(\vec{q}) =
    \int d^3 \vec{x}\, 
    e^{i \vec{q} \cdot \vec{x}} \delta(\vec{x});
\end{align}
since $\delta(\vec{x})$ is real, conjugating both sides gives a negative sign in the complex exponential. We can take this to be applied to the $\vec{q}$, so
\begin{align}
    \tilde{\delta}^*(\vec{q}) = \tilde{\delta}(-\vec{q}).
\end{align}
Hence, we see that 
\begin{align}
    P(\vec{k}) = \left< |\tilde{\delta}(\vec{k})|^2\right>;
    \label{eq:we_see}
\end{align}
there is no dependence on phase. 

Like the 2PCF, the power spectrum can depend on the direction of $\hat{k}$ with respect to the line of sight, due to RSD. It is often isotropically averaged, as $(4\pi)^{-1} \int d\Omega_k\; P(\vec{k})$, and then is a function only of $k$. It is also often binned in $k$, with order 50 bins typical in a modern analysis.

Also as is true with the 2PCF, one can expand the ($\hat{k}$-dependent) power spectrum in Legendre polynomials (multipoles) with respect to the line of sight, and these are related to the 2PCF multipoles by $j_{\ell}$ (Hankel) transforms. These transforms can be efficiently performed using the FFTLog algorithm (which by taking a log converts products to sums and thus renders the multiplicative Hankel transforms now convolutional) \citep{hamilton_fftlog}, discussed pedagogically in \citet{rotation}.\footnote{A package for this is here: \url{https://github.com/emsig/fftlog}} Typically the $\ell = 0, 2$, and sometimes $\ell = 4$ multipoles (monopole, quadrupole, hexadecapole) are measured, though the last is quite low signal-to-noise even in the largest current data-sets. There are applications where it is favorable to go to much higher multipole, for instance if one wishes to reconstruct a narrow wedge in angle to the line of sight to remove a systematic \citep{hand_2017}.

\subsection{Estimators}
One important difference between power spectrum and 2PCF is that in the former, we do not remove survey geometry effects at the estimator level; rather, the survey geometry (often called the ``window function'' when one is discussing Fourier-space analyses) is convolved with the theory model instead, prior to parameter fitting. 

However, one does need to convert the galaxy counts to a density fluctuation field. This is done by using a catalog of random particles. One has
\begin{align}
    \delta(\vec{x}) = n_{\rm g}(\vec{x}) - \alpha n_{\rm r}(\vec{x})
    \label{eq:power_spec_first_eq}
\end{align}
where $n$ is the number density of respectively galaxies (subscript g) and randoms (subscript r) at a given point, and typically more randoms are used than galaxies, so $\alpha$ is a scaling factor to counter this, $\alpha \equiv N_{\rm g}/N_{\rm r}$, with $N$ the number of each species.\footnote{In detail $\alpha$ is actually evaluated using the \textit{weighted} number of galaxies and randoms, \textit{e.g.} \citet{beutler2017clustering} \S3.} Each term in Eq. (\ref{eq:power_spec_first_eq}) is then weighted by the Feldman-Kaiser-Peacock (FKP) weight \citep{fkp_1994}, which is the assumed signal-to-noise ratio for the power spectrum:
\begin{align}
    w_{\rm FKP} \equiv \frac{1}{1 + n(z) P_0}
\end{align}
with $P_0 = P(k_0)$ the theory power spectrum evaluated at some fiducial wave-number $k_0$, and $n(z)$ the galaxy number density at redshift $z$.

Often additional weights are applied to the galaxy density, such as weights to correct collisions of the optical fibers used on the telescope plate to collect light and channel it to the spectrograph. Weights can also correct redshift failures (if a redshift was not obtained), stellar contamination, and seeing variation; \citet{gilmarin_bi}, \S2.1, nicely describes the weights used in SDSS BOSS. Often these weights are obtained by regressing some measured statistic (such as the galaxy density) against one of these variables and adjusting the weight until there is no remaining trend. 

With this in hand, one takes the FT of Eq. (\ref{eq:power_spec_first_eq}) and computes Eq. (\ref{eq:we_see}). We note that the power spectrum will contain a shot noise contribution, from any two points that are coincident in position space, and will also be sensitive to the geometry of the survey (known as the ``window function'' in Fourier space). The shot noise contribution is typically computed using the randoms and subtracted. As we have already mentioned, the window function cannot be so easily removed and instead is convolved with the theory model before comparing to the measurement (\textit{e.g.} \citealt{gilmarin_bi} \S3 and Appendix A). \citet{phil_no_win_p} is a recent effort towards an alternative approach to dealing with the window function. 

\subsection{Algorithms}
\subsubsection{Fourier Transform (FT)}
The power spectrum intrinsically is in Fourier space, so computing it requires the FT of the data. By far the most common approach to obtaining this is casting the data to a regular Cartesian grid and then using a Fast FT (FFT) algorithm; the cost is then $N_{\rm g} \log N_{\rm g}$, with $N_{\rm g}$ the number of grid points. The cost of the FT dominates over the subsequent step, of binning into shells in $k$. Historically grids where each dimension's size was a power of 2 were most efficient because the FFT algorithm used recursive bisection, and so one would zero-pad if this condition were not met. However more recent implementations, such as Fastest FT in the West (FFTW) \citep{frigo}, work well as long as the prime factorization of the grid size is compact by using the Rader or prime-factor algorithms; Bluestein's algorithm can be used for a non-prime grid size.\footnote{FFTW is publicly available at \url{https://www.fftw.org/}. For Bluestein's algorithm, a useful resource is \url{https://ccrma.stanford.edu/~jos/st/Bluestein_s_FFT_Algorithm.html}; \textsc{python's} \texttt{scipy.fft} uses this for non-prime grids.} The FFT was popularized by Cooley and Tukey in the 1960s \citep{cooley} but originally discovered by Gauss \citep{heideman}.


\subsubsection{Interpolation to a Grid}
There are various methods of casting an irregularly-spaced distribution, such as galaxies' 3D positions, to a regular grid, such as cloud-in-cell (CIC), triangular-shaped cloud (TSC), Lanczos kernel, and symmetric Daubechies wavelet. CIC or TSC are the more commonly used.\footnote{A public code exploring these options is here: \url{https://nbodykit.readthedocs.io/en/latest/cookbook/interpolation-windows.html}} In general, the grid spacing $\Delta$ sets the $k_{\rm max}$ one can probe as of order $1/\Delta$, and the minimum wavenumber $k_{\rm min}$ one can probe is set by $1/L_{\rm box}$ with $L_{\rm box}$ the box side length. For recent discussion of gridding effects and how to correct them, see \citet{sefu_2016_fourier_est}.

\subsubsection{Varying Line of Sight}
Much as in the 2PCF, there are a number of choices possible for the line of sight used in computation of the anisotropic power spectrum. This issue is discussed in more detail in \S\ref{subsec:los_choice}, and here we simply provide a brief pr\'ecis of some important references. \citet{bianchi} showed that one could use a single-pair-member line of sight (``Yamamoto'') but enjoy the speed advantage of FFTs and compute multipoles; this was done by writing the relevant Legendre polynomials as dot products of unit vectors, and then expanding these latter into Cartesian components. This led to a set of Cartesian tensors of the form $k_x^a k_y^b k_z^c n_x^a n_y^b n_z^c$, where $k_x$ \textit{etc.} are the components of $\vec{k}$ and $n_x$, \textit{etc.} are the components of the line of sight, $\hat{n}$. These tensors are separable and so can be efficiently integrated against the Fourier-space density field. 

However these tensors are not orthogonal to each other, and \citet{hand_2017}, motivated by \citet{se_3pcf_ft}, showed that instead using spherical harmonics, which are orthogonal, to split the Legendres, led to significant cost savings, for instance 40\% faster to $\ell_{\rm max} = 4$, and a factor of 3 fewer FFTs than the Cartesian method to go to $\ell_{\rm max} = 16$. This approach is publicly available in the \textsc{nbodykit}\footnote{\url{https://github.com/bccp/nbodykit}} \citep{nbodykit} and, more recently, \textsc{pypower}\footnote{\url{https://github.com/cosmodesi/pypower/tree/ae9a53446f299f64338e22352d0ced3068d3309c/pypower}} packages. Some recent work has also explored using a direct evaluation of the FT on small scales to obtain the power spectrum and bispectrum \citep{philcox_small_scale_p_config_space}, rather than gridding; this can be advantageous in avoiding artifacts without the need for a fine grid.

\subsection{Theory}
\label{sec:P_The}
In the standard inflationary paradigm, the primordial quantum fluctuations of the inflaton field are believed to form an almost scale-invariant GRF. This field is fully characterized by its 2-point statistics, or equivalently, by its power spectrum in Fourier space \citep{Bernardeau_2002}. During inflation, as the comoving horizon shrinks, perturbative modes cross the horizon \citep{Ma_1995}. Once they exit, their amplitude remains nearly constant, preserving the statistical properties of the fluctuations. As the Universe expands, the comoving horizon grows, and these modes re-enter and can ultimately begin to grow under gravity.
 
The matter transfer function, 
$T_{\rm m}(k, z)$ \citep{eisenstein1998baryonic}, describes how the matter power spectrum evolves from the primordial (density) power spectrum $P_{\rm pri} = A k^{n_{\rm s}}$, to the linear matter power spectrum at a given redshift $z$, via $P_{\rm lin}(k, z) = P_{\rm pri}(k) T_{\rm m}^2(k, z)$. $A$ is an amplitude, usually fixed by normalizing to late-time measurements of small-scale clustering (often, the root-mean-square fluctuations over an 8 Mpc/$h$ sphere, $\sigma_8$), while $n_{\rm s} \approx 0.9611$ is the scalar spectral tilt \citep{planck_final}.

The linear matter power spectrum is sensitive to cosmological parameters such as \( \Omega_{\rm m} \) (matter density), \( h \) (Hubble constant in units of 100 km/s/Mpc), \( n_{\rm s} \) (spectral index of the primordial power spectrum), \( \Omega_{\rm b} \) (baryon density), and the total neutrino mass, \( \sum m_{\nu} \) (the sum is over the three neutrino species). As a result, the linear matter power spectrum encodes a wealth of information about the fundamental physics and composition of the Universe \citep{mcquinn_linear_info}. 

However, we do not observe \( P_{\rm lin} \) directly. Rather, we observe galaxies, which are also biased tracers of the underlying (non-linear) matter field \citep{Desjacques_2018}. Additionally, RSD, such as the \citet{kaiser_1987} effect (see Eq. \ref{eqn:kaiser}), modify the observed clustering. Moreover, since gravitational growth is inherently nonlinear, structures (at least ignoring baryonic physics) can grow indefinitely in density. Consequently, linear PT breaks down on scales smaller than \( 10\,{\rm Mpc}/h \) at present, roughly the size of superclusters of galaxies \citep{carrasco2012effective}.

Therefore, there have been major efforts to extend the validity of power spectrum modeling to smaller scales (higher $k$). Standard PT (SPT) expands the density field around the linear solution, leading to the emergence of higher-order correction terms \citep{Bernardeau_2002, carrasco2012effective}. Similarly to how in particle physics, loop integrals are introduced, the 1-loop matter power spectrum is obtained by adding the higher-order loop integrals, commonly known as \( P_{22}(k) \) and \( P_{13}(k) \). The subscripts \( 22 \) and \( 13 \) refer to the order of the expansion in the density field; that is, for the \( 22 \) contribution, two second-order density fields must be contracted with each other, and for the \( 13 \) term, a linear density field must be contracted with a third-order density field. These corrections can be written as:
\begin{align}
    P_{13}(k) &= 6\, P_{\rm lin}(k)\int \frac{d^3\mathbf{q}}{(2\pi)^3}\,F_3^{({\rm s})}(\mathbf{k},\mathbf{q},-\mathbf{q})\, P_{\rm lin}(q),\label{Eq:13}\\
    P_{22}(k) &=2 \int\frac{d^3\mathbf{q}}{(2\pi)^3}\,[F_2^{({\rm s})}(\mathbf{k}-\mathbf{q},\mathbf{q})]^2\, P_{\rm lin}(|\mathbf{k}-\mathbf{q}|) \,P_{\rm lin}(q),\label{Eq:22}
\end{align}
and the 1-loop power spectrum is then:
\begin{align}
    P^{1-\rm loop}(k, z) = D^2(z)\, P_{\rm lin}(k)+ D^4(z) \,\left[P_{13}(k) + P_{22}(k)\right],
\end{align}
where $D(z)$ is the linear growth rate, which is $D(z)\propto a \propto 1/(1+z)$ in a matter-dominated Universe (as is assumed to obtain the kernels; this approximation is good to about 0.25\%). The superscript $s$ means that the $F$ kernels have been symmetrized over their arguments. 

How do we evaluate these integrals? One might think that we could numerically integrate over \( q \), but this is not always possible. The first reason is that it is computationally expensive. We are integrating from zero to infinity over \( q \), which means we need to extrapolate the power spectrum over this region (which is costly). Additionally, we need to finely bin the \( q \)-space because the integrals are sensitive to the bin width, as the power spectrum is scale-dependent. Most importantly, the linear power spectrum fails to describe the data on small scales, so how are we integrating the linear power spectrum all the way to infinity \citep{Schmittfull_2016, simonovic2018cosmological}?


This is a fundamental issue of SPT, and may be solved by introducing so-called ``counter-terms'' in the Effective Field Theory (EFT) of LSS (also written EFTofLSS) \citep{carrasco2012effective}. These capture the ultra-violet (UV) physics (large \( k \)) and are calibrated using $N$-body simulations, typically taking the form \( c_{\rm ctr} k^2 P_{\rm lin}(k) \).
\begin{figure}[ht]
    \centering
    {\includegraphics[width=0.8\textwidth]{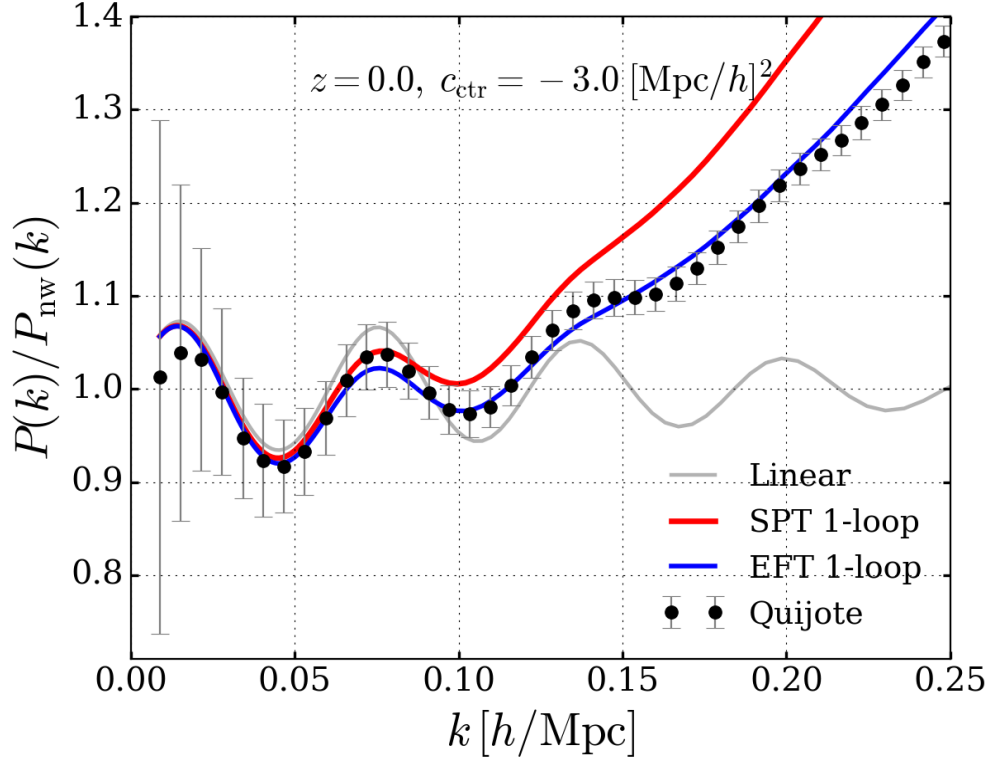} }
    \caption{The 1-loop matter power spectrum in real space normalized by the ``no-wiggle'' contribution to the power spectrum, $P_{\rm nw}(k)$. We have obtained the data points from 500 boxes of the \textsc{Quijote} \( N \)-body simulation suite \citep{villaescusa2020quijote} at redshift $z=0.0$, where each box has a side length of 1 Gpc/\( h \). The shaded region represents the \( 1\sigma \) uncertainty in the power spectrum measurement. The red curve shows the SPT prediction of the 1-loop power spectrum, while the blue curve represents the EFT prediction with a single counter-term, visually set to \( c_{\rm ctr} = -3.0\;[{\rm Mpc}/h]^2 \), which extends the validity of the model to $k_{\rm max}$$\sim$$0.2 \;h/{\rm Mpc}$. We have included IR resummation (IR means infrared) for both models in this plot.}
    \label{matterpk}
\end{figure}
The addition of counter-terms addresses the failure of the linear power spectrum on small scales by correcting for the ultraviolet (UV) behavior. 

However, the computational cost of these integrals remains a challenge. To tackle this, a method \citep{simonovic2018cosmological} harnessing FFTLog \citep{hamilton_fftlog} was developed. This approach expands the linear power spectrum as a sum of complex power laws (the expansion coefficients are computed fast using FFTLog). Loop integrals for these power laws are then computed using dimensional regularization techniques, and added up weighted by the coefficients from the expansion of the linear power spectrum. This method enables non-linear corrections to be computed very efficiently, as it requires sampling the power spectrum at only around 100 points (typically from \( q_{\rm min} = 10^{-5} \, h / {\rm Mpc} \) to \( q_{\rm max} = 10^{5} \, h / {\rm Mpc} \)) \citep{class_pt, simonovic2018cosmological}. For previous approaches to loop integrals in PT, \citet{slepian_decoup} offers a succinct review, and also see recent work on an expansion of the linear power spectrum in a basis of Gaussian functions \citep{bakx}.

Long-wavelength (IR) modes can also impact the  matter power spectrum. They can smear the BAO features, leading to a broadening of the peak recovered in the 2PCF (see Fig. \ref{fig:desi_2pcf}, and \citealt{senatore2018ir, ivanov2018infrared, blas2016time}). These IR modes primarily induce bulk motions, which damps the BAO wiggles in Fourier space---a feature not fully captured in the SPT framework. IR resummation systematically separates the smooth, ``no-wiggle'' contribution of the matter power spectrum \citep{eisenstein1998baryonic},  \( P_{\rm nw}(k) \), from the oscillatory wiggle contribution, \( P_{\rm w}(k) \), and accounts for the suppression of the BAO feature by a factor of \( e^{-\Sigma^2 k^2} \). We have
\begin{align}
    P_{\rm lin}(k) = P_{\rm nw}(k)+e^{-\Sigma^2 k^2} P_{\rm w}(k),
\end{align}
where $\Sigma^2$ is the smoothing scale. We have \citep{class_pt, senatore2018ir, blas2016time}:
\begin{align}
    \Sigma^2 = \frac{1}{6\pi^2} \int_0^{k_{\rm s}} dq \, P_{{\rm lin}}(q) \left[1 - j_0\left(\frac{q}{k_{{\rm BAO}}}\right) + 2 j_2\left(\frac{q}{k_{{\rm BAO}}}\right)\right].
    \label{Eq:sigma}
\end{align}
For a $\Lambda$CDM cosmology motivated by \textit{Planck} \citep{planck_final}, $\Sigma$$\sim$$5.5\;{\rm Mpc}/h$, also roughly the root-mean-square displacement one obtains in the Zel'dovich approximation. The scale \( k_{\rm s} \) separates the UV and IR modes, and it is recommended to set it to \( k_{\rm s} = 0.2\;h/{\rm Mpc} \) \citep{class_pt, ivanov2018infrared}. The quantity \( k_{\rm BAO} \) represents the BAO scale in Fourier space. IR resummation is implemented in almost all recent \textsc{Python} packages developed for large-scale structure analysis. 

The algorithm proceeds as follows \citep{class_pt}. First, we apply a Fast Sine Transform (FST) to the power spectrum and separate the odd and even harmonics. Next, we remove the BAO peaks in the harmonics using interpolation. Finally, we apply the inverse FST, suppress \( P_{\rm w}(k) \) according to Eq.~(\ref{Eq:sigma}), and obtain the IR-resummed power spectrum.

Fig. \ref{matterpk} shows the matter power spectra normalized by the ``no-wiggle'' power spectrum for a fiducial cosmology of the \textsc{Quijote} $N$-body simulation suite \citep{villaescusa2020quijote} with parameters $\theta=\{\sum m_{\nu}=0.0\;{\rm eV}, \Omega_{\rm m} = 0.3175,\Omega_{\rm b} = 0.049, \sigma_8 = 0.834, h=0.6711, n_{\rm s} = 0.9624\}$. The red curve represents the 1-loop SPT power spectrum, while the blue curve corresponds to the EFT 1-loop with the counter-term \( c_{\rm ctr} = -3 \;[{\rm Mpc}/h]^2 \). The data points are the averages from 15,000 realizations of the \textsc{Quijote} $N$-body simulation suite. As expected, both the linear theory and the SPT 1-loop power spectra fail to describe the data well. However, EFT successfully fits the data with only one free parameter up to $k_{\rm max}$$\sim$$0.2 \;h/{\rm Mpc}$.

\begin{figure}[ht]
    \centering
    {\includegraphics[width=0.8\textwidth]{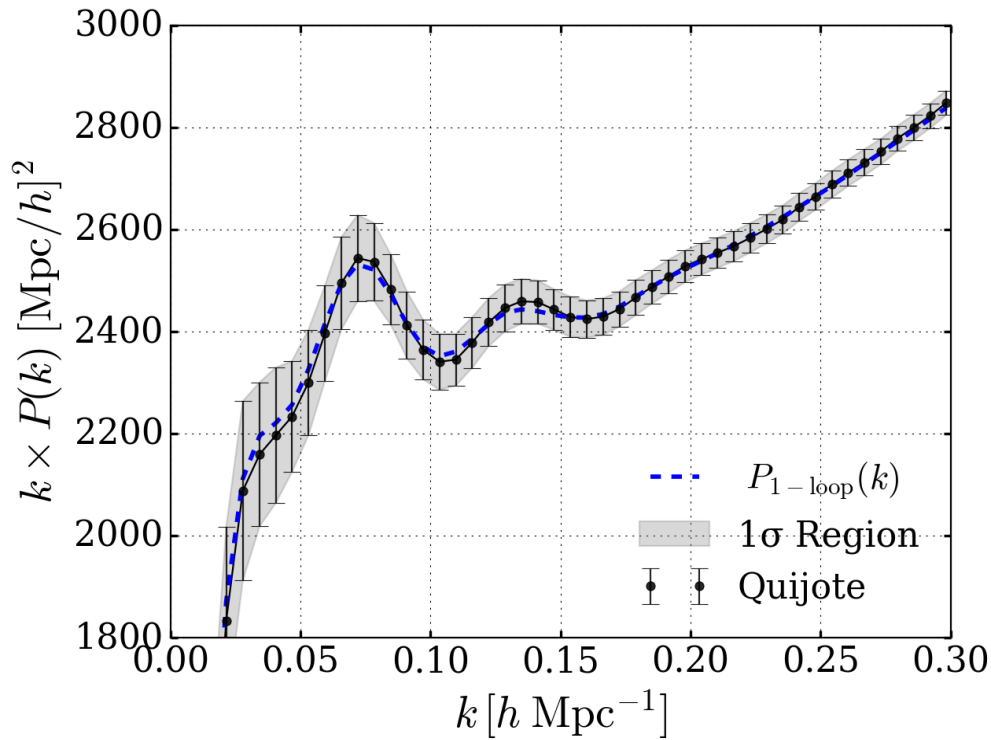} }
    \caption{The 1-loop halo power spectrum in real space obtained from 500 boxes from the \textsc{Quijote} $N$-body simulation suite \citep{villaescusa2020quijote}; each box is 1 Gpc/$h$ on each side. The shaded region shows the $1\sigma$ region of the power spectrum measurement. This power spectrum model is sensitive to three bias parameters $b_1$, $b_2$ and $b_{\rm tidal}$ and also one counter-term coefficient, $c_{\rm ctr}$, which originates from the EFTofLSS.}
    \label{halopk}s
\end{figure}

As we mentioned earlier, we do not observe the matter field directly; instead, galaxies (and halos) are biased tracers of the underlying density field. Describing the measurements from galaxy surveys requires a bias model that accurately reflects the observations. We have already introduced such a model in Eq. (\ref{Eq:bias}). Fig. \ref{halopk} shows that, with only three bias parameters and one counter-term, we can accurately predict the halo power spectrum of the \textsc{Quijote} $N$-body simulation suite up to $k_{\rm max}$$\sim$$0.3 \;h/{\rm Mpc}$.

Observing galaxies in redshift space adds another layer of complexity. In addition to the nonlinear evolution of structure, we must also account for galaxy biasing, RSD, Finger-of-God effects (from galaxies' thermal velocities), and IR resummation. The power spectrum now depends on $\mu$, the cosine of the angle of $\hat{k}$ to the line of sight, so we write \( P(k, \mu) \). All of these effects have been properly modeled and included in computationally efficient codes such as \textsc{velocileptors}
\citep{velocileptors_1, velocileptors_2},\footnote{\url{https://github.com/sfschen/velocileptors}} \textsc{pybird} \citep{py_bird}
,\footnote{\url{https://github.com/pierrexyz/pybird}} and \textsc{class-pt} \citep{class_pt},\footnote{\url{https://github.com/Michalychforever/CLASS-PT}} which produce the galaxy power spectrum in redshift space. Additionally, \textsc{folps-nu} \citep{folps_nu}\footnote{\url{https://github.com/henoriega/FOLPS-nu}} incorporates the impact of neutrino free-streaming on the galaxy power spectrum (derived fully from an analytical, two-fluid standpoint in \citealt{kamali_2fluid}).


\subsection{Recent Measurements and Applications}

The early measurements of the power spectrum using a redshift survey go back to the 2 degree Field Galaxy Redshift Survey (2dFGRS) \citep{2dfsurvey} and SDSS \citep{Tegmark_2004}, where the latter reduced the uncertainties in the cosmological parameters such as $h \equiv H_0/(100 $ km/s/Mpc) and $\Omega_{\rm m}$. They also reduced the 95\% upper bounds on the neutrino mass $\sum m_{\nu}$ from $11\; \rm eV$ to $0.6\; \rm eV$.

More recently, full-shape measurements of the redshift-space power spectrum from the Baryon Oscillation Spectroscopic Survey (BOSS) \citep{dawson2012baryon} and its extension to higher redshifts, eBOSS \citep{dawson2016sdss, alam2021completed}, within SDSS-III \citep{eisenstein2011sdss}, have reduced the error-bars on all parameters. The BOSS and eBOSS collaborations have made extensive efforts to perform these measurements using full-shape power spectrum analysis. BOSS probed a vast cosmic volume of approximately $18\;{\rm Gpc}^3$ \citep{alam2017clustering, beutler2017clustering}. 

In addition to the BOSS collaboration, other groups have explored the BOSS galaxy survey data using various models and methods \citep{ivanov2020cosmological, markovic2023cosmological, troster2020cosmology, chen2022new, kobayashi2022full, philcox2022boss, brieden2022model, noriega2024unveiling, chudaykin2023cosmological, donald2023analysis}, including EFT \citep{carrasco2012effective, senatore2015bias, baldauf2020effective}, the Halo Occupation Distribution (HOD) framework \citep{cooray2002halo, HODmodel, berlind2002halo}, and PT including massive neutrinos \citep{aviles2020lagrangian, aviles2021clustering, Blas_2014, wong2008higher, kamalinejad2020non}.

The most recent measurement of the galaxy power spectrum comes from Data Release 1 (DR1) of DESI \citep{levi2013desi, aghamousa2016desi, adame2024early} as well as from work with Year 3 data that will be included in DR2 \citep{DESI:2025zgx, DESI:2025zpo}. These works have reduced the constraint on the sum of the neutrino masses to $\sum m_{\nu} < 0.072\;{\rm eV}$ at the 95\% confidence level \citep{adame2024desi, DESI:2025zgx}. DESI also set new limits on the dynamical dark energy equation of state parameters $w_a$ and $w_0$, showing an intriguing possible departure from the $\Lambda$CDM predictions of $w_a = 0$ and $w_0 = -1$.\footnote{These are parameters within the Chevallier-Polarski-Linder (CPL) parametrization for $w(a)$, as $w(a) = w_0 + w_a ( 1 - a)$ \citep{chev_pol, linder}.}

\section{3-Point Correlation Function}
\label{sec:3pcf}
\subsection{Motivation and Definition}
The 3PCF measures excess clustering of triplets over and above what a spatially random field would have. One major motivation for it is that non-linear structure formation produces it from an initially very-nearly GRF density field (at the end of inflation), as discussed in \S\ref{subsub:nonlin}. Another motivation is PNG (\S\ref{subsub:png}). A third is galaxy biasing (\S\ref{subsub:bias}); in particular a classic paper by \citet{fry_1993} showed that adding the 3PCF can break the degeneracy between the clustering amplitude $\sigma_8$ and the linear bias $b_1$ that exists if the 2PCF (or power spectrum) alone is used.

The 3PCF has a robust history, with early work measuring the projected 3PCF in the 1970s on surveys done with photographic plates, such as the Automated Plate Measurement (APM) survey, and subsequent work on early redshift surveys in the 1990s. Primarily the 3PCF has been measured for galaxies, with one work for quasars \citep{Borderia_1991} and recent measurements also on galaxy clusters \citep{moresco_bao},  Ly-$\alpha$ forest data \citep{Tie_Lya_3}, and MHD turbulence simulations \citep{portillo, saydjari}.

\subsection{Estimators}
The current state-of-the art 3PCF estimator is due to \citet{szapudi_szalay}, and is 
\begin{align}
    \hat{\zeta}(\vec{r}_1, \vec{r}_2; \vec{x}) = \frac{N(\vec{x}) N(\vec{x} + \vec{r}_1) N(\vec{x} + \vec{r}_2)}{R(\vec{x}) R(\vec{x} + \vec{r}_1) R(\vec{x} + \vec{r}_2)}
\end{align}
with $N(\vec{x}) \equiv D(\vec{x}) - R(\vec{x})$, etc. This estimates the 3PCF around a point $\vec{x}$. Analogously to the 2PCF estimator, one then optimally weights this assuming the shot-noise limit, in other words, multiplies by the 3PCF of the randoms. We then have \citep{se_3pt_alg}
\begin{align}
    \zeta_{\rm opt}(\vec{r}_1, \vec{r}_2) = 
    \frac{ \int d^3 \vec{x} \; \frac{N(\vec{x}) N(\vec{x} + \vec{r}_1) N(\vec{x} + \vec{r}_2)}{R(\vec{x}) R(\vec{x} + \vec{r}_1) R(\vec{x} + \vec{r}_2)} \times R(\vec{x}) R(\vec{x} + \vec{r}_1) R(\vec{x} + \vec{r}_2)}{\int d^3 \vec{x} \;R(\vec{x}) R(\vec{x} + \vec{r}_1) R(\vec{x} + \vec{r}_2) }
\end{align}    
The denominator is a normalization of the weights. We notice that this has six degrees of freedom, but in fact the 3PCF (assuming full isotropy) depends only on triangles, and thus should have three. This means one can select a basis; the most common recent choice has been to expand the 3PCF's dependence on $\hat{r}_1 \cdot \hat{r}_2$ into Legendre polynomials $\mathcal{L}_{\ell}$ \citep{se_3pt_alg, szapudi_3pcf, pan_3pcf} or isotropic basis functions \citep{cahn_iso, iso_gen}, which, for the two-argument case, are equivalent to Legendres up to a normalization and phase.

One then expands the 3PCF as:
\begin{align}
    \zeta(\vec{r}_1, \vec{r}_2) = \sum_{\ell} \zeta_{\ell}(r_1, r_2) \mathcal{L}_{\ell}(\hat{r}_1 \cdot \hat{r}_2).
    \label{eqn:3pcf_multip}
\end{align}

\subsection{Algorithms}
An efficient algorithm \citep{se_3pt_alg} to compute $\zeta_{\ell}$ takes advantage of the spherical harmonic addition theorem to split the dependence on $\hat{r}_1$ and $\hat{r}_2$, so that one, about each point $\vec{x}$, bins neighbors into spherical shells, obtains the harmonic expansion coefficients $a_{\ell m}$ on each shell, and then forms cross-correlations of shells rather than objects. This reduces the combinatorics, around each point, to $N_{\rm bins}^2$ rather than $(nV_{\rm max})^2$. The cost of obtaining the harmonic coefficients is as $n V_{\rm max}$, as this is the number of relevant neighbors around each point and evaluating their projection onto harmonics scales linearly as their number. Overall, this algorithm scales as $N (nV_{\rm max})$, where the factor of $N$ is because the harmonic expansion and bin combination must be done around each point.

Code implementing this algorithm is publicly available in the \textsc{encore} package \citep{encore} (in \textsc{C++}).\footnote{\url{https://github.com/oliverphilcox/encore}} The algorithm can also be framed as a convolution and implemented with FTs, as outlined in \citet{se_3pcf_ft} and publicly available in the \textsc{sarabande} package in \textsc{python}  \citep{sarabande}.\footnote{\url{https://github.com/James11222/sarabande}} In detail, both of these codes measure the 3PCF in the isotropic basis \citep{cahn_iso, iso_gen}, which for two arguments, is equivalent to Legendres up to a normalization and phase.\footnote{The isotropic basis functions of two arguments are orthonormal over $d\Omega_1 d\Omega_2$, while the Legendres are orthogonal, but not orthonormal, over this integration measure.} The ideas behind this algorithm can also be extended to measure the 3PCF in arbitrary spatial dimension \citep{phil_nd}. Another related 3PCF approach is \textsc{conker}, which also exploits the FT \citep{conker}.

Algorithms to measure the line-of-sight dependence of the 3PCF that extend these spherical harmonic methods have also been proposed, taking advantage of the presumed symmetry under rotation about the line of sight \citep{se_aniso_3pcf, friesen}. One can compute the 3PCF using either the triangle centroid, or the average of the triplet members' lines of sight, with the same algorithmic efficiency, by using a Taylor series about the results for a single-triplet-member line of sight \citep{garcia}; these choices have the advantage of being symmetric under interchange of the three triplet members. 

There have been efforts to measure the 3PCF without a Legendre decomposition, for instance using $kd$-trees; these are very efficient for highly clustered data \citep{gardner, march_phd} but work less well on large scales ($\gtrsim 100$ Mpc), when the data becomes nearly unclustered so the tree's space partitioning offers little advantage. This method was used in early SDSS 3PCF measurements \citep{mcbride_1, mcbride_2}.

Notably, recent work, building on a method developed to produce simulated CMB maps with prescribed bispectra \citep{smith}, showed how to produce a density field with a specified 3PCF around each point \citep{slepian_const_npcf}; this is useful in testing pipelines and assessing the impact of observational systematics on a sample with realistic clustering, at lower cost than running full or approximate simulations.

\subsection{Theory}
As with the 2PCF, the primary source of 3PCF theory has been to obtain the theory model for the bispectrum and then take the inverse FT. Thus, we reserve some theory discussion for when we turn to bispectrum. Here we simply outline what has been done regarding modeling the 3PCF. 

Modeling work using both the hierarchical ansatz and PT was done by Peebles, Fry, and collaborators throughout the 1980s: Monte Carlo modeling \citep{fry_monte_carlo}, hierarchical amplitudes \citep{fry_hier_in_pt, fry_1984_Npoint}, relationship of NPCFs to the Probability Distribution Function (PDF) \citep{fry_pdf}, 3PCF for biasing \citep{melott}, from $N$-body simulations \citep{fry_melott_sims, Fry_Melott_3d_sims}, and for separating gravitational clustering and galaxy biasing \citep{fry_bias_1994}. Work on the hierarchical ansatz\footnote{A phenomenological form for the higher-order functions that takes them to be products of lower-order functions; for instance, that the 3PCF is a cyclic sum of products of pairs of 2PCFs around the triangle, \textit{i.e.} $\zeta$$\sim$$\xi(r_1)\xi(r_2) + \xi(r_2)\xi(r_3) + \xi(r_3)\xi(r_1)$.} and related ideas is reviewed in \citet{suto_1993}, and the scale dependence of the 3PCF was explored with $N$-body simulations in \citet{matsubara_1994}. In the 2000s, \citet{buch_2000} modeled the angular 3PCF, with further work on the full 3PCF in this period being \citet{takada, wang_2004_3pcf, marin}.

\begin{figure}[ht]
    \centering
    {\includegraphics[width=0.6\textwidth]{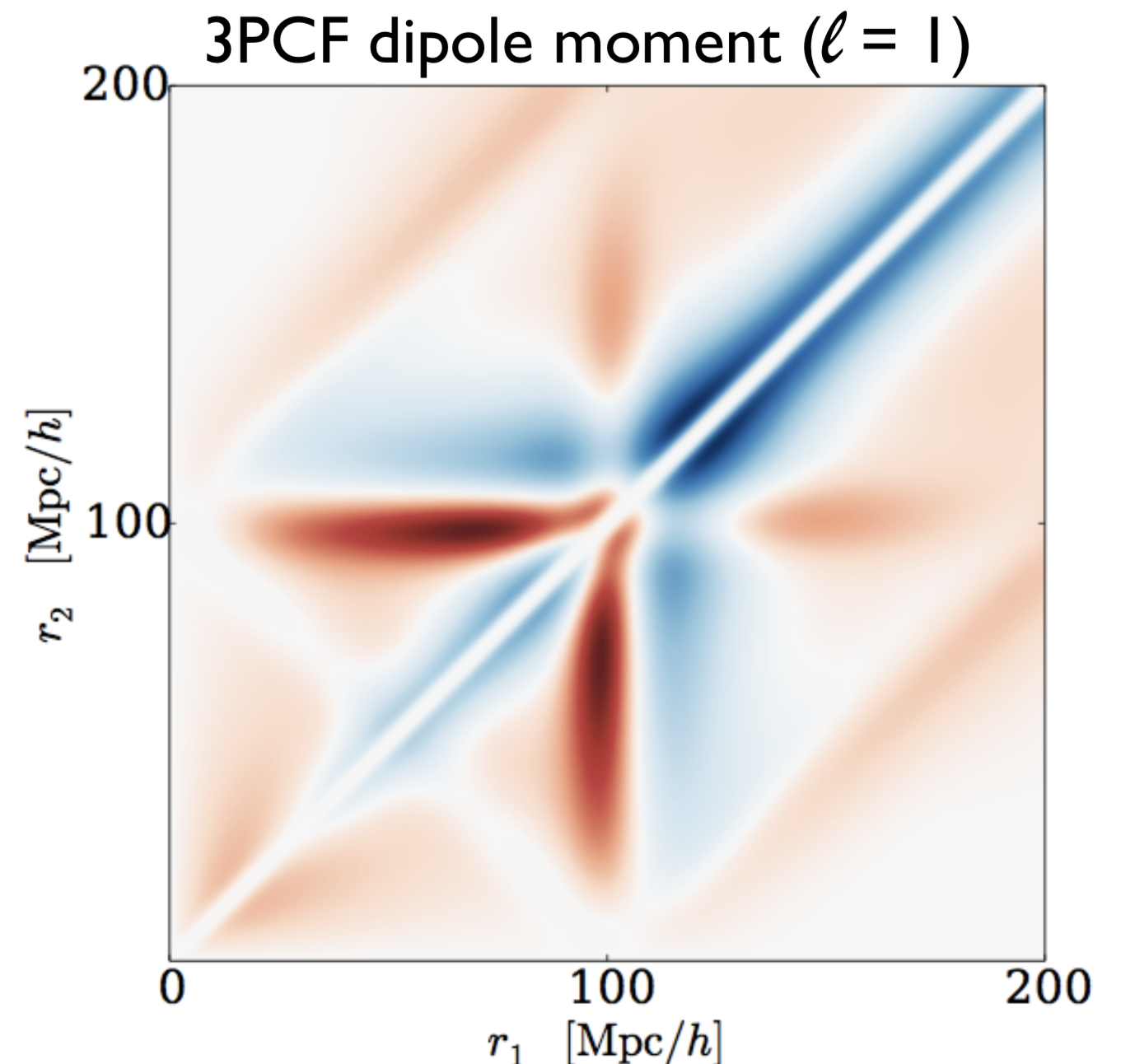} }
    \caption{Here we show the 3PCF with its internal angle dependence projected onto $\cos \theta_{12}$ (the dipole, $\ell = 1$ in Eq. \ref{eqn:3pcf_multip}), vs. the triangle side lengths $r_1$ and $r_2$. Red represents an excess of triangles over random, blue a deficit. The plot has been weighted by $r_1^2 r_2^2$ to take out the large-scale fall-off of the 3PCF, and the diagonal is intentionally artificially suppressed, as it is dominated by triangles where $r_3 \to 0$ and thus PT should break down. The sharp transition from red to blue at $r_1 = 100$ Mpc/$h$ (and same for $r_2$) is due to the BAO, as are the diagonal creases connecting 100 Mpc/$h$ on each axis. The diagonal creases that go from 100 to 200 Mpc/$h$ are also due to BAO. This structure simply reflects different triangle sides' exploring the BAO scale \citep{se_3pcf_rsd}. This BAO signature was used to make the first detection of BAO in the 3PCF in \citet{se_3pcf_bao} on 777,202 BOSS CMASS LRGs, enabling a 1.7\% precision measurement of the distance to $z = 0.57$ from the 3PCF alone. For further details see \S\ref{subsub:bao_3pcf}.}
    \label{fig:3pcf}
\end{figure}

\citet{se_rsd_3pcf} presented a large-scale, PT-based model of the 3PCF including Kaiser formula RSD, quadratic, and tidal tensor bias, and showed that this model could be quickly evaluated with 1D Hankel transforms (and with some 2D transforms for terms that contribute negligibly to the 3PCF values). This work also showed that redshift space largely results in a simple, constant rescaling of the real-space 3PCF. It obtained the 3PCF in the Legendre basis of \citet{se_3pt_alg} and split the 3PCF by bias coefficient and multipole; it also identified a clear BAO signature in the 3PCF. It exploited mathematical ideas first developed in \citet{se_rv}, which showed that galaxy biasing due to the baryon-dark matter relative velocity (sourced by their differing behaviors prior to decoupling) would have a unique signature in the 3PCF (see also \citealt{yoo_dalal}). An efficient implementation of the sub-leading order 3PCF model using FFTLog \citep{hamilton_fftlog} was suggested in \citet{guidi_3pcf}, an idea earlier explored for the leading-order anisotropic 3PCF in \citet{umeh}.

Regarding robustness of 3PCF analyses, \citet{hoffman_18} used simulations to analyze consistency of the parameter constraints one would infer from 3PCF vs. from bispectrum. \citet{veropalumbo} used simulations to further explore the impact of covariance and modeling on 3PCF parameter constraints. An important work on covariance was \citet{se_3pt_alg}, which presented an analytic template based on assuming the density to be a GRF. This template enabled analysis of the 3PCF even with limited number of mocks relative to degrees of freedom, and ultimately led to the further templates of \citet{hou_covar} for NPCF covariances, as well as to the first detection of BAO in the 3PCF \citep{se_3pcf_bao}, with the largest 3PCF measurement to date (the 777,202 LRGs of BOSS CMASS).

\subsection{Applications}
\citet{peebles_groth-3pcf_1975} computed the 3PCF on the Zwicky galaxy catalog in 2D using counts-in-cells to avoid direct counting of triplets, and \citet{peebles_groth_1977} on Shane-Wirtaanen catalog data. \citet{fry_cluster_galaxy_3pcf} then measured the cluster-galaxy-galaxy 3PCF. 

Measurements of the quasar 3PCF were made by \citet{Borderia_1991}, and for galaxies, by \citet{jing_3pcf} (Las Campanas redshift survey, 1998), \citet{kayo_2004} (SDSS, 2004), \citet{pan_3pcf} (2dFGS), \citet{nichol_2006} (SDSS, 2006; used $kd$-tree algorithm), \citet{marin_2011} (SDSS DR7 LRGs), \citet{mcbride_1, mcbride_2} (SDSS, 2011; used $kd$-tree algorithm), \citet{Guo_2013} (SDSS DR7, 2013), and \citet{Guo_2015}. VIPERS' 3PCF was measured by \citet{vipers_2017}.

\subsubsection{BAO in the 3PCF}
\label{subsub:bao_3pcf}
Some of the figures in \citet{szapudi_3pcf} figures (\textit{e.g.} Fig. 2) show clear evidence of BAO wiggles in the theory model for the 3PCF in the Legendre basis.  Some evidence was found for BAO in the 3PCF in a measurement on SDSS DR9 LRGs in 2009 \citep{gaztanaga_2009_3pcf}.

\citet{se_3pcf_bao} made the first detection of BAO in the 3PCF, measuring the large-scale SDSS BOSS CMASS 3PCF in a series of two papers, first in a compressed basis where one triangle side was integrated out \citep{se_boss_3pcf}, with no strong BAO detection, but some evidence, and then the full basis \citep{se_3pcf_bao}. We show the 3PCF BAO signature in Fig. \ref{fig:3pcf}.

This first detection used the Legendre polynomial/spherical harmonic-based algorithm \citep{se_3pt_alg}, which has unique efficiency on large scales, enabling probing the BAO scale and environs, and remains the largest 3PCF measurement to date. \citet{se_rv_boss} also constrained the baryon-dark matter relative velocity term in galaxy biasing using these 3PCF measurements, following the model of \citet{se_rv, se_3pcf_rsd}. A first detection of BAO in the bispectrum, using SDSS BOSS, was made around the same time \citep{pearson_bao_bispec}; for additional discussion of bispectrum BAO, see \citet{child1, child2, behera}. A first detection of BAO in the 3PCF of galaxy clusters was made by \citet{moresco_3pcf_bao}.

\subsubsection{3PCF Beyond Galaxies}
Application of the 3PCF to MHD turbulence was made in \citet{portillo}, 
sensitivity of the Lyman-alpha forest 3PCF to a fluctuating UV background was found in simulations in 2019 \citep{Tie_Lya_3}, and the use of 3PCF to probe reionization was suggested in \citet{jennings_2020}.

\section{Bispectrum}
\label{sec:bispec}
\subsection{Motivation and Definition}
Though the distribution of matter in the Universe at the end of inflation is thought to be a GRF, as discussed in \S\ref{subsec:inflation}, and pair statistics would be sufficient in that case, there are significant motivations to go beyond pairs, as outlined in \S\ref{subsec:beyond_pairs}. To recap succinctly: inflationary models can produce Primordial Non-Gaussianity, which can be measured with the bispectrum. Second, non-linear gravitational evolution of the cosmic web induces non-Gaussianity (and a bispectrum). Thus, the bispectrum contains a wealth of information on the cosmological model and parameters. It also has been shown that the bispectrum can improve the parameter constraints by at least $10\%$. For instance, \citet{Hahn_2020, ruggeri2018demnuni, yankelevich2019cosmological, kamalinejad2020non} show that the bispectrum monopole (with respect to the line of sight, not internal triangle angle) can be used to break parameter degeneracies in the power spectrum and improve the bounds on the parameters significantly.

The bispectrum has also been proposed as a tool to constrain cosmic reionization (from the 21-cm signal) \citep{watkins_22, noble_24},  MHD turbulence \citep{turbulence_bispectrum_1976, burkhart, obrien}, and X-ray variability \citep{arur_xray}. A related statistic is the biphase, which correlates triples of the Fourier-space field's phase \citep{maccarone_biphase}; the bicoherence is a very similar measure to the biphase \citep{maccarone_bicoherence}.

The bispectrum is defined as
\begin{align}
    (2\pi)^3 \delta_{\rm D}^{[3]} (\vec{k}_1 + \vec{k}_2 + \vec{k}_3) B(\vec{k}_1, \vec{k}_2,\vec{k}_3) = \left<\tilde{\delta}(\vec{k}_1) \tilde{\delta}(\vec{k}_2) \tilde{\delta}(\vec{k}_3)  \right>. 
    \label{EQ:Bispec}
\end{align}
We note that, while here we have written $\vec{k}_1, \vec{k}_2$ as arguments, one could as easily use $\vec{k}_3$ in place of one of them, due to the Delta function. Further, while $\vec{k}_1, \vec{k}_2$ have six degrees of freedom, under the assumption of isotropy the bispectrum must be invariant under joint rotations of these. Now, averaging over the three Euler angles subtracts three degrees of freedom, leaving three remaining---exactly as needed to describe a triangle whose sides are $k_1, k_2, k_3$. The bispectrum can equivalently be parametrized by $k_1, k_2, \hat{k}_1\cdot \hat{k}_2 = \cos \theta_{12}$ with $\theta_{12}$ the angle enclosed by $k_1$ and $k_2$.

Now, in detail, due to RSD, the bispectrum is not actually isotropic, but depends on the orientation of each wave-vector to the line of sight. In full detail, one has three Fourier space density fields, and one can think of them each as having an associated line of sight $\hat{n}_i$ (conceptually, imagine taking the FTs of three position-space densities but retaining what the l.o.s. was to each), so the full wide-angle redshift-space bispectrum depends on $\hat{k}_i \cdot \hat{n}_i$ as well as the three ``internal'' parameters (\textit{e.g.} $k_1, k_2, k_3$).

However, one has to be cautious, because since, for instance, $\vec{k}_3$ is constrained by the Delta function, its angle to $\hat{n}_3$ is not completely free to be independent of those for $\vec{k}_1$ and $\vec{k}_2$. One solution to this is to work in a spherical Fourier-Bessel basis \citep{benabou}. This is intended to address wide-angle effects, which are most relevant for analyses on the largest spatial scales, specifically searches for primordial non-Gaussianity from inflation. 

More commonly, such as in typical LSS analyses where the bispectrum is used to probe gravity, dark energy, and galaxy bias, a single line of sight is adopted to each galaxy triplet (taking the $\hat{n}_i$ of the paragraph above to be $\equiv \hat{n}$, but allowing this $\hat{n}$ to be defined by each given set of three position-space densities). This approach is used in \citet{sugi_1}, but with additional averaging over rotation of the triplet about this $\hat{n}$. \citet{scf_99} was an early, influential work that presented the bispectrum with a single line of sight, expanding in spherical harmonics the dependence on the azimuthal  and polar angles between $\hat{k}_1$ and $\hat{n}$. \citet{scf_99} suggested further averaging over azimuth, corresponding to setting $m = 0$ in the harmonic. \citet{gagrani_2017, yankelevich2019cosmological} study the information content of the bispectrum coefficients in this expansion; see also \citet{gualdi_2020}.

\subsection{Estimators}
\subsubsection{Overview}
Much like the power spectrum, the bispectrum is constructed from forming correlations of Fourier space density fields (in this case, three), where the density field is formed by differencing the galaxy and random number densities and weighting this difference appropriately to optimize signal to noise and correct systematics. One only forms correlators where the three wave-vectors form a closed triangle. One typically bins in the lengths of each relevant wave-vector (either two, or three, depending on the basis used). 

We note that the FKP weight is generally used for the bispectrum calculation even though it is not strictly optimal (an optimal S/N weighting here would require instead using a theoretical bispectrum template as the signal, and accounting for the bispectrum's variance). The same fundamental construction as in Eq. (\ref{eq:power_spec_first_eq}) is used to obtain the position-space density fluctuation field that then has its FT taken. To obtain the bispectrum we simply then form triplets of the Fourier-space density at three different wave-vectors, subject to the constraint that they form a closed triangle, as Eq. (\ref{EQ:Bispec}) dictates. \citet{gilmarin_bi} \S3.3 nicely presents the details in the context of the SDSS BOSS bispectrum measurement.

\subsubsection{Window Function}
As with the power spectrum, for the bispectrum the window function is not removed at the estimator level, but rather convolved with the theory model prior to fitting. Unlike with the power spectrum, due to the dimensionality of the bispectrum this convolution is computationally expensive and so is usually performed only approximately, by applying the window function only to the power spectra entering the bispectrum theoretical model (\textit{e.g.} \citealt{gilmarin_bi}). More recently, \citet{wang_window} shows that full window convolution is possible for the bispectrum as measured in the tripolar spherical harmonic basis of \citet{sugi_1} (further detailed in \S\ref{subsec:algs}), using the window coefficients computed in this same basis; this scheme also is not exact as one must truncate the window expansion at some $\ell_{\rm max}$. \citet{phil_no_win_b} presents an alternative approach in an effort to remove the window.

\subsubsection{Shot Noise}
Unlike the 3PCF, the bispectrum measured from data will include a shot noise term at the signal level. In position space, one can simply set a minimum triangle side, and this naturally excludes contributions from two or more objects' being coincident. However, in Fourier space, two coincident points (essentially, a Delta function correlation) spread their signal o all $k$-modes (an example of the uncertainty relation), and so there is an unavoidable bispectrum contribution at all $k$ from coincident pairs or coincident triplets (\textit{e.g.} \citealt{hoffman_18}). This can be removed simply by subtracting terms in $1/n$ or $1/n^2$, with $n$ the number density (\textit{e.g.} \citealt{gilmarin_bi}).

\subsection{Algorithms}
\label{subsec:algs}
One can measure the (isotropic) bispectrum as a function of $k_1, k_2, k_3$, or of $k_1, k_2, \cos \theta_{12} \equiv \hat{k}_1 \cdot \hat{k}_2$, and the latter is often decomposed as a series in spherical harmonics of each $\hat{k}_i$, with an additional spherical harmonic capturing dependence on the line of sight, $\hat{n}$. This latter basis is called ``tripolar spherical harmonics'' \citep{sugi_1}.

The current state-of-the art algorithm to measure the bispectrum as a function of $k_1, k_2, k_3$ is \citet{scocc_2015}, which rewrites the Fourier-space Delta function as an FT of plane waves about $\vec{x}$. This splits up the $k_i$ dependence; one may then compute bin the density field onto shells in $k_i$, about each $\vec{x}$ on a grid, multiply them, and finally integrate over $d^3 \vec{x}$ (which enforces the Delta function constraint). This has cost $N_{\rm bins}^3$ about each $\vec{x}$ and then requires an $\mathcal{O}(N_{\rm g})$ sum, with $N_{\rm g}$ the number of grid points in $\vec{x}$. It is more efficient than the naive $N_k^3$ computation (with $N_k$ the number of Fourier modes), since it allows one to bin in $k_i$ rather than considering triplets of $k_i$. The acceleration is significant since there can be $\mathcal{O}(1,000)$ $k$-modes but only $\mathcal{O}(50)$ bins in each $k_i$. This approach also can generalize to include some information about the dependence on line of sight. The algorithm is implemented in the \textsc{rustico} package, which also measures the power spectrum.\footnote{\url{https://github.com/hectorgil/Rustico}} 

Regarding the $k_1, k_2, k_3$ basis, an opening-angle-averaged (equivalent by law of cosines to averaging over $k_3$) CPU/GPU code, \textsc{spatialstats} \citep{obrien}, which uses Monte Carlo to do the averaging, is also publicly available.\footnote{\url{https://github.com/mjo22/spatialstats}} In the tripolar spherical harmonics basis, the bispectrum can be measured using the algorithm of \citet{sugi_1}. This approach is publicly available in the \textsc{triumvirate} package \citep{triumvirate}. Recent work also shows how to produce a density field with arbitrary desired bispectrum \citep{fergusson_separable, regan_bispectrum_real}, building on an approach originally developed for the CMB \citep{smith}.

\subsection{Theory}

The galaxy bispectrum is one of the key higher-order statistics of the density field. Beginning with its definition in Eq. (\ref{EQ:Bispec}), we expand the density field using PT and express it in terms of the kernels $F_2^{({\rm s})}$ (density) and $G_2^{({\rm s})}$ (velocity) to obtain the tree-level bispectrum \citep{Bernardeau_2002, scoccimarro1999bispectrum}. In redshift space, the galaxy bispectrum can be further expanded as:
\begin{align}
    B_{\rm ggg}(\mathbf{k_1}, \mathbf{k_2}, \mathbf{k_3}) &= 2 \,Z_1(\mathbf{k}_1) \,Z_1(\mathbf{k}_2) \,Z_2(\mathbf{k}_1, \mathbf{k}_2)\, P_{\rm lin}(k_1) \,P_{\rm lin}(k_2) + {\rm cyc.} 
    \label{eq:Bcb}
\end{align}
where $P_{\rm lin}$ is the linear power spectrum, the $Z$ are the redshift-space kernels that include the biases ($b_1$, $b_2$ and $b_{\mathrm{tidal}}$) and the logarithmic growth rate, $f$, as:
\begin{align}
    Z_1(\mathbf{k}) &= b_1+f\mu^2,\\
    Z_2(\mathbf{k}_1, \mathbf{k}_2) &= b_1 {F}_2(\mathbf{k}_1, \mathbf{k}_2)+f\mu^2 {G}_2(\mathbf{k}_1, \mathbf{k}_2)+\frac{f\mu k}{2}\left[\frac{\mu_1}{k_1}(b_1+f \mu_2^2)+\frac{\mu_2}{k_2}(b_1+f \mu_1^2)\right]\\&\qquad\qquad+\frac{b_2}{2}+ b_{\mathrm{tidal}}\left(\left(\frac{\mathbf{k}_1\cdot\mathbf{k}_2}{k_1 k_2}\right)^2-1\right),\nonumber
\end{align}
where $\mu \equiv \hat{k}\cdot\hat{z}$ and $\mathbf{k}\equiv \mathbf{k}_1+\mathbf{k}_2$, and we have set the line of sight to be in the $\hat{z}$ direction. The $Z$ kernels as above are symmetrized. 

The redshift-space bispectrum in Eq. (\ref{eq:Bcb}) depends on five parameters: the three triangle sides, \( k_1 \), \( k_2 \), and \( k_3 \); the cosine of the angle between \( k_1 \) and the line of sight, \( \mu_1 \); and the azimuthal angle of \( k_2 \) about \( k_1 \), which we denote by \( \phi \) following \citet{scoccimarro1999bispectrum}. The cyclic summation in Eq. (\ref{eq:Bcb}) involves calculating \( \mu_2 \) and \( \mu_3 \), the cosines of the angles between \( \mathbf{k}_2 \) and \( \mathbf{k}_3 \) and the line of sight, respectively. These are given by:

\begin{align}
    x &\equiv \frac{\mathbf{k}_1\cdot\mathbf{k}_2}{k_1k_2},\\
    \mu_2 &= \mu x - \sqrt{1-\mu^2}\sqrt{1-x^2}\cos{\phi},\\
    \mu_3 &= -\frac{k_1}{k_3}\mu-\frac{k_2}{k_3}\mu_2.
\end{align}
However, the bispectrum is typically averaged over the azimuthal angle \( \phi \) and the angle with respect to the line of sight, \( \mu \). This averaging yields the spherically averaged bispectrum, known as the bispectrum monopole, \( B_0 \), which is independent of the line of sight. Similar to the power spectrum, the bispectrum monopole is the least noisy projection and, consequently, the only one widely used in the community so far.
\begin{figure}[ht]
    \centering
    {\includegraphics[width=0.8\textwidth]{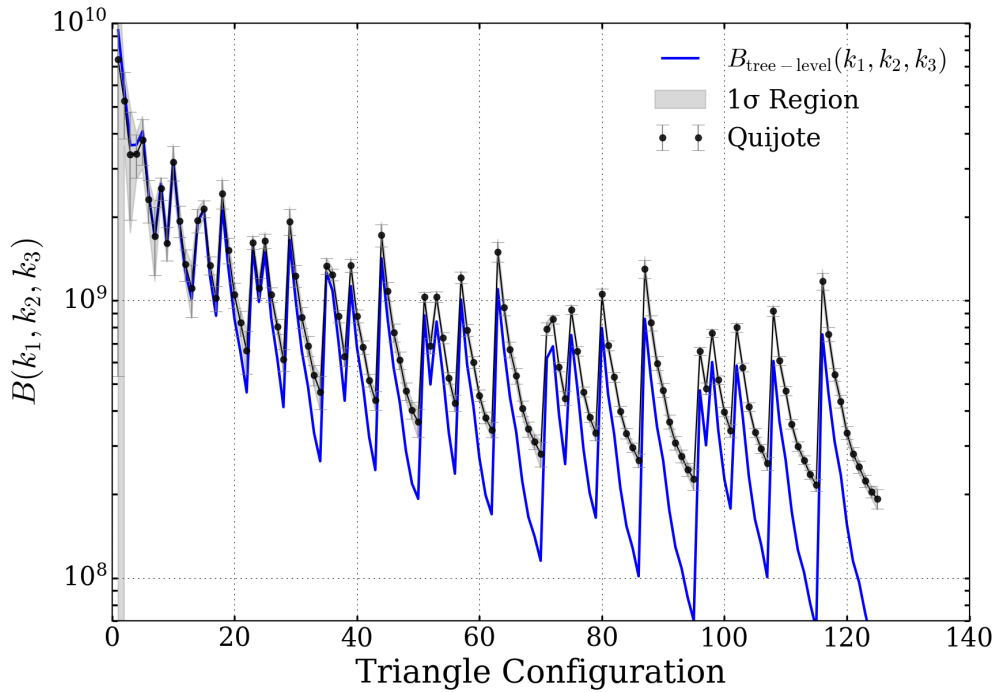} }
    \caption{The halo bispectrum (black points) is obtained by averaging over 500 boxes from the \textsc{Quijote} $N$-body suite. Here we compare with a PT-based theoretical model with three bias parameters, $b_1, b_2$, and $b_{\rm tidal}$ up to \( k_{\rm max} = 0.2\;h/{\rm Mpc} \). This tree-level bispectrum model (blue curve) fails to accurately describe the $N$-body simulations on scales smaller than \( k > 0.08\;h/{\rm Mpc} \). The triangle configuration refers to the sets of \( k_1 \), \( k_2 \), and \( k_3 \) that form a closed triangle. We map each set of $k_i$ to a single index, labeled ``triangle configuration'' above, by first varying \( k_3 \) while keeping \( k_1 \) and \( k_2 \) fixed, then varying \( k_2 \), and finally over \( k_1 \).  
}
    \label{syntheticdata}
\end{figure}
Similar to the power spectrum, the tree-level bispectrum is only valid on linear scales. 

However, unlike the power spectrum, the tree-level bispectrum remains accurate up to larger scales, around $k_{\rm max}$$\sim$$0.08\;h/{\rm Mpc}$ \citep{Ivanov_2022}. Higher-order corrections (1-loop and 2-loop corrections) can extend its validity to smaller scales, reaching $k_{\rm max}$$\sim$$0.2\;h/{\rm Mpc}$ \citep{angulo2015one, twoloopbis, eft_exmt}. Additionally, the bispectrum requires IR resummation to accurately capture the BAO features \citep{Rampf_2012}; see also \cite{behera}.

There have also been recent efforts to model the bispectrum beyond PT. For instance, \citet{geoftp} employs a functional form derived from PT to describe the bispectrum in the mildly nonlinear regime. \citet{simbigchang} utilizes simulation-based inference to study the bispectrum at highly nonlinear scales. \citet{lado_peak_position} further simplifies the redshift-space power spectrum and bispectrum by introducing efficient analytical formulas to account for the effect of Alcock-Paczynski \citep{ap_effect} distortions.









\subsection{Recent Measurements and Applications}

The bispectrum can directly probe Primordial Non-Gaussianity (PNG). Since a Gaussian field does not produce a bispectrum at the linear level, measuring any excess bispectrum signal beyond the non-Gaussianity induced by the nonlinear growth of matter provides a direct indication of primordial non-Gaussianity \citep{bisp_png,emiliano2006cosmology, scoccimarro2000bispectrum, dizgah2021primordial,baldauf2011primordial}.

Aside from its usefulness to probe PNG, the bispectrum serves as a test for theories of gravity. The standard $\Lambda$CDM model assumes General Relativity (GR) governs gravitational interactions, so any observational deviation from $\Lambda$CDM could indicate physics beyond GR \citep{bose2018one, hashimoto2017precision}. \citet{Aviles_2023} models the 3PCF of biased tracers at tree-level using SPT for GR, $f(R)$ \citep{hu2007models, sotiriou2010f}, and the Dvali-Gababadze-Porrati (DGP) model \citep{dvali20004d}. A joint analysis of the 2- and 3-point correlation functions confirms BOSS DR12's consistency with GR \citep{Sugiyama_BOSS_MG}. In preparation for DESI, \citet{alam2021towards} examined multiple summary statistics, including 2-point and 3-point statistics in real and Fourier space, to assess DESI's ability to test gravity theories. \citet{Bose_GR_Bispec} compares halo model-based approaches and fitting formulae to $N$-body simulations, finding that the halo-model-corrected fitting formula performs best. \citet{Takahashi_bihalofit} provides a fitting formula for the nonlinear matter bispectrum calibrated on high-resolution $N$-body simulations.

The bispectrum can also help break degeneracies in the power spectrum parameter space \citep{Hahn_2020, yankelevich2019cosmological, kamalinejad2020non}. The linear bias \( b_1 \) and the amplitude of density fluctuations \( \sigma_8 \) are completely degenerate in the linear power spectrum since the halo (and galaxy) power spectrum scales as \( b_1^2 \sigma_8^2 \), making their simultaneous measurement impossible \citep{fry_1993}. Adding nonlinear corrections as well as higher-order biases ($b_2$ and $b_{\rm tidal}$ which were explained in \S\ref{sec:P_The}) to the power spectrum partially alleviates this degeneracy, as non-linear terms originated from the $P_{22}$ (\ref{Eq:22}) and $P_{13}$ (\ref{Eq:13}) scale as $b_1^2 \sigma_8^4$ (which contains different powers of $b_1$ and $\sigma_8$). Higher-order biases also introduce extra terms in $b_2$ and $b_{\rm tidal}$ which will further break this degeneracy \citep{class_pt, simonovic2018cosmological}. However, \( b_1 \) and \( \sigma_8 \) remain highly correlated. The halo bispectrum helps break this degeneracy because it scales as \( b_1^3 \sigma_8^4 \), introducing different powers of \( b_1 \) and \( \sigma_8 \). \citet{b1Ombreakingdeg} also proposed a method to measure the linear bias \( b_1 \) independently of the matter density, \( \Omega_{\rm m} \), by using second-order PT in redshift-space. A  joint analysis of the power spectrum and bispectrum will therefore tighten constraints on the recovered parameters, as we will discuss further.

The first measurement of cosmological parameters from the galaxy bispectrum was performed by \citet{feldman2001constraints}. They measured the bispectrum of the IRAS PSC$z$ galaxy redshift survey and constrain the biases $1/b_1$ and $b_2/b_1^2$ and showed that there is no sign of a scale-dependent bias up to $k=0.2\;h\;{\rm Mpc}^{-1}$. \citet{verde20022df} later used the data from 2dF Galaxy Redshift Survey (2dFGRS) to measure the cosmological parameters from the bispectrum on scales $<0.1 \;h\;{\rm Mpc}^{-1}<k<0.5 \;h\;{\rm Mpc}^{-1}$. They measured the matter density contrast as $\Omega_{\rm m} = 0.27\pm 0.06$ and also measured the linear and non-linear biases. The linear bias of the sample was consistent with one and the non-linear bias with zero.

In  the era of precision cosmology, as the galaxy redshift surveys probe larger volumes and fainter objects, we can set tighter bounds on the cosmological parameters. \citet{gil2016clustering} measured the Baryonic Oscillation Spectroscopic Survey (BOSS) Data Release (DR) 12 cosmological parameters with a joint analysis of the power spectrum and bispectrum. They could set constraints on the logarithmic growth-rate $f$, the amplitude of the clustering fluctuations $\sigma_8$ and few other cosmological parameters. The BOSS DR12 sample consists of two samples: LOWZ galaxy sample which has 361,762 galaxies at an effective redshift of $z = 0.32$ and CMASS which has  777,202 galaxies at redshift of $z=0.57$. The scales considered in this analysis were $0.03 \;h\;{\rm Mpc}^{-1}<k<0.18 \;h\;{\rm Mpc}^{-1}$ for the LOWZ and $0.03 \;h\;{\rm Mpc}^{-1}<k<0.22 \;h\;{\rm Mpc}^{-1}$ for the CMASS sample. 

Higher-order corrections to the tree-level bispectrum from PT extends the validity of the bispectrum to even smaller scales. This will give more accurate description of the structure formation, enabling us to probe smaller scales. \citet{d2022boss} used EFTofLSS and found out that the using this framework can enhance the cosmological parameter constraints by at least $10\%$ over the power spectrum-only analysis. For some parameters, this enhancement is up to $30\%$.

\citet{ivanov2023cosmology} used the BOSS DR12 CMASS sample to measure the window-free bispectrum \citep{philcox2021cosmology}. Other than the theoretical methods for analyzing the power spectrum and bispectrum of galaxy surveys, we can use simulations to estimate the cosmological parameters. In this approach, $N$-body simulations are used instead of theory models that often struggle to give precise predictions on small scales. \citet{hahn2023rmsscriptsizeimbigcosmological} used this method to measure the BOSS DR12 bispectrum monopole to $k_{\rm max} = 0.5\;h\;{\rm Mpc}^{-1}$. \citet{hahn2023rmsscriptsizeimbigcosmological} demonstrated that the use of the bispectrum can enhance the parameter constraints by about a factor of two over the power spectrum. \citet{rezaei_mahdi_BAO} showed that when the BAO feature of the tree-level bispectrum is isolated and damped by a factor of $\exp{(-\Sigma^2 k^2/2)}$, the bispectrum can more accurately describe the data on smaller scales.

We can demonstrate the power of the redshift-space summary statistics in estimating cosmological parameters through a Fisher forecast analysis. Fisher forecasting has been a crucial tool in cosmology over the past decades \citep{Tegmark_1998}. Here, we highlight the strength of the redshift-space power spectrum and bispectrum in constraining cosmological parameters.  Let us first assume that the effective volume we probe is equivalent to the DESI Year 5 (Y5) LRG sample, with an effective volume of \( 25\;[{\rm Gpc}/h]^3 \) for the power spectrum and \( 14\;[{\rm Gpc}/h]^3 \) for the bispectrum  \citep{DESI:2016, grove2022desi}. These volumes are not the exact values describing the survey but are approximately 5 times the volume of the BOSS LRG CMASS sample \citep{alam2021completed}.

The fiducial cosmology we consider is a $\Lambda$CDM Universe with $\Omega_{\rm m} = 0.315$, $h=0.67$, $\Omega_{\rm b} = 0.049$, $n_{\rm s} = 0.962$, and $\sigma_8 = 0.811$. Since large-scale structure (LSS) data cannot tightly constrain the baryon density $\Omega_{\rm b}$, we fix it to its fiducial value. We assume three degenerate neutrinos with a total mass sum of $\sum m_{\nu} = 0.26\;{\rm eV}$, which corresponds to the $2\sigma$ upper bound from \textit{Planck} \citep{Planck:2018vyg}. The galaxy sample is described by three bias parameters with fiducial values: $b_1 = 2$, $b_2 = -1$, and $b_{\rm tidal} = -0.35$, all chosen to be within $1\sigma$ of the BOSS measurements \citep{ivanov2023cosmology, eggemeier2020testing}. We then construct the theoretical covariance matrix for the redshift-space bispectrum following \citet{hector_bis_cov, biagetti2022covariance, gualdi2020galaxy, gualdi2019enhancing}. The Fisher matrix is defined as \citep{Tegmark_1998}:
\begin{align}
    F_{\rm ij} = \frac{\partial \mathbf{D}}{\partial \theta_{\rm i}}\mathbf{C}^{-1}\frac{\partial \mathbf{D}}{\partial \theta_{\rm j}}
\end{align}
where $F$ is the Fisher matrix, with $i$ and $j$ indexing the $i$-th and $j$-th elements. $\theta$ represents the parameter of interest, $\mathbf{D}$ is the data vector corresponding to either the power spectrum or bispectrum, and $\mathbf{C}$ represents the covariance matrix of the data vector. Here we note that we have neglected the cross-covariance between the power spectrum and bispectrum. Next, we calculate the power spectrum using the \textsc{velocileptors} package \citep{velocileptors_1, velocileptors_2}, and obtain a ``triangle'' plot by using the \textsc{getdist} package \citep{lewis2019getdistpythonpackageanalysing}. The inverse of the Fisher matrix will then provide the error-bars on the parameters as well as the correlations (degeneracies) between them.

\begin{figure}[ht]
    \centering
    {\includegraphics[width=0.8\textwidth]{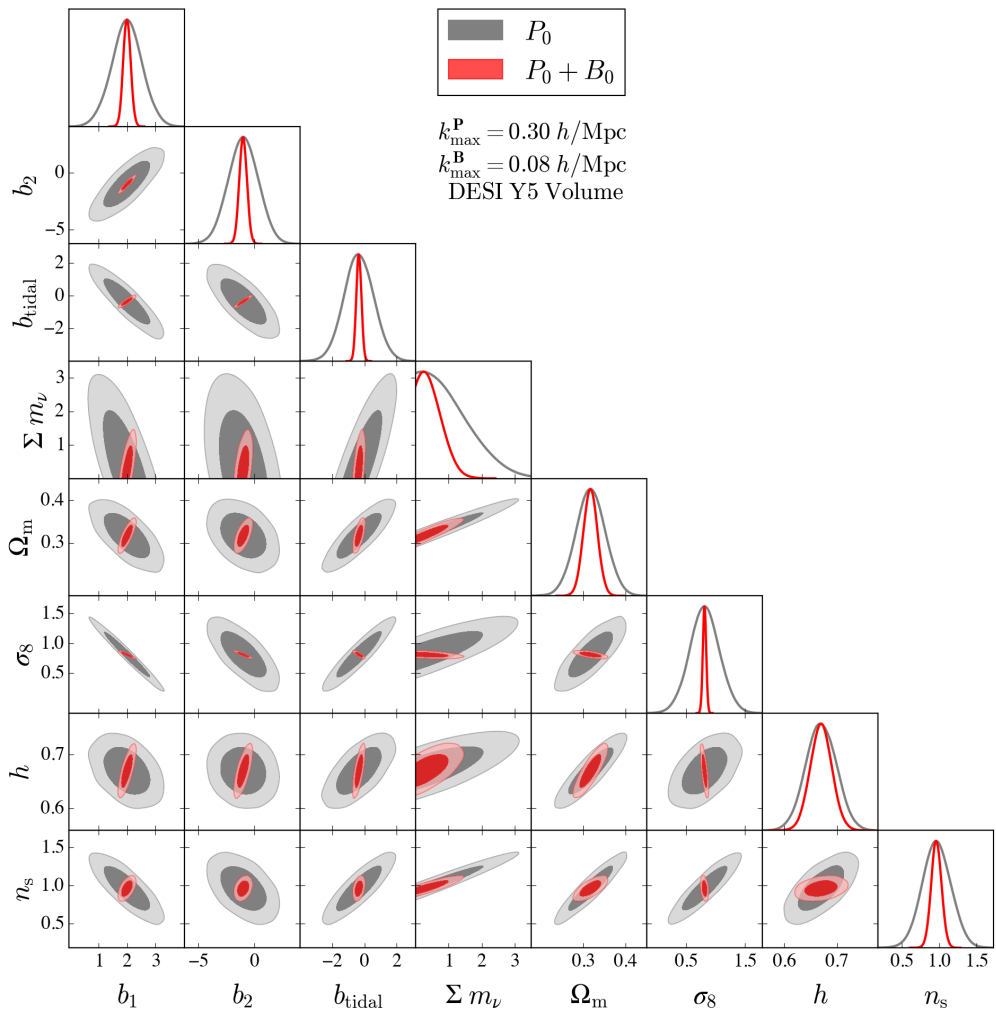} }
    \caption{A Fisher forecast of the 1-loop redshift-space galaxy power spectrum (gray, $P_0$) and the joint redshift-space power spectrum + redshift-space tree-level bispectrum ($B_0$) (red). For each, we use the monopole with respect to the line of sight, hence the subscript zero. The bispectrum helps reducing the error-bars in estimation of the cosmological parameters and galaxy biases by a noticeable amount. In this forecast, we have used theoretical covariance matrices for both power spectrum and bispectrum. The sample is DESI Y5-like, with effective volume of $25\;[{\rm Gpc}/h]^3$ for the power spectrum and $14\;[{\rm Gpc}/h]^3$ for the bispectrum (effective volume can differ depending on the observable). The maximum wave-number used for each is indicated below the legend.}
    \label{forecast}
\end{figure}

Fig. \ref{forecast} shows the $68\%$ and $95\%$ confidence level error-bars on the parameters, marginalized over the counter-term of the power spectrum, $c_{\rm ctr}$. For the power spectrum, we considered scales up to $k_{\rm max} = 0.3\;h/{\rm Mpc}$, and for the bispectrum, we assumed that the tree-level bispectrum is valid up to $k_{\rm max} = 0.08\;h/{\rm Mpc}$ \citep{Ivanov_2022}. The bispectrum alone up to $k_{\rm max} = 0.08\;h/{\rm Mpc}$ does not provide much information on the cosmological parameters (not shown in the figure). However, since it affects the correlations between the parameters, it can significantly reduce the error-bars by breaking parameter degeneracies. We note that in this forecast, no external dataset is added. We have only fixed the baryon density. Including other datasets, such as the CMB measurements from \textit{Planck} \citep{Planck:2018vyg}, supernovae \citep{roman_sn}, or BAO geometric distance measurements, would further reduce the error-bars \citep{adame2024desi}. As we see, the bispectrum has especially succeeded in reducing the degeneracy between $\sigma_8$ and the linear bias $b_1$, which in turn will further reduce the error-bars on the other parameters.

\section{Beyond Triplets: N-Point Correlation Functions and Polyspectra}
\label{sec:beyond}
\subsection{History}
We now very briefly discuss NPCFs, $N>3$, and their Fourier-space counterparts, polyspectra. Here, we do give a little history as there is not much of it, and it is more difficult to find, and less well-reviewed, than the history of the 2PCF and power spectrum. 

\subsubsection{Earliest Years}
The earliest work on NPCFs goes back to measurements in the 1970s of the 4PCF on 2D, photometric survey data from the Lick and Zwicky catalogs \citep{fry_peebles_4pcf}, which also in the same series of papers had their 3PCF measured \citep{peebles_groth_3pcf}; early 2+3PCF measurements were also made on the Shane-Wirtanen catalog \citep{groth_2_plus_3}. \citet{bonometto_1, bonometto_2} proposed a method to measure coefficients of the 4PCF on 2D, counts-in-cells data without measuring the full 4PCF, a useful computational acceleration.

There was some work on theoretical models for the 4PCF in the 1980s, based on the idea that there are consistency relations between the number of pairs, triplets, and quadruplets, and using the BBGKY hierarchy \citep{fry_bbgky, inagaki, fall_1976, fry_1984_Npoint}; the hierarchy among correlation functions themselves is explored in \citet{white_hier}, with slightly more recent work exploring the 2, 3, and 4PCF hierarchy with $N$-body simulations in \citet{suto_1994_hierarchical_up_to_4}. In the 1990s, a number of works explored the skewness (limit of the 3PCF as all triangle sides go to zero) \citep{fry_skew_1994} and kurtosis  (analogous limit of the 4PCF) \citep{catelan_kurt_1994, lokas_kurt_1995, chodo_kurt} or both \citep{bern_skew_kurt_1993, luo_skew_kurt_1993, kim_1998_skew_kurt}.

\subsubsection{More Recent Work}
There was another flowering of work on the 4PCF in the late 1990s and early 2000s, more focused on the trispectrum (Fourier-space analog of the 4PCF) than 4PCF \citep{verde_2001, sefusatti_trispec_2005}, which has very recently been pushed forward further \citep{gualdi_2021_trispec, gualdi_joint_bi_tri}, culminating in a measurement on BOSS data \citep{gualdi_2022_data_trispec}. Theoretical development of the trispectrum in redshift-space at 1-loop (using the second and third-order PT kernels) and including EFT and galaxy biasing is in \citet{bertolini, steele_2021}; some of the interest in this is because certain limits of the trispectrum constitute the higher-order corrections to the 2PCF or power spectrum covariance matrix \citep{mohamed} and also enter corrections to the bispectrum covariance matrix \citep{sugi_covar}. \citet{lee_2020_ang_trispec} also recently explored the angular (\textit{i.e.} projected) trispectrum. The model of \citet{gualdi_2021_trispec}
has also been translated into position space to give a model for the 4PCF \citep{william_4pcf}. The connected (GRF contribution-removed) 4PCF has also been measured, for the first time, on BOSS CMASS \citep{phil_4pcf}, exploiting the ability to remove the disconnected (\textit{i.e.} due to the GRF) piece as outlined in \citet{encore}. The 4PCF was also, for the first time, recently applied to MHD turbulence in \citet{williamson_mhd}.

Notably, \citet{schmitt_npcf_recon} connects the 3 and 4-point statistics to the standardly-used technique of density field reconstruction \citep{eis_recon, pad_recon}, which moves galaxies backwards by inferring the gravitational forces from their current positions and then reversing the forces. This makes the density field more linear and thus puts much of the 3 and 4-point statistics' information back into the 2-point statistics. However, as \citet{sam_npcf} shows, this is only the case for number densities below a certain threshold, which turns out, for the BOSS CMASS sample, to be roughly that sample's number density. For higher-number-density samples, 3 point-statistics can offer more standard ruler information than even the initial matter power spectrum.

\subsection{Algorithms for 4 Points and Beyond}
Early work on the 4PCF measured just the 2D version, generally using counts in cells on photographic plates; this was not of severe computational expense. More recently, \citet{sefusatti_trispec_2005} proposed an averaged trispectrum estimator of reasonable speed (but smoothing away some signal by averaging); \citet{tomlinson_polyspectra} proposed a polyspectrum estimator. In position space, \citet{sabiu}
suggested a solution using a graph spatial database. Slightly later, the \textsc{ENCORE} code became available, computing the 4PCF in a basis of isotropic functions \citep{cahn_iso, iso_gen}; this computation is fundamentally a convolution and so can be done via FFTs, as presented in the package \textsc{sarabande} \citep{sarabande}. Mathematically, the problem of a 4PCF is the same as that of the CMB bispectrum \citep{spergel_goldberg_bispec}, but with multiple radial shells rather than just one, and using each galaxy in the survey as the ``observer'' in sequence. An algorithm to measure the compressed, ``Parity-Odd Power Spectrum'', which is similar in spirit to the 4PCF, is \citet{jamieson_pops}.

\subsection{Parity Violation}
Recently it was pointed out that in 3D LSS the 4PCF is the lowest-order statistic sensitive to parity violation after rotation averaging \citep{cahn_parity}; earlier work had suggested using spatially offset pairs of power spectra (similar to a 4PCF) for this purpose \citep{jeong_fossil}. The division of the 4PCF in the isotropic basis into even and odd parity modes \citep{cahn_iso} was central to \citet{cahn_parity}. 

The odd 4PCF was first measured on SDSS BOSS CMASS and LOWZ data by \citet{hou_parity}, finding high statistical significance evidence in CMASS and lower-signficance evidence in the smaller LOWZ sample. Using a covariance matrix based on \citet{hou_covar} and produced by Hou, and the idea presented in \citet{cahn_parity}, \citet{phil_parity} also measured the parity-odd 4PCF, finding a lower significance, but due to a coarser choice of binning for the separations between the galaxies involved in the 4PCF; see footnote 1 in \citet{iso_gen} for further discussion. Using a simple model, \citet{cahn_parity} forecasts that if genuine, parity violation can be seen at very high significance by the time of DESI year 5 data, with the best sample being Luminous Red Galaxies. We caution that this statement could be model-dependent, however.

Gravitational waves (GW) have also been considered for the parity-odd trispectrum \citep{moretti_trispec_odd}, as well as, by lensing the LSS, a possible explanation for the BOSS results noted above \citep{inomata}. If the parity-odd 4PCF is genuine, it would most likely stem from some inflationary mechanism, for instance a tachyonic instability amplifying one helicity mode relative to another in axion inflation \citep{barnaby_2011, niu_axion_trispectrum, reinhard_axion}, or other beyond-Standard Model inflationary physics, \textit{e.g.} \citet{cabass}. Several works have investigated no-go theorems for this area \citep{cabass_no_go, thavanesan_no_go} or explored other general properties of the primordial trispectrum, such as factorization \citep{stefanyszyn}. \citet{coulton, jamieson_pops} suggested a simple toy parity-violating model useful for exploring the behavior of estimators and pipelines.\footnote{Available here: \url{https://github.com/dsjamieson/parity_violating_templates}} Searching for BAO in the odd 4PCF has also been suggested as a test of its genuine cosmological origin \citep{bao_odd}. 

While we are in early days observationally regarding parity violation, it is a strong motivator for use of the 4PCF on future data-sets designed to probe inflation, such as Spherex \citep{spherex}, DESI-II \citep{spec_roadmap} or the (proposed) Megamapper instrument \citep{megamapper}, as well as more general missions such as Roman \citep{roman_wang_spec} that will have spectroscopy over a large volume.\footnote{The Roman Galaxy Redshift Survey infrastructure site: \url{https://roman-grs-pit.caltech.edu/}}

\subsection{Beyond the 4PCF: 5 and 6PCF}
To our knowledge, no work has yet measured the 5 or 6PCF or their Fourier-space analogs the quad or pentaspectrum. However, both are natively 3D and hence could probe parity. The \textsc{encore} code \citep{encore} does have the capability to measure them efficiently, in the basis of \citet{cahn_iso}, and an analytic covariance template (assuming a GRF) has been obtained \citep{hou_covar}, which would enable analysis of the very large number of degrees of freedom these observables entail. Measurement of them could also be used to verify how well mock catalogs agree with data, as relevant for using mocks to estimate the covariance matrix. The 3PCF and 4PCF covariance matrix both contain contributions 
from the 5PCF and 6PCF, and finding agreement between these latter for data and for mocks would strengthen confidence that the mocks are accurately capturing the physics entering the covariance matrix.

\section{Outlook}
Since they naturally encode the presumed homogeneity (translation-invariance) and isotropy (rotation-invariance) of the Universe (and the laws of physics!), correlations of pairs, triplets, and beyond have become a core tool for probing the spatial clustering of galaxies. The 2PCF and power spectrum have been in broad use for a long time, with more recent inroads bringing the 3PCF and bispectrum into standard practice as well. With novel prospects for constraining parity violation in the early Universe, the interest in 4PCF and trispectrum is also now at a new height. 

Two major remaining challenges of making regular use of all these statistics together are: dealing with the large number of degrees of freedom, and estimating covariance matrices that include cross-talk between the different statistics. GRF templates are often helpful regarding the first, as they can deal with a large number of degrees of freedom either directly, or by use in a first, compression step (where only the highest signal-to-noise eigenvectors are chosen, and then mock catalogs are used to get the covariance of this small space). However, the GRF template predicts zero cross covariance between any pair of statistics with an odd total number of points (\textit{e.g.} 2PCF $\times$ 3PCF), and so cannot be helpful; at present, only simulations (or higher-order perturbative calculations that quickly become complicated) can capture cross-covariance in this scenario. Compression to a smaller number of degrees of freedom is not a silver bullet here, as one already needs a ``guess'' covariance matrix to perform this compression in the first place---exactly what one lacks as regards cross-covariance of different NPCFs or polyspectra. This is perhaps one area where Machine Learning can productively be hybridized with the more traditional techniques outlined in this article.

Overall, however, cosmology has (arguably) always been data-driven, and the wealth of present and future 3D spectroscopic surveys, such as DESI, DESI-II, Euclid, Spherex, and Roman, will guarantee that further development of the methods outlined here will be well-rewarded with rich targets for their application.

\bibliographystyle{harvard}

\bibliography{references.bib}

\begin{thebibliography*}{340}
\providecommand{\bibtype}[1]{}
\providecommand{\natexlab}[1]{#1}
{\catcode`\|=0\catcode`\#=12\catcode`\@=11\catcode`\\=12
|immediate|write|@auxout{\expandafter\ifx\csname natexlab\endcsname\relax\gdef\natexlab#1{#1}\fi}}
\renewcommand{\url}[1]{{\tt #1}}
\providecommand{\urlprefix}{URL }
\expandafter\ifx\csname urlstyle\endcsname\relax
  \providecommand{\doi}[1]{doi:\discretionary{}{}{}#1}\else
  \providecommand{\doi}{doi:\discretionary{}{}{}\begingroup \urlstyle{rm}\Url}\fi
\providecommand{\bibinfo}[2]{#2}
\providecommand{\eprint}[2][]{\url{#2}}

\bibtype{Article}%
\bibitem[Abdul~Karim et al.(2025{\natexlab{a}})]{desi_dr1}
\bibinfo{author}{Abdul~Karim M} and  et al. (\bibinfo{collaboration}{DESI}) (\bibinfo{year}{2025}{\natexlab{a}}), \bibinfo{month}{3}.
\bibinfo{title}{{Data Release 1 of the Dark Energy Spectroscopic Instrument}} \eprint{2503.14745}.

\bibtype{Article}%
\bibitem[Abdul~Karim et al.(2025{\natexlab{b}})]{DESI:2025zpo}
\bibinfo{author}{Abdul~Karim M} and  et al. (\bibinfo{collaboration}{DESI}) (\bibinfo{year}{2025}{\natexlab{b}}), \bibinfo{month}{3}.
\bibinfo{title}{{DESI DR2 Results I: Baryon Acoustic Oscillations from the Lyman Alpha Forest}} \eprint{2503.14739}.

\bibtype{Article}%
\bibitem[Abdul~Karim et al.(2025{\natexlab{c}})]{DESI:2025zgx}
\bibinfo{author}{Abdul~Karim M} and  et al. (\bibinfo{collaboration}{DESI}) (\bibinfo{year}{2025}{\natexlab{c}}), \bibinfo{month}{3}.
\bibinfo{title}{{DESI DR2 Results II: Measurements of Baryon Acoustic Oscillations and Cosmological Constraints}} \eprint{2503.14738}.

\bibtype{Article}%
\bibitem[{Ach{\'u}carro} et al.(2022)]{achu_2022}
\bibinfo{author}{{Ach{\'u}carro} A}, \bibinfo{author}{{Biagetti} M}, \bibinfo{author}{{Braglia} M}, \bibinfo{author}{{Cabass} G}, \bibinfo{author}{{Caldwell} R}, \bibinfo{author}{{Castorina} E}, \bibinfo{author}{{Chen} X}, \bibinfo{author}{{Coulton} W}, \bibinfo{author}{{Flauger} R}, \bibinfo{author}{{Fumagalli} J}, \bibinfo{author}{{Ivanov} MM}, \bibinfo{author}{{Lee} H}, \bibinfo{author}{{Maleknejad} A}, \bibinfo{author}{{Meerburg} PD}, \bibinfo{author}{{Moradinezhad Dizgah} A}, \bibinfo{author}{{Palma} GA}, \bibinfo{author}{{Pimentel} GL}, \bibinfo{author}{{Renaux-Petel} S}, \bibinfo{author}{{Wallisch} B}, \bibinfo{author}{{Wandelt} BD}, \bibinfo{author}{{Witkowski} LT} and  \bibinfo{author}{{Kimmy Wu} WL} (\bibinfo{year}{2022}), \bibinfo{month}{Mar.}
\bibinfo{title}{{Inflation: Theory and Observations}}.
\bibinfo{journal}{{\em arXiv e-prints}} , \bibinfo{eid}{arXiv:2203.08128}\bibinfo{doi}{\doi{10.48550/arXiv.2203.08128}}.
\eprint{2203.08128}.

\bibtype{Article}%
\bibitem[Adame et al.(2024{\natexlab{a}})]{adame2024desi}
\bibinfo{author}{Adame AG} and  et al. (\bibinfo{collaboration}{DESI}) (\bibinfo{year}{2024}{\natexlab{a}}), \bibinfo{month}{11}.
\bibinfo{title}{{DESI 2024 VII: Cosmological Constraints from the Full-Shape Modeling of Clustering Measurements}} \eprint{2411.12022}.

\bibtype{Article}%
\bibitem[Adame et al.(2024{\natexlab{b}})]{adame2024early}
\bibinfo{author}{Adame AG} and  et al. (\bibinfo{collaboration}{DESI}) (\bibinfo{year}{2024}{\natexlab{b}}).
\bibinfo{title}{{The Early Data Release of the Dark Energy Spectroscopic Instrument}}.
\bibinfo{journal}{{\em Astron. J.}} \bibinfo{volume}{168} (\bibinfo{number}{2}): \bibinfo{pages}{58}. \bibinfo{doi}{\doi{10.3847/1538-3881/ad3217}}.
\eprint{2306.06308}.

\bibtype{Article}%
\bibitem[Aghamousa et al.(2016)]{aghamousa2016desi}
\bibinfo{author}{Aghamousa A}, \bibinfo{author}{Aguilar J}, \bibinfo{author}{Ahlen S}, \bibinfo{author}{Alam S}, \bibinfo{author}{Allen LE}, \bibinfo{author}{Prieto CA}, \bibinfo{author}{Annis J}, \bibinfo{author}{Bailey S}, \bibinfo{author}{Balland C}, \bibinfo{author}{Ballester O} and  et al. (\bibinfo{year}{2016}).
\bibinfo{title}{The desi experiment part i: science, targeting, and survey design}.
\bibinfo{journal}{{\em arXiv preprint arXiv:1611.00036}} .

\bibtype{Article}%
\bibitem[Aghanim et al.(2020)]{Planck:2018vyg}
\bibinfo{author}{Aghanim N} and  et al. (\bibinfo{collaboration}{Planck}) (\bibinfo{year}{2020}).
\bibinfo{title}{{Planck 2018 results. VI. Cosmological parameters}}.
\bibinfo{journal}{{\em Astron. Astrophys.}} \bibinfo{volume}{641}: \bibinfo{pages}{A6}. \bibinfo{doi}{\doi{10.1051/0004-6361/201833910}}.
\bibinfo{note}{[Erratum: Astron.Astrophys. 652, C4 (2021)]}, \eprint{1807.06209}.

\bibtype{Article}%
\bibitem[Alam et al.(2017{\natexlab{a}})]{alam_2017_boss_2ps}
\bibinfo{author}{Alam S} and  et al. (\bibinfo{collaboration}{BOSS}) (\bibinfo{year}{2017}{\natexlab{a}}).
\bibinfo{title}{{The clustering of galaxies in the completed SDSS-III Baryon Oscillation Spectroscopic Survey: cosmological analysis of the DR12 galaxy sample}}.
\bibinfo{journal}{{\em Mon. Not. Roy. Astron. Soc.}} \bibinfo{volume}{470} (\bibinfo{number}{3}): \bibinfo{pages}{2617--2652}. \bibinfo{doi}{\doi{10.1093/mnras/stx721}}.
\eprint{1607.03155}.

\bibtype{Article}%
\bibitem[Alam et al.(2017{\natexlab{b}})]{alam2017clustering}
\bibinfo{author}{Alam S} and  et al. (\bibinfo{collaboration}{BOSS}) (\bibinfo{year}{2017}{\natexlab{b}}).
\bibinfo{title}{{The clustering of galaxies in the completed SDSS-III Baryon Oscillation Spectroscopic Survey: cosmological analysis of the DR12 galaxy sample}}.
\bibinfo{journal}{{\em Mon. Not. Roy. Astron. Soc.}} \bibinfo{volume}{470} (\bibinfo{number}{3}): \bibinfo{pages}{2617--2652}. \bibinfo{doi}{\doi{10.1093/mnras/stx721}}.
\eprint{1607.03155}.

\bibtype{Article}%
\bibitem[Alam et al.(2021{\natexlab{a}})]{alam_all_sdss}
\bibinfo{author}{Alam S} and  et al. (\bibinfo{collaboration}{eBOSS}) (\bibinfo{year}{2021}{\natexlab{a}}).
\bibinfo{title}{{Completed SDSS-IV extended Baryon Oscillation Spectroscopic Survey: Cosmological implications from two decades of spectroscopic surveys at the Apache Point Observatory}}.
\bibinfo{journal}{{\em Phys. Rev. D}} \bibinfo{volume}{103} (\bibinfo{number}{8}): \bibinfo{pages}{083533}. \bibinfo{doi}{\doi{10.1103/PhysRevD.103.083533}}.
\eprint{2007.08991}.

\bibtype{Article}%
\bibitem[Alam et al.(2021{\natexlab{b}})]{alam2021completed}
\bibinfo{author}{Alam S} and  et al. (\bibinfo{collaboration}{eBOSS}) (\bibinfo{year}{2021}{\natexlab{b}}).
\bibinfo{title}{{Completed SDSS-IV extended Baryon Oscillation Spectroscopic Survey: Cosmological implications from two decades of spectroscopic surveys at the Apache Point Observatory}}.
\bibinfo{journal}{{\em Phys. Rev. D}} \bibinfo{volume}{103} (\bibinfo{number}{8}): \bibinfo{pages}{083533}. \bibinfo{doi}{\doi{10.1103/PhysRevD.103.083533}}.
\eprint{2007.08991}.

\bibtype{Article}%
\bibitem[Alam et al.(2021{\natexlab{c}})]{alam2021towards}
\bibinfo{author}{Alam S} and  et al. (\bibinfo{year}{2021}{\natexlab{c}}).
\bibinfo{title}{{Towards testing the theory of gravity with DESI: summary statistics, model predictions and future simulation requirements}}.
\bibinfo{journal}{{\em JCAP}} \bibinfo{volume}{11} (\bibinfo{number}{11}): \bibinfo{pages}{050}. \bibinfo{doi}{\doi{10.1088/1475-7516/2021/11/050}}.
\eprint{2011.05771}.

\bibtype{Article}%
\bibitem[Alcock and Paczynski(1979)]{ap_effect}
\bibinfo{author}{Alcock C} and  \bibinfo{author}{Paczynski B} (\bibinfo{year}{1979}).
\bibinfo{title}{{An evolution free test for non-zero cosmological constant}}.
\bibinfo{journal}{{\em Nature}} \bibinfo{volume}{281}: \bibinfo{pages}{358--359}. \bibinfo{doi}{\doi{10.1038/281358a0}}.

\bibtype{Article}%
\bibitem[Angulo et al.(2015)]{angulo2015one}
\bibinfo{author}{Angulo RE}, \bibinfo{author}{Foreman S}, \bibinfo{author}{Schmittfull M} and  \bibinfo{author}{Senatore L} (\bibinfo{year}{2015}).
\bibinfo{title}{The one-loop matter bispectrum in the effective field theory of large scale structures}.
\bibinfo{journal}{{\em Journal of Cosmology and Astroparticle Physics}} \bibinfo{volume}{2015} (\bibinfo{number}{10}): \bibinfo{pages}{039}.

\bibtype{Article}%
\bibitem[{Arur} and {Maccarone}(2022)]{arur_xray}
\bibinfo{author}{{Arur} K} and  \bibinfo{author}{{Maccarone} TJ} (\bibinfo{year}{2022}), \bibinfo{month}{Aug.}
\bibinfo{title}{{Using the bispectrum to probe radio X-ray correlations in GRS 1915+105}}.
\bibinfo{journal}{{\em \mnras}} \bibinfo{volume}{514} (\bibinfo{number}{2}): \bibinfo{pages}{1720--1732}. \bibinfo{doi}{\doi{10.1093/mnras/stac1463}}.
\eprint{2205.12981}.

\bibtype{Article}%
\bibitem[Aviles and Banerjee(2020)]{aviles2020lagrangian}
\bibinfo{author}{Aviles A} and  \bibinfo{author}{Banerjee A} (\bibinfo{year}{2020}).
\bibinfo{title}{A lagrangian perturbation theory in the presence of massive neutrinos}.
\bibinfo{journal}{{\em Journal of Cosmology and Astroparticle Physics}} \bibinfo{volume}{2020} (\bibinfo{number}{10}): \bibinfo{pages}{034}.

\bibtype{Article}%
\bibitem[Aviles and Niz(2023)]{Aviles_2023}
\bibinfo{author}{Aviles A} and  \bibinfo{author}{Niz G} (\bibinfo{year}{2023}), \bibinfo{month}{Mar.}
\bibinfo{title}{Galaxy three-point correlation function in modified gravity}.
\bibinfo{journal}{{\em Physical Review D}} \bibinfo{volume}{107} (\bibinfo{number}{6}).
ISSN \bibinfo{issn}{2470-0029}. \bibinfo{doi}{\doi{10.1103/physrevd.107.063525}}.
\bibinfo{url}{\url{http://dx.doi.org/10.1103/PhysRevD.107.063525}}.

\bibtype{Article}%
\bibitem[Aviles et al.(2021)]{aviles2021clustering}
\bibinfo{author}{Aviles A}, \bibinfo{author}{Banerjee A}, \bibinfo{author}{Niz G} and  \bibinfo{author}{Slepian Z} (\bibinfo{year}{2021}).
\bibinfo{title}{Clustering in massive neutrino cosmologies via eulerian perturbation theory}.
\bibinfo{journal}{{\em Journal of Cosmology and Astroparticle Physics}} \bibinfo{volume}{2021} (\bibinfo{number}{11}): \bibinfo{pages}{028}.

\bibtype{Article}%
\bibitem[{Bakx} et al.(2024)]{bakx}
\bibinfo{author}{{Bakx} T}, \bibinfo{author}{{Chisari} NE} and  \bibinfo{author}{{Vlah} Z} (\bibinfo{year}{2024}), \bibinfo{month}{Jul.}
\bibinfo{title}{{COBRA: Optimal Factorization of Cosmological Observables}}.
\bibinfo{journal}{{\em arXiv e-prints}} , \bibinfo{eid}{arXiv:2407.04660}\bibinfo{doi}{\doi{10.48550/arXiv.2407.04660}}.
\eprint{2407.04660}.

\bibtype{Article}%
\bibitem[Baldauf(2020)]{baldauf2020effective}
\bibinfo{author}{Baldauf T} (\bibinfo{year}{2020}).
\bibinfo{title}{Effective field theory of large-scale structure}.
\bibinfo{journal}{{\em Effective Field Theory in Particle Physics and Cosmology: Lecture Notes of the Les Houches Summer School: Volume 108, July 2017}} \bibinfo{volume}{108}: \bibinfo{pages}{415}.

\bibtype{Article}%
\bibitem[Baldauf et al.(2011)]{baldauf2011primordial}
\bibinfo{author}{Baldauf T}, \bibinfo{author}{Seljak U} and  \bibinfo{author}{Senatore L} (\bibinfo{year}{2011}).
\bibinfo{title}{Primordial non-gaussianity in the bispectrum of the halo density field}.
\bibinfo{journal}{{\em Journal of Cosmology and Astroparticle Physics}} \bibinfo{volume}{2011} (\bibinfo{number}{04}): \bibinfo{pages}{006}.

\bibtype{Article}%
\bibitem[{Baldauf} et al.(2012)]{baldauf}
\bibinfo{author}{{Baldauf} T}, \bibinfo{author}{{Seljak} U}, \bibinfo{author}{{Desjacques} V} and  \bibinfo{author}{{McDonald} P} (\bibinfo{year}{2012}), \bibinfo{month}{Oct.}
\bibinfo{title}{{Evidence for quadratic tidal tensor bias from the halo bispectrum}}.
\bibinfo{journal}{{\em \prd}} \bibinfo{volume}{86} (\bibinfo{number}{8}), \bibinfo{eid}{083540}. \bibinfo{doi}{\doi{10.1103/PhysRevD.86.083540}}.
\eprint{1201.4827}.

\bibtype{Article}%
\bibitem[{Baldauf} et al.(2021)]{twoloopbis}
\bibinfo{author}{{Baldauf} T}, \bibinfo{author}{{Garny} M}, \bibinfo{author}{{Taule} P} and  \bibinfo{author}{{Steele} T} (\bibinfo{year}{2021}), \bibinfo{month}{Dec.}
\bibinfo{title}{{Two-loop bispectrum of large-scale structure}}.
\bibinfo{journal}{{\em \prd}} \bibinfo{volume}{104} (\bibinfo{number}{12}), \bibinfo{eid}{123551}. \bibinfo{doi}{\doi{10.1103/PhysRevD.104.123551}}.
\eprint{2110.13930}.

\bibtype{Article}%
\bibitem[{Barnaby} and {Peloso}(2011)]{barnaby_2011}
\bibinfo{author}{{Barnaby} N} and  \bibinfo{author}{{Peloso} M} (\bibinfo{year}{2011}), \bibinfo{month}{May}.
\bibinfo{title}{{Large Non-Gaussianity in Axion Inflation}}.
\bibinfo{journal}{{\em \prl}} \bibinfo{volume}{106} (\bibinfo{number}{18}), \bibinfo{eid}{181301}. \bibinfo{doi}{\doi{10.1103/PhysRevLett.106.181301}}.
\eprint{1011.1500}.

\bibtype{Article}%
\bibitem[{Bartolo} et al.(2001)]{bartolo_isocurv}
\bibinfo{author}{{Bartolo} N}, \bibinfo{author}{{Matarrese} S} and  \bibinfo{author}{{Riotto} A} (\bibinfo{year}{2001}), \bibinfo{month}{Dec.}
\bibinfo{title}{{Adiabatic and isocurvature perturbations from inflation: Power spectra and consistency relations}}.
\bibinfo{journal}{{\em \prd}} \bibinfo{volume}{64} (\bibinfo{number}{12}): \bibinfo{pages}{123504}. \bibinfo{doi}{\doi{10.1103/PhysRevD.64.123504}}.
\eprint{astro-ph/0107502}.

\bibtype{Article}%
\bibitem[{Bartolo} et al.(2004)]{bartolo}
\bibinfo{author}{{Bartolo} N}, \bibinfo{author}{{Komatsu} E}, \bibinfo{author}{{Matarrese} S} and  \bibinfo{author}{{Riotto} A} (\bibinfo{year}{2004}), \bibinfo{month}{Nov.}
\bibinfo{title}{{Non-Gaussianity from inflation: theory and observations}}.
\bibinfo{journal}{{\em \physrep}} \bibinfo{volume}{402} (\bibinfo{number}{3-4}): \bibinfo{pages}{103--266}. \bibinfo{doi}{\doi{10.1016/j.physrep.2004.08.022}}.
\eprint{astro-ph/0406398}.

\bibtype{Article}%
\bibitem[{Bashinsky} and {Bertschinger}(2002)]{bashinsky}
\bibinfo{author}{{Bashinsky} S} and  \bibinfo{author}{{Bertschinger} E} (\bibinfo{year}{2002}), \bibinfo{month}{Jun.}
\bibinfo{title}{{Dynamics of cosmological perturbations in position space}}.
\bibinfo{journal}{{\em \prd}} \bibinfo{volume}{65} (\bibinfo{number}{12}), \bibinfo{eid}{123008}. \bibinfo{doi}{\doi{10.1103/PhysRevD.65.123008}}.
\eprint{astro-ph/0202215}.

\bibtype{Article}%
\bibitem[{Baumann}(2009)]{baumann_tasi}
\bibinfo{author}{{Baumann} D} (\bibinfo{year}{2009}), \bibinfo{month}{Jul.}
\bibinfo{title}{{TASI Lectures on Inflation}}.
\bibinfo{journal}{{\em arXiv e-prints}} , \bibinfo{eid}{arXiv:0907.5424}\bibinfo{doi}{\doi{10.48550/arXiv.0907.5424}}.
\eprint{0907.5424}.

\bibtype{Article}%
\bibitem[{Bautista} et al.(2021)]{bautista}
\bibinfo{author}{{Bautista} JE}, \bibinfo{author}{{Paviot} R}, \bibinfo{author}{{Vargas Maga{\~n}a} M}, \bibinfo{author}{{de la Torre} S}, \bibinfo{author}{{Fromenteau} S}, \bibinfo{author}{{Gil-Mar{\'\i}n} H}, \bibinfo{author}{{Ross} AJ}, \bibinfo{author}{{Burtin} E}, \bibinfo{author}{{Dawson} KS}, \bibinfo{author}{{Hou} J}, \bibinfo{author}{{Kneib} JP}, \bibinfo{author}{{de Mattia} A}, \bibinfo{author}{{Percival} WJ}, \bibinfo{author}{{Rossi} G}, \bibinfo{author}{{Tojeiro} R}, \bibinfo{author}{{Zhao} C}, \bibinfo{author}{{Zhao} GB}, \bibinfo{author}{{Alam} S}, \bibinfo{author}{{Brownstein} J}, \bibinfo{author}{{Chapman} MJ}, \bibinfo{author}{{Choi} PD}, \bibinfo{author}{{Chuang} CH}, \bibinfo{author}{{Escoffier} S}, \bibinfo{author}{{de la Macorra} A}, \bibinfo{author}{{du Mas des Bourboux} H}, \bibinfo{author}{{Mohammad} FG}, \bibinfo{author}{{Moon} J}, \bibinfo{author}{{M{\"u}ller} EM}, \bibinfo{author}{{Nadathur} S}, \bibinfo{author}{{Newman} JA}, \bibinfo{author}{{Schneider} D}, \bibinfo{author}{{Seo}
  HJ} and  \bibinfo{author}{{Wang} Y} (\bibinfo{year}{2021}), \bibinfo{month}{Jan.}
\bibinfo{title}{{The completed SDSS-IV extended Baryon Oscillation Spectroscopic Survey: measurement of the BAO and growth rate of structure of the luminous red galaxy sample from the anisotropic correlation function between redshifts 0.6 and 1}}.
\bibinfo{journal}{{\em \mnras}} \bibinfo{volume}{500} (\bibinfo{number}{1}): \bibinfo{pages}{736--762}. \bibinfo{doi}{\doi{10.1093/mnras/staa2800}}.
\eprint{2007.08993}.

\bibtype{Article}%
\bibitem[{Behera} et al.(2024{\natexlab{a}})]{behera}
\bibinfo{author}{{Behera} J}, \bibinfo{author}{{Rezaie} M}, \bibinfo{author}{{Samushia} L} and  \bibinfo{author}{{Ereza} J} (\bibinfo{year}{2024}{\natexlab{a}}), \bibinfo{month}{Jul.}
\bibinfo{title}{{Modelling the BAO feature in bispectrum}}.
\bibinfo{journal}{{\em \mnras}} \bibinfo{volume}{531} (\bibinfo{number}{3}): \bibinfo{pages}{3326--3335}. \bibinfo{doi}{\doi{10.1093/mnras/stae1161}}.
\eprint{2312.05942}.

\bibtype{Article}%
\bibitem[{Behera} et al.(2024{\natexlab{b}})]{rezaei_mahdi_BAO}
\bibinfo{author}{{Behera} J}, \bibinfo{author}{{Rezaie} M}, \bibinfo{author}{{Samushia} L} and  \bibinfo{author}{{Ereza} J} (\bibinfo{year}{2024}{\natexlab{b}}), \bibinfo{month}{Jul.}
\bibinfo{title}{{Modelling the BAO feature in bispectrum}}.
\bibinfo{journal}{{\em \mnras}} \bibinfo{volume}{531} (\bibinfo{number}{3}): \bibinfo{pages}{3326--3335}. \bibinfo{doi}{\doi{10.1093/mnras/stae1161}}.
\eprint{2312.05942}.

\bibtype{Article}%
\bibitem[{Benabou} et al.(2024)]{benabou}
\bibinfo{author}{{Benabou} JN}, \bibinfo{author}{{Testa} A}, \bibinfo{author}{{Heinrich} C}, \bibinfo{author}{{Gebhardt} HSG} and  \bibinfo{author}{{Dor{\'e}} O} (\bibinfo{year}{2024}), \bibinfo{month}{May}.
\bibinfo{title}{{Galaxy bispectrum in the spherical Fourier-Bessel basis}}.
\bibinfo{journal}{{\em \prd}} \bibinfo{volume}{109} (\bibinfo{number}{10}), \bibinfo{eid}{103507}. \bibinfo{doi}{\doi{10.1103/PhysRevD.109.103507}}.
\eprint{2312.15992}.

\bibtype{Article}%
\bibitem[{Bennett} et al.(2013)]{wmap_9}
\bibinfo{author}{{Bennett} CL}, \bibinfo{author}{{Larson} D}, \bibinfo{author}{{Weiland} JL}, \bibinfo{author}{{Jarosik} N}, \bibinfo{author}{{Hinshaw} G}, \bibinfo{author}{{Odegard} N}, \bibinfo{author}{{Smith} KM}, \bibinfo{author}{{Hill} RS}, \bibinfo{author}{{Gold} B}, \bibinfo{author}{{Halpern} M}, \bibinfo{author}{{Komatsu} E}, \bibinfo{author}{{Nolta} MR}, \bibinfo{author}{{Page} L}, \bibinfo{author}{{Spergel} DN}, \bibinfo{author}{{Wollack} E}, \bibinfo{author}{{Dunkley} J}, \bibinfo{author}{{Kogut} A}, \bibinfo{author}{{Limon} M}, \bibinfo{author}{{Meyer} SS}, \bibinfo{author}{{Tucker} GS} and  \bibinfo{author}{{Wright} EL} (\bibinfo{year}{2013}), \bibinfo{month}{Oct.}
\bibinfo{title}{{Nine-year Wilkinson Microwave Anisotropy Probe (WMAP) Observations: Final Maps and Results}}.
\bibinfo{journal}{{\em \apjs}} \bibinfo{volume}{208} (\bibinfo{number}{2}), \bibinfo{eid}{20}. \bibinfo{doi}{\doi{10.1088/0067-0049/208/2/20}}.
\eprint{1212.5225}.

\bibtype{Article}%
\bibitem[Berlind and Weinberg(2002)]{berlind2002halo}
\bibinfo{author}{Berlind AA} and  \bibinfo{author}{Weinberg DH} (\bibinfo{year}{2002}).
\bibinfo{title}{The halo occupation distribution: Toward an empirical determination of the relation between galaxies and mass}.
\bibinfo{journal}{{\em The Astrophysical Journal}} \bibinfo{volume}{575} (\bibinfo{number}{2}): \bibinfo{pages}{587}.

\bibtype{Article}%
\bibitem[{Bernardeau}(1994)]{bern_skew_kurt_1993}
\bibinfo{author}{{Bernardeau} F} (\bibinfo{year}{1994}), \bibinfo{month}{Sep.}
\bibinfo{title}{{Skewness and Kurtosis in Large-Scale Cosmic Fields}}.
\bibinfo{journal}{{\em \apj}} \bibinfo{volume}{433}: \bibinfo{pages}{1}. \bibinfo{doi}{\doi{10.1086/174620}}.
\eprint{astro-ph/9312026}.

\bibtype{Article}%
\bibitem[Bernardeau et al.(2002)]{Bernardeau_2002}
\bibinfo{author}{Bernardeau F}, \bibinfo{author}{Colombi S}, \bibinfo{author}{Gaztañaga E} and  \bibinfo{author}{Scoccimarro R} (\bibinfo{year}{2002}), \bibinfo{month}{Sep.}
\bibinfo{title}{Large-scale structure of the universe and cosmological perturbation theory}.
\bibinfo{journal}{{\em Physics Reports}} \bibinfo{volume}{367} (\bibinfo{number}{1–3}): \bibinfo{pages}{1–248}.
ISSN \bibinfo{issn}{0370-1573}. \bibinfo{doi}{\doi{10.1016/s0370-1573(02)00135-7}}.
\bibinfo{url}{\url{http://dx.doi.org/10.1016/S0370-1573(02)00135-7}}.

\bibtype{Article}%
\bibitem[{Bertolini} et al.(2016)]{bertolini}
\bibinfo{author}{{Bertolini} D}, \bibinfo{author}{{Schutz} K}, \bibinfo{author}{{Solon} MP} and  \bibinfo{author}{{Zurek} KM} (\bibinfo{year}{2016}), \bibinfo{month}{Jun.}
\bibinfo{title}{{The trispectrum in the Effective Field Theory of Large Scale Structure}}.
\bibinfo{journal}{{\em \jcap}} \bibinfo{volume}{2016} (\bibinfo{number}{6}), \bibinfo{eid}{052}. \bibinfo{doi}{\doi{10.1088/1475-7516/2016/06/052}}.
\eprint{1604.01770}.

\bibtype{Article}%
\bibitem[Beutler et al.(2017)]{beutler2017clustering}
\bibinfo{author}{Beutler F}, \bibinfo{author}{Seo HJ}, \bibinfo{author}{Saito S}, \bibinfo{author}{Chuang CH}, \bibinfo{author}{Cuesta AJ}, \bibinfo{author}{Eisenstein DJ}, \bibinfo{author}{Gil-Mar{\'\i}n H}, \bibinfo{author}{Grieb JN}, \bibinfo{author}{Hand N}, \bibinfo{author}{Kitaura FS} and  et al. (\bibinfo{year}{2017}).
\bibinfo{title}{The clustering of galaxies in the completed sdss-iii baryon oscillation spectroscopic survey: anisotropic galaxy clustering in fourier space}.
\bibinfo{journal}{{\em Monthly Notices of the Royal Astronomical Society}} \bibinfo{volume}{466} (\bibinfo{number}{2}): \bibinfo{pages}{2242--2260}.

\bibtype{Article}%
\bibitem[Biagetti et al.(2022)]{biagetti2022covariance}
\bibinfo{author}{Biagetti M}, \bibinfo{author}{Castiblanco L}, \bibinfo{author}{Nore{\~n}a J} and  \bibinfo{author}{Sefusatti E} (\bibinfo{year}{2022}).
\bibinfo{title}{The covariance of squeezed bispectrum configurations}.
\bibinfo{journal}{{\em Journal of Cosmology and Astroparticle Physics}} \bibinfo{volume}{2022} (\bibinfo{number}{09}): \bibinfo{pages}{009}.

\bibtype{Article}%
\bibitem[{Bianchi} et al.(2015)]{bianchi}
\bibinfo{author}{{Bianchi} D}, \bibinfo{author}{{Gil-Mar{\'\i}n} H}, \bibinfo{author}{{Ruggeri} R} and  \bibinfo{author}{{Percival} WJ} (\bibinfo{year}{2015}), \bibinfo{month}{Oct.}
\bibinfo{title}{{Measuring line-of-sight-dependent Fourier-space clustering using FFTs}}.
\bibinfo{journal}{{\em \mnras}} \bibinfo{volume}{453} (\bibinfo{number}{1}): \bibinfo{pages}{L11--L15}. \bibinfo{doi}{\doi{10.1093/mnrasl/slv090}}.
\eprint{1505.05341}.

\bibtype{Article}%
\bibitem[{Bianchi} et al.(2024)]{bianchi_fiber_2024}
\bibinfo{author}{{Bianchi} D}, \bibinfo{author}{{Hanif} MMS}, \bibinfo{author}{{Carnero Rosell} A}, \bibinfo{author}{{Lasker} J}, \bibinfo{author}{{Ross} AJ}, \bibinfo{author}{{Pinon} M}, \bibinfo{author}{{de Mattia} A}, \bibinfo{author}{{White} M}, \bibinfo{author}{{Ahlen} S}, \bibinfo{author}{{Bailey} S}, \bibinfo{author}{{Brooks} D}, \bibinfo{author}{{Burtin} E}, \bibinfo{author}{{Chaussidon} E}, \bibinfo{author}{{Claybaugh} T}, \bibinfo{author}{{Cole} S}, \bibinfo{author}{{de la Macorra} A}, \bibinfo{author}{{Ferraro} S}, \bibinfo{author}{{Font-Ribera} A}, \bibinfo{author}{{Forero-Romero} JE}, \bibinfo{author}{{Gazta{\~n}aga} E}, \bibinfo{author}{{Gontcho} SGA}, \bibinfo{author}{{Gutierrez} G}, \bibinfo{author}{{Guy} J}, \bibinfo{author}{{Hahn} C}, \bibinfo{author}{{Honscheid} K}, \bibinfo{author}{{Howlett} C}, \bibinfo{author}{{Juneau} S}, \bibinfo{author}{{Kirkby} D}, \bibinfo{author}{{Kisner} T}, \bibinfo{author}{{Kremin} A}, \bibinfo{author}{{Landriau} M}, \bibinfo{author}{{Le Guillou} L},
  \bibinfo{author}{{Levi} ME}, \bibinfo{author}{{McDonald} P}, \bibinfo{author}{{Meisner} A}, \bibinfo{author}{{Miquel} R}, \bibinfo{author}{{Moustakas} J}, \bibinfo{author}{{Palanque-Delabrouille} N}, \bibinfo{author}{{Percival} WJ}, \bibinfo{author}{{Prada} F}, \bibinfo{author}{{P{\'e}rez-R{\`a}fols} I}, \bibinfo{author}{{Raichoor} A}, \bibinfo{author}{{Rossi} G}, \bibinfo{author}{{Sanchez} E}, \bibinfo{author}{{Schlegel} D}, \bibinfo{author}{{Schubnell} M}, \bibinfo{author}{{Sharples} R}, \bibinfo{author}{{Silber} J}, \bibinfo{author}{{Sprayberry} D}, \bibinfo{author}{{Tarl{\'e}} G}, \bibinfo{author}{{Vargas-Maga{\~n}a} M}, \bibinfo{author}{{Weaver} BA}, \bibinfo{author}{{Zarrouk} P}, \bibinfo{author}{{Zhou} R} and  \bibinfo{author}{{Zou} H} (\bibinfo{year}{2024}), \bibinfo{month}{Nov.}
\bibinfo{title}{{Characterization of DESI fiber assignment incompleteness effect on 2-point clustering and mitigation methods for DR1 analysis}}.
\bibinfo{journal}{{\em arXiv e-prints}} , \bibinfo{eid}{arXiv:2411.12025}\bibinfo{doi}{\doi{10.48550/arXiv.2411.12025}}.
\eprint{2411.12025}.

\bibtype{Article}%
\bibitem[Blas et al.(2014)]{Blas_2014}
\bibinfo{author}{Blas D}, \bibinfo{author}{Garny M}, \bibinfo{author}{Konstandin T} and  \bibinfo{author}{Lesgourgues J} (\bibinfo{year}{2014}), \bibinfo{month}{Nov.}
\bibinfo{title}{Structure formation with massive neutrinos: going beyond linear theory}.
\bibinfo{journal}{{\em Journal of Cosmology and Astroparticle Physics}} \bibinfo{volume}{2014} (\bibinfo{number}{11}): \bibinfo{pages}{039–039}.
ISSN \bibinfo{issn}{1475-7516}. \bibinfo{doi}{\doi{10.1088/1475-7516/2014/11/039}}.
\bibinfo{url}{\url{http://dx.doi.org/10.1088/1475-7516/2014/11/039}}.

\bibtype{Article}%
\bibitem[Blas et al.(2016)]{blas2016time}
\bibinfo{author}{Blas D}, \bibinfo{author}{Garny M}, \bibinfo{author}{Ivanov MM} and  \bibinfo{author}{Sibiryakov S} (\bibinfo{year}{2016}).
\bibinfo{title}{Time-sliced perturbation theory ii: baryon acoustic oscillations and infrared resummation}.
\bibinfo{journal}{{\em Journal of Cosmology and Astroparticle Physics}} \bibinfo{volume}{2016} (\bibinfo{number}{07}): \bibinfo{pages}{028}.

\bibtype{Article}%
\bibitem[{Bonometto} and {Lucchin}(1980)]{bonometto_1}
\bibinfo{author}{{Bonometto} SA} and  \bibinfo{author}{{Lucchin} F} (\bibinfo{year}{1980}), \bibinfo{month}{Feb.}
\bibinfo{title}{{A new statistical test for galaxy clustering}}.
\bibinfo{journal}{{\em \aap}} \bibinfo{volume}{82} (\bibinfo{number}{3}): \bibinfo{pages}{287}.

\bibtype{Article}%
\bibitem[{Bonometto} and {Sharp}(1980)]{bonometto_2}
\bibinfo{author}{{Bonometto} SA} and  \bibinfo{author}{{Sharp} NA} (\bibinfo{year}{1980}), \bibinfo{month}{Dec.}
\bibinfo{title}{{On the derivation of higher order correlation functions}}.
\bibinfo{journal}{{\em \aap}} \bibinfo{volume}{92} (\bibinfo{number}{1-2}): \bibinfo{pages}{222--224}.

\bibtype{Article}%
\bibitem[{Borderia} et al.(1991)]{Borderia_1991}
\bibinfo{author}{{Borderia} MJP}, \bibinfo{author}{{Iovino} A} and  \bibinfo{author}{{Bonometto} SA} (\bibinfo{year}{1991}), \bibinfo{month}{Aug.}
\bibinfo{title}{{Search For the 3-Point Correlation Function for Quasars}}.
\bibinfo{journal}{{\em \aj}} \bibinfo{volume}{102}: \bibinfo{pages}{495}. \bibinfo{doi}{\doi{10.1086/115888}}.

\bibtype{Article}%
\bibitem[Bose and Taruya(2018)]{bose2018one}
\bibinfo{author}{Bose B} and  \bibinfo{author}{Taruya A} (\bibinfo{year}{2018}).
\bibinfo{title}{The one-loop matter bispectrum as a probe of gravity and dark energy}.
\bibinfo{journal}{{\em Journal of Cosmology and Astroparticle Physics}} \bibinfo{volume}{2018} (\bibinfo{number}{10}): \bibinfo{pages}{019}.

\bibtype{Article}%
\bibitem[{Bose} et al.(2020)]{Bose_GR_Bispec}
\bibinfo{author}{{Bose} B}, \bibinfo{author}{{Byun} J}, \bibinfo{author}{{Lacasa} F}, \bibinfo{author}{{Moradinezhad Dizgah} A} and  \bibinfo{author}{{Lombriser} L} (\bibinfo{year}{2020}), \bibinfo{month}{Feb.}
\bibinfo{title}{{Modelling the matter bispectrum at small scales in modified gravity}}.
\bibinfo{journal}{{\em \jcap}} \bibinfo{volume}{2020} (\bibinfo{number}{2}), \bibinfo{eid}{025}. \bibinfo{doi}{\doi{10.1088/1475-7516/2020/02/025}}.
\eprint{1909.02504}.

\bibtype{Article}%
\bibitem[Brieden et al.(2022)]{brieden2022model}
\bibinfo{author}{Brieden S}, \bibinfo{author}{Gil-Mar{\'\i}n H} and  \bibinfo{author}{Verde L} (\bibinfo{year}{2022}).
\bibinfo{title}{Model-agnostic interpretation of 10 billion years of cosmic evolution traced by boss and eboss data}.
\bibinfo{journal}{{\em Journal of Cosmology and Astroparticle Physics}} \bibinfo{volume}{2022} (\bibinfo{number}{08}): \bibinfo{pages}{024}.

\bibtype{Article}%
\bibitem[Brout et al.(1978)]{brout_1978}
\bibinfo{author}{Brout R}, \bibinfo{author}{Englert F} and  \bibinfo{author}{Gunzig E} (\bibinfo{year}{1978}).
\bibinfo{title}{The creation of the universe as a quantum phenomenon}.
\bibinfo{journal}{{\em Annals of Physics}} \bibinfo{volume}{115} (\bibinfo{number}{1}): \bibinfo{pages}{78--106}.
ISSN \bibinfo{issn}{0003-4916}. \bibinfo{doi}{\doi{https://doi.org/10.1016/0003-4916(78)90176-8}}.
\bibinfo{url}{\url{https://www.sciencedirect.com/science/article/pii/0003491678901768}}.

\bibtype{Article}%
\bibitem[{Brout} et al.(1979)]{brout_1979}
\bibinfo{author}{{Brout} R}, \bibinfo{author}{{Englert} F} and  \bibinfo{author}{{Gunzig} E} (\bibinfo{year}{1979}), \bibinfo{month}{Jan.}
\bibinfo{title}{{The causal universe.}}
\bibinfo{journal}{{\em General Relativity and Gravitation}} \bibinfo{volume}{10} (\bibinfo{number}{1}): \bibinfo{pages}{1--6}. \bibinfo{doi}{\doi{10.1007/BF00757018}}.

\bibtype{Article}%
\bibitem[{Brown} et al.(2022)]{conker}
\bibinfo{author}{{Brown} Z}, \bibinfo{author}{{Mishtaku} G} and  \bibinfo{author}{{Demina} R} (\bibinfo{year}{2022}), \bibinfo{month}{Nov.}
\bibinfo{title}{{ConKer: An algorithm for evaluating correlations of arbitrary order}}.
\bibinfo{journal}{{\em \aap}} \bibinfo{volume}{667}, \bibinfo{eid}{A129}. \bibinfo{doi}{\doi{10.1051/0004-6361/202141917}}.

\bibtype{Article}%
\bibitem[{Buchalter} et al.(2000)]{buch_2000}
\bibinfo{author}{{Buchalter} A}, \bibinfo{author}{{Kamionkowski} M} and  \bibinfo{author}{{Jaffe} AH} (\bibinfo{year}{2000}), \bibinfo{month}{Feb.}
\bibinfo{title}{{The Angular Three-Point Correlation Function in the Quasi-linear Regime}}.
\bibinfo{journal}{{\em \apj}} \bibinfo{volume}{530} (\bibinfo{number}{1}): \bibinfo{pages}{36--52}. \bibinfo{doi}{\doi{10.1086/308339}}.
\eprint{astro-ph/9903486}.

\bibtype{Article}%
\bibitem[{Burden} et al.(2017)]{burden}
\bibinfo{author}{{Burden} A}, \bibinfo{author}{{Padmanabhan} N}, \bibinfo{author}{{Cahn} RN}, \bibinfo{author}{{White} MJ} and  \bibinfo{author}{{Samushia} L} (\bibinfo{year}{2017}), \bibinfo{month}{Mar.}
\bibinfo{title}{{Mitigating the impact of the DESI fiber assignment on galaxy clustering}}.
\bibinfo{journal}{{\em \jcap}} \bibinfo{volume}{2017} (\bibinfo{number}{3}), \bibinfo{eid}{001}. \bibinfo{doi}{\doi{10.1088/1475-7516/2017/03/001}}.
\eprint{1611.04635}.

\bibtype{Article}%
\bibitem[{Burkhart} et al.(2009)]{burkhart}
\bibinfo{author}{{Burkhart} B}, \bibinfo{author}{{Falceta-Gon{\c{c}}alves} D}, \bibinfo{author}{{Kowal} G} and  \bibinfo{author}{{Lazarian} A} (\bibinfo{year}{2009}), \bibinfo{month}{Mar.}
\bibinfo{title}{{Density Studies of MHD Interstellar Turbulence: Statistical Moments, Correlations and Bispectrum}}.
\bibinfo{journal}{{\em \apj}} \bibinfo{volume}{693} (\bibinfo{number}{1}): \bibinfo{pages}{250--266}. \bibinfo{doi}{\doi{10.1088/0004-637X/693/1/250}}.
\eprint{0811.0822}.

\bibtype{Article}%
\bibitem[{Cabass} et al.(2023)]{cabass}
\bibinfo{author}{{Cabass} G}, \bibinfo{author}{{Ivanov} MM} and  \bibinfo{author}{{Philcox} OHE} (\bibinfo{year}{2023}), \bibinfo{month}{Jan.}
\bibinfo{title}{{Colliders and ghosts: Constraining inflation with the parity-odd galaxy four-point function}}.
\bibinfo{journal}{{\em \prd}} \bibinfo{volume}{107} (\bibinfo{number}{2}), \bibinfo{eid}{023523}. \bibinfo{doi}{\doi{10.1103/PhysRevD.107.023523}}.
\eprint{2210.16320}.

\bibtype{Article}%
\bibitem[Cabass et al.(2023)]{cabass_no_go}
\bibinfo{author}{Cabass G} and  et al. (\bibinfo{year}{2023}).
\bibinfo{title}{{Parity violation in the scalar trispectrum: no-go theorems and yes-go examples}}.
\bibinfo{journal}{{\em JHEP}} \bibinfo{volume}{02}: \bibinfo{pages}{021}. \bibinfo{doi}{\doi{10.1007/JHEP02(2023)021}}.
\eprint{2210.02907}.

\bibtype{Article}%
\bibitem[Cahn and Slepian(2023)]{cahn_iso}
\bibinfo{author}{Cahn RN} and  \bibinfo{author}{Slepian Z} (\bibinfo{year}{2023}), \bibinfo{month}{jul}.
\bibinfo{title}{Isotropic n-point basis functions and their properties}.
\bibinfo{journal}{{\em Journal of Physics A: Mathematical and Theoretical}} \bibinfo{volume}{56} (\bibinfo{number}{32}): \bibinfo{pages}{325204}. \bibinfo{doi}{\doi{10.1088/1751-8121/acdfc4}}.
\bibinfo{url}{\url{https://dx.doi.org/10.1088/1751-8121/acdfc4}}.

\bibtype{Article}%
\bibitem[{Cahn} et al.(2023)]{cahn_parity}
\bibinfo{author}{{Cahn} RN}, \bibinfo{author}{{Slepian} Z} and  \bibinfo{author}{{Hou} J} (\bibinfo{year}{2023}), \bibinfo{month}{May}.
\bibinfo{title}{{Test for Cosmological Parity Violation Using the 3D Distribution of Galaxies}}.
\bibinfo{journal}{{\em \prl}} \bibinfo{volume}{130} (\bibinfo{number}{20}), \bibinfo{eid}{201002}. \bibinfo{doi}{\doi{10.1103/PhysRevLett.130.201002}}.
\eprint{2110.12004}.

\bibtype{Article}%
\bibitem[Carrasco et al.(2012)]{carrasco2012effective}
\bibinfo{author}{Carrasco JJM}, \bibinfo{author}{Hertzberg MP} and  \bibinfo{author}{Senatore L} (\bibinfo{year}{2012}).
\bibinfo{title}{The effective field theory of cosmological large scale structures}.
\bibinfo{journal}{{\em Journal of High Energy Physics}} \bibinfo{volume}{2012} (\bibinfo{number}{9}): \bibinfo{pages}{1--40}.

\bibtype{Article}%
\bibitem[{Catelan} and {Moscardini}(1994)]{catelan_kurt_1994}
\bibinfo{author}{{Catelan} P} and  \bibinfo{author}{{Moscardini} L} (\bibinfo{year}{1994}), \bibinfo{month}{Nov.}
\bibinfo{title}{{Kurtosis as a Non-Gaussian Signature of the Large-Scale Velocity Field}}.
\bibinfo{journal}{{\em \apj}} \bibinfo{volume}{436}: \bibinfo{pages}{5}. \bibinfo{doi}{\doi{10.1086/174875}}.
\eprint{astro-ph/9403035}.

\bibtype{Article}%
\bibitem[{Chan} et al.(2012)]{chan}
\bibinfo{author}{{Chan} KC}, \bibinfo{author}{{Scoccimarro} R} and  \bibinfo{author}{{Sheth} RK} (\bibinfo{year}{2012}), \bibinfo{month}{Apr.}
\bibinfo{title}{{Gravity and large-scale nonlocal bias}}.
\bibinfo{journal}{{\em \prd}} \bibinfo{volume}{85} (\bibinfo{number}{8}), \bibinfo{eid}{083509}. \bibinfo{doi}{\doi{10.1103/PhysRevD.85.083509}}.
\eprint{1201.3614}.

\bibtype{Article}%
\bibitem[{Chen} et al.(2020)]{velocileptors_1}
\bibinfo{author}{{Chen} SF}, \bibinfo{author}{{Vlah} Z} and  \bibinfo{author}{{White} M} (\bibinfo{year}{2020}), \bibinfo{month}{Jul.}
\bibinfo{title}{{Consistent modeling of velocity statistics and redshift-space distortions in one-loop perturbation theory}}.
\bibinfo{journal}{{\em \jcap}} \bibinfo{volume}{2020} (\bibinfo{number}{7}), \bibinfo{eid}{062}. \bibinfo{doi}{\doi{10.1088/1475-7516/2020/07/062}}.
\eprint{2005.00523}.

\bibtype{Article}%
\bibitem[{Chen} et al.(2021)]{velocileptors_2}
\bibinfo{author}{{Chen} SF}, \bibinfo{author}{{Vlah} Z}, \bibinfo{author}{{Castorina} E} and  \bibinfo{author}{{White} M} (\bibinfo{year}{2021}), \bibinfo{month}{Mar.}
\bibinfo{title}{{Redshift-space distortions in Lagrangian perturbation theory}}.
\bibinfo{journal}{{\em \jcap}} \bibinfo{volume}{2021} (\bibinfo{number}{3}), \bibinfo{eid}{100}. \bibinfo{doi}{\doi{10.1088/1475-7516/2021/03/100}}.
\eprint{2012.04636}.

\bibtype{Article}%
\bibitem[{Chen} et al.(2022)]{chen_boss}
\bibinfo{author}{{Chen} SF}, \bibinfo{author}{{Vlah} Z} and  \bibinfo{author}{{White} M} (\bibinfo{year}{2022}), \bibinfo{month}{Feb.}
\bibinfo{title}{{A new analysis of galaxy 2-point functions in the BOSS survey, including full-shape information and post-reconstruction BAO}}.
\bibinfo{journal}{{\em \jcap}} \bibinfo{volume}{2022} (\bibinfo{number}{2}), \bibinfo{eid}{008}. \bibinfo{doi}{\doi{10.1088/1475-7516/2022/02/008}}.
\eprint{2110.05530}.

\bibtype{Article}%
\bibitem[Chen et al.(2022)]{chen2022new}
\bibinfo{author}{Chen SF}, \bibinfo{author}{Vlah Z} and  \bibinfo{author}{White M} (\bibinfo{year}{2022}).
\bibinfo{title}{A new analysis of galaxy 2-point functions in the boss survey, including full-shape information and post-reconstruction bao}.
\bibinfo{journal}{{\em Journal of Cosmology and Astroparticle Physics}} \bibinfo{volume}{2022} (\bibinfo{number}{02}): \bibinfo{pages}{008}.

\bibtype{Article}%
\bibitem[{Chevallier} and {Polarski}(2001)]{chev_pol}
\bibinfo{author}{{Chevallier} M} and  \bibinfo{author}{{Polarski} D} (\bibinfo{year}{2001}), \bibinfo{month}{Jan.}
\bibinfo{title}{{Accelerating Universes with Scaling Dark Matter}}.
\bibinfo{journal}{{\em International Journal of Modern Physics D}} \bibinfo{volume}{10} (\bibinfo{number}{2}): \bibinfo{pages}{213--223}. \bibinfo{doi}{\doi{10.1142/S0218271801000822}}.
\eprint{gr-qc/0009008}.

\bibtype{Article}%
\bibitem[Child et al.(2018{\natexlab{a}})]{child2}
\bibinfo{author}{Child HL} and  et al. (\bibinfo{year}{2018}{\natexlab{a}}), \bibinfo{month}{11}.
\bibinfo{title}{{A Physical Picture of Bispectrum Baryon Acoustic Oscillations in the Interferometric Basis}} \eprint{1811.12396}.

\bibtype{Article}%
\bibitem[Child et al.(2018{\natexlab{b}})]{child1}
\bibinfo{author}{Child HL} and  et al. (\bibinfo{year}{2018}{\natexlab{b}}).
\bibinfo{title}{{Bispectrum as Baryon Acoustic Oscillation Interferometer}}.
\bibinfo{journal}{{\em Phys. Rev. D}} \bibinfo{volume}{98} (\bibinfo{number}{12}): \bibinfo{pages}{123521}. \bibinfo{doi}{\doi{10.1103/PhysRevD.98.123521}}.
\eprint{1806.11147}.

\bibtype{Article}%
\bibitem[{Chodorowski} and {Bouchet}(1996)]{chodo_kurt}
\bibinfo{author}{{Chodorowski} MJ} and  \bibinfo{author}{{Bouchet} FR} (\bibinfo{year}{1996}), \bibinfo{month}{Mar.}
\bibinfo{title}{{Kurtosis in large-scale structure as a constraint on non-Gaussian initial conditions}}.
\bibinfo{journal}{{\em \mnras}} \bibinfo{volume}{279} (\bibinfo{number}{2}): \bibinfo{pages}{557--563}. \bibinfo{doi}{\doi{10.1093/mnras/279.2.557}}.
\eprint{astro-ph/9507038}.

\bibtype{Article}%
\bibitem[{Chuang} et al.(2015)]{ezmock}
\bibinfo{author}{{Chuang} CH}, \bibinfo{author}{{Kitaura} FS}, \bibinfo{author}{{Prada} F}, \bibinfo{author}{{Zhao} C} and  \bibinfo{author}{{Yepes} G} (\bibinfo{year}{2015}), \bibinfo{month}{Jan.}
\bibinfo{title}{{EZmocks: extending the Zel'dovich approximation to generate mock galaxy catalogues with accurate clustering statistics}}.
\bibinfo{journal}{{\em \mnras}} \bibinfo{volume}{446} (\bibinfo{number}{3}): \bibinfo{pages}{2621--2628}. \bibinfo{doi}{\doi{10.1093/mnras/stu2301}}.
\eprint{1409.1124}.

\bibtype{Article}%
\bibitem[Chudaykin and Ivanov(2023)]{chudaykin2023cosmological}
\bibinfo{author}{Chudaykin A} and  \bibinfo{author}{Ivanov MM} (\bibinfo{year}{2023}).
\bibinfo{title}{Cosmological constraints from the power spectrum of eboss quasars}.
\bibinfo{journal}{{\em Physical Review D}} \bibinfo{volume}{107} (\bibinfo{number}{4}): \bibinfo{pages}{043518}.

\bibtype{Article}%
\bibitem[{Chudaykin} et al.(2020)]{class_pt}
\bibinfo{author}{{Chudaykin} A}, \bibinfo{author}{{Ivanov} MM}, \bibinfo{author}{{Philcox} OHE} and  \bibinfo{author}{{Simonovi{\'c}} M} (\bibinfo{year}{2020}), \bibinfo{month}{Sep.}
\bibinfo{title}{{Nonlinear perturbation theory extension of the Boltzmann code CLASS}}.
\bibinfo{journal}{{\em \prd}} \bibinfo{volume}{102} (\bibinfo{number}{6}), \bibinfo{eid}{063533}. \bibinfo{doi}{\doi{10.1103/PhysRevD.102.063533}}.
\eprint{2004.10607}.

\bibtype{Article}%
\bibitem[Cole et al.(2005)]{cole_2005}
\bibinfo{author}{Cole S}, \bibinfo{author}{Percival WJ}, \bibinfo{author}{Peacock JA}, \bibinfo{author}{Norberg P}, \bibinfo{author}{Baugh CM}, \bibinfo{author}{Frenk CS}, \bibinfo{author}{Baldry I}, \bibinfo{author}{Bland-Hawthorn J}, \bibinfo{author}{Bridges T}, \bibinfo{author}{Cannon R}, \bibinfo{author}{Colless M}, \bibinfo{author}{Collins C}, \bibinfo{author}{Couch W}, \bibinfo{author}{Cross NJG}, \bibinfo{author}{Dalton G}, \bibinfo{author}{Eke VR}, \bibinfo{author}{de~Propris R}, \bibinfo{author}{Driver SP}, \bibinfo{author}{Efstathiou G}, \bibinfo{author}{Ellis RS}, \bibinfo{author}{Glazebrook K}, \bibinfo{author}{Jackson C}, \bibinfo{author}{Jenkins A}, \bibinfo{author}{Lahav O}, \bibinfo{author}{Lewis I}, \bibinfo{author}{Lumsden S}, \bibinfo{author}{Maddox S}, \bibinfo{author}{Madgwick D}, \bibinfo{author}{Peterson BA}, \bibinfo{author}{Sutherland W}, \bibinfo{author}{Taylor K} and  \bibinfo{author}{2dFGRS Team T} (\bibinfo{year}{2005}), \bibinfo{month}{09}.
\bibinfo{title}{The 2df galaxy redshift survey: power-spectrum analysis of the final data set and cosmological implications}.
\bibinfo{journal}{{\em Monthly Notices of the Royal Astronomical Society}} \bibinfo{volume}{362} (\bibinfo{number}{2}): \bibinfo{pages}{505--534}.
ISSN \bibinfo{issn}{0035-8711}. \bibinfo{doi}{\doi{10.1111/j.1365-2966.2005.09318.x}}.
\eprint{https://academic.oup.com/mnras/article-pdf/362/2/505/6155670/362-2-505.pdf}, \bibinfo{url}{\url{https://doi.org/10.1111/j.1365-2966.2005.09318.x}}.

\bibtype{Article}%
\bibitem[{Colless} et al.(2001)]{2dfsurvey}
\bibinfo{author}{{Colless} M}, \bibinfo{author}{{Dalton} G}, \bibinfo{author}{{Maddox} S}, \bibinfo{author}{{Sutherland} W}, \bibinfo{author}{{Norberg} P}, \bibinfo{author}{{Cole} S}, \bibinfo{author}{{Bland-Hawthorn} J}, \bibinfo{author}{{Bridges} T}, \bibinfo{author}{{Cannon} R}, \bibinfo{author}{{Collins} C}, \bibinfo{author}{{Couch} W}, \bibinfo{author}{{Cross} N}, \bibinfo{author}{{Deeley} K}, \bibinfo{author}{{De Propris} R}, \bibinfo{author}{{Driver} SP}, \bibinfo{author}{{Efstathiou} G}, \bibinfo{author}{{Ellis} RS}, \bibinfo{author}{{Frenk} CS}, \bibinfo{author}{{Glazebrook} K}, \bibinfo{author}{{Jackson} C}, \bibinfo{author}{{Lahav} O}, \bibinfo{author}{{Lewis} I}, \bibinfo{author}{{Lumsden} S}, \bibinfo{author}{{Madgwick} D}, \bibinfo{author}{{Peacock} JA}, \bibinfo{author}{{Peterson} BA}, \bibinfo{author}{{Price} I}, \bibinfo{author}{{Seaborne} M} and  \bibinfo{author}{{Taylor} K} (\bibinfo{year}{2001}), \bibinfo{month}{Dec.}
\bibinfo{title}{{The 2dF Galaxy Redshift Survey: spectra and redshifts}}.
\bibinfo{journal}{{\em \mnras}} \bibinfo{volume}{328} (\bibinfo{number}{4}): \bibinfo{pages}{1039--1063}. \bibinfo{doi}{\doi{10.1046/j.1365-8711.2001.04902.x}}.
\eprint{astro-ph/0106498}.

\bibtype{Misc}%
\bibitem[Cooley and Tukey(1965)]{cooley}
\bibinfo{author}{Cooley J} and  \bibinfo{author}{Tukey J} (\bibinfo{year}{1965}).
\bibinfo{title}{An algorithm for the machine calculation of complex fourier series, 1965}.
\eprint{https://www.ams.org/journals/mcom/1965-19-090/S0025-5718-1965-0178586-1/}, \bibinfo{url}{\url{https://www.ams.org/journals/mcom/1965-19-090/S0025-5718-1965-0178586-1/}}.

\bibtype{Article}%
\bibitem[Cooray and Sheth(2002)]{cooray2002halo}
\bibinfo{author}{Cooray A} and  \bibinfo{author}{Sheth R} (\bibinfo{year}{2002}).
\bibinfo{title}{Halo models of large scale structure}.
\bibinfo{journal}{{\em Physics reports}} \bibinfo{volume}{372} (\bibinfo{number}{1}): \bibinfo{pages}{1--129}.

\bibtype{Article}%
\bibitem[{Copeland} et al.(2006)]{copeland}
\bibinfo{author}{{Copeland} EJ}, \bibinfo{author}{{Sami} M} and  \bibinfo{author}{{Tsujikawa} S} (\bibinfo{year}{2006}), \bibinfo{month}{Jan.}
\bibinfo{title}{{Dynamics of Dark Energy}}.
\bibinfo{journal}{{\em International Journal of Modern Physics D}} \bibinfo{volume}{15} (\bibinfo{number}{11}): \bibinfo{pages}{1753--1935}. \bibinfo{doi}{\doi{10.1142/S021827180600942X}}.
\eprint{hep-th/0603057}.

\bibtype{Article}%
\bibitem[{Coulton} et al.(2024)]{coulton}
\bibinfo{author}{{Coulton} WR}, \bibinfo{author}{{Philcox} OHE} and  \bibinfo{author}{{Villaescusa-Navarro} F} (\bibinfo{year}{2024}), \bibinfo{month}{Jan.}
\bibinfo{title}{{Signatures of a parity-violating universe}}.
\bibinfo{journal}{{\em \prd}} \bibinfo{volume}{109} (\bibinfo{number}{2}), \bibinfo{eid}{023531}. \bibinfo{doi}{\doi{10.1103/PhysRevD.109.023531}}.
\eprint{2306.11782}.

\bibtype{Inproceedings}%
\bibitem[{Crill} et al.(2020)]{spherex}
\bibinfo{author}{{Crill} BP}, \bibinfo{author}{{Werner} M}, \bibinfo{author}{{Akeson} R}, \bibinfo{author}{{Ashby} M}, \bibinfo{author}{{Bleem} L}, \bibinfo{author}{{Bock} JJ}, \bibinfo{author}{{Bryan} S}, \bibinfo{author}{{Burnham} J}, \bibinfo{author}{{Byunh} J}, \bibinfo{author}{{Chang} TC}, \bibinfo{author}{{Chiang} YK}, \bibinfo{author}{{Cook} W}, \bibinfo{author}{{Cooray} A}, \bibinfo{author}{{Davis} A}, \bibinfo{author}{{Dor{\'e}} O}, \bibinfo{author}{{Dowell} CD}, \bibinfo{author}{{Dubois-Felsmann} G}, \bibinfo{author}{{Eifler} T}, \bibinfo{author}{{Faisst} A}, \bibinfo{author}{{Habib} S}, \bibinfo{author}{{Heinrich} C}, \bibinfo{author}{{Heitmann} K}, \bibinfo{author}{{Heaton} G}, \bibinfo{author}{{Hirata} C}, \bibinfo{author}{{Hristov} V}, \bibinfo{author}{{Hui} H}, \bibinfo{author}{{Jeong} WS}, \bibinfo{author}{{Kang} JH}, \bibinfo{author}{{Kecman} B}, \bibinfo{author}{{Kirkpatrick} JD}, \bibinfo{author}{{Korngut} PM}, \bibinfo{author}{{Krause} E}, \bibinfo{author}{{Lee} B},
  \bibinfo{author}{{Lisse} C}, \bibinfo{author}{{Masters} D}, \bibinfo{author}{{Mauskopf} P}, \bibinfo{author}{{Melnick} G}, \bibinfo{author}{{Miyasaka} H}, \bibinfo{author}{{Nayyeri} H}, \bibinfo{author}{{Nguyen} H}, \bibinfo{author}{{{\"O}berg} K}, \bibinfo{author}{{Padin} S}, \bibinfo{author}{{Paladini} R}, \bibinfo{author}{{Pourrahmani} M}, \bibinfo{author}{{Pyo} J}, \bibinfo{author}{{Smith} R}, \bibinfo{author}{{Song} YS}, \bibinfo{author}{{Symons} T}, \bibinfo{author}{{Teplitz} H}, \bibinfo{author}{{Tolls} V}, \bibinfo{author}{{Unwin} S}, \bibinfo{author}{{Windhorst} R}, \bibinfo{author}{{Yang} Y} and  \bibinfo{author}{{Zemcov} M} (\bibinfo{year}{2020}), \bibinfo{month}{Dec.}, \bibinfo{title}{{SPHEREx: NASA's near-infrared spectrophotometric all-sky survey}}, \bibinfo{editor}{{Lystrup} M} and  \bibinfo{editor}{{Perrin} MD}, (Eds.), \bibinfo{booktitle}{Space Telescopes and Instrumentation 2020: Optical, Infrared, and Millimeter Wave}, \bibinfo{series}{Society of Photo-Optical Instrumentation Engineers
  (SPIE) Conference Series}, \bibinfo{volume}{11443}, pp. \bibinfo{pages}{114430I}, \eprint{2404.11017}.

\bibtype{Article}%
\bibitem[{D'Amico} et al.(2021)]{py_bird}
\bibinfo{author}{{D'Amico} G}, \bibinfo{author}{{Senatore} L} and  \bibinfo{author}{{Zhang} P} (\bibinfo{year}{2021}), \bibinfo{month}{Jan.}
\bibinfo{title}{{Limits on wCDM from the EFTofLSS with the PyBird code}}.
\bibinfo{journal}{{\em \jcap}} \bibinfo{volume}{2021} (\bibinfo{number}{1}), \bibinfo{eid}{006}. \bibinfo{doi}{\doi{10.1088/1475-7516/2021/01/006}}.
\eprint{2003.07956}.

\bibtype{Article}%
\bibitem[D'Amico et al.(2022)]{d2022boss}
\bibinfo{author}{D'Amico G}, \bibinfo{author}{Donath Y}, \bibinfo{author}{Lewandowski M}, \bibinfo{author}{Senatore L} and  \bibinfo{author}{Zhang P} (\bibinfo{year}{2022}).
\bibinfo{title}{The boss bispectrum analysis at one loop from the effective field theory of large-scale structure}.
\bibinfo{journal}{{\em arXiv preprint arXiv:2206.08327}} .

\bibtype{Article}%
\bibitem[{Davidson} et al.(2008)]{lepto}
\bibinfo{author}{{Davidson} S}, \bibinfo{author}{{Nardi} E} and  \bibinfo{author}{{Nir} Y} (\bibinfo{year}{2008}), \bibinfo{month}{Sep.}
\bibinfo{title}{{Leptogenesis}}.
\bibinfo{journal}{{\em \physrep}} \bibinfo{volume}{466} (\bibinfo{number}{4-5}): \bibinfo{pages}{105--177}. \bibinfo{doi}{\doi{10.1016/j.physrep.2008.06.002}}.
\eprint{0802.2962}.

\bibtype{Article}%
\bibitem[Dawson et al.(2012)]{dawson2012baryon}
\bibinfo{author}{Dawson KS}, \bibinfo{author}{Schlegel DJ}, \bibinfo{author}{Ahn CP}, \bibinfo{author}{Anderson SF}, \bibinfo{author}{Aubourg {\'E}}, \bibinfo{author}{Bailey S}, \bibinfo{author}{Barkhouser RH}, \bibinfo{author}{Bautista JE}, \bibinfo{author}{Beifiori A}, \bibinfo{author}{Berlind AA} and  et al. (\bibinfo{year}{2012}).
\bibinfo{title}{The baryon oscillation spectroscopic survey of sdss-iii}.
\bibinfo{journal}{{\em The Astronomical Journal}} \bibinfo{volume}{145} (\bibinfo{number}{1}): \bibinfo{pages}{10}.

\bibtype{Article}%
\bibitem[Dawson et al.(2016)]{dawson2016sdss}
\bibinfo{author}{Dawson KS}, \bibinfo{author}{Kneib JP}, \bibinfo{author}{Percival WJ}, \bibinfo{author}{Alam S}, \bibinfo{author}{Albareti FD}, \bibinfo{author}{Anderson SF}, \bibinfo{author}{Armengaud E}, \bibinfo{author}{Aubourg {\'E}}, \bibinfo{author}{Bailey S}, \bibinfo{author}{Bautista JE} and  et al. (\bibinfo{year}{2016}).
\bibinfo{title}{The sdss-iv extended baryon oscillation spectroscopic survey: overview and early data}.
\bibinfo{journal}{{\em The Astronomical Journal}} \bibinfo{volume}{151} (\bibinfo{number}{2}): \bibinfo{pages}{44}.

\bibtype{Article}%
\bibitem[{DESI Collaboration} et al.(2016)]{DESI:2016}
\bibinfo{author}{{DESI Collaboration}}, \bibinfo{author}{{Aghamousa} A}, \bibinfo{author}{{Aguilar} J}, \bibinfo{author}{{Ahlen} S}, \bibinfo{author}{{Alam} S}, \bibinfo{author}{{Allen} LE}, \bibinfo{author}{{Allende Prieto} C}, \bibinfo{author}{{Annis} J}, \bibinfo{author}{{Bailey} S}, \bibinfo{author}{{Balland} C} and  \bibinfo{author}{et~al.} (\bibinfo{year}{2016}), \bibinfo{month}{Oct.}
\bibinfo{title}{{The DESI Experiment Part I: Science,Targeting, and Survey Design}}.
\bibinfo{journal}{{\em ArXiv e-prints}} \eprint{1611.00036}.

\bibtype{Article}%
\bibitem[{DESI Collaboration} et al.(2024)]{desi_y1_2ps}
\bibinfo{author}{{DESI Collaboration}}, \bibinfo{author}{{Adame} AG}, \bibinfo{author}{{Aguilar} J}, \bibinfo{author}{{Ahlen} S}, \bibinfo{author}{{Alam} S}, \bibinfo{author}{{Alexander} DM}, \bibinfo{author}{{Alvarez} M}, \bibinfo{author}{{Alves} O}, \bibinfo{author}{{Anand} A}, \bibinfo{author}{{Andrade} U}, \bibinfo{author}{{Armengaud} E}, \bibinfo{author}{{Avila} S}, \bibinfo{author}{{Aviles} A}, \bibinfo{author}{{Awan} H}, \bibinfo{author}{{Bailey} S}, \bibinfo{author}{{Baltay} C}, \bibinfo{author}{{Bault} A}, \bibinfo{author}{{Behera} J}, \bibinfo{author}{{BenZvi} S}, \bibinfo{author}{{Beutler} F}, \bibinfo{author}{{Bianchi} D}, \bibinfo{author}{{Blake} C}, \bibinfo{author}{{Blum} R}, \bibinfo{author}{{Brieden} S}, \bibinfo{author}{{Brodzeller} A}, \bibinfo{author}{{Brooks} D}, \bibinfo{author}{{Brown} Z}, \bibinfo{author}{{Buckley-Geer} E}, \bibinfo{author}{{Burtin} E}, \bibinfo{author}{{Calderon} R}, \bibinfo{author}{{Canning} R}, \bibinfo{author}{{Carnero Rosell} A}, \bibinfo{author}{{Cereskaite} R},
  \bibinfo{author}{{Cervantes-Cota} JL}, \bibinfo{author}{{Chabanier} S}, \bibinfo{author}{{Chaussidon} E}, \bibinfo{author}{{Chaves-Montero} J}, \bibinfo{author}{{Chen} S}, \bibinfo{author}{{Chen} X}, \bibinfo{author}{{Claybaugh} T}, \bibinfo{author}{{Cole} S}, \bibinfo{author}{{Cuceu} A}, \bibinfo{author}{{Davis} TM}, \bibinfo{author}{{Dawson} K}, \bibinfo{author}{{de la Macorra} A}, \bibinfo{author}{{de Mattia} A}, \bibinfo{author}{{Deiosso} N}, \bibinfo{author}{{Demina} R}, \bibinfo{author}{{Dey} A}, \bibinfo{author}{{Dey} B}, \bibinfo{author}{{Ding} Z}, \bibinfo{author}{{Doel} P}, \bibinfo{author}{{Edelstein} J}, \bibinfo{author}{{Eftekharzadeh} S}, \bibinfo{author}{{Eisenstein} DJ}, \bibinfo{author}{{Elliott} A}, \bibinfo{author}{{Fagrelius} P}, \bibinfo{author}{{Fanning} K}, \bibinfo{author}{{Ferraro} S}, \bibinfo{author}{{Ereza} J}, \bibinfo{author}{{Findlay} N}, \bibinfo{author}{{Flaugher} B}, \bibinfo{author}{{Font-Ribera} A}, \bibinfo{author}{{Forero-S{\'a}nchez} D},
  \bibinfo{author}{{Forero-Romero} JE}, \bibinfo{author}{{Frenk} CS}, \bibinfo{author}{{Garcia-Quintero} C}, \bibinfo{author}{{Gazta{\~n}aga} E}, \bibinfo{author}{{Gil-Mar{\'\i}n} H}, \bibinfo{author}{{Gontcho} SGA}, \bibinfo{author}{{Gonzalez-Morales} AX}, \bibinfo{author}{{Gonzalez-Perez} V}, \bibinfo{author}{{Gordon} C}, \bibinfo{author}{{Green} D}, \bibinfo{author}{{Gruen} D}, \bibinfo{author}{{Gsponer} R}, \bibinfo{author}{{Gutierrez} G}, \bibinfo{author}{{Guy} J}, \bibinfo{author}{{Hadzhiyska} B}, \bibinfo{author}{{Hahn} C}, \bibinfo{author}{{Hanif} MMS}, \bibinfo{author}{{Herrera-Alcantar} HK}, \bibinfo{author}{{Honscheid} K}, \bibinfo{author}{{Hou} J}, \bibinfo{author}{{Howlett} C}, \bibinfo{author}{{Huterer} D}, \bibinfo{author}{{Ir{\v{s}}i{\v{c}}} V}, \bibinfo{author}{{Ishak} M}, \bibinfo{author}{{Juneau} S}, \bibinfo{author}{{Kara{\c{c}}ayl{\i}} NG}, \bibinfo{author}{{Kehoe} R}, \bibinfo{author}{{Kent} S}, \bibinfo{author}{{Kirkby} D}, \bibinfo{author}{{Kitaura} FS}, \bibinfo{author}{{Kong} H},
  \bibinfo{author}{{Kremin} A}, \bibinfo{author}{{Krolewski} A}, \bibinfo{author}{{Lai} Y}, \bibinfo{author}{{Lan} TW}, \bibinfo{author}{{Landriau} M}, \bibinfo{author}{{Lang} D}, \bibinfo{author}{{Lasker} J}, \bibinfo{author}{{Le Goff} JM}, \bibinfo{author}{{Le Guillou} L}, \bibinfo{author}{{Leauthaud} A}, \bibinfo{author}{{Levi} ME}, \bibinfo{author}{{Li} TS}, \bibinfo{author}{{Lodha} K}, \bibinfo{author}{{Magneville} C}, \bibinfo{author}{{Manera} M}, \bibinfo{author}{{Margala} D}, \bibinfo{author}{{Martini} P}, \bibinfo{author}{{Maus} M}, \bibinfo{author}{{McDonald} P}, \bibinfo{author}{{Medina-Varela} L}, \bibinfo{author}{{Meisner} A}, \bibinfo{author}{{Mena-Fern{\'a}ndez} J}, \bibinfo{author}{{Miquel} R}, \bibinfo{author}{{Moon} J}, \bibinfo{author}{{Moore} S}, \bibinfo{author}{{Moustakas} J}, \bibinfo{author}{{Mudur} N}, \bibinfo{author}{{Mueller} E}, \bibinfo{author}{{Mu{\~n}oz-Guti{\'e}rrez} A}, \bibinfo{author}{{Myers} AD}, \bibinfo{author}{{Nadathur} S}, \bibinfo{author}{{Napolitano} L},
  \bibinfo{author}{{Neveux} R}, \bibinfo{author}{{Newman} JA}, \bibinfo{author}{{Nguyen} NM}, \bibinfo{author}{{Nie} J}, \bibinfo{author}{{Niz} G}, \bibinfo{author}{{Noriega} HE}, \bibinfo{author}{{Padmanabhan} N}, \bibinfo{author}{{Paillas} E}, \bibinfo{author}{{Palanque-Delabrouille} N}, \bibinfo{author}{{Pan} J}, \bibinfo{author}{{Penmetsa} S}, \bibinfo{author}{{Percival} WJ}, \bibinfo{author}{{Pieri} MM}, \bibinfo{author}{{Pinon} M}, \bibinfo{author}{{Poppett} C}, \bibinfo{author}{{Porredon} A}, \bibinfo{author}{{Prada} F}, \bibinfo{author}{{P{\'e}rez-Fern{\'a}ndez} A}, \bibinfo{author}{{P{\'e}rez-R{\`a}fols} I}, \bibinfo{author}{{Rabinowitz} D}, \bibinfo{author}{{Raichoor} A}, \bibinfo{author}{{Ram{\'\i}rez-P{\'e}rez} C}, \bibinfo{author}{{Ramirez-Solano} S}, \bibinfo{author}{{Rashkovetskyi} M}, \bibinfo{author}{{Ravoux} C}, \bibinfo{author}{{Rezaie} M}, \bibinfo{author}{{Rich} J}, \bibinfo{author}{{Rocher} A}, \bibinfo{author}{{Rockosi} C}, \bibinfo{author}{{Roe} NA}, \bibinfo{author}{{Rosado-Marin} A},
  \bibinfo{author}{{Ross} AJ}, \bibinfo{author}{{Rossi} G}, \bibinfo{author}{{Ruggeri} R}, \bibinfo{author}{{Ruhlmann-Kleider} V}, \bibinfo{author}{{Samushia} L}, \bibinfo{author}{{Sanchez} E}, \bibinfo{author}{{Saulder} C}, \bibinfo{author}{{Schlafly} EF}, \bibinfo{author}{{Schlegel} D}, \bibinfo{author}{{Scholte} D}, \bibinfo{author}{{Schubnell} M}, \bibinfo{author}{{Seo} H}, \bibinfo{author}{{Sharples} R}, \bibinfo{author}{{Silber} J}, \bibinfo{author}{{Slosar} A}, \bibinfo{author}{{Smith} A}, \bibinfo{author}{{Sprayberry} D}, \bibinfo{author}{{Tan} T}, \bibinfo{author}{{Tarl{\'e}} G}, \bibinfo{author}{{Trusov} S}, \bibinfo{author}{{Vaisakh} R}, \bibinfo{author}{{Valcin} D}, \bibinfo{author}{{Valdes} F}, \bibinfo{author}{{Vargas-Maga{\~n}a} M}, \bibinfo{author}{{Verde} L}, \bibinfo{author}{{Walther} M}, \bibinfo{author}{{Wang} B}, \bibinfo{author}{{Wang} MS}, \bibinfo{author}{{Weaver} BA}, \bibinfo{author}{{Weaverdyck} N}, \bibinfo{author}{{Wechsler} RH}, \bibinfo{author}{{Weinberg} DH},
  \bibinfo{author}{{White} M}, \bibinfo{author}{{Wilson} MJ}, \bibinfo{author}{{Yu} J}, \bibinfo{author}{{Yu} Y}, \bibinfo{author}{{Yuan} S}, \bibinfo{author}{{Y{\`e}che} C}, \bibinfo{author}{{Zaborowski} EA}, \bibinfo{author}{{Zarrouk} P}, \bibinfo{author}{{Zhang} H} and  \bibinfo{author}{{Zhao} C} (\bibinfo{year}{2024}), \bibinfo{month}{Nov.}
\bibinfo{title}{{DESI 2024 II: Sample Definitions, Characteristics, and Two-point Clustering Statistics}}.
\bibinfo{journal}{{\em arXiv e-prints}} , \bibinfo{eid}{arXiv:2411.12020}\bibinfo{doi}{\doi{10.48550/arXiv.2411.12020}}.
\eprint{2411.12020}.

\bibtype{Article}%
\bibitem[{Desjacques} et al.(2018)]{desjacques}
\bibinfo{author}{{Desjacques} V}, \bibinfo{author}{{Jeong} D} and  \bibinfo{author}{{Schmidt} F} (\bibinfo{year}{2018}), \bibinfo{month}{Feb.}
\bibinfo{title}{{Large-scale galaxy bias}}.
\bibinfo{journal}{{\em \physrep}} \bibinfo{volume}{733}: \bibinfo{pages}{1--193}. \bibinfo{doi}{\doi{10.1016/j.physrep.2017.12.002}}.
\eprint{1611.09787}.

\bibtype{Article}%
\bibitem[Desjacques et al.(2018)]{Desjacques_2018}
\bibinfo{author}{Desjacques V}, \bibinfo{author}{Jeong D} and  \bibinfo{author}{Schmidt F} (\bibinfo{year}{2018}), \bibinfo{month}{Feb.}
\bibinfo{title}{Large-scale galaxy bias}.
\bibinfo{journal}{{\em Physics Reports}} \bibinfo{volume}{733}: \bibinfo{pages}{1–193}.
ISSN \bibinfo{issn}{0370-1573}. \bibinfo{doi}{\doi{10.1016/j.physrep.2017.12.002}}.
\bibinfo{url}{\url{http://dx.doi.org/10.1016/j.physrep.2017.12.002}}.

\bibtype{Article}%
\bibitem[Dizgah et al.(2021)]{dizgah2021primordial}
\bibinfo{author}{Dizgah AM}, \bibinfo{author}{Biagetti M}, \bibinfo{author}{Sefusatti E}, \bibinfo{author}{Desjacques V} and  \bibinfo{author}{Nore{\~n}a J} (\bibinfo{year}{2021}).
\bibinfo{title}{Primordial non-gaussianity from biased tracers: likelihood analysis of real-space power spectrum and bispectrum}.
\bibinfo{journal}{{\em Journal of Cosmology and Astroparticle Physics}} \bibinfo{volume}{2021} (\bibinfo{number}{05}): \bibinfo{pages}{015}.

\bibtype{Article}%
\bibitem[Donald-McCann et al.(2023)]{donald2023analysis}
\bibinfo{author}{Donald-McCann J}, \bibinfo{author}{Gsponer R}, \bibinfo{author}{Zhao R}, \bibinfo{author}{Koyama K} and  \bibinfo{author}{Beutler F} (\bibinfo{year}{2023}).
\bibinfo{title}{Analysis of unified galaxy power spectrum multipole measurements}.
\bibinfo{journal}{{\em Monthly Notices of the Royal Astronomical Society}} \bibinfo{volume}{526} (\bibinfo{number}{3}): \bibinfo{pages}{3461--3481}.

\bibtype{Article}%
\bibitem[Dvali et al.(2000)]{dvali20004d}
\bibinfo{author}{Dvali G}, \bibinfo{author}{Gabadadze G} and  \bibinfo{author}{Porrati M} (\bibinfo{year}{2000}).
\bibinfo{title}{4d gravity on a brane in 5d minkowski space}.
\bibinfo{journal}{{\em Physics Letters B}} \bibinfo{volume}{485} (\bibinfo{number}{1-3}): \bibinfo{pages}{208--214}.

\bibtype{Article}%
\bibitem[Eggemeier et al.(2020)]{eggemeier2020testing}
\bibinfo{author}{Eggemeier A}, \bibinfo{author}{Scoccimarro R}, \bibinfo{author}{Crocce M}, \bibinfo{author}{Pezzotta A} and  \bibinfo{author}{S{\'a}nchez AG} (\bibinfo{year}{2020}).
\bibinfo{title}{Testing one-loop galaxy bias: Power spectrum}.
\bibinfo{journal}{{\em Physical Review D}} \bibinfo{volume}{102} (\bibinfo{number}{10}): \bibinfo{pages}{103530}.

\bibtype{Article}%
\bibitem[Eisenstein and Hu(1998)]{eisenstein1998baryonic}
\bibinfo{author}{Eisenstein DJ} and  \bibinfo{author}{Hu W} (\bibinfo{year}{1998}).
\bibinfo{title}{Baryonic features in the matter transfer function}.
\bibinfo{journal}{{\em The Astrophysical Journal}} \bibinfo{volume}{496} (\bibinfo{number}{2}): \bibinfo{pages}{605}.

\bibtype{Article}%
\bibitem[{Eisenstein} et al.(2005)]{Eisenstein_05}
\bibinfo{author}{{Eisenstein} DJ}, \bibinfo{author}{{Zehavi} I}, \bibinfo{author}{{Hogg} DW}, \bibinfo{author}{{Scoccimarro} R}, \bibinfo{author}{{Blanton} MR}, \bibinfo{author}{{Nichol} RC}, \bibinfo{author}{{Scranton} R}, \bibinfo{author}{{Seo} HJ}, \bibinfo{author}{{Tegmark} M}, \bibinfo{author}{{Zheng} Z}, \bibinfo{author}{{Anderson} SF}, \bibinfo{author}{{Annis} J}, \bibinfo{author}{{Bahcall} N}, \bibinfo{author}{{Brinkmann} J}, \bibinfo{author}{{Burles} S}, \bibinfo{author}{{Castander} FJ}, \bibinfo{author}{{Connolly} A}, \bibinfo{author}{{Csabai} I}, \bibinfo{author}{{Doi} M}, \bibinfo{author}{{Fukugita} M}, \bibinfo{author}{{Frieman} JA}, \bibinfo{author}{{Glazebrook} K}, \bibinfo{author}{{Gunn} JE}, \bibinfo{author}{{Hendry} JS}, \bibinfo{author}{{Hennessy} G}, \bibinfo{author}{{Ivezi{\'c}} Z}, \bibinfo{author}{{Kent} S}, \bibinfo{author}{{Knapp} GR}, \bibinfo{author}{{Lin} H}, \bibinfo{author}{{Loh} YS}, \bibinfo{author}{{Lupton} RH}, \bibinfo{author}{{Margon} B}, \bibinfo{author}{{McKay} TA},
  \bibinfo{author}{{Meiksin} A}, \bibinfo{author}{{Munn} JA}, \bibinfo{author}{{Pope} A}, \bibinfo{author}{{Richmond} MW}, \bibinfo{author}{{Schlegel} D}, \bibinfo{author}{{Schneider} DP}, \bibinfo{author}{{Shimasaku} K}, \bibinfo{author}{{Stoughton} C}, \bibinfo{author}{{Strauss} MA}, \bibinfo{author}{{SubbaRao} M}, \bibinfo{author}{{Szalay} AS}, \bibinfo{author}{{Szapudi} I}, \bibinfo{author}{{Tucker} DL}, \bibinfo{author}{{Yanny} B} and  \bibinfo{author}{{York} DG} (\bibinfo{year}{2005}), \bibinfo{month}{Nov.}
\bibinfo{title}{{Detection of the Baryon Acoustic Peak in the Large-Scale Correlation Function of SDSS Luminous Red Galaxies}}.
\bibinfo{journal}{{\em \apj}} \bibinfo{volume}{633} (\bibinfo{number}{2}): \bibinfo{pages}{560--574}. \bibinfo{doi}{\doi{10.1086/466512}}.
\eprint{astro-ph/0501171}.

\bibtype{Article}%
\bibitem[{Eisenstein} et al.(2007{\natexlab{a}})]{eis_recon}
\bibinfo{author}{{Eisenstein} DJ}, \bibinfo{author}{{Seo} HJ}, \bibinfo{author}{{Sirko} E} and  \bibinfo{author}{{Spergel} DN} (\bibinfo{year}{2007}{\natexlab{a}}), \bibinfo{month}{Aug.}
\bibinfo{title}{{Improving Cosmological Distance Measurements by Reconstruction of the Baryon Acoustic Peak}}.
\bibinfo{journal}{{\em \apj}} \bibinfo{volume}{664} (\bibinfo{number}{2}): \bibinfo{pages}{675--679}. \bibinfo{doi}{\doi{10.1086/518712}}.
\eprint{astro-ph/0604362}.

\bibtype{Article}%
\bibitem[{Eisenstein} et al.(2007{\natexlab{b}})]{esw_07}
\bibinfo{author}{{Eisenstein} DJ}, \bibinfo{author}{{Seo} HJ} and  \bibinfo{author}{{White} M} (\bibinfo{year}{2007}{\natexlab{b}}), \bibinfo{month}{Aug.}
\bibinfo{title}{{On the Robustness of the Acoustic Scale in the Low-Redshift Clustering of Matter}}.
\bibinfo{journal}{{\em \apj}} \bibinfo{volume}{664} (\bibinfo{number}{2}): \bibinfo{pages}{660--674}. \bibinfo{doi}{\doi{10.1086/518755}}.
\eprint{astro-ph/0604361}.

\bibtype{Article}%
\bibitem[Eisenstein et al.(2011)]{eisenstein2011sdss}
\bibinfo{author}{Eisenstein DJ}, \bibinfo{author}{Weinberg DH}, \bibinfo{author}{Agol E}, \bibinfo{author}{Aihara H}, \bibinfo{author}{Prieto CA}, \bibinfo{author}{Anderson SF}, \bibinfo{author}{Arns JA}, \bibinfo{author}{Aubourg {\'E}}, \bibinfo{author}{Bailey S}, \bibinfo{author}{Balbinot E} and  et al. (\bibinfo{year}{2011}).
\bibinfo{title}{Sdss-iii: Massive spectroscopic surveys of the distant universe, the milky way, and extra-solar planetary systems}.
\bibinfo{journal}{{\em The Astronomical Journal}} \bibinfo{volume}{142} (\bibinfo{number}{3}): \bibinfo{pages}{72}.

\bibtype{Article}%
\bibitem[Emiliano et al.(2006)]{emiliano2006cosmology}
\bibinfo{author}{Emiliano S}, \bibinfo{author}{Martin C} and  \bibinfo{author}{Sebastian P} (\bibinfo{year}{2006}).
\bibinfo{title}{Cosmology and the bispectrum}.
\bibinfo{journal}{{\em Phys. Rev. D}} \bibinfo{volume}{74}: \bibinfo{pages}{023522}.

\bibtype{Article}%
\bibitem[{Fall} and {Saslaw}(1976)]{fall_1976}
\bibinfo{author}{{Fall} SM} and  \bibinfo{author}{{Saslaw} WC} (\bibinfo{year}{1976}), \bibinfo{month}{Mar.}
\bibinfo{title}{{The growth of correlations in an expanding universe and the clustering of galaxies.}}
\bibinfo{journal}{{\em \apj}} \bibinfo{volume}{204}: \bibinfo{pages}{631--641}. \bibinfo{doi}{\doi{10.1086/154212}}.

\bibtype{Article}%
\bibitem[{Feldman} et al.(1994)]{fkp_1994}
\bibinfo{author}{{Feldman} HA}, \bibinfo{author}{{Kaiser} N} and  \bibinfo{author}{{Peacock} JA} (\bibinfo{year}{1994}), \bibinfo{month}{May}.
\bibinfo{title}{{Power-Spectrum Analysis of Three-dimensional Redshift Surveys}}.
\bibinfo{journal}{{\em \apj}} \bibinfo{volume}{426}: \bibinfo{pages}{23}. \bibinfo{doi}{\doi{10.1086/174036}}.
\eprint{astro-ph/9304022}.

\bibtype{Article}%
\bibitem[Feldman et al.(2001)]{feldman2001constraints}
\bibinfo{author}{Feldman HA}, \bibinfo{author}{Frieman JA}, \bibinfo{author}{Fry JN} and  \bibinfo{author}{Scoccimarro R} (\bibinfo{year}{2001}).
\bibinfo{title}{Constraints on galaxy bias, matter density, and primordial non-gaussianity from the psc z galaxy redshift survey}.
\bibinfo{journal}{{\em Physical Review Letters}} \bibinfo{volume}{86} (\bibinfo{number}{8}): \bibinfo{pages}{1434}.

\bibtype{Article}%
\bibitem[{Fergusson} et al.(2012{\natexlab{a}})]{fergusson_separable}
\bibinfo{author}{{Fergusson} JR}, \bibinfo{author}{{Regan} DM} and  \bibinfo{author}{{Shellard} EPS} (\bibinfo{year}{2012}{\natexlab{a}}), \bibinfo{month}{Sep.}
\bibinfo{title}{{Rapid separable analysis of higher order correlators in large-scale structure}}.
\bibinfo{journal}{{\em \prd}} \bibinfo{volume}{86} (\bibinfo{number}{6}), \bibinfo{eid}{063511}. \bibinfo{doi}{\doi{10.1103/PhysRevD.86.063511}}.
\eprint{1008.1730}.

\bibtype{Article}%
\bibitem[{Fergusson} et al.(2012{\natexlab{b}})]{regan_bispectrum_real}
\bibinfo{author}{{Fergusson} JR}, \bibinfo{author}{{Regan} DM} and  \bibinfo{author}{{Shellard} EPS} (\bibinfo{year}{2012}{\natexlab{b}}), \bibinfo{month}{Sep.}
\bibinfo{title}{{Rapid separable analysis of higher order correlators in large-scale structure}}.
\bibinfo{journal}{{\em \prd}} \bibinfo{volume}{86} (\bibinfo{number}{6}), \bibinfo{eid}{063511}. \bibinfo{doi}{\doi{10.1103/PhysRevD.86.063511}}.
\eprint{1008.1730}.

\bibtype{Article}%
\bibitem[{Fisher} et al.(1995)]{fisher}
\bibinfo{author}{{Fisher} KB}, \bibinfo{author}{{Lahav} O}, \bibinfo{author}{{Hoffman} Y}, \bibinfo{author}{{Lynden-Bell} D} and  \bibinfo{author}{{Zaroubi} S} (\bibinfo{year}{1995}), \bibinfo{month}{Feb.}
\bibinfo{title}{{Wiener reconstruction of density, velocity and potential fields from all-sky galaxy redshift surveys}}.
\bibinfo{journal}{{\em \mnras}} \bibinfo{volume}{272} (\bibinfo{number}{4}): \bibinfo{pages}{885--908}. \bibinfo{doi}{\doi{10.1093/mnras/272.4.885}}.
\eprint{astro-ph/9406009}.

\bibtype{Article}%
\bibitem[{Fixsen} et al.(1996)]{cobe}
\bibinfo{author}{{Fixsen} DJ}, \bibinfo{author}{{Cheng} ES}, \bibinfo{author}{{Gales} JM}, \bibinfo{author}{{Mather} JC}, \bibinfo{author}{{Shafer} RA} and  \bibinfo{author}{{Wright} EL} (\bibinfo{year}{1996}), \bibinfo{month}{Dec.}
\bibinfo{title}{{The Cosmic Microwave Background Spectrum from the Full COBE FIRAS Data Set}}.
\bibinfo{journal}{{\em \apj}} \bibinfo{volume}{473}: \bibinfo{pages}{576}. \bibinfo{doi}{\doi{10.1086/178173}}.
\eprint{astro-ph/9605054}.

\bibtype{Inproceedings}%
\bibitem[Friesen et al.(2017)]{friesen}
\bibinfo{author}{Friesen B}, \bibinfo{author}{Patwary MMA}, \bibinfo{author}{Austin B}, \bibinfo{author}{Satish N}, \bibinfo{author}{Slepian Z}, \bibinfo{author}{Sundaram N}, \bibinfo{author}{Bard D}, \bibinfo{author}{Eisenstein DJ}, \bibinfo{author}{Deslippe J}, \bibinfo{author}{Dubey P} and  \bibinfo{author}{Prabhat} (\bibinfo{year}{2017}), \bibinfo{title}{Galactos: computing the anisotropic 3-point correlation function for 2 billion galaxies}, \bibinfo{booktitle}{Proceedings of the International Conference for High Performance Computing, Networking, Storage and Analysis}, \bibinfo{series}{SC '17}, \bibinfo{publisher}{Association for Computing Machinery}, \bibinfo{address}{New York, NY, USA}, \bibinfo{url}{\url{https://doi.org/10.1145/3126908.3126927}}.

\bibtype{Article}%
\bibitem[Frigo and Johnson(2005)]{frigo}
\bibinfo{author}{Frigo M} and  \bibinfo{author}{Johnson S} (\bibinfo{year}{2005}).
\bibinfo{title}{The design and implementation of fftw3}.
\bibinfo{journal}{{\em Proceedings of the IEEE}} \bibinfo{volume}{93} (\bibinfo{number}{2}): \bibinfo{pages}{216--231}. \bibinfo{doi}{\doi{10.1109/JPROC.2004.840301}}.

\bibtype{Article}%
\bibitem[{Fry}(1982)]{fry_bbgky}
\bibinfo{author}{{Fry} JN} (\bibinfo{year}{1982}), \bibinfo{month}{Nov.}
\bibinfo{title}{{The four-point function in the BBGKY hierarchy}}.
\bibinfo{journal}{{\em \apj}} \bibinfo{volume}{262}: \bibinfo{pages}{424--431}. \bibinfo{doi}{\doi{10.1086/160437}}.

\bibtype{Article}%
\bibitem[{Fry}(1984{\natexlab{a}})]{fry_1984_Npoint}
\bibinfo{author}{{Fry} JN} (\bibinfo{year}{1984}{\natexlab{a}}), \bibinfo{month}{Feb.}
\bibinfo{title}{{Galaxy N-point correlation functions - Theoretical amplitudes for arbitrary N}}.
\bibinfo{journal}{{\em \apjl}} \bibinfo{volume}{277}: \bibinfo{pages}{L5--L8}. \bibinfo{doi}{\doi{10.1086/184189}}.

\bibtype{Article}%
\bibitem[{Fry}(1984{\natexlab{b}})]{fry_hier_in_pt}
\bibinfo{author}{{Fry} JN} (\bibinfo{year}{1984}{\natexlab{b}}), \bibinfo{month}{Apr.}
\bibinfo{title}{{The Galaxy correlation hierarchy in perturbation theory}}.
\bibinfo{journal}{{\em \apj}} \bibinfo{volume}{279}: \bibinfo{pages}{499--510}. \bibinfo{doi}{\doi{10.1086/161913}}.

\bibtype{Article}%
\bibitem[{Fry}(1985)]{fry_pdf}
\bibinfo{author}{{Fry} JN} (\bibinfo{year}{1985}), \bibinfo{month}{Feb.}
\bibinfo{title}{{Cosmological density fluctuations and large-scale structure From N-point correlation functions to the probability distribution}}.
\bibinfo{journal}{{\em \apj}} \bibinfo{volume}{289}: \bibinfo{pages}{10--17}. \bibinfo{doi}{\doi{10.1086/162859}}.

\bibtype{Article}%
\bibitem[Fry(1994)]{fry_bias_1994}
\bibinfo{author}{Fry JN} (\bibinfo{year}{1994}), \bibinfo{month}{Jul}.
\bibinfo{title}{Gravity, bias, and the galaxy three-point correlation function}.
\bibinfo{journal}{{\em Phys. Rev. Lett.}} \bibinfo{volume}{73}: \bibinfo{pages}{215--219}. \bibinfo{doi}{\doi{10.1103/PhysRevLett.73.215}}.
\bibinfo{url}{\url{https://link.aps.org/doi/10.1103/PhysRevLett.73.215}}.

\bibtype{Article}%
\bibitem[{Fry}(1996)]{fry_bias_evol}
\bibinfo{author}{{Fry} JN} (\bibinfo{year}{1996}), \bibinfo{month}{Apr.}
\bibinfo{title}{{The Evolution of Bias}}.
\bibinfo{journal}{{\em \apjl}} \bibinfo{volume}{461}: \bibinfo{pages}{L65}. \bibinfo{doi}{\doi{10.1086/310006}}.

\bibtype{Article}%
\bibitem[{Fry} and {Gaztanaga}(1993)]{fry_1993}
\bibinfo{author}{{Fry} JN} and  \bibinfo{author}{{Gaztanaga} E} (\bibinfo{year}{1993}), \bibinfo{month}{Aug.}
\bibinfo{title}{{Biasing and Hierarchical Statistics in Large-Scale Structure}}.
\bibinfo{journal}{{\em \apj}} \bibinfo{volume}{413}: \bibinfo{pages}{447}. \bibinfo{doi}{\doi{10.1086/173015}}.
\eprint{astro-ph/9302009}.

\bibtype{Article}%
\bibitem[{Fry} and {Peebles}(1978)]{fry_peebles_4pcf}
\bibinfo{author}{{Fry} JN} and  \bibinfo{author}{{Peebles} PJE} (\bibinfo{year}{1978}), \bibinfo{month}{Apr.}
\bibinfo{title}{{Statistical analysis of catalogs of extragalactic objects. IX. The four-point galaxy correlation function.}}
\bibinfo{journal}{{\em \apj}} \bibinfo{volume}{221}: \bibinfo{pages}{19--33}. \bibinfo{doi}{\doi{10.1086/156001}}.

\bibtype{Article}%
\bibitem[{Fry} and {Peebles}(1980{\natexlab{a}})]{fry_monte_carlo}
\bibinfo{author}{{Fry} JN} and  \bibinfo{author}{{Peebles} PJE} (\bibinfo{year}{1980}{\natexlab{a}}), \bibinfo{month}{Mar.}
\bibinfo{title}{{Monte-Carlo Model for the Evolution of the Galaxy Correlation Functions}}.
\bibinfo{journal}{{\em \apj}} \bibinfo{volume}{236}: \bibinfo{pages}{343}. \bibinfo{doi}{\doi{10.1086/157752}}.

\bibtype{Article}%
\bibitem[{Fry} and {Peebles}(1980{\natexlab{b}})]{fry_cluster_galaxy_3pcf}
\bibinfo{author}{{Fry} JN} and  \bibinfo{author}{{Peebles} PJE} (\bibinfo{year}{1980}{\natexlab{b}}), \bibinfo{month}{Jun.}
\bibinfo{title}{{Statistical analysis of catalogs of extragalactic objects. XII - The cluster-galaxy-galaxy three-point correlation function}}.
\bibinfo{journal}{{\em \apj}} \bibinfo{volume}{238}: \bibinfo{pages}{785--792}. \bibinfo{doi}{\doi{10.1086/158037}}.

\bibtype{Article}%
\bibitem[{Fry} and {Scherrer}(1994)]{fry_skew_1994}
\bibinfo{author}{{Fry} JN} and  \bibinfo{author}{{Scherrer} RJ} (\bibinfo{year}{1994}), \bibinfo{month}{Jul.}
\bibinfo{title}{{Skewness in Large-Scale Structure and Non-Gaussian Initial Conditions}}.
\bibinfo{journal}{{\em \apj}} \bibinfo{volume}{429}: \bibinfo{pages}{36}. \bibinfo{doi}{\doi{10.1086/174300}}.

\bibtype{Article}%
\bibitem[{Fry} et al.(1992)]{fry_melott_sims}
\bibinfo{author}{{Fry} JN}, \bibinfo{author}{{Melott} AL} and  \bibinfo{author}{{Shandarin} SF} (\bibinfo{year}{1992}), \bibinfo{month}{Jul.}
\bibinfo{title}{{The Three-Point Function in an Ensemble of Numerical Simulations}}.
\bibinfo{journal}{{\em \apj}} \bibinfo{volume}{393}: \bibinfo{pages}{431}. \bibinfo{doi}{\doi{10.1086/171516}}.

\bibtype{Article}%
\bibitem[{Fry} et al.(1993)]{Fry_Melott_3d_sims}
\bibinfo{author}{{Fry} JN}, \bibinfo{author}{{Melott} AL} and  \bibinfo{author}{{Shandarin} SF} (\bibinfo{year}{1993}), \bibinfo{month}{Aug.}
\bibinfo{title}{{The Three-Point Correlation Function in an Ensemble of Three-dimensional Simulations}}.
\bibinfo{journal}{{\em \apj}} \bibinfo{volume}{412}: \bibinfo{pages}{504}. \bibinfo{doi}{\doi{10.1086/172938}}.

\bibtype{Article}%
\bibitem[{Gagrani} and {Samushia}(2017)]{gagrani_2017}
\bibinfo{author}{{Gagrani} P} and  \bibinfo{author}{{Samushia} L} (\bibinfo{year}{2017}), \bibinfo{month}{May}.
\bibinfo{title}{{Information Content of the Angular Multipoles of Redshift-Space Galaxy Bispectrum}}.
\bibinfo{journal}{{\em \mnras}} \bibinfo{volume}{467} (\bibinfo{number}{1}): \bibinfo{pages}{928--935}. \bibinfo{doi}{\doi{10.1093/mnras/stx135}}.
\eprint{1610.03488}.

\bibtype{Article}%
\bibitem[Garcia and Slepian(2022)]{garcia}
\bibinfo{author}{Garcia K} and  \bibinfo{author}{Slepian Z} (\bibinfo{year}{2022}), \bibinfo{month}{06}.
\bibinfo{title}{Improving the line of sight for the anisotropic 3-point correlation function of galaxies: Centroid and unit-vector-average methods scaling as $\mathcal{O}(n^2)$}.
\bibinfo{journal}{{\em Monthly Notices of the Royal Astronomical Society}} \bibinfo{volume}{515} (\bibinfo{number}{1}): \bibinfo{pages}{1199--1217}.
ISSN \bibinfo{issn}{0035-8711}. \bibinfo{doi}{\doi{10.1093/mnras/stac1540}}.
\eprint{https://academic.oup.com/mnras/article-pdf/515/1/1199/45045172/stac1540.pdf}, \bibinfo{url}{\url{https://doi.org/10.1093/mnras/stac1540}}.

\bibtype{Inproceedings}%
\bibitem[{Gardner} et al.(2007)]{gardner}
\bibinfo{author}{{Gardner} JP}, \bibinfo{author}{{Connolly} A} and  \bibinfo{author}{{McBride} C} (\bibinfo{year}{2007}), \bibinfo{month}{Oct.}, \bibinfo{title}{{A Framework for Analyzing Massive Astrophysical Datasets on a Distributed Grid}}, \bibinfo{editor}{{Shaw} RA}, \bibinfo{editor}{{Hill} F} and  \bibinfo{editor}{{Bell} DJ}, (Eds.), \bibinfo{booktitle}{Astronomical Data Analysis Software and Systems XVI}, \bibinfo{series}{Astronomical Society of the Pacific Conference Series}, \bibinfo{volume}{376}, pp.~\bibinfo{pages}{69}.

\bibtype{Article}%
\bibitem[{Gazta{\~n}aga} et al.(2009)]{gaztanaga_2009_3pcf}
\bibinfo{author}{{Gazta{\~n}aga} E}, \bibinfo{author}{{Cabr{\'e}} A}, \bibinfo{author}{{Castander} F}, \bibinfo{author}{{Crocce} M} and  \bibinfo{author}{{Fosalba} P} (\bibinfo{year}{2009}), \bibinfo{month}{Oct.}
\bibinfo{title}{{Clustering of luminous red galaxies - III. Baryon acoustic peak in the three-point correlation}}.
\bibinfo{journal}{{\em \mnras}} \bibinfo{volume}{399} (\bibinfo{number}{2}): \bibinfo{pages}{801--811}. \bibinfo{doi}{\doi{10.1111/j.1365-2966.2009.15313.x}}.
\eprint{0807.2448}.

\bibtype{Article}%
\bibitem[{Gil-Mar{\'\i}n} et al.(2015)]{gilmarin_bi}
\bibinfo{author}{{Gil-Mar{\'\i}n} H}, \bibinfo{author}{{Nore{\~n}a} J}, \bibinfo{author}{{Verde} L}, \bibinfo{author}{{Percival} WJ}, \bibinfo{author}{{Wagner} C}, \bibinfo{author}{{Manera} M} and  \bibinfo{author}{{Schneider} DP} (\bibinfo{year}{2015}), \bibinfo{month}{Jul.}
\bibinfo{title}{{The power spectrum and bispectrum of SDSS DR11 BOSS galaxies - I. Bias and gravity}}.
\bibinfo{journal}{{\em \mnras}} \bibinfo{volume}{451} (\bibinfo{number}{1}): \bibinfo{pages}{539--580}. \bibinfo{doi}{\doi{10.1093/mnras/stv961}}.
\eprint{1407.5668}.

\bibtype{Article}%
\bibitem[Gil-Mar{\'\i}n et al.(2016)]{gil2016clustering}
\bibinfo{author}{Gil-Mar{\'\i}n H}, \bibinfo{author}{Percival WJ}, \bibinfo{author}{Brownstein JR}, \bibinfo{author}{Chuang CH}, \bibinfo{author}{Grieb JN}, \bibinfo{author}{Ho S}, \bibinfo{author}{Kitaura FS}, \bibinfo{author}{Maraston C}, \bibinfo{author}{Prada F}, \bibinfo{author}{Rodr{\'\i}guez-Torres S} and  et al. (\bibinfo{year}{2016}).
\bibinfo{title}{The clustering of galaxies in the sdss-iii baryon oscillation spectroscopic survey: Rsd measurement from the los-dependent power spectrum of dr12 boss galaxies}.
\bibinfo{journal}{{\em Monthly Notices of the Royal Astronomical Society}} \bibinfo{volume}{460} (\bibinfo{number}{4}): \bibinfo{pages}{4188--4209}.

\bibtype{Article}%
\bibitem[{Gon{\c{c}}alves} et al.(2021)]{goncalves}
\bibinfo{author}{{Gon{\c{c}}alves} RS}, \bibinfo{author}{{Carvalho} GC}, \bibinfo{author}{{Andrade} U}, \bibinfo{author}{{Bengaly} CAP}, \bibinfo{author}{{Carvalho} JC} and  \bibinfo{author}{{Alcaniz} J} (\bibinfo{year}{2021}), \bibinfo{month}{Mar.}
\bibinfo{title}{{Measuring the cosmic homogeneity scale with SDSS-IV DR16 quasars}}.
\bibinfo{journal}{{\em \jcap}} \bibinfo{volume}{2021} (\bibinfo{number}{3}), \bibinfo{eid}{029}. \bibinfo{doi}{\doi{10.1088/1475-7516/2021/03/029}}.
\eprint{2010.06635}.

\bibtype{Article}%
\bibitem[{Goroff} et al.(1986)]{goroff}
\bibinfo{author}{{Goroff} MH}, \bibinfo{author}{{Grinstein} B}, \bibinfo{author}{{Rey} SJ} and  \bibinfo{author}{{Wise} MB} (\bibinfo{year}{1986}), \bibinfo{month}{Dec.}
\bibinfo{title}{{Coupling of modes of cosmological mass density fluctuations}}.
\bibinfo{journal}{{\em \apj}} \bibinfo{volume}{311}: \bibinfo{pages}{6--14}. \bibinfo{doi}{\doi{10.1086/164749}}.

\bibtype{Article}%
\bibitem[{Gouyou Beauchamps} et al.(2025)]{gouyou}
\bibinfo{author}{{Gouyou Beauchamps} S}, \bibinfo{author}{{Baratta} P}, \bibinfo{author}{{Escoffier} S}, \bibinfo{author}{{Gillard} W}, \bibinfo{author}{{Bel} J}, \bibinfo{author}{{Bautista} J} and  \bibinfo{author}{{Carbone} C} (\bibinfo{year}{2025}), \bibinfo{month}{Jan.}
\bibinfo{title}{{Cosmological inference including massive neutrinos from the matter power spectrum: Biases induced by uncertainties in the covariance matrix}}.
\bibinfo{journal}{{\em \aap}} \bibinfo{volume}{693}, \bibinfo{eid}{A226}. \bibinfo{doi}{\doi{10.1051/0004-6361/202347164}}.
\eprint{2306.05988}.

\bibtype{Article}%
\bibitem[{Groth} and {Peebles}(1977{\natexlab{a}})]{peebles_groth_1977}
\bibinfo{author}{{Groth} EJ} and  \bibinfo{author}{{Peebles} PJE} (\bibinfo{year}{1977}{\natexlab{a}}), \bibinfo{month}{Oct.}
\bibinfo{title}{{Statistical analysis of catalogs of extragalactic objects. VII. Two- and three-point correlation functions for the high-resolution Shane-Wirtanen catalog of galaxies.}}
\bibinfo{journal}{{\em \apj}} \bibinfo{volume}{217}: \bibinfo{pages}{385--405}. \bibinfo{doi}{\doi{10.1086/155588}}.

\bibtype{Article}%
\bibitem[{Groth} and {Peebles}(1977{\natexlab{b}})]{groth_2_plus_3}
\bibinfo{author}{{Groth} EJ} and  \bibinfo{author}{{Peebles} PJE} (\bibinfo{year}{1977}{\natexlab{b}}), \bibinfo{month}{Oct.}
\bibinfo{title}{{Statistical analysis of catalogs of extragalactic objects. VII. Two- and three-point correlation functions for the high-resolution Shane-Wirtanen catalog of galaxies.}}
\bibinfo{journal}{{\em \apj}} \bibinfo{volume}{217}: \bibinfo{pages}{385--405}. \bibinfo{doi}{\doi{10.1086/155588}}.

\bibtype{Article}%
\bibitem[Grove et al.(2022)]{grove2022desi}
\bibinfo{author}{Grove C}, \bibinfo{author}{Chuang CH}, \bibinfo{author}{Devi NC}, \bibinfo{author}{Garrison L}, \bibinfo{author}{L’Huillier B}, \bibinfo{author}{Feng Y}, \bibinfo{author}{Helly J}, \bibinfo{author}{Hern{\'a}ndez-Aguayo C}, \bibinfo{author}{Alam S}, \bibinfo{author}{Zhang H} and  et al. (\bibinfo{year}{2022}).
\bibinfo{title}{The desi n-body simulation project--i. testing the robustness of simulations for the desi dark time survey}.
\bibinfo{journal}{{\em Monthly Notices of the Royal Astronomical Society}} \bibinfo{volume}{515} (\bibinfo{number}{2}): \bibinfo{pages}{1854--1870}.

\bibtype{Article}%
\bibitem[{Gualdi} and {Verde}(2020)]{gualdi_2020}
\bibinfo{author}{{Gualdi} D} and  \bibinfo{author}{{Verde} L} (\bibinfo{year}{2020}), \bibinfo{month}{Jun.}
\bibinfo{title}{{Galaxy redshift-space bispectrum: the importance of being anisotropic}}.
\bibinfo{journal}{{\em \jcap}} \bibinfo{volume}{2020} (\bibinfo{number}{6}), \bibinfo{eid}{041}. \bibinfo{doi}{\doi{10.1088/1475-7516/2020/06/041}}.
\eprint{2003.12075}.

\bibtype{Article}%
\bibitem[Gualdi and Verde(2020)]{gualdi2020galaxy}
\bibinfo{author}{Gualdi D} and  \bibinfo{author}{Verde L} (\bibinfo{year}{2020}).
\bibinfo{title}{Galaxy redshift-space bispectrum: the importance of being anisotropic}.
\bibinfo{journal}{{\em Journal of Cosmology and Astroparticle Physics}} \bibinfo{volume}{2020} (\bibinfo{number}{06}): \bibinfo{pages}{041}.

\bibtype{Article}%
\bibitem[{Gualdi} and {Verde}(2022)]{gualdi_2022_data_trispec}
\bibinfo{author}{{Gualdi} D} and  \bibinfo{author}{{Verde} L} (\bibinfo{year}{2022}), \bibinfo{month}{Sep.}
\bibinfo{title}{{Integrated trispectrum detection from BOSS DR12 NGC CMASS}}.
\bibinfo{journal}{{\em \jcap}} \bibinfo{volume}{2022} (\bibinfo{number}{9}), \bibinfo{eid}{050}. \bibinfo{doi}{\doi{10.1088/1475-7516/2022/09/050}}.
\eprint{2201.06932}.

\bibtype{Article}%
\bibitem[Gualdi et al.(2019)]{gualdi2019enhancing}
\bibinfo{author}{Gualdi D}, \bibinfo{author}{Gil-Mar{\'\i}n H}, \bibinfo{author}{Schuhmann RL}, \bibinfo{author}{Manera M}, \bibinfo{author}{Joachimi B} and  \bibinfo{author}{Lahav O} (\bibinfo{year}{2019}).
\bibinfo{title}{Enhancing boss bispectrum cosmological constraints with maximal compression}.
\bibinfo{journal}{{\em Monthly Notices of the Royal Astronomical Society}} \bibinfo{volume}{484} (\bibinfo{number}{3}): \bibinfo{pages}{3713--3730}.

\bibtype{Article}%
\bibitem[{Gualdi} et al.(2021{\natexlab{a}})]{gualdi_joint_bi_tri}
\bibinfo{author}{{Gualdi} D}, \bibinfo{author}{{Gil-Mar{\'\i}n} H} and  \bibinfo{author}{{Verde} L} (\bibinfo{year}{2021}{\natexlab{a}}), \bibinfo{month}{Jul.}
\bibinfo{title}{{Joint analysis of anisotropic power spectrum, bispectrum and trispectrum: application to N-body simulations}}.
\bibinfo{journal}{{\em \jcap}} \bibinfo{volume}{2021} (\bibinfo{number}{7}), \bibinfo{eid}{008}. \bibinfo{doi}{\doi{10.1088/1475-7516/2021/07/008}}.
\eprint{2104.03976}.

\bibtype{Article}%
\bibitem[{Gualdi} et al.(2021{\natexlab{b}})]{gualdi_2021_trispec}
\bibinfo{author}{{Gualdi} D}, \bibinfo{author}{{Novell} S}, \bibinfo{author}{{Gil-Mar{\'\i}n} H} and  \bibinfo{author}{{Verde} L} (\bibinfo{year}{2021}{\natexlab{b}}), \bibinfo{month}{Jan.}
\bibinfo{title}{{Matter trispectrum: theoretical modelling and comparison to N-body simulations}}.
\bibinfo{journal}{{\em \jcap}} \bibinfo{volume}{2021} (\bibinfo{number}{1}), \bibinfo{eid}{015}. \bibinfo{doi}{\doi{10.1088/1475-7516/2021/01/015}}.
\eprint{2009.02290}.

\bibtype{Article}%
\bibitem[{Guidi} et al.(2023)]{guidi_3pcf}
\bibinfo{author}{{Guidi} M}, \bibinfo{author}{{Veropalumbo} A}, \bibinfo{author}{{Branchini} E}, \bibinfo{author}{{Eggemeier} A} and  \bibinfo{author}{{Carbone} C} (\bibinfo{year}{2023}), \bibinfo{month}{Aug.}
\bibinfo{title}{{Modelling the next-to-leading order matter three-point correlation function using FFTLog}}.
\bibinfo{journal}{{\em \jcap}} \bibinfo{volume}{2023} (\bibinfo{number}{8}), \bibinfo{eid}{066}. \bibinfo{doi}{\doi{10.1088/1475-7516/2023/08/066}}.
\eprint{2212.07382}.

\bibtype{Article}%
\bibitem[Guo et al.(2014)]{Guo_2013}
\bibinfo{author}{Guo H}, \bibinfo{author}{Li C}, \bibinfo{author}{Jing YP} and  \bibinfo{author}{B\"orner G} (\bibinfo{year}{2014}).
\bibinfo{title}{{Stellar Mass and color dependence of the three-point correlation function of galaxies in the local universe}}.
\bibinfo{journal}{{\em Astrophys. J.}} \bibinfo{volume}{780}: \bibinfo{pages}{139}. \bibinfo{doi}{\doi{10.1088/0004-637X/780/2/139}}.
\eprint{1303.2609}.

\bibtype{Article}%
\bibitem[{Guo} et al.(2015)]{Guo_2015}
\bibinfo{author}{{Guo} H}, \bibinfo{author}{{Zheng} Z}, \bibinfo{author}{{Jing} YP}, \bibinfo{author}{{Zehavi} I}, \bibinfo{author}{{Li} C}, \bibinfo{author}{{Weinberg} DH}, \bibinfo{author}{{Skibba} RA}, \bibinfo{author}{{Nichol} RC}, \bibinfo{author}{{Rossi} G}, \bibinfo{author}{{Sabiu} CG}, \bibinfo{author}{{Schneider} DP} and  \bibinfo{author}{{McBride} CK} (\bibinfo{year}{2015}), \bibinfo{month}{Apr.}
\bibinfo{title}{{Modelling the redshift-space three-point correlation function in SDSS-III.}}
\bibinfo{journal}{{\em \mnras}} \bibinfo{volume}{449}: \bibinfo{pages}{L95--L99}. \bibinfo{doi}{\doi{10.1093/mnrasl/slv020}}.
\eprint{1409.7389}.

\bibtype{Article}%
\bibitem[{Guth}(2007)]{guth_2007}
\bibinfo{author}{{Guth} AH} (\bibinfo{year}{2007}), \bibinfo{month}{Jun.}
\bibinfo{title}{{Eternal inflation and its implications}}.
\bibinfo{journal}{{\em Journal of Physics A Mathematical General}} \bibinfo{volume}{40} (\bibinfo{number}{25}): \bibinfo{pages}{6811--6826}. \bibinfo{doi}{\doi{10.1088/1751-8113/40/25/S25}}.
\eprint{hep-th/0702178}.

\bibtype{Article}%
\bibitem[{Hahn} et al.(2017)]{hahn_fiber}
\bibinfo{author}{{Hahn} C}, \bibinfo{author}{{Scoccimarro} R}, \bibinfo{author}{{Blanton} MR}, \bibinfo{author}{{Tinker} JL} and  \bibinfo{author}{{Rodr{\'\i}guez-Torres} SA} (\bibinfo{year}{2017}), \bibinfo{month}{May}.
\bibinfo{title}{{The Effect of Fiber Collisions on the Galaxy Power Spectrum Multipoles}}.
\bibinfo{journal}{{\em \mnras}} \bibinfo{volume}{467} (\bibinfo{number}{2}): \bibinfo{pages}{1940--1956}. \bibinfo{doi}{\doi{10.1093/mnras/stx185}}.
\eprint{1609.01714}.

\bibtype{Article}%
\bibitem[Hahn et al.(2020)]{Hahn_2020}
\bibinfo{author}{Hahn C}, \bibinfo{author}{Villaescusa-Navarro F}, \bibinfo{author}{Castorina E} and  \bibinfo{author}{Scoccimarro R} (\bibinfo{year}{2020}), \bibinfo{month}{Mar.}
\bibinfo{title}{Constraining m$\nu$ with the bispectrum. part i. breaking parameter degeneracies}.
\bibinfo{journal}{{\em Journal of Cosmology and Astroparticle Physics}} \bibinfo{volume}{2020} (\bibinfo{number}{03}): \bibinfo{pages}{040–040}.
ISSN \bibinfo{issn}{1475-7516}. \bibinfo{doi}{\doi{10.1088/1475-7516/2020/03/040}}.
\bibinfo{url}{\url{http://dx.doi.org/10.1088/1475-7516/2020/03/040}}.

\bibtype{Misc}%
\bibitem[Hahn et al.(2023)]{hahn2023rmsscriptsizeimbigcosmological}
\bibinfo{author}{Hahn C}, \bibinfo{author}{Eickenberg M}, \bibinfo{author}{Ho S}, \bibinfo{author}{Hou J}, \bibinfo{author}{Lemos P}, \bibinfo{author}{Massara E}, \bibinfo{author}{Modi C}, \bibinfo{author}{Dizgah AM}, \bibinfo{author}{Parker L} and  \bibinfo{author}{Blancard BRS} (\bibinfo{year}{2023}).
\bibinfo{title}{\textsc{SimBIG}: The first cosmological constraints from the non-linear galaxy bispectrum}.
\eprint{2310.15243}, \bibinfo{url}{\url{https://arxiv.org/abs/2310.15243}}.

\bibtype{Article}%
\bibitem[{Hahn} et al.(2024)]{simbigchang}
\bibinfo{author}{{Hahn} C}, \bibinfo{author}{{Eickenberg} M}, \bibinfo{author}{{Ho} S}, \bibinfo{author}{{Hou} J}, \bibinfo{author}{{Lemos} P}, \bibinfo{author}{{Massara} E}, \bibinfo{author}{{Modi} C}, \bibinfo{author}{{Dizgah} AM}, \bibinfo{author}{{Parker} L}, \bibinfo{author}{{Blancard} BRS} and  \bibinfo{author}{{SimBIG Collaboration}} (\bibinfo{year}{2024}), \bibinfo{month}{Apr.}
\bibinfo{title}{{Cosmological constraints from the nonlinear galaxy bispectrum}}.
\bibinfo{journal}{{\em \prd}} \bibinfo{volume}{109} (\bibinfo{number}{8}), \bibinfo{eid}{083534}. \bibinfo{doi}{\doi{10.1103/PhysRevD.109.083534}}.
\eprint{2310.15243}.

\bibtype{Inproceedings}%
\bibitem[{Hamilton}(1998)]{hamilton_1998}
\bibinfo{author}{{Hamilton} AJS} (\bibinfo{year}{1998}), \bibinfo{month}{Jan.}, \bibinfo{title}{{Linear Redshift Distortions: a Review}}, \bibinfo{editor}{{Hamilton} D}, (Ed.), \bibinfo{booktitle}{The Evolving Universe}, \bibinfo{series}{Astrophysics and Space Science Library}, \bibinfo{volume}{231}, pp. \bibinfo{pages}{185}, \eprint{astro-ph/9708102}.

\bibtype{Article}%
\bibitem[{Hamilton}(2000)]{hamilton_fftlog}
\bibinfo{author}{{Hamilton} AJS} (\bibinfo{year}{2000}), \bibinfo{month}{Feb.}
\bibinfo{title}{{Uncorrelated modes of the non-linear power spectrum}}.
\bibinfo{journal}{{\em \mnras}} \bibinfo{volume}{312} (\bibinfo{number}{2}): \bibinfo{pages}{257--284}. \bibinfo{doi}{\doi{10.1046/j.1365-8711.2000.03071.x}}.
\eprint{astro-ph/9905191}.

\bibtype{Article}%
\bibitem[{Hamilton} and {Culhane}(1996)]{hamilton_culhane}
\bibinfo{author}{{Hamilton} AJS} and  \bibinfo{author}{{Culhane} M} (\bibinfo{year}{1996}), \bibinfo{month}{Jan.}
\bibinfo{title}{{Spherical redshift distortions}}.
\bibinfo{journal}{{\em \mnras}} \bibinfo{volume}{278}: \bibinfo{pages}{73}. \bibinfo{doi}{\doi{10.1093/mnras/278.1.73}}.
\eprint{astro-ph/9507021}.

\bibtype{Article}%
\bibitem[{Hand} et al.(2017{\natexlab{a}})]{hand_2017}
\bibinfo{author}{{Hand} N}, \bibinfo{author}{{Li} Y}, \bibinfo{author}{{Slepian} Z} and  \bibinfo{author}{{Seljak} U} (\bibinfo{year}{2017}{\natexlab{a}}), \bibinfo{month}{Jul.}
\bibinfo{title}{{An optimal FFT-based anisotropic power spectrum estimator}}.
\bibinfo{journal}{{\em \jcap}} \bibinfo{volume}{2017} (\bibinfo{number}{7}), \bibinfo{eid}{002}. \bibinfo{doi}{\doi{10.1088/1475-7516/2017/07/002}}.
\eprint{1704.02357}.

\bibtype{Article}%
\bibitem[{Hand} et al.(2017{\natexlab{b}})]{hand}
\bibinfo{author}{{Hand} N}, \bibinfo{author}{{Li} Y}, \bibinfo{author}{{Slepian} Z} and  \bibinfo{author}{{Seljak} U} (\bibinfo{year}{2017}{\natexlab{b}}), \bibinfo{month}{Jul.}
\bibinfo{title}{{An optimal FFT-based anisotropic power spectrum estimator}}.
\bibinfo{journal}{{\em \jcap}} \bibinfo{volume}{2017} (\bibinfo{number}{7}), \bibinfo{eid}{002}. \bibinfo{doi}{\doi{10.1088/1475-7516/2017/07/002}}.
\eprint{1704.02357}.

\bibtype{Article}%
\bibitem[{Hand} et al.(2018)]{nbodykit}
\bibinfo{author}{{Hand} N}, \bibinfo{author}{{Feng} Y}, \bibinfo{author}{{Beutler} F}, \bibinfo{author}{{Li} Y}, \bibinfo{author}{{Modi} C}, \bibinfo{author}{{Seljak} U} and  \bibinfo{author}{{Slepian} Z} (\bibinfo{year}{2018}), \bibinfo{month}{Oct.}
\bibinfo{title}{{nbodykit: An Open-source, Massively Parallel Toolkit for Large-scale Structure}}.
\bibinfo{journal}{{\em \aj}} \bibinfo{volume}{156} (\bibinfo{number}{4}), \bibinfo{eid}{160}. \bibinfo{doi}{\doi{10.3847/1538-3881/aadae0}}.
\eprint{1712.05834}.

\bibtype{Article}%
\bibitem[{Hartlap} et al.(2007)]{hartlap}
\bibinfo{author}{{Hartlap} J}, \bibinfo{author}{{Simon} P} and  \bibinfo{author}{{Schneider} P} (\bibinfo{year}{2007}), \bibinfo{month}{Mar.}
\bibinfo{title}{{Why your model parameter confidences might be too optimistic. Unbiased estimation of the inverse covariance matrix}}.
\bibinfo{journal}{{\em \aap}} \bibinfo{volume}{464} (\bibinfo{number}{1}): \bibinfo{pages}{399--404}. \bibinfo{doi}{\doi{10.1051/0004-6361:20066170}}.
\eprint{astro-ph/0608064}.

\bibtype{Article}%
\bibitem[Hashimoto et al.(2017)]{hashimoto2017precision}
\bibinfo{author}{Hashimoto I}, \bibinfo{author}{Rasera Y} and  \bibinfo{author}{Taruya A} (\bibinfo{year}{2017}).
\bibinfo{title}{Precision cosmology with redshift-space bispectrum: a perturbation theory based model at one-loop order}.
\bibinfo{journal}{{\em Physical Review D}} \bibinfo{volume}{96} (\bibinfo{number}{4}): \bibinfo{pages}{043526}.

\bibtype{Article}%
\bibitem[{Heavens} and {Taylor}(1995)]{heavens_taylor}
\bibinfo{author}{{Heavens} AF} and  \bibinfo{author}{{Taylor} AN} (\bibinfo{year}{1995}), \bibinfo{month}{Jul.}
\bibinfo{title}{{A spherical harmonic analysis of redshift space}}.
\bibinfo{journal}{{\em \mnras}} \bibinfo{volume}{275} (\bibinfo{number}{2}): \bibinfo{pages}{483--497}. \bibinfo{doi}{\doi{10.1093/mnras/275.2.483}}.
\eprint{astro-ph/9409027}.

\bibtype{Article}%
\bibitem[Heideman et al.(1984)]{heideman}
\bibinfo{author}{Heideman M}, \bibinfo{author}{Johnson D} and  \bibinfo{author}{Burrus C} (\bibinfo{year}{1984}).
\bibinfo{title}{Gauss and the history of the fast fourier transform}.
\bibinfo{journal}{{\em IEEE ASSP Magazine}} \bibinfo{volume}{1} (\bibinfo{number}{4}): \bibinfo{pages}{14--21}. \bibinfo{doi}{\doi{10.1109/MASSP.1984.1162257}}.

\bibtype{Article}%
\bibitem[{Hoffmann} et al.(2018)]{hoffman_18}
\bibinfo{author}{{Hoffmann} K}, \bibinfo{author}{{Gazta{\~n}aga} E}, \bibinfo{author}{{Scoccimarro} R} and  \bibinfo{author}{{Crocce} M} (\bibinfo{year}{2018}), \bibinfo{month}{May}.
\bibinfo{title}{{Testing the consistency of three-point halo clustering in Fourier and configuration space}}.
\bibinfo{journal}{{\em \mnras}} \bibinfo{volume}{476} (\bibinfo{number}{1}): \bibinfo{pages}{814--829}. \bibinfo{doi}{\doi{10.1093/mnras/sty187}}.
\eprint{1708.08941}.

\bibtype{Article}%
\bibitem[{Hou} et al.(2021)]{hou_eboss}
\bibinfo{author}{{Hou} J}, \bibinfo{author}{{S{\'a}nchez} AG}, \bibinfo{author}{{Ross} AJ}, \bibinfo{author}{{Smith} A}, \bibinfo{author}{{Neveux} R}, \bibinfo{author}{{Bautista} J}, \bibinfo{author}{{Burtin} E}, \bibinfo{author}{{Zhao} C}, \bibinfo{author}{{Scoccimarro} R}, \bibinfo{author}{{Dawson} KS}, \bibinfo{author}{{de Mattia} A}, \bibinfo{author}{{de la Macorra} A}, \bibinfo{author}{{du Mas des Bourboux} H}, \bibinfo{author}{{Eisenstein} DJ}, \bibinfo{author}{{Gil-Mar{\'\i}n} H}, \bibinfo{author}{{Lyke} BW}, \bibinfo{author}{{Mohammad} FG}, \bibinfo{author}{{Mueller} EM}, \bibinfo{author}{{Percival} WJ}, \bibinfo{author}{{Rossi} G}, \bibinfo{author}{{Vargas Maga{\~n}a} M}, \bibinfo{author}{{Zarrouk} P}, \bibinfo{author}{{Zhao} GB}, \bibinfo{author}{{Brinkmann} J}, \bibinfo{author}{{Brownstein} JR}, \bibinfo{author}{{Chuang} CH}, \bibinfo{author}{{Myers} AD}, \bibinfo{author}{{Newman} JA}, \bibinfo{author}{{Schneider} DP} and  \bibinfo{author}{{Vivek} M} (\bibinfo{year}{2021}), \bibinfo{month}{Jan.}
\bibinfo{title}{{The completed SDSS-IV extended Baryon Oscillation Spectroscopic Survey: BAO and RSD measurements from anisotropic clustering analysis of the quasar sample in configuration space between redshift 0.8 and 2.2}}.
\bibinfo{journal}{{\em \mnras}} \bibinfo{volume}{500} (\bibinfo{number}{1}): \bibinfo{pages}{1201--1221}. \bibinfo{doi}{\doi{10.1093/mnras/staa3234}}.
\eprint{2007.08998}.

\bibtype{Article}%
\bibitem[{Hou} et al.(2022)]{hou_covar}
\bibinfo{author}{{Hou} J}, \bibinfo{author}{{Cahn} RN}, \bibinfo{author}{{Philcox} OHE} and  \bibinfo{author}{{Slepian} Z} (\bibinfo{year}{2022}), \bibinfo{month}{Aug.}
\bibinfo{title}{{Analytic Gaussian covariance matrices for galaxy N -point correlation functions}}.
\bibinfo{journal}{{\em \prd}} \bibinfo{volume}{106} (\bibinfo{number}{4}), \bibinfo{eid}{043515}. \bibinfo{doi}{\doi{10.1103/PhysRevD.106.043515}}.
\eprint{2108.01714}.

\bibtype{Article}%
\bibitem[{Hou} et al.(2023)]{hou_parity}
\bibinfo{author}{{Hou} J}, \bibinfo{author}{{Slepian} Z} and  \bibinfo{author}{{Cahn} RN} (\bibinfo{year}{2023}), \bibinfo{month}{May}.
\bibinfo{title}{{Measurement of parity-odd modes in the large-scale 4-point correlation function of Sloan Digital Sky Survey Baryon Oscillation Spectroscopic Survey twelfth data release CMASS and LOWZ galaxies}}.
\bibinfo{journal}{{\em \mnras}} \bibinfo{volume}{522} (\bibinfo{number}{4}): \bibinfo{pages}{5701--5739}. \bibinfo{doi}{\doi{10.1093/mnras/stad1062}}.
\eprint{2206.03625}.

\bibtype{Article}%
\bibitem[{Hou} et al.(2024)]{bao_odd}
\bibinfo{author}{{Hou} J}, \bibinfo{author}{{Slepian} Z} and  \bibinfo{author}{{Jamieson} D} (\bibinfo{year}{2024}), \bibinfo{month}{Oct.}
\bibinfo{title}{{Can Baryon Acoustic Oscillations Illuminate the Parity-Violating Galaxy 4PCF?}}
\bibinfo{journal}{{\em arXiv e-prints}} , \bibinfo{eid}{arXiv:2410.05230}\bibinfo{doi}{\doi{10.48550/arXiv.2410.05230}}.
\eprint{2410.05230}.

\bibtype{Article}%
\bibitem[Hu and Sawicki(2007)]{hu2007models}
\bibinfo{author}{Hu W} and  \bibinfo{author}{Sawicki I} (\bibinfo{year}{2007}).
\bibinfo{title}{Models of f (r) cosmic acceleration that evade solar system tests}.
\bibinfo{journal}{{\em Physical Review D—Particles, Fields, Gravitation, and Cosmology}} \bibinfo{volume}{76} (\bibinfo{number}{6}): \bibinfo{pages}{064004}.

\bibtype{Article}%
\bibitem[{Hu} and {Sugiyama}(1996)]{hs_96}
\bibinfo{author}{{Hu} W} and  \bibinfo{author}{{Sugiyama} N} (\bibinfo{year}{1996}), \bibinfo{month}{Nov.}
\bibinfo{title}{{Small-Scale Cosmological Perturbations: an Analytic Approach}}.
\bibinfo{journal}{{\em \apj}} \bibinfo{volume}{471}: \bibinfo{pages}{542}. \bibinfo{doi}{\doi{10.1086/177989}}.
\eprint{astro-ph/9510117}.

\bibtype{Article}%
\bibitem[{Inagaki}(1976)]{inagaki}
\bibinfo{author}{{Inagaki} S} (\bibinfo{year}{1976}), \bibinfo{month}{Jan.}
\bibinfo{title}{{On the density correlations of fluctuations in an expanding universe. I.}}
\bibinfo{journal}{{\em \pasj}} \bibinfo{volume}{28} (\bibinfo{number}{1}): \bibinfo{pages}{77--87}.

\bibtype{Article}%
\bibitem[{Inomata} et al.(2025)]{inomata}
\bibinfo{author}{{Inomata} K}, \bibinfo{author}{{Jenks} L} and  \bibinfo{author}{{Kamionkowski} M} (\bibinfo{year}{2025}), \bibinfo{month}{Feb.}
\bibinfo{title}{{Parity-breaking galaxy 4-point function from lensing by chiral gravitational waves}}.
\bibinfo{journal}{{\em \prd}} \bibinfo{volume}{111} (\bibinfo{number}{4}), \bibinfo{eid}{043504}. \bibinfo{doi}{\doi{10.1103/PhysRevD.111.043504}}.
\eprint{2408.03994}.

\bibtype{Article}%
\bibitem[Isserlis(1918)]{isserlis}
\bibinfo{author}{Isserlis L} (\bibinfo{year}{1918}), \bibinfo{month}{11}.
\bibinfo{title}{On a formula for the product-moment coefficient of any order of a normal frequency distribution in any number of variables}.
\bibinfo{journal}{{\em Biometrika}} \bibinfo{volume}{12} (\bibinfo{number}{1-2}): \bibinfo{pages}{134--139}.
ISSN \bibinfo{issn}{0006-3444}. \bibinfo{doi}{\doi{10.1093/biomet/12.1-2.134}}.
\eprint{https://academic.oup.com/biomet/article-pdf/12/1-2/134/481266/12-1-2-134.pdf}, \bibinfo{url}{\url{https://doi.org/10.1093/biomet/12.1-2.134}}.

\bibtype{Article}%
\bibitem[Ivanov and Sibiryakov(2018)]{ivanov2018infrared}
\bibinfo{author}{Ivanov MM} and  \bibinfo{author}{Sibiryakov S} (\bibinfo{year}{2018}).
\bibinfo{title}{Infrared resummation for biased tracers in redshift space}.
\bibinfo{journal}{{\em Journal of Cosmology and Astroparticle Physics}} \bibinfo{volume}{2018} (\bibinfo{number}{07}): \bibinfo{pages}{053}.

\bibtype{Article}%
\bibitem[Ivanov et al.(2020)]{ivanov2020cosmological}
\bibinfo{author}{Ivanov MM}, \bibinfo{author}{Simonovi{\'c} M} and  \bibinfo{author}{Zaldarriaga M} (\bibinfo{year}{2020}).
\bibinfo{title}{Cosmological parameters from the boss galaxy power spectrum}.
\bibinfo{journal}{{\em Journal of Cosmology and Astroparticle Physics}} \bibinfo{volume}{2020} (\bibinfo{number}{05}): \bibinfo{pages}{042}.

\bibtype{Article}%
\bibitem[Ivanov et al.(2022)]{Ivanov_2022}
\bibinfo{author}{Ivanov MM}, \bibinfo{author}{Philcox OH}, \bibinfo{author}{Nishimichi T}, \bibinfo{author}{Simonović M}, \bibinfo{author}{Takada M} and  \bibinfo{author}{Zaldarriaga M} (\bibinfo{year}{2022}), \bibinfo{month}{Mar.}
\bibinfo{title}{Precision analysis of the redshift-space galaxy bispectrum}.
\bibinfo{journal}{{\em Physical Review D}} \bibinfo{volume}{105} (\bibinfo{number}{6}).
ISSN \bibinfo{issn}{2470-0029}. \bibinfo{doi}{\doi{10.1103/physrevd.105.063512}}.
\bibinfo{url}{\url{http://dx.doi.org/10.1103/PhysRevD.105.063512}}.

\bibtype{Article}%
\bibitem[Ivanov et al.(2023)]{ivanov2023cosmology}
\bibinfo{author}{Ivanov MM}, \bibinfo{author}{Philcox OH}, \bibinfo{author}{Cabass G}, \bibinfo{author}{Nishimichi T}, \bibinfo{author}{Simonovi{\'c} M} and  \bibinfo{author}{Zaldarriaga M} (\bibinfo{year}{2023}).
\bibinfo{title}{Cosmology with the galaxy bispectrum multipoles: Optimal estimation and application to boss data}.
\bibinfo{journal}{{\em Physical Review D}} \bibinfo{volume}{107} (\bibinfo{number}{8}): \bibinfo{pages}{083515}.

\bibtype{Article}%
\bibitem[{Jain} and {Bertschinger}(1994)]{bert_jain}
\bibinfo{author}{{Jain} B} and  \bibinfo{author}{{Bertschinger} E} (\bibinfo{year}{1994}), \bibinfo{month}{Aug.}
\bibinfo{title}{{Second-Order Power Spectrum and Nonlinear Evolution at High Redshift}}.
\bibinfo{journal}{{\em \apj}} \bibinfo{volume}{431}: \bibinfo{pages}{495}. \bibinfo{doi}{\doi{10.1086/174502}}.
\eprint{astro-ph/9311070}.

\bibtype{Article}%
\bibitem[{Jamieson} et al.(2024)]{jamieson_pops}
\bibinfo{author}{{Jamieson} D}, \bibinfo{author}{{Caravano} A}, \bibinfo{author}{{Hou} J}, \bibinfo{author}{{Slepian} Z} and  \bibinfo{author}{{Komatsu} E} (\bibinfo{year}{2024}), \bibinfo{month}{Sep.}
\bibinfo{title}{{Parity-odd power spectra: concise statistics for cosmological parity violation}}.
\bibinfo{journal}{{\em \mnras}} \bibinfo{volume}{533} (\bibinfo{number}{3}): \bibinfo{pages}{2582--2598}. \bibinfo{doi}{\doi{10.1093/mnras/stae1924}}.
\eprint{2406.15683}.

\bibtype{Article}%
\bibitem[{Jennings} et al.(2020)]{jennings_2020}
\bibinfo{author}{{Jennings} WD}, \bibinfo{author}{{Watkinson} CA} and  \bibinfo{author}{{Abdalla} FB} (\bibinfo{year}{2020}), \bibinfo{month}{Nov.}
\bibinfo{title}{{Analysing the Epoch of Reionization with three-point correlation functions and machine learning techniques}}.
\bibinfo{journal}{{\em \mnras}} \bibinfo{volume}{498} (\bibinfo{number}{3}): \bibinfo{pages}{4518--4532}. \bibinfo{doi}{\doi{10.1093/mnras/staa2598}}.
\eprint{2011.14157}.

\bibtype{Article}%
\bibitem[{Jeong} and {Kamionkowski}(2012)]{jeong_fossil}
\bibinfo{author}{{Jeong} D} and  \bibinfo{author}{{Kamionkowski} M} (\bibinfo{year}{2012}), \bibinfo{month}{Jun.}
\bibinfo{title}{{Clustering Fossils from the Early Universe}}.
\bibinfo{journal}{{\em \prl}} \bibinfo{volume}{108} (\bibinfo{number}{25}), \bibinfo{eid}{251301}. \bibinfo{doi}{\doi{10.1103/PhysRevLett.108.251301}}.
\eprint{1203.0302}.

\bibtype{Article}%
\bibitem[{Jing}(2005)]{Jing_05}
\bibinfo{author}{{Jing} YP} (\bibinfo{year}{2005}), \bibinfo{month}{Feb.}
\bibinfo{title}{{Correcting for the Alias Effect When Measuring the Power Spectrum Using a Fast Fourier Transform}}.
\bibinfo{journal}{{\em \apj}} \bibinfo{volume}{620} (\bibinfo{number}{2}): \bibinfo{pages}{559--563}. \bibinfo{doi}{\doi{10.1086/427087}}.
\eprint{astro-ph/0409240}.

\bibtype{Article}%
\bibitem[{Jing} and {B{\"o}rner}(1998)]{jing_3pcf}
\bibinfo{author}{{Jing} YP} and  \bibinfo{author}{{B{\"o}rner} G} (\bibinfo{year}{1998}), \bibinfo{month}{Aug.}
\bibinfo{title}{{The Three-Point Correlation Function of Galaxies Determined from the Las Campanas Redshift Survey}}.
\bibinfo{journal}{{\em \apj}} \bibinfo{volume}{503} (\bibinfo{number}{1}): \bibinfo{pages}{37--47}. \bibinfo{doi}{\doi{10.1086/305997}}.
\eprint{astro-ph/9802011}.

\bibtype{Article}%
\bibitem[{Kaiser}(1987)]{kaiser_1987}
\bibinfo{author}{{Kaiser} N} (\bibinfo{year}{1987}), \bibinfo{month}{Jul.}
\bibinfo{title}{{Clustering in real space and in redshift space}}.
\bibinfo{journal}{{\em \mnras}} \bibinfo{volume}{227}: \bibinfo{pages}{1--21}. \bibinfo{doi}{\doi{10.1093/mnras/227.1.1}}.

\bibtype{Article}%
\bibitem[Kamalinejad and Slepian(2020)]{kamalinejad2020non}
\bibinfo{author}{Kamalinejad F} and  \bibinfo{author}{Slepian Z} (\bibinfo{year}{2020}).
\bibinfo{title}{A non-degenerate neutrino mass signature in the galaxy bispectrum}.
\bibinfo{journal}{{\em arXiv preprint arXiv:2011.00899}} .

\bibtype{Article}%
\bibitem[{Kamalinejad} and {Slepian}(2022)]{kamali_2fluid}
\bibinfo{author}{{Kamalinejad} F} and  \bibinfo{author}{{Slepian} Z} (\bibinfo{year}{2022}), \bibinfo{month}{Mar.}
\bibinfo{title}{{A Simple Analytic Treatment of Neutrino Mass Impact on the Full Power Spectrum Shape via a Two-Fluid Approximation}}.
\bibinfo{journal}{{\em arXiv e-prints}} , \bibinfo{eid}{arXiv:2203.13103}\bibinfo{doi}{\doi{10.48550/arXiv.2203.13103}}.
\eprint{2203.13103}.

\bibtype{Article}%
\bibitem[{Kayo} et al.(2004)]{kayo_2004}
\bibinfo{author}{{Kayo} I}, \bibinfo{author}{{Suto} Y}, \bibinfo{author}{{Nichol} RC}, \bibinfo{author}{{Pan} J}, \bibinfo{author}{{Szapudi} I}, \bibinfo{author}{{Connolly} AJ}, \bibinfo{author}{{Gardner} J}, \bibinfo{author}{{Jain} B}, \bibinfo{author}{{Kulkarni} G}, \bibinfo{author}{{Matsubara} T}, \bibinfo{author}{{Sheth} R}, \bibinfo{author}{{Szalay} AS} and  \bibinfo{author}{{Brinkmann} J} (\bibinfo{year}{2004}), \bibinfo{month}{Jun.}
\bibinfo{title}{{Three-Point Correlation Functions of SDSS Galaxies in Redshift Space: Morphology, Color, and Luminosity Dependence}}.
\bibinfo{journal}{{\em \pasj}} \bibinfo{volume}{56} (\bibinfo{number}{3}): \bibinfo{pages}{415--423}. \bibinfo{doi}{\doi{10.1093/pasj/56.3.415}}.
\eprint{astro-ph/0403638}.

\bibtype{Article}%
\bibitem[{Kerscher} et al.(2000)]{kerscher_2000}
\bibinfo{author}{{Kerscher} M}, \bibinfo{author}{{Szapudi} I} and  \bibinfo{author}{{Szalay} AS} (\bibinfo{year}{2000}), \bibinfo{month}{May}.
\bibinfo{title}{{A Comparison of Estimators for the Two-Point Correlation Function}}.
\bibinfo{journal}{{\em \apjl}} \bibinfo{volume}{535} (\bibinfo{number}{1}): \bibinfo{pages}{L13--L16}. \bibinfo{doi}{\doi{10.1086/312702}}.
\eprint{astro-ph/9912088}.

\bibtype{Article}%
\bibitem[{Khomeriki} and {Samushia}(2024)]{lado_peak_position}
\bibinfo{author}{{Khomeriki} G} and  \bibinfo{author}{{Samushia} L} (\bibinfo{year}{2024}), \bibinfo{month}{Jan.}
\bibinfo{title}{{Mixing bispectrum multipoles under geometric distortions}}.
\bibinfo{journal}{{\em \mnras}} \bibinfo{volume}{527} (\bibinfo{number}{3}): \bibinfo{pages}{5886--5894}. \bibinfo{doi}{\doi{10.1093/mnras/stad3335}}.
\eprint{2307.16498}.

\bibtype{Article}%
\bibitem[{Kim} and {Strauss}(1998)]{kim_1998_skew_kurt}
\bibinfo{author}{{Kim} RSJ} and  \bibinfo{author}{{Strauss} MA} (\bibinfo{year}{1998}), \bibinfo{month}{Jan.}
\bibinfo{title}{{Measuring High-Order Moments of the Galaxy Distribution from Counts in Cells: The Edgeworth Approximation}}.
\bibinfo{journal}{{\em \apj}} \bibinfo{volume}{493} (\bibinfo{number}{1}): \bibinfo{pages}{39--51}. \bibinfo{doi}{\doi{10.1086/305095}}.

\bibtype{Article}%
\bibitem[Kobayashi et al.(2022)]{kobayashi2022full}
\bibinfo{author}{Kobayashi Y}, \bibinfo{author}{Nishimichi T}, \bibinfo{author}{Takada M} and  \bibinfo{author}{Miyatake H} (\bibinfo{year}{2022}).
\bibinfo{title}{Full-shape cosmology analysis of the sdss-iii boss galaxy power spectrum using an emulator-based halo model: A 5\% determination of $\sigma$ 8}.
\bibinfo{journal}{{\em Physical Review D}} \bibinfo{volume}{105} (\bibinfo{number}{8}): \bibinfo{pages}{083517}.

\bibtype{Article}%
\bibitem[{Kofman} et al.(1994)]{reheating_1994}
\bibinfo{author}{{Kofman} L}, \bibinfo{author}{{Linde} A} and  \bibinfo{author}{{Starobinsky} AA} (\bibinfo{year}{1994}), \bibinfo{month}{Dec.}
\bibinfo{title}{{Reheating after inflation}}.
\bibinfo{journal}{{\em \prl}} \bibinfo{volume}{73} (\bibinfo{number}{24}): \bibinfo{pages}{3195--3198}. \bibinfo{doi}{\doi{10.1103/PhysRevLett.73.3195}}.
\eprint{hep-th/9405187}.

\bibtype{Article}%
\bibitem[{Landy} and {Szalay}(1993)]{landy_szalay}
\bibinfo{author}{{Landy} SD} and  \bibinfo{author}{{Szalay} AS} (\bibinfo{year}{1993}), \bibinfo{month}{Jul.}
\bibinfo{title}{{Bias and Variance of Angular Correlation Functions}}.
\bibinfo{journal}{{\em \apj}} \bibinfo{volume}{412}: \bibinfo{pages}{64}. \bibinfo{doi}{\doi{10.1086/172900}}.

\bibtype{Article}%
\bibitem[{Laurent} et al.(2016)]{laurent}
\bibinfo{author}{{Laurent} P}, \bibinfo{author}{{Le Goff} JM}, \bibinfo{author}{{Burtin} E}, \bibinfo{author}{{Hamilton} JC}, \bibinfo{author}{{Hogg} DW}, \bibinfo{author}{{Myers} A}, \bibinfo{author}{{Ntelis} P}, \bibinfo{author}{{P{\^a}ris} I}, \bibinfo{author}{{Rich} J}, \bibinfo{author}{{Aubourg} E}, \bibinfo{author}{{Bautista} J}, \bibinfo{author}{{Delubac} T}, \bibinfo{author}{{du Mas des Bourboux} H}, \bibinfo{author}{{Eftekharzadeh} S}, \bibinfo{author}{{Palanque Delabrouille} N}, \bibinfo{author}{{Petitjean} P}, \bibinfo{author}{{Rossi} G}, \bibinfo{author}{{Schneider} DP} and  \bibinfo{author}{{Yeche} C} (\bibinfo{year}{2016}), \bibinfo{month}{Nov.}
\bibinfo{title}{{A 14 h$^{-3}$ Gpc$^{3}$ study of cosmic homogeneity using BOSS DR12 quasar sample}}.
\bibinfo{journal}{{\em \jcap}} \bibinfo{volume}{2016} (\bibinfo{number}{11}), \bibinfo{eid}{060}. \bibinfo{doi}{\doi{10.1088/1475-7516/2016/11/060}}.
\eprint{1602.09010}.

\bibtype{Article}%
\bibitem[{Lee} and {Dvorkin}(2020)]{lee_2020_ang_trispec}
\bibinfo{author}{{Lee} H} and  \bibinfo{author}{{Dvorkin} C} (\bibinfo{year}{2020}), \bibinfo{month}{May}.
\bibinfo{title}{{Cosmological angular trispectra and non-Gaussian covariance}}.
\bibinfo{journal}{{\em \jcap}} \bibinfo{volume}{2020} (\bibinfo{number}{5}), \bibinfo{eid}{044}. \bibinfo{doi}{\doi{10.1088/1475-7516/2020/05/044}}.
\eprint{2001.00584}.

\bibtype{Article}%
\bibitem[Levi et al.(2013)]{levi2013desi}
\bibinfo{author}{Levi M}, \bibinfo{author}{Bebek C}, \bibinfo{author}{Beers T}, \bibinfo{author}{Blum R}, \bibinfo{author}{Cahn R}, \bibinfo{author}{Eisenstein D}, \bibinfo{author}{Flaugher B}, \bibinfo{author}{Honscheid K}, \bibinfo{author}{Kron R}, \bibinfo{author}{Lahav O} and  et al. (\bibinfo{year}{2013}).
\bibinfo{title}{The desi experiment, a whitepaper for snowmass 2013}.
\bibinfo{journal}{{\em arXiv preprint arXiv:1308.0847}} .

\bibtype{Misc}%
\bibitem[Lewis(2019)]{lewis2019getdistpythonpackageanalysing}
\bibinfo{author}{Lewis A} (\bibinfo{year}{2019}).
\bibinfo{title}{Getdist: a python package for analysing monte carlo samples}.
\eprint{1910.13970}, \bibinfo{url}{\url{https://arxiv.org/abs/1910.13970}}.

\bibtype{incollection}%
\bibitem[Lifshitz(1992)]{lifshitz_repub}
\bibinfo{author}{Lifshitz E} (\bibinfo{year}{1992}), \bibinfo{title}{15 - on the gravitational stability of the expanding universe}, \bibinfo{editor}{PITAEVSKI L}, (Ed.), \bibinfo{booktitle}{Perspectives in Theoretical Physics}, \bibinfo{publisher}{Pergamon}, \bibinfo{address}{Amsterdam},  \bibinfo{pages}{219--239}, \bibinfo{url}{\url{https://www.sciencedirect.com/science/article/pii/B978008036364650020X}}.

\bibtype{Article}%
\bibitem[{Lii} et al.(1976)]{turbulence_bispectrum_1976}
\bibinfo{author}{{Lii} KS}, \bibinfo{author}{{Rosenblatt} M} and  \bibinfo{author}{{van Atta} C} (\bibinfo{year}{1976}), \bibinfo{month}{Sep.}
\bibinfo{title}{{Bispectral measurements in turbulence}}.
\bibinfo{journal}{{\em Journal of Fluid Mechanics}} \bibinfo{volume}{77}: \bibinfo{pages}{45--62}. \bibinfo{doi}{\doi{10.1017/S0022112076001122}}.

\bibtype{Article}%
\bibitem[Linder(2003)]{linder}
\bibinfo{author}{Linder EV} (\bibinfo{year}{2003}), \bibinfo{month}{Mar}.
\bibinfo{title}{Exploring the expansion history of the universe}.
\bibinfo{journal}{{\em Phys. Rev. Lett.}} \bibinfo{volume}{90}: \bibinfo{pages}{091301}. \bibinfo{doi}{\doi{10.1103/PhysRevLett.90.091301}}.
\bibinfo{url}{\url{https://link.aps.org/doi/10.1103/PhysRevLett.90.091301}}.

\bibtype{Article}%
\bibitem[{Lokas} et al.(1995)]{lokas_kurt_1995}
\bibinfo{author}{{Lokas} EL}, \bibinfo{author}{{Juszkiewicz} R}, \bibinfo{author}{{Weinberg} DH} and  \bibinfo{author}{{Bouchet} FR} (\bibinfo{year}{1995}), \bibinfo{month}{Jun.}
\bibinfo{title}{{Kurtosis of large-scale cosmic fields}}.
\bibinfo{journal}{{\em \mnras}} \bibinfo{volume}{274} (\bibinfo{number}{3}): \bibinfo{pages}{730--744}. \bibinfo{doi}{\doi{10.1093/mnras/274.3.730}}.
\eprint{astro-ph/9407095}.

\bibtype{Article}%
\bibitem[{Luo} and {Schramm}(1993)]{luo_skew_kurt_1993}
\bibinfo{author}{{Luo} X} and  \bibinfo{author}{{Schramm} DN} (\bibinfo{year}{1993}), \bibinfo{month}{May}.
\bibinfo{title}{{Kurtosis, Skewness, and Non-Gaussian Cosmological Density Perturbations}}.
\bibinfo{journal}{{\em \apj}} \bibinfo{volume}{408}: \bibinfo{pages}{33}. \bibinfo{doi}{\doi{10.1086/172567}}.

\bibtype{Article}%
\bibitem[{Ma} and {Bertschinger}(1995)]{ma_bertschinger}
\bibinfo{author}{{Ma} CP} and  \bibinfo{author}{{Bertschinger} E} (\bibinfo{year}{1995}), \bibinfo{month}{Dec.}
\bibinfo{title}{{Cosmological Perturbation Theory in the Synchronous and Conformal Newtonian Gauges}}.
\bibinfo{journal}{{\em \apj}} \bibinfo{volume}{455}: \bibinfo{pages}{7}. \bibinfo{doi}{\doi{10.1086/176550}}.
\eprint{astro-ph/9506072}.

\bibtype{Article}%
\bibitem[Ma and Bertschinger(1995)]{Ma_1995}
\bibinfo{author}{Ma CP} and  \bibinfo{author}{Bertschinger E} (\bibinfo{year}{1995}), \bibinfo{month}{Dec.}
\bibinfo{title}{Cosmological perturbation theory in the synchronous and conformal newtonian gauges}.
\bibinfo{journal}{{\em The Astrophysical Journal}} \bibinfo{volume}{455}: \bibinfo{pages}{7}.
ISSN \bibinfo{issn}{1538-4357}. \bibinfo{doi}{\doi{10.1086/176550}}.
\bibinfo{url}{\url{http://dx.doi.org/10.1086/176550}}.

\bibtype{Article}%
\bibitem[{Maccarone}(2013)]{maccarone_biphase}
\bibinfo{author}{{Maccarone} TJ} (\bibinfo{year}{2013}), \bibinfo{month}{Nov.}
\bibinfo{title}{{The biphase explained: understanding the asymmetries in coupled Fourier components of astronomical time series}}.
\bibinfo{journal}{{\em \mnras}} \bibinfo{volume}{435} (\bibinfo{number}{4}): \bibinfo{pages}{3547--3558}. \bibinfo{doi}{\doi{10.1093/mnras/stt1546}}.
\eprint{1308.3150}.

\bibtype{Article}%
\bibitem[{Maccarone} and {Schnittman}(2005)]{maccarone_bicoherence}
\bibinfo{author}{{Maccarone} TJ} and  \bibinfo{author}{{Schnittman} JD} (\bibinfo{year}{2005}), \bibinfo{month}{Feb.}
\bibinfo{title}{{The bicoherence as a diagnostic for models of high-frequency quasi-periodic oscillations}}.
\bibinfo{journal}{{\em \mnras}} \bibinfo{volume}{357} (\bibinfo{number}{1}): \bibinfo{pages}{12--16}. \bibinfo{doi}{\doi{10.1111/j.1365-2966.2004.08615.x}}.
\eprint{astro-ph/0411266}.

\bibtype{Misc}%
\bibitem[March(2013)]{march_phd}
\bibinfo{author}{March W} (\bibinfo{year}{2013}).
\bibinfo{title}{Multi-tree algorithms for computational statistics and physics: Ph.d. thesis}.
\eprint{http://hdl.handle.net/1853/49116}, \bibinfo{url}{\url{http://hdl.handle.net/1853/49116}}.

\bibtype{Article}%
\bibitem[{Mar{\'\i}n}(2011)]{marin_2011}
\bibinfo{author}{{Mar{\'\i}n} F} (\bibinfo{year}{2011}), \bibinfo{month}{Aug.}
\bibinfo{title}{{The Large-scale Three-point Correlation Function of Sloan Digital Sky Survey Luminous Red Galaxies}}.
\bibinfo{journal}{{\em \apj}} \bibinfo{volume}{737} (\bibinfo{number}{2}), \bibinfo{eid}{97}. \bibinfo{doi}{\doi{10.1088/0004-637X/737/2/97}}.
\eprint{1011.4530}.

\bibtype{Article}%
\bibitem[{Mar{\'\i}n} et al.(2008)]{marin}
\bibinfo{author}{{Mar{\'\i}n} FA}, \bibinfo{author}{{Wechsler} RH}, \bibinfo{author}{{Frieman} JA} and  \bibinfo{author}{{Nichol} RC} (\bibinfo{year}{2008}), \bibinfo{month}{Jan.}
\bibinfo{title}{{Modeling the Galaxy Three-Point Correlation Function}}.
\bibinfo{journal}{{\em \apj}} \bibinfo{volume}{672} (\bibinfo{number}{2}): \bibinfo{pages}{849--860}. \bibinfo{doi}{\doi{10.1086/523628}}.
\eprint{0704.0255}.

\bibtype{Article}%
\bibitem[Markovic(2023)]{markovic2023cosmological}
\bibinfo{author}{Markovic D} (\bibinfo{year}{2023}).
\bibinfo{title}{The cosmological analysis of the sdss/boss data from the effective field theory of large-scale structure}.
\bibinfo{journal}{{\em (No Title)}} .

\bibtype{Article}%
\bibitem[{Matarrese} et al.(1998)]{matarrese_1998}
\bibinfo{author}{{Matarrese} S}, \bibinfo{author}{{Mollerach} S} and  \bibinfo{author}{{Bruni} M} (\bibinfo{year}{1998}), \bibinfo{month}{Aug.}
\bibinfo{title}{{Relativistic second-order perturbations of the Einstein-de Sitter universe}}.
\bibinfo{journal}{{\em \prd}} \bibinfo{volume}{58} (\bibinfo{number}{4}), \bibinfo{eid}{043504}. \bibinfo{doi}{\doi{10.1103/PhysRevD.58.043504}}.
\eprint{astro-ph/9707278}.

\bibtype{Article}%
\bibitem[{Matsubara} and {Suto}(1994)]{matsubara_1994}
\bibinfo{author}{{Matsubara} T} and  \bibinfo{author}{{Suto} Y} (\bibinfo{year}{1994}), \bibinfo{month}{Jan.}
\bibinfo{title}{{Scale Dependence of Three-Point Correlation Functions: Model Predictions and Redshift-Space Contamination}}.
\bibinfo{journal}{{\em \apj}} \bibinfo{volume}{420}: \bibinfo{pages}{497}. \bibinfo{doi}{\doi{10.1086/173580}}.

\bibtype{Article}%
\bibitem[McBride et al.(2011{\natexlab{a}})]{mcbride_1}
\bibinfo{author}{McBride CK} and  et al. (\bibinfo{year}{2011}{\natexlab{a}}).
\bibinfo{title}{{Three-Point Correlation Functions of SDSS Galaxies: Constraining Galaxy-Mass Bias}}.
\bibinfo{journal}{{\em Astrophys. J.}} \bibinfo{volume}{739}: \bibinfo{pages}{85}. \bibinfo{doi}{\doi{10.1088/0004-637X/739/2/85}}.
\eprint{1012.3462}.

\bibtype{Article}%
\bibitem[McBride et al.(2011{\natexlab{b}})]{mcbride_2}
\bibinfo{author}{McBride CK} and  et al. (\bibinfo{year}{2011}{\natexlab{b}}).
\bibinfo{title}{{Three-Point Correlation Functions of SDSS Galaxies: Luminosity and Color Dependence in Redshift and Projected Space}}.
\bibinfo{journal}{{\em Astrophys. J.}} \bibinfo{volume}{726}: \bibinfo{pages}{13}. \bibinfo{doi}{\doi{10.1088/0004-637X/726/1/13}}.
\eprint{1007.2414}.

\bibtype{Article}%
\bibitem[{McQuinn}(2021)]{mcquinn_linear_info}
\bibinfo{author}{{McQuinn} M} (\bibinfo{year}{2021}), \bibinfo{month}{Jun.}
\bibinfo{title}{{On the primordial information available to galaxy redshift surveys}}.
\bibinfo{journal}{{\em \jcap}} \bibinfo{volume}{2021} (\bibinfo{number}{6}), \bibinfo{eid}{024}. \bibinfo{doi}{\doi{10.1088/1475-7516/2021/06/024}}.
\eprint{2008.12312}.

\bibtype{Article}%
\bibitem[{Melott} and {Fry}(1986)]{melott}
\bibinfo{author}{{Melott} AL} and  \bibinfo{author}{{Fry} JN} (\bibinfo{year}{1986}), \bibinfo{month}{Jun.}
\bibinfo{title}{{An Independent Test of Biased Galaxy Formation with Cold Particles: The Three-Point Function}}.
\bibinfo{journal}{{\em \apj}} \bibinfo{volume}{305}: \bibinfo{pages}{1}. \bibinfo{doi}{\doi{10.1086/164222}}.

\bibtype{Article}%
\bibitem[{Mohammed} et al.(2017{\natexlab{a}})]{mohamed}
\bibinfo{author}{{Mohammed} I}, \bibinfo{author}{{Seljak} U} and  \bibinfo{author}{{Vlah} Z} (\bibinfo{year}{2017}{\natexlab{a}}), \bibinfo{month}{Apr.}
\bibinfo{title}{{Perturbative approach to covariance matrix of the matter power spectrum}}.
\bibinfo{journal}{{\em \mnras}} \bibinfo{volume}{466} (\bibinfo{number}{1}): \bibinfo{pages}{780--797}. \bibinfo{doi}{\doi{10.1093/mnras/stw3196}}.
\eprint{1607.00043}.

\bibtype{Article}%
\bibitem[{Mohammed} et al.(2017{\natexlab{b}})]{sugi_covar}
\bibinfo{author}{{Mohammed} I}, \bibinfo{author}{{Seljak} U} and  \bibinfo{author}{{Vlah} Z} (\bibinfo{year}{2017}{\natexlab{b}}), \bibinfo{month}{Apr.}
\bibinfo{title}{{Perturbative approach to covariance matrix of the matter power spectrum}}.
\bibinfo{journal}{{\em \mnras}} \bibinfo{volume}{466} (\bibinfo{number}{1}): \bibinfo{pages}{780--797}. \bibinfo{doi}{\doi{10.1093/mnras/stw3196}}.
\eprint{1607.00043}.

\bibtype{Article}%
\bibitem[{Moresco} et al.(2017)]{vipers_2017}
\bibinfo{author}{{Moresco} M}, \bibinfo{author}{{Marulli} F}, \bibinfo{author}{{Moscardini} L}, \bibinfo{author}{{Branchini} E}, \bibinfo{author}{{Cappi} A}, \bibinfo{author}{{Davidzon} I}, \bibinfo{author}{{Granett} BR}, \bibinfo{author}{{de la Torre} S}, \bibinfo{author}{{Guzzo} L}, \bibinfo{author}{{Abbas} U}, \bibinfo{author}{{Adami} C}, \bibinfo{author}{{Arnouts} S}, \bibinfo{author}{{Bel} J}, \bibinfo{author}{{Bolzonella} M}, \bibinfo{author}{{Bottini} D}, \bibinfo{author}{{Carbone} C}, \bibinfo{author}{{Coupon} J}, \bibinfo{author}{{Cucciati} O}, \bibinfo{author}{{De Lucia} G}, \bibinfo{author}{{Franzetti} P}, \bibinfo{author}{{Fritz} A}, \bibinfo{author}{{Fumana} M}, \bibinfo{author}{{Garilli} B}, \bibinfo{author}{{Ilbert} O}, \bibinfo{author}{{Iovino} A}, \bibinfo{author}{{Krywult} J}, \bibinfo{author}{{Le Brun} V}, \bibinfo{author}{{Le F{\`e}vre} O}, \bibinfo{author}{{Ma{\l}ek} K}, \bibinfo{author}{{McCracken} HJ}, \bibinfo{author}{{Polletta} M}, \bibinfo{author}{{Pollo} A},
  \bibinfo{author}{{Scodeggio} M}, \bibinfo{author}{{Tasca} LAM}, \bibinfo{author}{{Tojeiro} R}, \bibinfo{author}{{Vergani} D} and  \bibinfo{author}{{Zanichelli} A} (\bibinfo{year}{2017}), \bibinfo{month}{Aug.}
\bibinfo{title}{{The VIMOS Public Extragalactic Redshift Survey (VIPERS) . Exploring the dependence of the three-point correlation function on stellar mass and luminosity at 0.5 <z < 1.1}}.
\bibinfo{journal}{{\em \aap}} \bibinfo{volume}{604}, \bibinfo{eid}{A133}. \bibinfo{doi}{\doi{10.1051/0004-6361/201628589}}.
\eprint{1603.08924}.

\bibtype{Article}%
\bibitem[{Moresco} et al.(2021{\natexlab{a}})]{moresco_bao}
\bibinfo{author}{{Moresco} M}, \bibinfo{author}{{Veropalumbo} A}, \bibinfo{author}{{Marulli} F}, \bibinfo{author}{{Moscardini} L} and  \bibinfo{author}{{Cimatti} A} (\bibinfo{year}{2021}{\natexlab{a}}), \bibinfo{month}{Oct.}
\bibinfo{title}{{C$^{3}$: Cluster Clustering Cosmology. II. First Detection of the Baryon Acoustic Oscillations Peak in the Three-point Correlation Function of Galaxy Clusters}}.
\bibinfo{journal}{{\em \apj}} \bibinfo{volume}{919} (\bibinfo{number}{2}), \bibinfo{eid}{144}. \bibinfo{doi}{\doi{10.3847/1538-4357/ac10c9}}.
\eprint{2011.04665}.

\bibtype{Article}%
\bibitem[{Moresco} et al.(2021{\natexlab{b}})]{moresco_3pcf_bao}
\bibinfo{author}{{Moresco} M}, \bibinfo{author}{{Veropalumbo} A}, \bibinfo{author}{{Marulli} F}, \bibinfo{author}{{Moscardini} L} and  \bibinfo{author}{{Cimatti} A} (\bibinfo{year}{2021}{\natexlab{b}}), \bibinfo{month}{Oct.}
\bibinfo{title}{{C$^{3}$: Cluster Clustering Cosmology. II. First Detection of the Baryon Acoustic Oscillations Peak in the Three-point Correlation Function of Galaxy Clusters}}.
\bibinfo{journal}{{\em \apj}} \bibinfo{volume}{919} (\bibinfo{number}{2}), \bibinfo{eid}{144}. \bibinfo{doi}{\doi{10.3847/1538-4357/ac10c9}}.
\eprint{2011.04665}.

\bibtype{Article}%
\bibitem[{Moretti} et al.(2024)]{moretti_trispec_odd}
\bibinfo{author}{{Moretti} T}, \bibinfo{author}{{Bartolo} N} and  \bibinfo{author}{{Greco} A} (\bibinfo{year}{2024}), \bibinfo{month}{Oct.}
\bibinfo{title}{{Breaking Parity: the case of the Trispectrum from Chiral Scalar-Tensor Theories of Gravity}}.
\bibinfo{journal}{{\em arXiv e-prints}} , \bibinfo{eid}{arXiv:2410.11801}\bibinfo{doi}{\doi{10.48550/arXiv.2410.11801}}.
\eprint{2410.11801}.

\bibtype{Article}%
\bibitem[{Munthe-Kaas} et al.(2025)]{iso_isserlis}
\bibinfo{author}{{Munthe-Kaas} HZ}, \bibinfo{author}{{Verdier} O} and  \bibinfo{author}{{Vilmart} G} (\bibinfo{year}{2025}), \bibinfo{month}{Mar.}
\bibinfo{title}{{A short proof of Isserlis' theorem}}.
\bibinfo{journal}{{\em arXiv e-prints}} , \bibinfo{eid}{arXiv:2503.01588}\bibinfo{doi}{\doi{10.48550/arXiv.2503.01588}}.
\eprint{2503.01588}.

\bibtype{Article}%
\bibitem[{Nichol} et al.(2006)]{nichol_2006}
\bibinfo{author}{{Nichol} RC}, \bibinfo{author}{{Sheth} RK}, \bibinfo{author}{{Suto} Y}, \bibinfo{author}{{Gray} AJ}, \bibinfo{author}{{Kayo} I}, \bibinfo{author}{{Wechsler} RH}, \bibinfo{author}{{Marin} F}, \bibinfo{author}{{Kulkarni} G}, \bibinfo{author}{{Blanton} M}, \bibinfo{author}{{Connolly} AJ}, \bibinfo{author}{{Gardner} JP}, \bibinfo{author}{{Jain} B}, \bibinfo{author}{{Miller} CJ}, \bibinfo{author}{{Moore} AW}, \bibinfo{author}{{Pope} A}, \bibinfo{author}{{Pun} J}, \bibinfo{author}{{Schneider} D}, \bibinfo{author}{{Schneider} J}, \bibinfo{author}{{Szalay} A}, \bibinfo{author}{{Szapudi} I}, \bibinfo{author}{{Zehavi} I}, \bibinfo{author}{{Bahcall} NA}, \bibinfo{author}{{Csabai} I} and  \bibinfo{author}{{Brinkmann} J} (\bibinfo{year}{2006}), \bibinfo{month}{Jun.}
\bibinfo{title}{{The effect of large-scale structure on the SDSS galaxy three-point correlation function}}.
\bibinfo{journal}{{\em \mnras}} \bibinfo{volume}{368} (\bibinfo{number}{4}): \bibinfo{pages}{1507--1514}. \bibinfo{doi}{\doi{10.1111/j.1365-2966.2006.10239.x}}.
\eprint{astro-ph/0602548}.

\bibtype{Article}%
\bibitem[{Niu} et al.(2023)]{niu_axion_trispectrum}
\bibinfo{author}{{Niu} X}, \bibinfo{author}{{Rahat} MH}, \bibinfo{author}{{Srinivasan} K} and  \bibinfo{author}{{Xue} W} (\bibinfo{year}{2023}), \bibinfo{month}{May}.
\bibinfo{title}{{Parity-odd and even trispectrum from axion inflation}}.
\bibinfo{journal}{{\em \jcap}} \bibinfo{volume}{2023} (\bibinfo{number}{5}), \bibinfo{eid}{018}. \bibinfo{doi}{\doi{10.1088/1475-7516/2023/05/018}}.
\eprint{2211.14324}.

\bibtype{Article}%
\bibitem[{Noble} et al.(2024)]{noble_24}
\bibinfo{author}{{Noble} L}, \bibinfo{author}{{Kamran} M}, \bibinfo{author}{{Majumdar} S}, \bibinfo{author}{{Murmu} CS}, \bibinfo{author}{{Ghara} R}, \bibinfo{author}{{Mellema} G}, \bibinfo{author}{{Iliev} IT} and  \bibinfo{author}{{Pritchard} JR} (\bibinfo{year}{2024}), \bibinfo{month}{Oct.}
\bibinfo{title}{{Impact of the Epoch of Reionization sources on the 21-cm bispectrum}}.
\bibinfo{journal}{{\em \jcap}} \bibinfo{volume}{2024} (\bibinfo{number}{10}), \bibinfo{eid}{003}. \bibinfo{doi}{\doi{10.1088/1475-7516/2024/10/003}}.
\eprint{2406.03118}.

\bibtype{Article}%
\bibitem[Noriega and Aviles(2024)]{noriega2024unveiling}
\bibinfo{author}{Noriega HE} and  \bibinfo{author}{Aviles A} (\bibinfo{year}{2024}).
\bibinfo{title}{Unveiling neutrino masses: Insights from robust (e) boss data analysis and prospects for desi and beyond}.
\bibinfo{journal}{{\em arXiv preprint arXiv:2407.06117}} .

\bibtype{Article}%
\bibitem[{Noriega} et al.(2022)]{folps_nu}
\bibinfo{author}{{Noriega} HE}, \bibinfo{author}{{Aviles} A}, \bibinfo{author}{{Fromenteau} S} and  \bibinfo{author}{{Vargas-Maga{\~n}a} M} (\bibinfo{year}{2022}), \bibinfo{month}{Nov.}
\bibinfo{title}{{Fast computation of non-linear power spectrum in cosmologies with massive neutrinos}}.
\bibinfo{journal}{{\em \jcap}} \bibinfo{volume}{2022} (\bibinfo{number}{11}), \bibinfo{eid}{038}. \bibinfo{doi}{\doi{10.1088/1475-7516/2022/11/038}}.
\eprint{2208.02791}.

\bibtype{Article}%
\bibitem[{Novell-Masot} et al.(2023)]{geoftp}
\bibinfo{author}{{Novell-Masot} S}, \bibinfo{author}{{Gualdi} D}, \bibinfo{author}{{Gil-Mar{\'\i}n} H} and  \bibinfo{author}{{Verde} L} (\bibinfo{year}{2023}), \bibinfo{month}{Nov.}
\bibinfo{title}{{GEO-FPT: a model of the galaxy bispectrum at mildly non-linear scales}}.
\bibinfo{journal}{{\em \jcap}} \bibinfo{volume}{2023} (\bibinfo{number}{11}), \bibinfo{eid}{044}. \bibinfo{doi}{\doi{10.1088/1475-7516/2023/11/044}}.
\eprint{2303.15510}.

\bibtype{Article}%
\bibitem[{Novell-Masot} et al.(2024)]{hector_bis_cov}
\bibinfo{author}{{Novell-Masot} S}, \bibinfo{author}{{Gil-Mar{\'\i}n} H} and  \bibinfo{author}{{Verde} L} (\bibinfo{year}{2024}), \bibinfo{month}{Jun.}
\bibinfo{title}{{On approximations of the redshift-space bispectrum and power spectrum multipoles covariance matrix}}.
\bibinfo{journal}{{\em \jcap}} \bibinfo{volume}{2024} (\bibinfo{number}{6}), \bibinfo{eid}{048}. \bibinfo{doi}{\doi{10.1088/1475-7516/2024/06/048}}.
\eprint{2306.03137}.

\bibtype{Article}%
\bibitem[{O'Brien} et al.(2022)]{obrien}
\bibinfo{author}{{O'Brien} MJ}, \bibinfo{author}{{Burkhart} B} and  \bibinfo{author}{{Shelley} MJ} (\bibinfo{year}{2022}), \bibinfo{month}{May}.
\bibinfo{title}{{Studying Interstellar Turbulence Driving Scales Using the Bispectrum}}.
\bibinfo{journal}{{\em \apj}} \bibinfo{volume}{930} (\bibinfo{number}{2}), \bibinfo{eid}{149}. \bibinfo{doi}{\doi{10.3847/1538-4357/ac6502}}.
\eprint{2203.13334}.

\bibtype{Article}%
\bibitem[{Ortol{\'a} Leonard} et al.(2025)]{william_4pcf}
\bibinfo{author}{{Ortol{\'a} Leonard} W}, \bibinfo{author}{{Slepian} Z} and  \bibinfo{author}{{Hou} J} (\bibinfo{year}{2025}), \bibinfo{month}{Jan.}
\bibinfo{title}{{A model for the redshift-space galaxy 4-point correlation function}}.
\bibinfo{journal}{{\em \jcap}} \bibinfo{volume}{2025} (\bibinfo{number}{1}), \bibinfo{eid}{090}. \bibinfo{doi}{\doi{10.1088/1475-7516/2025/01/090}}.
\eprint{2402.15510}.

\bibtype{Article}%
\bibitem[{Padmanabhan} et al.(2009)]{pad_recon}
\bibinfo{author}{{Padmanabhan} N}, \bibinfo{author}{{White} M} and  \bibinfo{author}{{Cohn} JD} (\bibinfo{year}{2009}), \bibinfo{month}{Mar.}
\bibinfo{title}{{Reconstructing baryon oscillations: A Lagrangian theory perspective}}.
\bibinfo{journal}{{\em \prd}} \bibinfo{volume}{79} (\bibinfo{number}{6}), \bibinfo{eid}{063523}. \bibinfo{doi}{\doi{10.1103/PhysRevD.79.063523}}.
\eprint{0812.2905}.

\bibtype{Article}%
\bibitem[{Pan} and {Szapudi}(2005)]{pan_3pcf}
\bibinfo{author}{{Pan} J} and  \bibinfo{author}{{Szapudi} I} (\bibinfo{year}{2005}), \bibinfo{month}{Oct.}
\bibinfo{title}{{The monopole moment of the three-point correlation function of the two-degree Field Galaxy Redshift Survey}}.
\bibinfo{journal}{{\em \mnras}} \bibinfo{volume}{362} (\bibinfo{number}{4}): \bibinfo{pages}{1363--1370}. \bibinfo{doi}{\doi{10.1111/j.1365-2966.2005.09407.x}}.
\eprint{astro-ph/0505422}.

\bibtype{Article}%
\bibitem[{P{\'a}pai} and {Szapudi}(2008)]{papai}
\bibinfo{author}{{P{\'a}pai} P} and  \bibinfo{author}{{Szapudi} I} (\bibinfo{year}{2008}), \bibinfo{month}{Sep.}
\bibinfo{title}{{Non-perturbative effects of geometry in wide-angle redshift distortions}}.
\bibinfo{journal}{{\em \mnras}} \bibinfo{volume}{389} (\bibinfo{number}{1}): \bibinfo{pages}{292--296}. \bibinfo{doi}{\doi{10.1111/j.1365-2966.2008.13572.x}}.
\eprint{0802.2940}.

\bibtype{Article}%
\bibitem[{Pearson} and {Samushia}(2018)]{pearson_bao_bispec}
\bibinfo{author}{{Pearson} DW} and  \bibinfo{author}{{Samushia} L} (\bibinfo{year}{2018}), \bibinfo{month}{Aug.}
\bibinfo{title}{{A Detection of the Baryon Acoustic Oscillation features in the SDSS BOSS DR12 Galaxy Bispectrum}}.
\bibinfo{journal}{{\em \mnras}} \bibinfo{volume}{478} (\bibinfo{number}{4}): \bibinfo{pages}{4500--4512}. \bibinfo{doi}{\doi{10.1093/mnras/sty1266}}.
\eprint{1712.04970}.

\bibtype{Inproceedings}%
\bibitem[{Peebles}(2001)]{peebles_2001}
\bibinfo{author}{{Peebles} PJE} (\bibinfo{year}{2001}), \bibinfo{month}{Jan.}, \bibinfo{title}{{The Galaxy and Mass N-Point Correlation Functions: a Blast from the Past}}, \bibinfo{editor}{{Mart{\'\i}nez} VJ}, \bibinfo{editor}{{Trimble} V} and  \bibinfo{editor}{{Pons-Border{\'\i}a} MJ}, (Eds.), \bibinfo{booktitle}{Historical Development of Modern Cosmology}, \bibinfo{series}{Astronomical Society of the Pacific Conference Series}, \bibinfo{volume}{252}, pp. \bibinfo{pages}{201}, \eprint{astro-ph/0103040}.

\bibtype{Article}%
\bibitem[{Peebles} and {Groth}(1975{\natexlab{a}})]{peebles_groth-3pcf_1975}
\bibinfo{author}{{Peebles} PJE} and  \bibinfo{author}{{Groth} EJ} (\bibinfo{year}{1975}{\natexlab{a}}), \bibinfo{month}{Feb.}
\bibinfo{title}{{Statistical analysis of catalogs of extragalactic objects. V. Three-point correlation function for the galaxy distribution in the Zwicky catalog.}}
\bibinfo{journal}{{\em \apj}} \bibinfo{volume}{196}: \bibinfo{pages}{1--11}. \bibinfo{doi}{\doi{10.1086/153390}}.

\bibtype{Article}%
\bibitem[{Peebles} and {Groth}(1975{\natexlab{b}})]{peebles_groth_3pcf}
\bibinfo{author}{{Peebles} PJE} and  \bibinfo{author}{{Groth} EJ} (\bibinfo{year}{1975}{\natexlab{b}}), \bibinfo{month}{Feb.}
\bibinfo{title}{{Statistical analysis of catalogs of extragalactic objects. V. Three-point correlation function for the galaxy distribution in the Zwicky catalog.}}
\bibinfo{journal}{{\em \apj}} \bibinfo{volume}{196}: \bibinfo{pages}{1--11}. \bibinfo{doi}{\doi{10.1086/153390}}.

\bibtype{Article}%
\bibitem[{Peebles} and {Yu}(1970)]{peebles_1970}
\bibinfo{author}{{Peebles} PJE} and  \bibinfo{author}{{Yu} JT} (\bibinfo{year}{1970}), \bibinfo{month}{Dec.}
\bibinfo{title}{{Primeval Adiabatic Perturbation in an Expanding Universe}}.
\bibinfo{journal}{{\em \apj}} \bibinfo{volume}{162}: \bibinfo{pages}{815}. \bibinfo{doi}{\doi{10.1086/150713}}.

\bibtype{Article}%
\bibitem[{Percival} et al.(2014)]{percival_covar}
\bibinfo{author}{{Percival} WJ}, \bibinfo{author}{{Ross} AJ}, \bibinfo{author}{{S{\'a}nchez} AG}, \bibinfo{author}{{Samushia} L}, \bibinfo{author}{{Burden} A}, \bibinfo{author}{{Crittenden} R}, \bibinfo{author}{{Cuesta} AJ}, \bibinfo{author}{{Magana} MV}, \bibinfo{author}{{Manera} M}, \bibinfo{author}{{Beutler} F}, \bibinfo{author}{{Chuang} CH}, \bibinfo{author}{{Eisenstein} DJ}, \bibinfo{author}{{Ho} S}, \bibinfo{author}{{McBride} CK}, \bibinfo{author}{{Montesano} F}, \bibinfo{author}{{Padmanabhan} N}, \bibinfo{author}{{Reid} B}, \bibinfo{author}{{Saito} S}, \bibinfo{author}{{Schneider} DP}, \bibinfo{author}{{Seo} HJ}, \bibinfo{author}{{Tojeiro} R} and  \bibinfo{author}{{Weaver} BA} (\bibinfo{year}{2014}), \bibinfo{month}{Apr.}
\bibinfo{title}{{The clustering of Galaxies in the SDSS-III Baryon Oscillation Spectroscopic Survey: including covariance matrix errors}}.
\bibinfo{journal}{{\em \mnras}} \bibinfo{volume}{439} (\bibinfo{number}{3}): \bibinfo{pages}{2531--2541}. \bibinfo{doi}{\doi{10.1093/mnras/stu112}}.
\eprint{1312.4841}.

\bibtype{Article}%
\bibitem[Philcox(2021)]{philcox2021cosmology}
\bibinfo{author}{Philcox OH} (\bibinfo{year}{2021}).
\bibinfo{title}{Cosmology without window functions. ii. cubic estimators for the galaxy bispectrum}.
\bibinfo{journal}{{\em Physical Review D}} \bibinfo{volume}{104} (\bibinfo{number}{12}): \bibinfo{pages}{123529}.

\bibtype{Article}%
\bibitem[{Philcox}(2021{\natexlab{a}})]{phil_no_win_b}
\bibinfo{author}{{Philcox} OHE} (\bibinfo{year}{2021}{\natexlab{a}}), \bibinfo{month}{Dec.}
\bibinfo{title}{{Cosmology without window functions. II. Cubic estimators for the galaxy bispectrum}}.
\bibinfo{journal}{{\em \prd}} \bibinfo{volume}{104} (\bibinfo{number}{12}), \bibinfo{eid}{123529}. \bibinfo{doi}{\doi{10.1103/PhysRevD.104.123529}}.
\eprint{2107.06287}.

\bibtype{Article}%
\bibitem[{Philcox}(2021{\natexlab{b}})]{phil_no_win_p}
\bibinfo{author}{{Philcox} OHE} (\bibinfo{year}{2021}{\natexlab{b}}), \bibinfo{month}{May}.
\bibinfo{title}{{Cosmology without window functions: Quadratic estimators for the galaxy power spectrum}}.
\bibinfo{journal}{{\em \prd}} \bibinfo{volume}{103} (\bibinfo{number}{10}), \bibinfo{eid}{103504}. \bibinfo{doi}{\doi{10.1103/PhysRevD.103.103504}}.
\eprint{2012.09389}.

\bibtype{Article}%
\bibitem[{Philcox}(2022)]{phil_parity}
\bibinfo{author}{{Philcox} OHE} (\bibinfo{year}{2022}), \bibinfo{month}{Sep.}
\bibinfo{title}{{Probing parity violation with the four-point correlation function of BOSS galaxies}}.
\bibinfo{journal}{{\em \prd}} \bibinfo{volume}{106} (\bibinfo{number}{6}), \bibinfo{eid}{063501}. \bibinfo{doi}{\doi{10.1103/PhysRevD.106.063501}}.
\eprint{2206.04227}.

\bibtype{Article}%
\bibitem[{Philcox} and {Eisenstein}(2020)]{philcox_small_scale_p_config_space}
\bibinfo{author}{{Philcox} OHE} and  \bibinfo{author}{{Eisenstein} DJ} (\bibinfo{year}{2020}), \bibinfo{month}{Feb.}
\bibinfo{title}{{Computing the small-scale galaxy power spectrum and bispectrum in configuration space}}.
\bibinfo{journal}{{\em \mnras}} \bibinfo{volume}{492} (\bibinfo{number}{1}): \bibinfo{pages}{1214--1242}. \bibinfo{doi}{\doi{10.1093/mnras/stz3335}}.
\eprint{1912.01010}.

\bibtype{Article}%
\bibitem[Philcox and Ivanov(2022)]{philcox2022boss}
\bibinfo{author}{Philcox OH} and  \bibinfo{author}{Ivanov MM} (\bibinfo{year}{2022}).
\bibinfo{title}{Boss dr12 full-shape cosmology: $\lambda$ cdm constraints from the large-scale galaxy power spectrum and bispectrum monopole}.
\bibinfo{journal}{{\em Physical Review D}} \bibinfo{volume}{105} (\bibinfo{number}{4}): \bibinfo{pages}{043517}.

\bibtype{Article}%
\bibitem[{Philcox} and {Slepian}(2021)]{phil_slep_yama}
\bibinfo{author}{{Philcox} OHE} and  \bibinfo{author}{{Slepian} Z} (\bibinfo{year}{2021}), \bibinfo{month}{Jun.}
\bibinfo{title}{{Beyond the Yamamoto approximation: Anisotropic power spectra and correlation functions with pairwise lines of sight}}.
\bibinfo{journal}{{\em \prd}} \bibinfo{volume}{103} (\bibinfo{number}{12}), \bibinfo{eid}{123509}. \bibinfo{doi}{\doi{10.1103/PhysRevD.103.123509}}.
\eprint{2102.08384}.

\bibtype{Article}%
\bibitem[{Philcox} and {Slepian}(2022)]{phil_nd}
\bibinfo{author}{{Philcox} OHE} and  \bibinfo{author}{{Slepian} Z} (\bibinfo{year}{2022}), \bibinfo{month}{Aug.}
\bibinfo{title}{{Efficient computation of N-point correlation functions in D dimensions}}.
\bibinfo{journal}{{\em Proceedings of the National Academy of Science}} \bibinfo{volume}{119} (\bibinfo{number}{33}), \bibinfo{eid}{e2111366119}. \bibinfo{doi}{\doi{10.1073/pnas.2111366119}}.
\eprint{2106.10278}.

\bibtype{Article}%
\bibitem[{Philcox} et al.(2021)]{phil_4pcf}
\bibinfo{author}{{Philcox} OHE}, \bibinfo{author}{{Hou} J} and  \bibinfo{author}{{Slepian} Z} (\bibinfo{year}{2021}), \bibinfo{month}{Aug.}
\bibinfo{title}{{A First Detection of the Connected 4-Point Correlation Function of Galaxies Using the BOSS CMASS Sample}}.
\bibinfo{journal}{{\em arXiv e-prints}} , \bibinfo{eid}{arXiv:2108.01670}\bibinfo{doi}{\doi{10.48550/arXiv.2108.01670}}.
\eprint{2108.01670}.

\bibtype{Article}%
\bibitem[{Philcox} et al.(2022)]{encore}
\bibinfo{author}{{Philcox} OHE}, \bibinfo{author}{{Slepian} Z}, \bibinfo{author}{{Hou} J}, \bibinfo{author}{{Warner} C}, \bibinfo{author}{{Cahn} RN} and  \bibinfo{author}{{Eisenstein} DJ} (\bibinfo{year}{2022}), \bibinfo{month}{Jan.}
\bibinfo{title}{{ENCORE: an O (N$_{g}$$^{2}$) estimator for galaxy N-point correlation functions}}.
\bibinfo{journal}{{\em \mnras}} \bibinfo{volume}{509} (\bibinfo{number}{2}): \bibinfo{pages}{2457--2481}. \bibinfo{doi}{\doi{10.1093/mnras/stab3025}}.
\eprint{2105.08722}.

\bibtype{Article}%
\bibitem[{Pinol} et al.(2017)]{pinol}
\bibinfo{author}{{Pinol} L}, \bibinfo{author}{{Cahn} RN}, \bibinfo{author}{{Hand} N}, \bibinfo{author}{{Seljak} U} and  \bibinfo{author}{{White} M} (\bibinfo{year}{2017}), \bibinfo{month}{Apr.}
\bibinfo{title}{{Imprint of DESI fiber assignment on the anisotropic power spectrum of emission line galaxies}}.
\bibinfo{journal}{{\em \jcap}} \bibinfo{volume}{2017} (\bibinfo{number}{4}), \bibinfo{eid}{008}. \bibinfo{doi}{\doi{10.1088/1475-7516/2017/04/008}}.
\eprint{1611.05007}.

\bibtype{Article}%
\bibitem[{Planck Collaboration} et al.(2020)]{planck_final}
\bibinfo{author}{{Planck Collaboration}}, \bibinfo{author}{{Aghanim} N}, \bibinfo{author}{{Akrami} Y}, \bibinfo{author}{{Arroja} F}, \bibinfo{author}{{Ashdown} M}, \bibinfo{author}{{Aumont} J}, \bibinfo{author}{{Baccigalupi} C}, \bibinfo{author}{{Ballardini} M}, \bibinfo{author}{{Banday} AJ}, \bibinfo{author}{{Barreiro} RB}, \bibinfo{author}{{Bartolo} N}, \bibinfo{author}{{Basak} S}, \bibinfo{author}{{Battye} R}, \bibinfo{author}{{Benabed} K}, \bibinfo{author}{{Bernard} JP}, \bibinfo{author}{{Bersanelli} M}, \bibinfo{author}{{Bielewicz} P}, \bibinfo{author}{{Bock} JJ}, \bibinfo{author}{{Bond} JR}, \bibinfo{author}{{Borrill} J}, \bibinfo{author}{{Bouchet} FR}, \bibinfo{author}{{Boulanger} F}, \bibinfo{author}{{Bucher} M}, \bibinfo{author}{{Burigana} C}, \bibinfo{author}{{Butler} RC}, \bibinfo{author}{{Calabrese} E}, \bibinfo{author}{{Cardoso} JF}, \bibinfo{author}{{Carron} J}, \bibinfo{author}{{Casaponsa} B}, \bibinfo{author}{{Challinor} A}, \bibinfo{author}{{Chiang} HC}, \bibinfo{author}{{Colombo} LPL},
  \bibinfo{author}{{Combet} C}, \bibinfo{author}{{Contreras} D}, \bibinfo{author}{{Crill} BP}, \bibinfo{author}{{Cuttaia} F}, \bibinfo{author}{{de Bernardis} P}, \bibinfo{author}{{de Zotti} G}, \bibinfo{author}{{Delabrouille} J}, \bibinfo{author}{{Delouis} JM}, \bibinfo{author}{{D{\'e}sert} FX}, \bibinfo{author}{{Di Valentino} E}, \bibinfo{author}{{Dickinson} C}, \bibinfo{author}{{Diego} JM}, \bibinfo{author}{{Donzelli} S}, \bibinfo{author}{{Dor{\'e}} O}, \bibinfo{author}{{Douspis} M}, \bibinfo{author}{{Ducout} A}, \bibinfo{author}{{Dupac} X}, \bibinfo{author}{{Efstathiou} G}, \bibinfo{author}{{Elsner} F}, \bibinfo{author}{{En{\ss}lin} TA}, \bibinfo{author}{{Eriksen} HK}, \bibinfo{author}{{Falgarone} E}, \bibinfo{author}{{Fantaye} Y}, \bibinfo{author}{{Fergusson} J}, \bibinfo{author}{{Fernandez-Cobos} R}, \bibinfo{author}{{Finelli} F}, \bibinfo{author}{{Forastieri} F}, \bibinfo{author}{{Frailis} M}, \bibinfo{author}{{Franceschi} E}, \bibinfo{author}{{Frolov} A}, \bibinfo{author}{{Galeotta} S},
  \bibinfo{author}{{Galli} S}, \bibinfo{author}{{Ganga} K}, \bibinfo{author}{{G{\'e}nova-Santos} RT}, \bibinfo{author}{{Gerbino} M}, \bibinfo{author}{{Ghosh} T}, \bibinfo{author}{{Gonz{\'a}lez-Nuevo} J}, \bibinfo{author}{{G{\'o}rski} KM}, \bibinfo{author}{{Gratton} S}, \bibinfo{author}{{Gruppuso} A}, \bibinfo{author}{{Gudmundsson} JE}, \bibinfo{author}{{Hamann} J}, \bibinfo{author}{{Handley} W}, \bibinfo{author}{{Hansen} FK}, \bibinfo{author}{{Helou} G}, \bibinfo{author}{{Herranz} D}, \bibinfo{author}{{Hildebrandt} SR}, \bibinfo{author}{{Hivon} E}, \bibinfo{author}{{Huang} Z}, \bibinfo{author}{{Jaffe} AH}, \bibinfo{author}{{Jones} WC}, \bibinfo{author}{{Karakci} A}, \bibinfo{author}{{Keih{\"a}nen} E}, \bibinfo{author}{{Keskitalo} R}, \bibinfo{author}{{Kiiveri} K}, \bibinfo{author}{{Kim} J}, \bibinfo{author}{{Kisner} TS}, \bibinfo{author}{{Knox} L}, \bibinfo{author}{{Krachmalnicoff} N}, \bibinfo{author}{{Kunz} M}, \bibinfo{author}{{Kurki-Suonio} H}, \bibinfo{author}{{Lagache} G}, \bibinfo{author}{{Lamarre}
  JM}, \bibinfo{author}{{Langer} M}, \bibinfo{author}{{Lasenby} A}, \bibinfo{author}{{Lattanzi} M}, \bibinfo{author}{{Lawrence} CR}, \bibinfo{author}{{Le Jeune} M}, \bibinfo{author}{{Leahy} JP}, \bibinfo{author}{{Lesgourgues} J}, \bibinfo{author}{{Levrier} F}, \bibinfo{author}{{Lewis} A}, \bibinfo{author}{{Liguori} M}, \bibinfo{author}{{Lilje} PB}, \bibinfo{author}{{Lilley} M}, \bibinfo{author}{{Lindholm} V}, \bibinfo{author}{{L{\'o}pez-Caniego} M}, \bibinfo{author}{{Lubin} PM}, \bibinfo{author}{{Ma} YZ}, \bibinfo{author}{{Mac{\'\i}as-P{\'e}rez} JF}, \bibinfo{author}{{Maggio} G}, \bibinfo{author}{{Maino} D}, \bibinfo{author}{{Mandolesi} N}, \bibinfo{author}{{Mangilli} A}, \bibinfo{author}{{Marcos-Caballero} A}, \bibinfo{author}{{Maris} M}, \bibinfo{author}{{Martin} PG}, \bibinfo{author}{{Martinelli} M}, \bibinfo{author}{{Mart{\'\i}nez-Gonz{\'a}lez} E}, \bibinfo{author}{{Matarrese} S}, \bibinfo{author}{{Mauri} N}, \bibinfo{author}{{McEwen} JD}, \bibinfo{author}{{Meerburg} PD}, \bibinfo{author}{{Meinhold} PR},
  \bibinfo{author}{{Melchiorri} A}, \bibinfo{author}{{Mennella} A}, \bibinfo{author}{{Migliaccio} M}, \bibinfo{author}{{Millea} M}, \bibinfo{author}{{Mitra} S}, \bibinfo{author}{{Miville-Desch{\^e}nes} MA}, \bibinfo{author}{{Molinari} D}, \bibinfo{author}{{Moneti} A}, \bibinfo{author}{{Montier} L}, \bibinfo{author}{{Morgante} G}, \bibinfo{author}{{Moss} A}, \bibinfo{author}{{Mottet} S}, \bibinfo{author}{{M{\"u}nchmeyer} M}, \bibinfo{author}{{Natoli} P}, \bibinfo{author}{{N{\o}rgaard-Nielsen} HU}, \bibinfo{author}{{Oxborrow} CA}, \bibinfo{author}{{Pagano} L}, \bibinfo{author}{{Paoletti} D}, \bibinfo{author}{{Partridge} B}, \bibinfo{author}{{Patanchon} G}, \bibinfo{author}{{Pearson} TJ}, \bibinfo{author}{{Peel} M}, \bibinfo{author}{{Peiris} HV}, \bibinfo{author}{{Perrotta} F}, \bibinfo{author}{{Pettorino} V}, \bibinfo{author}{{Piacentini} F}, \bibinfo{author}{{Polastri} L}, \bibinfo{author}{{Polenta} G}, \bibinfo{author}{{Puget} JL}, \bibinfo{author}{{Rachen} JP}, \bibinfo{author}{{Reinecke} M},
  \bibinfo{author}{{Remazeilles} M}, \bibinfo{author}{{Renault} C}, \bibinfo{author}{{Renzi} A}, \bibinfo{author}{{Rocha} G}, \bibinfo{author}{{Rosset} C}, \bibinfo{author}{{Roudier} G}, \bibinfo{author}{{Rubi{\~n}o-Mart{\'\i}n} JA}, \bibinfo{author}{{Ruiz-Granados} B}, \bibinfo{author}{{Salvati} L}, \bibinfo{author}{{Sandri} M}, \bibinfo{author}{{Savelainen} M}, \bibinfo{author}{{Scott} D}, \bibinfo{author}{{Shellard} EPS}, \bibinfo{author}{{Shiraishi} M}, \bibinfo{author}{{Sirignano} C}, \bibinfo{author}{{Sirri} G}, \bibinfo{author}{{Spencer} LD}, \bibinfo{author}{{Sunyaev} R}, \bibinfo{author}{{Suur-Uski} AS}, \bibinfo{author}{{Tauber} JA}, \bibinfo{author}{{Tavagnacco} D}, \bibinfo{author}{{Tenti} M}, \bibinfo{author}{{Terenzi} L}, \bibinfo{author}{{Toffolatti} L}, \bibinfo{author}{{Tomasi} M}, \bibinfo{author}{{Trombetti} T}, \bibinfo{author}{{Valiviita} J}, \bibinfo{author}{{Van Tent} B}, \bibinfo{author}{{Vibert} L}, \bibinfo{author}{{Vielva} P}, \bibinfo{author}{{Villa} F}, \bibinfo{author}{{Vittorio}
  N}, \bibinfo{author}{{Wandelt} BD}, \bibinfo{author}{{Wehus} IK}, \bibinfo{author}{{White} M}, \bibinfo{author}{{White} SDM}, \bibinfo{author}{{Zacchei} A} and  \bibinfo{author}{{Zonca} A} (\bibinfo{year}{2020}), \bibinfo{month}{Sep.}
\bibinfo{title}{{Planck 2018 results. I. Overview and the cosmological legacy of Planck}}.
\bibinfo{journal}{{\em \aap}} \bibinfo{volume}{641}, \bibinfo{eid}{A1}. \bibinfo{doi}{\doi{10.1051/0004-6361/201833880}}.
\eprint{1807.06205}.

\bibtype{Article}%
\bibitem[{Portillo} et al.(2018)]{portillo}
\bibinfo{author}{{Portillo} SKN}, \bibinfo{author}{{Slepian} Z}, \bibinfo{author}{{Burkhart} B}, \bibinfo{author}{{Kahraman} S} and  \bibinfo{author}{{Finkbeiner} DP} (\bibinfo{year}{2018}), \bibinfo{month}{Aug.}
\bibinfo{title}{{Developing the 3-point Correlation Function for the Turbulent Interstellar Medium}}.
\bibinfo{journal}{{\em \apj}} \bibinfo{volume}{862} (\bibinfo{number}{2}), \bibinfo{eid}{119}. \bibinfo{doi}{\doi{10.3847/1538-4357/aacb80}}.
\eprint{1711.09907}.

\bibtype{Article}%
\bibitem[{Press} and {Schechter}(1974)]{press_schechter}
\bibinfo{author}{{Press} WH} and  \bibinfo{author}{{Schechter} P} (\bibinfo{year}{1974}), \bibinfo{month}{Feb.}
\bibinfo{title}{{Formation of Galaxies and Clusters of Galaxies by Self-Similar Gravitational Condensation}}.
\bibinfo{journal}{{\em \apj}} \bibinfo{volume}{187}: \bibinfo{pages}{425--438}. \bibinfo{doi}{\doi{10.1086/152650}}.

\bibtype{Article}%
\bibitem[{Press} and {Vishniac}(1980)]{press_vishniac}
\bibinfo{author}{{Press} WH} and  \bibinfo{author}{{Vishniac} ET} (\bibinfo{year}{1980}), \bibinfo{month}{Mar.}
\bibinfo{title}{{Propagation of adiabatic cosmological perturbations through the ERA of matter-radiation decoupling}}.
\bibinfo{journal}{{\em \apj}} \bibinfo{volume}{236}: \bibinfo{pages}{323--334}. \bibinfo{doi}{\doi{10.1086/157749}}.

\bibtype{Article}%
\bibitem[Rampf and Wong(2012)]{Rampf_2012}
\bibinfo{author}{Rampf C} and  \bibinfo{author}{Wong YY} (\bibinfo{year}{2012}), \bibinfo{month}{Jun.}
\bibinfo{title}{Lagrangian perturbations and the matter bispectrum ii: the resummed one-loop correction to the matter bispectrum}.
\bibinfo{journal}{{\em Journal of Cosmology and Astroparticle Physics}} \bibinfo{volume}{2012} (\bibinfo{number}{06}): \bibinfo{pages}{018–018}.
ISSN \bibinfo{issn}{1475-7516}. \bibinfo{doi}{\doi{10.1088/1475-7516/2012/06/018}}.
\bibinfo{url}{\url{http://dx.doi.org/10.1088/1475-7516/2012/06/018}}.

\bibtype{Article}%
\bibitem[{Reinhard} et al.(2024)]{reinhard_axion}
\bibinfo{author}{{Reinhard} M}, \bibinfo{author}{{Slepian} Z}, \bibinfo{author}{{Hou} J} and  \bibinfo{author}{{Greco} A} (\bibinfo{year}{2024}), \bibinfo{month}{Dec.}
\bibinfo{title}{{Full Parity-Violating Trispectrum in Axion Inflation: Reduction to Low-D Integrals}}.
\bibinfo{journal}{{\em arXiv e-prints}} , \bibinfo{eid}{arXiv:2412.16037}\bibinfo{doi}{\doi{10.48550/arXiv.2412.16037}}.
\eprint{2412.16037}.

\bibtype{Inproceedings}%
\bibitem[{Riotto}(1999)]{riotto}
\bibinfo{author}{{Riotto} A} (\bibinfo{year}{1999}), \bibinfo{month}{Jan.}, \bibinfo{title}{{Theories of Baryogenesis}}, \bibinfo{editor}{{Masiero} A}, \bibinfo{editor}{{Senjanovic} G} and  \bibinfo{editor}{{Smirnov} A}, (Eds.), \bibinfo{booktitle}{High Energy Physics and Cosmology, 1998 Summer School}, pp. \bibinfo{pages}{326}, \eprint{hep-ph/9807454}.

\bibtype{Article}%
\bibitem[{Rose} et al.(2021)]{roman_sn}
\bibinfo{author}{{Rose} BM}, \bibinfo{author}{{Baltay} C}, \bibinfo{author}{{Hounsell} R}, \bibinfo{author}{{Macias} P}, \bibinfo{author}{{Rubin} D}, \bibinfo{author}{{Scolnic} D}, \bibinfo{author}{{Aldering} G}, \bibinfo{author}{{Bohlin} R}, \bibinfo{author}{{Dai} M}, \bibinfo{author}{{Deustua} SE}, \bibinfo{author}{{Foley} RJ}, \bibinfo{author}{{Fruchter} A}, \bibinfo{author}{{Galbany} L}, \bibinfo{author}{{Jha} SW}, \bibinfo{author}{{Jones} DO}, \bibinfo{author}{{Joshi} BA}, \bibinfo{author}{{Kelly} PL}, \bibinfo{author}{{Kessler} R}, \bibinfo{author}{{Kirshner} RP}, \bibinfo{author}{{Mandel} KS}, \bibinfo{author}{{Perlmutter} S}, \bibinfo{author}{{Pierel} J}, \bibinfo{author}{{Qu} H}, \bibinfo{author}{{Rabinowitz} D}, \bibinfo{author}{{Rest} A}, \bibinfo{author}{{Riess} AG}, \bibinfo{author}{{Rodney} S}, \bibinfo{author}{{Sako} M}, \bibinfo{author}{{Siebert} MR}, \bibinfo{author}{{Strolger} L}, \bibinfo{author}{{Suzuki} N}, \bibinfo{author}{{Thorp} S}, \bibinfo{author}{{Van Dyk} SD},
  \bibinfo{author}{{Wang} K}, \bibinfo{author}{{Ward} SM} and  \bibinfo{author}{{Wood-Vasey} WM} (\bibinfo{year}{2021}), \bibinfo{month}{Nov.}
\bibinfo{title}{{A Reference Survey for Supernova Cosmology with the Nancy Grace Roman Space Telescope}}.
\bibinfo{journal}{{\em arXiv e-prints}} , \bibinfo{eid}{arXiv:2111.03081}\bibinfo{doi}{\doi{10.48550/arXiv.2111.03081}}.
\eprint{2111.03081}.

\bibtype{Article}%
\bibitem[Ruggeri et al.(2018)]{ruggeri2018demnuni}
\bibinfo{author}{Ruggeri R}, \bibinfo{author}{Castorina E}, \bibinfo{author}{Carbone C} and  \bibinfo{author}{Sefusatti E} (\bibinfo{year}{2018}).
\bibinfo{title}{Demnuni: Massive neutrinos and the bispectrum of large scale structures}.
\bibinfo{journal}{{\em Journal of Cosmology and Astroparticle Physics}} \bibinfo{volume}{2018} (\bibinfo{number}{03}): \bibinfo{pages}{003}.

\bibtype{Article}%
\bibitem[{Sabiu} et al.(2019)]{sabiu}
\bibinfo{author}{{Sabiu} CG}, \bibinfo{author}{{Hoyle} B}, \bibinfo{author}{{Kim} J} and  \bibinfo{author}{{Li} XD} (\bibinfo{year}{2019}), \bibinfo{month}{Jun.}
\bibinfo{title}{{Graph Database Solution for Higher-order Spatial Statistics in the Era of Big Data}}.
\bibinfo{journal}{{\em \apjs}} \bibinfo{volume}{242} (\bibinfo{number}{2}), \bibinfo{eid}{29}. \bibinfo{doi}{\doi{10.3847/1538-4365/ab22b5}}.
\eprint{1901.00296}.

\bibtype{Article}%
\bibitem[{Sakharov}(1966)]{sakh_1966}
\bibinfo{author}{{Sakharov} AD} (\bibinfo{year}{1966}), \bibinfo{month}{Jan.}
\bibinfo{title}{{The Initial Stage of an Expanding Universe and the Appearance of a Nonuniform Distribution of Matter}}.
\bibinfo{journal}{{\em Soviet Journal of Experimental and Theoretical Physics}} \bibinfo{volume}{22}: \bibinfo{pages}{241}.

\bibtype{Article}%
\bibitem[Sakharov(1967)]{sakh_1967}
\bibinfo{author}{Sakharov AD} (\bibinfo{year}{1967}).
\bibinfo{title}{{Violation of CP Invariance, C asymmetry, and baryon asymmetry of the universe}}.
\bibinfo{journal}{{\em Pisma Zh. Eksp. Teor. Fiz.}} \bibinfo{volume}{5}: \bibinfo{pages}{32--35}. \bibinfo{doi}{\doi{10.1070/PU1991v034n05ABEH002497}}.

\bibtype{Article}%
\bibitem[{Samushia}(2019)]{samushia_sfb}
\bibinfo{author}{{Samushia} L} (\bibinfo{year}{2019}), \bibinfo{month}{Jun.}
\bibinfo{title}{{Proper Fourier decomposition formalism for cosmological fields in spherical shells}}.
\bibinfo{journal}{{\em arXiv e-prints}} , \bibinfo{eid}{arXiv:1906.05866}\bibinfo{doi}{\doi{10.48550/arXiv.1906.05866}}.
\eprint{1906.05866}.

\bibtype{Article}%
\bibitem[{Samushia} et al.(2021)]{sam_npcf}
\bibinfo{author}{{Samushia} L}, \bibinfo{author}{{Slepian} Z} and  \bibinfo{author}{{Villaescusa-Navarro} F} (\bibinfo{year}{2021}), \bibinfo{month}{Jul.}
\bibinfo{title}{{Information content of higher order galaxy correlation functions}}.
\bibinfo{journal}{{\em \mnras}} \bibinfo{volume}{505} (\bibinfo{number}{1}): \bibinfo{pages}{628--641}. \bibinfo{doi}{\doi{10.1093/mnras/stab1199}}.
\eprint{2102.01696}.

\bibtype{Article}%
\bibitem[{Saydjari} et al.(2021)]{saydjari}
\bibinfo{author}{{Saydjari} AK}, \bibinfo{author}{{Portillo} SKN}, \bibinfo{author}{{Slepian} Z}, \bibinfo{author}{{Kahraman} S}, \bibinfo{author}{{Burkhart} B} and  \bibinfo{author}{{Finkbeiner} DP} (\bibinfo{year}{2021}), \bibinfo{month}{Apr.}
\bibinfo{title}{{Classification of Magnetohydrodynamic Simulations Using Wavelet Scattering Transforms}}.
\bibinfo{journal}{{\em \apj}} \bibinfo{volume}{910} (\bibinfo{number}{2}), \bibinfo{eid}{122}. \bibinfo{doi}{\doi{10.3847/1538-4357/abe46d}}.
\eprint{2010.11963}.

\bibtype{Article}%
\bibitem[Schlegel et al.(2022{\natexlab{a}})]{spec_roadmap}
\bibinfo{author}{Schlegel DJ} and  et al. (\bibinfo{collaboration}{DESI}) (\bibinfo{year}{2022}{\natexlab{a}}), \bibinfo{month}{9}.
\bibinfo{title}{{A Spectroscopic Road Map for Cosmic Frontier: DESI, DESI-II, Stage-5}}.
\bibinfo{journal}{{\em arXiv e-prints}} \eprint{2209.03585}.

\bibtype{Article}%
\bibitem[Schlegel et al.(2022{\natexlab{b}})]{megamapper}
\bibinfo{author}{Schlegel DJ} and  et al. (\bibinfo{year}{2022}{\natexlab{b}}), \bibinfo{month}{9}.
\bibinfo{title}{{The MegaMapper: A Stage-5 Spectroscopic Instrument Concept for the Study of Inflation and Dark Energy}}.
\bibinfo{journal}{{\em arXiv e-prints}} \bibinfo{doi}{\doi{10.48550/arXiv.2209.04322}}.
\eprint{2209.04322}.

\bibtype{Article}%
\bibitem[Schmittfull and Vlah(2016)]{Schmittfull_2016}
\bibinfo{author}{Schmittfull M} and  \bibinfo{author}{Vlah Z} (\bibinfo{year}{2016}), \bibinfo{month}{Nov.}
\bibinfo{title}{Reducing the two-loop large-scale structure power spectrum to low-dimensional, radial integrals}.
\bibinfo{journal}{{\em Physical Review D}} \bibinfo{volume}{94} (\bibinfo{number}{10}).
ISSN \bibinfo{issn}{2470-0029}. \bibinfo{doi}{\doi{10.1103/physrevd.94.103530}}.
\bibinfo{url}{\url{http://dx.doi.org/10.1103/PhysRevD.94.103530}}.

\bibtype{Article}%
\bibitem[{Schmittfull} et al.(2015)]{schmitt_npcf_recon}
\bibinfo{author}{{Schmittfull} M}, \bibinfo{author}{{Feng} Y}, \bibinfo{author}{{Beutler} F}, \bibinfo{author}{{Sherwin} B} and  \bibinfo{author}{{Chu} MY} (\bibinfo{year}{2015}), \bibinfo{month}{Dec.}
\bibinfo{title}{{Eulerian BAO reconstructions and N -point statistics}}.
\bibinfo{journal}{{\em \prd}} \bibinfo{volume}{92} (\bibinfo{number}{12}), \bibinfo{eid}{123522}. \bibinfo{doi}{\doi{10.1103/PhysRevD.92.123522}}.
\eprint{1508.06972}.

\bibtype{Article}%
\bibitem[Scoccimarro(2000)]{scoccimarro2000bispectrum}
\bibinfo{author}{Scoccimarro R} (\bibinfo{year}{2000}).
\bibinfo{title}{The bispectrum: from theory to observations}.
\bibinfo{journal}{{\em The Astrophysical Journal}} \bibinfo{volume}{544} (\bibinfo{number}{2}): \bibinfo{pages}{597}.

\bibtype{Article}%
\bibitem[{Scoccimarro}(2015)]{scocc_2015}
\bibinfo{author}{{Scoccimarro} R} (\bibinfo{year}{2015}), \bibinfo{month}{Oct.}
\bibinfo{title}{{Fast estimators for redshift-space clustering}}.
\bibinfo{journal}{{\em \prd}} \bibinfo{volume}{92} (\bibinfo{number}{8}), \bibinfo{eid}{083532}. \bibinfo{doi}{\doi{10.1103/PhysRevD.92.083532}}.
\eprint{1506.02729}.

\bibtype{Article}%
\bibitem[Scoccimarro et al.(1999)]{scoccimarro1999bispectrum}
\bibinfo{author}{Scoccimarro R}, \bibinfo{author}{Couchman H} and  \bibinfo{author}{Frieman JA} (\bibinfo{year}{1999}).
\bibinfo{title}{The bispectrum as a signature of gravitational instability in redshift space}.
\bibinfo{journal}{{\em The Astrophysical Journal}} \bibinfo{volume}{517} (\bibinfo{number}{2}): \bibinfo{pages}{531}.

\bibtype{Article}%
\bibitem[{Scoccimarro} et al.(1999)]{scf_99}
\bibinfo{author}{{Scoccimarro} R}, \bibinfo{author}{{Couchman} HMP} and  \bibinfo{author}{{Frieman} JA} (\bibinfo{year}{1999}), \bibinfo{month}{Jun.}
\bibinfo{title}{{The Bispectrum as a Signature of Gravitational Instability in Redshift Space}}.
\bibinfo{journal}{{\em \apj}} \bibinfo{volume}{517} (\bibinfo{number}{2}): \bibinfo{pages}{531--540}. \bibinfo{doi}{\doi{10.1086/307220}}.
\eprint{astro-ph/9808305}.

\bibtype{Article}%
\bibitem[{Scoccimarro} et al.(2004)]{bisp_png}
\bibinfo{author}{{Scoccimarro} R}, \bibinfo{author}{{Sefusatti} E} and  \bibinfo{author}{{Zaldarriaga} M} (\bibinfo{year}{2004}), \bibinfo{month}{May}.
\bibinfo{title}{{Probing primordial non-Gaussianity with large-scale structure}}.
\bibinfo{journal}{{\em \prd}} \bibinfo{volume}{69} (\bibinfo{number}{10}), \bibinfo{eid}{103513}. \bibinfo{doi}{\doi{10.1103/PhysRevD.69.103513}}.
\eprint{astro-ph/0312286}.

\bibtype{Article}%
\bibitem[{Sefusatti} and {Scoccimarro}(2005)]{sefusatti_trispec_2005}
\bibinfo{author}{{Sefusatti} E} and  \bibinfo{author}{{Scoccimarro} R} (\bibinfo{year}{2005}), \bibinfo{month}{Mar.}
\bibinfo{title}{{Galaxy bias and halo-occupation numbers from large-scale clustering}}.
\bibinfo{journal}{{\em \prd}} \bibinfo{volume}{71} (\bibinfo{number}{6}), \bibinfo{eid}{063001}. \bibinfo{doi}{\doi{10.1103/PhysRevD.71.063001}}.
\eprint{astro-ph/0412626}.

\bibtype{Article}%
\bibitem[{Sefusatti} et al.(2016)]{sefu_2016_fourier_est}
\bibinfo{author}{{Sefusatti} E}, \bibinfo{author}{{Crocce} M}, \bibinfo{author}{{Scoccimarro} R} and  \bibinfo{author}{{Couchman} HMP} (\bibinfo{year}{2016}), \bibinfo{month}{Aug.}
\bibinfo{title}{{Accurate estimators of correlation functions in Fourier space}}.
\bibinfo{journal}{{\em \mnras}} \bibinfo{volume}{460} (\bibinfo{number}{4}): \bibinfo{pages}{3624--3636}. \bibinfo{doi}{\doi{10.1093/mnras/stw1229}}.
\eprint{1512.07295}.

\bibtype{Article}%
\bibitem[{Seljak}(1994)]{seljak_1994}
\bibinfo{author}{{Seljak} U} (\bibinfo{year}{1994}), \bibinfo{month}{Nov.}
\bibinfo{title}{{A Two-Fluid Approximation for Calculating the Cosmic Microwave Background Anisotropies}}.
\bibinfo{journal}{{\em \apjl}} \bibinfo{volume}{435}: \bibinfo{pages}{L87}. \bibinfo{doi}{\doi{10.1086/187601}}.
\eprint{astro-ph/9406050}.

\bibtype{Article}%
\bibitem[{Sellentin} and {Heavens}(2016)]{sellentin}
\bibinfo{author}{{Sellentin} E} and  \bibinfo{author}{{Heavens} AF} (\bibinfo{year}{2016}), \bibinfo{month}{Feb.}
\bibinfo{title}{{Parameter inference with estimated covariance matrices}}.
\bibinfo{journal}{{\em \mnras}} \bibinfo{volume}{456} (\bibinfo{number}{1}): \bibinfo{pages}{L132--L136}. \bibinfo{doi}{\doi{10.1093/mnrasl/slv190}}.
\eprint{1511.05969}.

\bibtype{Article}%
\bibitem[Senatore(2015)]{senatore2015bias}
\bibinfo{author}{Senatore L} (\bibinfo{year}{2015}).
\bibinfo{title}{Bias in the effective field theory of large scale structures}.
\bibinfo{journal}{{\em Journal of Cosmology and Astroparticle Physics}} \bibinfo{volume}{2015} (\bibinfo{number}{11}): \bibinfo{pages}{007}.

\bibtype{Article}%
\bibitem[Senatore and Trevisan(2018)]{senatore2018ir}
\bibinfo{author}{Senatore L} and  \bibinfo{author}{Trevisan G} (\bibinfo{year}{2018}).
\bibinfo{title}{On the ir-resummation in the eftoflss}.
\bibinfo{journal}{{\em Journal of Cosmology and Astroparticle Physics}} \bibinfo{volume}{2018} (\bibinfo{number}{05}): \bibinfo{pages}{019}.

\bibtype{Article}%
\bibitem[Simonovi{\'c} et al.(2018)]{simonovic2018cosmological}
\bibinfo{author}{Simonovi{\'c} M}, \bibinfo{author}{Baldauf T}, \bibinfo{author}{Zaldarriaga M}, \bibinfo{author}{Carrasco JJ} and  \bibinfo{author}{Kollmeier JA} (\bibinfo{year}{2018}).
\bibinfo{title}{Cosmological perturbation theory using the fftlog: formalism and connection to qft loop integrals}.
\bibinfo{journal}{{\em Journal of Cosmology and Astroparticle Physics}} \bibinfo{volume}{2018} (\bibinfo{number}{04}): \bibinfo{pages}{030}.

\bibtype{Article}%
\bibitem[{Slepian}(2021)]{slepian_decoup}
\bibinfo{author}{{Slepian} Z} (\bibinfo{year}{2021}), \bibinfo{month}{Oct.}
\bibinfo{title}{{On decoupling the integrals of cosmological perturbation theory}}.
\bibinfo{journal}{{\em \mnras}} \bibinfo{volume}{507} (\bibinfo{number}{1}): \bibinfo{pages}{1337--1360}. \bibinfo{doi}{\doi{10.1093/mnras/staa1789}}.

\bibtype{Article}%
\bibitem[{Slepian}(2024)]{slepian_const_npcf}
\bibinfo{author}{{Slepian} Z} (\bibinfo{year}{2024}), \bibinfo{month}{Jan.}
\bibinfo{title}{{Algorithm to produce a density field with given two-, three-, and four-point correlation functions}}.
\bibinfo{journal}{{\em RAS Techniques and Instruments}} \bibinfo{volume}{3} (\bibinfo{number}{1}): \bibinfo{pages}{584--592}. \bibinfo{doi}{\doi{10.1093/rasti/rzae028}}.
\eprint{2306.05383}.

\bibtype{Article}%
\bibitem[{Slepian} and {Eisenstein}(2015{\natexlab{a}})]{se_wa}
\bibinfo{author}{{Slepian} Z} and  \bibinfo{author}{{Eisenstein} DJ} (\bibinfo{year}{2015}{\natexlab{a}}), \bibinfo{month}{Oct.}
\bibinfo{title}{{A new look at lines of sight: using Fourier methods for the wide-angle anisotropic 2-point correlation function}}.
\bibinfo{journal}{{\em arXiv e-prints}} , \bibinfo{eid}{arXiv:1510.04809}\bibinfo{doi}{\doi{10.48550/arXiv.1510.04809}}.
\eprint{1510.04809}.

\bibtype{Article}%
\bibitem[{Slepian} and {Eisenstein}(2015{\natexlab{b}})]{se_3pt_alg}
\bibinfo{author}{{Slepian} Z} and  \bibinfo{author}{{Eisenstein} DJ} (\bibinfo{year}{2015}{\natexlab{b}}), \bibinfo{month}{Dec.}
\bibinfo{title}{{Computing the three-point correlation function of galaxies in O(N\^2) time}}.
\bibinfo{journal}{{\em \mnras}} \bibinfo{volume}{454} (\bibinfo{number}{4}): \bibinfo{pages}{4142--4158}. \bibinfo{doi}{\doi{10.1093/mnras/stv2119}}.
\eprint{1506.02040}.

\bibtype{Article}%
\bibitem[{Slepian} and {Eisenstein}(2015{\natexlab{c}})]{se_rv}
\bibinfo{author}{{Slepian} Z} and  \bibinfo{author}{{Eisenstein} DJ} (\bibinfo{year}{2015}{\natexlab{c}}), \bibinfo{month}{Mar.}
\bibinfo{title}{{On the signature of the baryon-dark matter relative velocity in the two- and three-point galaxy correlation functions}}.
\bibinfo{journal}{{\em \mnras}} \bibinfo{volume}{448} (\bibinfo{number}{1}): \bibinfo{pages}{9--26}. \bibinfo{doi}{\doi{10.1093/mnras/stu2627}}.
\eprint{1411.4052}.

\bibtype{Article}%
\bibitem[{Slepian} and {Eisenstein}(2016{\natexlab{a}})]{SE_BAO_2016}
\bibinfo{author}{{Slepian} Z} and  \bibinfo{author}{{Eisenstein} DJ} (\bibinfo{year}{2016}{\natexlab{a}}), \bibinfo{month}{Mar.}
\bibinfo{title}{{A simple analytic treatment of linear growth of structure with baryon acoustic oscillations}}.
\bibinfo{journal}{{\em \mnras}} \bibinfo{volume}{457} (\bibinfo{number}{1}): \bibinfo{pages}{24--37}. \bibinfo{doi}{\doi{10.1093/mnras/stv2889}}.
\eprint{1509.08199}.

\bibtype{Article}%
\bibitem[{Slepian} and {Eisenstein}(2016{\natexlab{b}})]{se_3pcf_ft}
\bibinfo{author}{{Slepian} Z} and  \bibinfo{author}{{Eisenstein} DJ} (\bibinfo{year}{2016}{\natexlab{b}}), \bibinfo{month}{Jan.}
\bibinfo{title}{{Accelerating the two-point and three-point galaxy correlation functions using Fourier transforms}}.
\bibinfo{journal}{{\em \mnras}} \bibinfo{volume}{455} (\bibinfo{number}{1}): \bibinfo{pages}{L31--L35}. \bibinfo{doi}{\doi{10.1093/mnrasl/slv133}}.
\eprint{1506.04746}.

\bibtype{Article}%
\bibitem[{Slepian} and {Eisenstein}(2017{\natexlab{a}})]{se_3pcf_rsd}
\bibinfo{author}{{Slepian} Z} and  \bibinfo{author}{{Eisenstein} DJ} (\bibinfo{year}{2017}{\natexlab{a}}), \bibinfo{month}{Aug.}
\bibinfo{title}{{Modelling the large-scale redshift-space 3-point correlation function of galaxies}}.
\bibinfo{journal}{{\em \mnras}} \bibinfo{volume}{469} (\bibinfo{number}{2}): \bibinfo{pages}{2059--2076}. \bibinfo{doi}{\doi{10.1093/mnras/stx490}}.
\eprint{1607.03109}.

\bibtype{Article}%
\bibitem[{Slepian} and {Eisenstein}(2017{\natexlab{b}})]{se_rsd_3pcf}
\bibinfo{author}{{Slepian} Z} and  \bibinfo{author}{{Eisenstein} DJ} (\bibinfo{year}{2017}{\natexlab{b}}), \bibinfo{month}{Aug.}
\bibinfo{title}{{Modelling the large-scale redshift-space 3-point correlation function of galaxies}}.
\bibinfo{journal}{{\em \mnras}} \bibinfo{volume}{469} (\bibinfo{number}{2}): \bibinfo{pages}{2059--2076}. \bibinfo{doi}{\doi{10.1093/mnras/stx490}}.
\eprint{1607.03109}.

\bibtype{Article}%
\bibitem[{Slepian} and {Eisenstein}(2018)]{se_aniso_3pcf}
\bibinfo{author}{{Slepian} Z} and  \bibinfo{author}{{Eisenstein} DJ} (\bibinfo{year}{2018}), \bibinfo{month}{Aug.}
\bibinfo{title}{{A practical computational method for the anisotropic redshift-space three-point correlation function}}.
\bibinfo{journal}{{\em \mnras}} \bibinfo{volume}{478} (\bibinfo{number}{2}): \bibinfo{pages}{1468--1483}. \bibinfo{doi}{\doi{10.1093/mnras/sty1063}}.
\eprint{1709.10150}.

\bibtype{Article}%
\bibitem[Slepian et al.(2017{\natexlab{a}})]{se_3pcf_bao}
\bibinfo{author}{Slepian Z} and  et al. (\bibinfo{year}{2017}{\natexlab{a}}).
\bibinfo{title}{{Detection of baryon acoustic oscillation features in the large-scale three-point correlation function of SDSS BOSS DR12 CMASS galaxies}}.
\bibinfo{journal}{{\em Mon. Not. Roy. Astron. Soc.}} \bibinfo{volume}{469} (\bibinfo{number}{2}): \bibinfo{pages}{1738--1751}. \bibinfo{doi}{\doi{10.1093/mnras/stx488}}.
\eprint{1607.06097}.

\bibtype{Article}%
\bibitem[Slepian et al.(2017{\natexlab{b}})]{se_boss_3pcf}
\bibinfo{author}{Slepian Z} and  et al. (\bibinfo{year}{2017}{\natexlab{b}}).
\bibinfo{title}{{The large-scale three-point correlation function of the SDSS BOSS DR12 CMASS galaxies}}.
\bibinfo{journal}{{\em Mon. Not. Roy. Astron. Soc.}} \bibinfo{volume}{468} (\bibinfo{number}{1}): \bibinfo{pages}{1070--1083}. \bibinfo{doi}{\doi{10.1093/mnras/stw3234}}.
\eprint{1512.02231}.

\bibtype{Article}%
\bibitem[Slepian et al.(2017{\natexlab{c}})]{se_rv_boss}
\bibinfo{author}{Slepian Z} and  et al. (\bibinfo{year}{2017}{\natexlab{c}}).
\bibinfo{title}{{The large-scale three-point correlation function of the SDSS BOSS DR12 CMASS galaxies}}.
\bibinfo{journal}{{\em Mon. Not. Roy. Astron. Soc.}} \bibinfo{volume}{468} (\bibinfo{number}{1}): \bibinfo{pages}{1070--1083}. \bibinfo{doi}{\doi{10.1093/mnras/stw3234}}.
\eprint{1512.02231}.

\bibtype{Article}%
\bibitem[{Slepian} et al.(2019)]{rotation}
\bibinfo{author}{{Slepian} Z}, \bibinfo{author}{{Li} Y}, \bibinfo{author}{{Schmittfull} M} and  \bibinfo{author}{{Vlah} Z} (\bibinfo{year}{2019}), \bibinfo{month}{Nov.}
\bibinfo{title}{{Rotation method for accelerating multiple-spherical Bessel function integrals against a numerical source function}}.
\bibinfo{journal}{{\em arXiv e-prints}} , \bibinfo{eid}{arXiv:1912.00065}\bibinfo{doi}{\doi{10.48550/arXiv.1912.00065}}.
\eprint{1912.00065}.

\bibtype{Article}%
\bibitem[{Slepian} et al.(2024)]{iso_gen}
\bibinfo{author}{{Slepian} Z}, \bibinfo{author}{{Chellino} J}, \bibinfo{author}{{Hou} J} and  \bibinfo{author}{{Greco} A} (\bibinfo{year}{2024}), \bibinfo{month}{Dec.}
\bibinfo{title}{{On a generating function for the isotropic basis functions and other connected results}}.
\bibinfo{journal}{{\em Journal of Physics A Mathematical General}} \bibinfo{volume}{57} (\bibinfo{number}{50}), \bibinfo{eid}{505203}. \bibinfo{doi}{\doi{10.1088/1751-8121/ad955c}}.
\eprint{2406.15385}.

\bibtype{Article}%
\bibitem[{Smith} and {Zaldarriaga}(2011)]{smith}
\bibinfo{author}{{Smith} KM} and  \bibinfo{author}{{Zaldarriaga} M} (\bibinfo{year}{2011}), \bibinfo{month}{Oct.}
\bibinfo{title}{{Algorithms for bispectra: forecasting, optimal analysis and simulation}}.
\bibinfo{journal}{{\em \mnras}} \bibinfo{volume}{417} (\bibinfo{number}{1}): \bibinfo{pages}{2--19}. \bibinfo{doi}{\doi{10.1111/j.1365-2966.2010.18175.x}}.
\eprint{astro-ph/0612571}.

\bibtype{Article}%
\bibitem[Sotiriou and Faraoni(2010)]{sotiriou2010f}
\bibinfo{author}{Sotiriou TP} and  \bibinfo{author}{Faraoni V} (\bibinfo{year}{2010}).
\bibinfo{title}{f (r) theories of gravity}.
\bibinfo{journal}{{\em Reviews of Modern Physics}} \bibinfo{volume}{82} (\bibinfo{number}{1}): \bibinfo{pages}{451--497}.

\bibtype{Article}%
\bibitem[{Spergel} and {Goldberg}(1999)]{spergel_goldberg_bispec}
\bibinfo{author}{{Spergel} DN} and  \bibinfo{author}{{Goldberg} DM} (\bibinfo{year}{1999}), \bibinfo{month}{May}.
\bibinfo{title}{{Microwave background bispectrum. I. Basic formalism}}.
\bibinfo{journal}{{\em \prd}} \bibinfo{volume}{59} (\bibinfo{number}{10}), \bibinfo{eid}{103001}. \bibinfo{doi}{\doi{10.1103/PhysRevD.59.103001}}.
\eprint{astro-ph/9811252}.

\bibtype{Article}%
\bibitem[Starobinsky(1982)]{Starobinsky:1982ee}
\bibinfo{author}{Starobinsky AA} (\bibinfo{year}{1982}).
\bibinfo{title}{{Dynamics of Phase Transition in the New Inflationary Universe Scenario and Generation of Perturbations}}.
\bibinfo{journal}{{\em Phys. Lett. B}} \bibinfo{volume}{117}: \bibinfo{pages}{175--178}. \bibinfo{doi}{\doi{10.1016/0370-2693(82)90541-X}}.

\bibtype{Article}%
\bibitem[{Steele} and {Baldauf}(2021)]{steele_2021}
\bibinfo{author}{{Steele} T} and  \bibinfo{author}{{Baldauf} T} (\bibinfo{year}{2021}), \bibinfo{month}{May}.
\bibinfo{title}{{Precise calibration of the one-loop trispectrum in the effective field theory of large scale structure}}.
\bibinfo{journal}{{\em \prd}} \bibinfo{volume}{103} (\bibinfo{number}{10}), \bibinfo{eid}{103518}. \bibinfo{doi}{\doi{10.1103/PhysRevD.103.103518}}.
\eprint{2101.10289}.

\bibtype{Article}%
\bibitem[{Stefanyszyn} et al.(2024)]{stefanyszyn}
\bibinfo{author}{{Stefanyszyn} D}, \bibinfo{author}{{Tong} X} and  \bibinfo{author}{{Zhu} Y} (\bibinfo{year}{2024}), \bibinfo{month}{May}.
\bibinfo{title}{{Cosmological correlators through the looking glass: reality, parity, and factorisation}}.
\bibinfo{journal}{{\em Journal of High Energy Physics}} \bibinfo{volume}{2024} (\bibinfo{number}{5}), \bibinfo{eid}{196}. \bibinfo{doi}{\doi{10.1007/JHEP05(2024)196}}.
\eprint{2309.07769}.

\bibtype{Article}%
\bibitem[{Sugiyama} et al.(2019)]{sugi_1}
\bibinfo{author}{{Sugiyama} NS}, \bibinfo{author}{{Saito} S}, \bibinfo{author}{{Beutler} F} and  \bibinfo{author}{{Seo} HJ} (\bibinfo{year}{2019}), \bibinfo{month}{Mar.}
\bibinfo{title}{{A complete FFT-based decomposition formalism for the redshift-space bispectrum}}.
\bibinfo{journal}{{\em \mnras}} \bibinfo{volume}{484} (\bibinfo{number}{1}): \bibinfo{pages}{364--384}. \bibinfo{doi}{\doi{10.1093/mnras/sty3249}}.
\eprint{1803.02132}.

\bibtype{Article}%
\bibitem[{Sugiyama} et al.(2023)]{Sugiyama_BOSS_MG}
\bibinfo{author}{{Sugiyama} NS}, \bibinfo{author}{{Yamauchi} D}, \bibinfo{author}{{Kobayashi} T}, \bibinfo{author}{{Fujita} T}, \bibinfo{author}{{Arai} S}, \bibinfo{author}{{Hirano} S}, \bibinfo{author}{{Saito} S}, \bibinfo{author}{{Beutler} F} and  \bibinfo{author}{{Seo} HJ} (\bibinfo{year}{2023}), \bibinfo{month}{Aug.}
\bibinfo{title}{{New constraints on cosmological modified gravity theories from anisotropic three-point correlation functions of BOSS DR12 galaxies}}.
\bibinfo{journal}{{\em \mnras}} \bibinfo{volume}{523} (\bibinfo{number}{2}): \bibinfo{pages}{3133--3191}. \bibinfo{doi}{\doi{10.1093/mnras/stad1505}}.
\eprint{2302.06808}.

\bibtype{Article}%
\bibitem[{Sunseri} et al.(2023)]{sarabande}
\bibinfo{author}{{Sunseri} J}, \bibinfo{author}{{Slepian} Z}, \bibinfo{author}{{Portillo} S}, \bibinfo{author}{{Hou} J}, \bibinfo{author}{{Kahraman} S} and  \bibinfo{author}{{Finkbeiner} DP} (\bibinfo{year}{2023}), \bibinfo{month}{Jan.}
\bibinfo{title}{{SARABANDE: 3/4 point correlation functions with fast Fourier transforms}}.
\bibinfo{journal}{{\em RAS Techniques and Instruments}} \bibinfo{volume}{2} (\bibinfo{number}{1}): \bibinfo{pages}{62--77}. \bibinfo{doi}{\doi{10.1093/rasti/rzad003}}.
\eprint{2210.10206}.

\bibtype{Article}%
\bibitem[Sunyaev and Zeldovich(1970)]{sunyaev_1970}
\bibinfo{author}{Sunyaev RA} and  \bibinfo{author}{Zeldovich YB} (\bibinfo{year}{1970}).
\bibinfo{title}{{Small-scale fluctuations of relic radiation}}.
\bibinfo{journal}{{\em Astrophys. Space Sci.}} \bibinfo{volume}{7} (\bibinfo{number}{1}): \bibinfo{pages}{3--19}. \bibinfo{doi}{\doi{10.1007/BF00653471}}.

\bibtype{Article}%
\bibitem[{Suto}(1993)]{suto_1993}
\bibinfo{author}{{Suto} Y} (\bibinfo{year}{1993}), \bibinfo{month}{Feb.}
\bibinfo{title}{{Three-point correlation functions and the hierarchical clustering ansatz in low-density cold dark matter universes}}.
\bibinfo{journal}{{\em \apjl}} \bibinfo{volume}{404} (\bibinfo{number}{1}): \bibinfo{pages}{L1--L4}. \bibinfo{doi}{\doi{10.1086/186729}}.

\bibtype{Article}%
\bibitem[{Suto} and {Matsubara}(1994)]{suto_1994_hierarchical_up_to_4}
\bibinfo{author}{{Suto} Y} and  \bibinfo{author}{{Matsubara} T} (\bibinfo{year}{1994}), \bibinfo{month}{Jan.}
\bibinfo{title}{{Departure from Hierarchical Clustering Relations for Two-, Three-, and Four-Point Correlation Functions: Analysis of Cosmological N-Body Simulations}}.
\bibinfo{journal}{{\em \apj}} \bibinfo{volume}{420}: \bibinfo{pages}{504}. \bibinfo{doi}{\doi{10.1086/173581}}.

\bibtype{Article}%
\bibitem[{Szapudi}(2004{\natexlab{a}})]{szapudi_3pcf}
\bibinfo{author}{{Szapudi} I} (\bibinfo{year}{2004}{\natexlab{a}}), \bibinfo{month}{Apr.}
\bibinfo{title}{{Three-Point Statistics from a New Perspective}}.
\bibinfo{journal}{{\em \apjl}} \bibinfo{volume}{605} (\bibinfo{number}{2}): \bibinfo{pages}{L89--L92}. \bibinfo{doi}{\doi{10.1086/420894}}.
\eprint{astro-ph/0404476}.

\bibtype{Article}%
\bibitem[{Szapudi}(2004{\natexlab{b}})]{szapudi_wa}
\bibinfo{author}{{Szapudi} I} (\bibinfo{year}{2004}{\natexlab{b}}), \bibinfo{month}{Oct.}
\bibinfo{title}{{Wide-Angle Redshift Distortions Revisited}}.
\bibinfo{journal}{{\em \apj}} \bibinfo{volume}{614} (\bibinfo{number}{1}): \bibinfo{pages}{51--55}. \bibinfo{doi}{\doi{10.1086/423168}}.
\eprint{astro-ph/0404477}.

\bibtype{Article}%
\bibitem[Szapudi and Szalay(1998)]{szapudi_szalay}
\bibinfo{author}{Szapudi I} and  \bibinfo{author}{Szalay AS} (\bibinfo{year}{1998}), \bibinfo{month}{jan}.
\bibinfo{title}{A new class of estimators for the n-point correlations}.
\bibinfo{journal}{{\em The Astrophysical Journal}} \bibinfo{volume}{494} (\bibinfo{number}{1}): \bibinfo{pages}{L41}. \bibinfo{doi}{\doi{10.1086/311146}}.
\bibinfo{url}{\url{https://dx.doi.org/10.1086/311146}}.

\bibtype{Article}%
\bibitem[{Takada} and {Jain}(2003)]{takada}
\bibinfo{author}{{Takada} M} and  \bibinfo{author}{{Jain} B} (\bibinfo{year}{2003}), \bibinfo{month}{Apr.}
\bibinfo{title}{{The three-point correlation function in cosmology}}.
\bibinfo{journal}{{\em \mnras}} \bibinfo{volume}{340} (\bibinfo{number}{2}): \bibinfo{pages}{580--608}. \bibinfo{doi}{\doi{10.1046/j.1365-8711.2003.06321.x}}.
\eprint{astro-ph/0209167}.

\bibtype{Article}%
\bibitem[{Takahashi} et al.(2020)]{Takahashi_bihalofit}
\bibinfo{author}{{Takahashi} R}, \bibinfo{author}{{Nishimichi} T}, \bibinfo{author}{{Namikawa} T}, \bibinfo{author}{{Taruya} A}, \bibinfo{author}{{Kayo} I}, \bibinfo{author}{{Osato} K}, \bibinfo{author}{{Kobayashi} Y} and  \bibinfo{author}{{Shirasaki} M} (\bibinfo{year}{2020}), \bibinfo{month}{Jun.}
\bibinfo{title}{{Fitting the Nonlinear Matter Bispectrum by the Halofit Approach}}.
\bibinfo{journal}{{\em \apj}} \bibinfo{volume}{895} (\bibinfo{number}{2}), \bibinfo{eid}{113}. \bibinfo{doi}{\doi{10.3847/1538-4357/ab908d}}.
\eprint{1911.07886}.

\bibtype{Article}%
\bibitem[Taylor and Valentine(1999)]{taylor}
\bibinfo{author}{Taylor A} and  \bibinfo{author}{Valentine H} (\bibinfo{year}{1999}), \bibinfo{month}{06}.
\bibinfo{title}{The inverse redshift-space operator: reconstructing cosmological density and velocity fields}.
\bibinfo{journal}{{\em Monthly Notices of the Royal Astronomical Society}} \bibinfo{volume}{306} (\bibinfo{number}{2}): \bibinfo{pages}{491--503}.
ISSN \bibinfo{issn}{0035-8711}. \bibinfo{doi}{\doi{10.1046/j.1365-8711.1999.02535.x}}.
\eprint{https://academic.oup.com/mnras/article-pdf/306/2/491/18635595/306-2-491.pdf}, \bibinfo{url}{\url{https://doi.org/10.1046/j.1365-8711.1999.02535.x}}.

\bibtype{Article}%
\bibitem[{Tegmark} and {Bromley}(1995)]{tegmark_bromley}
\bibinfo{author}{{Tegmark} M} and  \bibinfo{author}{{Bromley} BC} (\bibinfo{year}{1995}), \bibinfo{month}{Nov.}
\bibinfo{title}{{Real-Space Cosmic Fields from Redshift-Space Distributions: A Green's Function Approach}}.
\bibinfo{journal}{{\em \apj}} \bibinfo{volume}{453}: \bibinfo{pages}{533}. \bibinfo{doi}{\doi{10.1086/176415}}.
\eprint{astro-ph/9409038}.

\bibtype{Article}%
\bibitem[Tegmark et al.(1998)]{Tegmark_1998}
\bibinfo{author}{Tegmark M}, \bibinfo{author}{Hamilton AJS}, \bibinfo{author}{Strauss MA}, \bibinfo{author}{Vogeley MS} and  \bibinfo{author}{Szalay AS} (\bibinfo{year}{1998}), \bibinfo{month}{Jun.}
\bibinfo{title}{Measuring the galaxy power spectrum with future redshift surveys}.
\bibinfo{journal}{{\em The Astrophysical Journal}} \bibinfo{volume}{499} (\bibinfo{number}{2}): \bibinfo{pages}{555–576}.
ISSN \bibinfo{issn}{1538-4357}. \bibinfo{doi}{\doi{10.1086/305663}}.
\bibinfo{url}{\url{http://dx.doi.org/10.1086/305663}}.

\bibtype{Article}%
\bibitem[Tegmark et al.(2004)]{Tegmark_2004}
\bibinfo{author}{Tegmark M}, \bibinfo{author}{Strauss MA}, \bibinfo{author}{Blanton MR}, \bibinfo{author}{Abazajian K}, \bibinfo{author}{Dodelson S}, \bibinfo{author}{Sandvik H}, \bibinfo{author}{Wang X}, \bibinfo{author}{Weinberg DH}, \bibinfo{author}{Zehavi I}, \bibinfo{author}{Bahcall NA}, \bibinfo{author}{Hoyle F}, \bibinfo{author}{Schlegel D}, \bibinfo{author}{Scoccimarro R}, \bibinfo{author}{Vogeley MS}, \bibinfo{author}{Berlind A}, \bibinfo{author}{Budavari T}, \bibinfo{author}{Connolly A}, \bibinfo{author}{Eisenstein DJ}, \bibinfo{author}{Finkbeiner D}, \bibinfo{author}{Frieman JA}, \bibinfo{author}{Gunn JE}, \bibinfo{author}{Hui L}, \bibinfo{author}{Jain B}, \bibinfo{author}{Johnston D}, \bibinfo{author}{Kent S}, \bibinfo{author}{Lin H}, \bibinfo{author}{Nakajima R}, \bibinfo{author}{Nichol RC}, \bibinfo{author}{Ostriker JP}, \bibinfo{author}{Pope A}, \bibinfo{author}{Scranton R}, \bibinfo{author}{Seljak U}, \bibinfo{author}{Sheth RK}, \bibinfo{author}{Stebbins A}, \bibinfo{author}{Szalay AS},
  \bibinfo{author}{Szapudi I}, \bibinfo{author}{Xu Y}, \bibinfo{author}{Annis J}, \bibinfo{author}{Brinkmann J}, \bibinfo{author}{Burles S}, \bibinfo{author}{Castander FJ}, \bibinfo{author}{Csabai I}, \bibinfo{author}{Loveday J}, \bibinfo{author}{Doi M}, \bibinfo{author}{Fukugita M}, \bibinfo{author}{Gillespie B}, \bibinfo{author}{Hennessy G}, \bibinfo{author}{Hogg DW}, \bibinfo{author}{Ivezić {\v{Z}}}, \bibinfo{author}{Knapp GR}, \bibinfo{author}{Lamb DQ}, \bibinfo{author}{Lee BC}, \bibinfo{author}{Lupton RH}, \bibinfo{author}{McKay TA}, \bibinfo{author}{Kunszt P}, \bibinfo{author}{Munn JA}, \bibinfo{author}{O’Connell L}, \bibinfo{author}{Peoples J}, \bibinfo{author}{Pier JR}, \bibinfo{author}{Richmond M}, \bibinfo{author}{Rockosi C}, \bibinfo{author}{Schneider DP}, \bibinfo{author}{Stoughton C}, \bibinfo{author}{Tucker DL}, \bibinfo{author}{Vanden~Berk DE}, \bibinfo{author}{Yanny B} and  \bibinfo{author}{York DG} (\bibinfo{year}{2004}), \bibinfo{month}{May}.
\bibinfo{title}{Cosmological parameters from sdss and wmap}.
\bibinfo{journal}{{\em Physical Review D}} \bibinfo{volume}{69} (\bibinfo{number}{10}).
ISSN \bibinfo{issn}{1550-2368}. \bibinfo{doi}{\doi{10.1103/physrevd.69.103501}}.
\bibinfo{url}{\url{http://dx.doi.org/10.1103/PhysRevD.69.103501}}.

\bibtype{Article}%
\bibitem[{Thavanesan}(2025)]{thavanesan_no_go}
\bibinfo{author}{{Thavanesan} A} (\bibinfo{year}{2025}), \bibinfo{month}{Jan.}
\bibinfo{title}{{No-go Theorem for Cosmological Parity Violation}}.
\bibinfo{journal}{{\em arXiv e-prints}} , \bibinfo{eid}{arXiv:2501.06383}\bibinfo{doi}{\doi{10.48550/arXiv.2501.06383}}.
\eprint{2501.06383}.

\bibtype{Article}%
\bibitem[{Tie} et al.(2019)]{Tie_Lya_3}
\bibinfo{author}{{Tie} SS}, \bibinfo{author}{{Weinberg} DH}, \bibinfo{author}{{Martini} P}, \bibinfo{author}{{Zhu} W}, \bibinfo{author}{{Peirani} S}, \bibinfo{author}{{Suarez} T} and  \bibinfo{author}{{Colombi} S} (\bibinfo{year}{2019}), \bibinfo{month}{Aug.}
\bibinfo{title}{{UV background fluctuations and three-point correlations in the large-scale clustering of the Lyman {\ensuremath{\alpha}} forest}}.
\bibinfo{journal}{{\em \mnras}} \bibinfo{volume}{487} (\bibinfo{number}{4}): \bibinfo{pages}{5346--5362}. \bibinfo{doi}{\doi{10.1093/mnras/stz1632}}.
\eprint{1905.02208}.

\bibtype{Article}%
\bibitem[{Tomlinson} and {Jeong}(2023)]{eft_exmt}
\bibinfo{author}{{Tomlinson} J} and  \bibinfo{author}{{Jeong} D} (\bibinfo{year}{2023}), \bibinfo{month}{Aug.}
\bibinfo{title}{{Spherical bispectrum: a novel visualization scheme for facilitating comparisons}}.
\bibinfo{journal}{{\em \jcap}} \bibinfo{volume}{2023} (\bibinfo{number}{8}), \bibinfo{eid}{040}. \bibinfo{doi}{\doi{10.1088/1475-7516/2023/08/040}}.
\eprint{2204.00668}.

\bibtype{Article}%
\bibitem[{Tomlinson} et al.(2019)]{tomlinson_polyspectra}
\bibinfo{author}{{Tomlinson} J}, \bibinfo{author}{{Jeong} D} and  \bibinfo{author}{{Kim} J} (\bibinfo{year}{2019}), \bibinfo{month}{Sep.}
\bibinfo{title}{{Efficient Parallel Algorithm for Estimating Higher-order Polyspectra}}.
\bibinfo{journal}{{\em \aj}} \bibinfo{volume}{158} (\bibinfo{number}{3}), \bibinfo{eid}{116}. \bibinfo{doi}{\doi{10.3847/1538-3881/ab3223}}.
\eprint{1904.11055}.

\bibtype{Article}%
\bibitem[{Trodden}(1999)]{trodden}
\bibinfo{author}{{Trodden} M} (\bibinfo{year}{1999}), \bibinfo{month}{Oct.}
\bibinfo{title}{{Electroweak baryogenesis}}.
\bibinfo{journal}{{\em Reviews of Modern Physics}} \bibinfo{volume}{71} (\bibinfo{number}{5}): \bibinfo{pages}{1463--1500}. \bibinfo{doi}{\doi{10.1103/RevModPhys.71.1463}}.
\eprint{hep-ph/9803479}.

\bibtype{Article}%
\bibitem[Tr{\"o}ster et al.(2020)]{troster2020cosmology}
\bibinfo{author}{Tr{\"o}ster T}, \bibinfo{author}{S{\'a}nchez AG}, \bibinfo{author}{Asgari M}, \bibinfo{author}{Blake C}, \bibinfo{author}{Crocce M}, \bibinfo{author}{Heymans C}, \bibinfo{author}{Hildebrandt H}, \bibinfo{author}{Joachimi B}, \bibinfo{author}{Joudaki S}, \bibinfo{author}{Kannawadi A} and  et al. (\bibinfo{year}{2020}).
\bibinfo{title}{Cosmology from large-scale structure-constraining $\lambda$cdm with boss}.
\bibinfo{journal}{{\em Astronomy \& Astrophysics}} \bibinfo{volume}{633}: \bibinfo{pages}{L10}.

\bibtype{Article}%
\bibitem[{Umeh}(2021)]{umeh}
\bibinfo{author}{{Umeh} O} (\bibinfo{year}{2021}), \bibinfo{month}{May}.
\bibinfo{title}{{Optimal computation of anisotropic galaxy three point correlation function multipoles using 2DFFTLOG formalism}}.
\bibinfo{journal}{{\em \jcap}} \bibinfo{volume}{2021} (\bibinfo{number}{5}), \bibinfo{eid}{035}. \bibinfo{doi}{\doi{10.1088/1475-7516/2021/05/035}}.
\eprint{2011.05889}.

\bibtype{Article}%
\bibitem[{Verde} and {Heavens}(2001)]{verde_2001}
\bibinfo{author}{{Verde} L} and  \bibinfo{author}{{Heavens} AF} (\bibinfo{year}{2001}), \bibinfo{month}{May}.
\bibinfo{title}{{On the Trispectrum as a Gaussian Test for Cosmology}}.
\bibinfo{journal}{{\em \apj}} \bibinfo{volume}{553} (\bibinfo{number}{1}): \bibinfo{pages}{14--24}. \bibinfo{doi}{\doi{10.1086/320656}}.
\eprint{astro-ph/0101143}.

\bibtype{Article}%
\bibitem[{Verde} et al.(1998)]{b1Ombreakingdeg}
\bibinfo{author}{{Verde} L}, \bibinfo{author}{{Heavens} AF}, \bibinfo{author}{{Matarrese} S} and  \bibinfo{author}{{Moscardini} L} (\bibinfo{year}{1998}), \bibinfo{month}{Nov.}
\bibinfo{title}{{Large-scale bias in the Universe - II. Redshift-space bispectrum}}.
\bibinfo{journal}{{\em \mnras}} \bibinfo{volume}{300} (\bibinfo{number}{3}): \bibinfo{pages}{747--756}. \bibinfo{doi}{\doi{10.1046/j.1365-8711.1998.01937.x}}.
\eprint{astro-ph/9806028}.

\bibtype{Article}%
\bibitem[Verde et al.(2002)]{verde20022df}
\bibinfo{author}{Verde L}, \bibinfo{author}{Heavens AF}, \bibinfo{author}{Percival WJ}, \bibinfo{author}{Matarrese S}, \bibinfo{author}{Baugh CM}, \bibinfo{author}{Bland-Hawthorn J}, \bibinfo{author}{Bridges T}, \bibinfo{author}{Cannon R}, \bibinfo{author}{Cole S}, \bibinfo{author}{Colless M} and  et al. (\bibinfo{year}{2002}).
\bibinfo{title}{The 2df galaxy redshift survey: the bias of galaxies and the density of the universe}.
\bibinfo{journal}{{\em Monthly Notices of the Royal Astronomical Society}} \bibinfo{volume}{335} (\bibinfo{number}{2}): \bibinfo{pages}{432--440}.

\bibtype{Article}%
\bibitem[{Veropalumbo} et al.(2022)]{veropalumbo}
\bibinfo{author}{{Veropalumbo} A}, \bibinfo{author}{{Binetti} A}, \bibinfo{author}{{Branchini} E}, \bibinfo{author}{{Moresco} M}, \bibinfo{author}{{Monaco} P}, \bibinfo{author}{{Oddo} A}, \bibinfo{author}{{S{\'a}nchez} AG} and  \bibinfo{author}{{Sefusatti} E} (\bibinfo{year}{2022}), \bibinfo{month}{Sep.}
\bibinfo{title}{{The halo 3-point correlation function: a methodological analysis}}.
\bibinfo{journal}{{\em \jcap}} \bibinfo{volume}{2022} (\bibinfo{number}{9}), \bibinfo{eid}{033}. \bibinfo{doi}{\doi{10.1088/1475-7516/2022/09/033}}.
\eprint{2206.00672}.

\bibtype{Article}%
\bibitem[Villaescusa-Navarro et al.(2020)]{villaescusa2020quijote}
\bibinfo{author}{Villaescusa-Navarro F}, \bibinfo{author}{Hahn C}, \bibinfo{author}{Massara E}, \bibinfo{author}{Banerjee A}, \bibinfo{author}{Delgado AM}, \bibinfo{author}{Ramanah DK}, \bibinfo{author}{Charnock T}, \bibinfo{author}{Giusarma E}, \bibinfo{author}{Li Y}, \bibinfo{author}{Allys E} and  et al. (\bibinfo{year}{2020}).
\bibinfo{title}{The quijote simulations}.
\bibinfo{journal}{{\em The Astrophysical Journal Supplement Series}} \bibinfo{volume}{250} (\bibinfo{number}{1}): \bibinfo{pages}{2}.

\bibtype{Article}%
\bibitem[{Wang} et al.(2004)]{wang_2004_3pcf}
\bibinfo{author}{{Wang} Y}, \bibinfo{author}{{Yang} X}, \bibinfo{author}{{Mo} HJ}, \bibinfo{author}{{van den Bosch} FC} and  \bibinfo{author}{{Chu} Y} (\bibinfo{year}{2004}), \bibinfo{month}{Sep.}
\bibinfo{title}{{The three-point correlation function of galaxies: comparing halo occupation models with observations}}.
\bibinfo{journal}{{\em \mnras}} \bibinfo{volume}{353} (\bibinfo{number}{1}): \bibinfo{pages}{287--300}. \bibinfo{doi}{\doi{10.1111/j.1365-2966.2004.08141.x}}.
\eprint{astro-ph/0404143}.

\bibtype{Article}%
\bibitem[{Wang} et al.(2014)]{white_wang}
\bibinfo{author}{{Wang} L}, \bibinfo{author}{{Reid} B} and  \bibinfo{author}{{White} M} (\bibinfo{year}{2014}), \bibinfo{month}{Jan.}
\bibinfo{title}{{An analytic model for redshift-space distortions}}.
\bibinfo{journal}{{\em \mnras}} \bibinfo{volume}{437} (\bibinfo{number}{1}): \bibinfo{pages}{588--599}. \bibinfo{doi}{\doi{10.1093/mnras/stt1916}}.
\eprint{1306.1804}.

\bibtype{Article}%
\bibitem[{Wang} et al.(2022)]{roman_wang_spec}
\bibinfo{author}{{Wang} Y}, \bibinfo{author}{{Zhai} Z}, \bibinfo{author}{{Alavi} A}, \bibinfo{author}{{Massara} E}, \bibinfo{author}{{Pisani} A}, \bibinfo{author}{{Benson} A}, \bibinfo{author}{{Hirata} CM}, \bibinfo{author}{{Samushia} L}, \bibinfo{author}{{Weinberg} DH}, \bibinfo{author}{{Colbert} J}, \bibinfo{author}{{Dor{\'e}} O}, \bibinfo{author}{{Eifler} T}, \bibinfo{author}{{Heinrich} C}, \bibinfo{author}{{Ho} S}, \bibinfo{author}{{Krause} E}, \bibinfo{author}{{Padmanabhan} N}, \bibinfo{author}{{Spergel} D} and  \bibinfo{author}{{Teplitz} HI} (\bibinfo{year}{2022}), \bibinfo{month}{Mar.}
\bibinfo{title}{{The High Latitude Spectroscopic Survey on the Nancy Grace Roman Space Telescope}}.
\bibinfo{journal}{{\em \apj}} \bibinfo{volume}{928} (\bibinfo{number}{1}), \bibinfo{eid}{1}. \bibinfo{doi}{\doi{10.3847/1538-4357/ac4973}}.
\eprint{2110.01829}.

\bibtype{Article}%
\bibitem[{Wang} et al.(2023)]{triumvirate}
\bibinfo{author}{{Wang} M}, \bibinfo{author}{{Beutler} F} and  \bibinfo{author}{{Sugiyama} N} (\bibinfo{year}{2023}), \bibinfo{month}{Nov.}
\bibinfo{title}{{Triumvirate: A Python/C++ package for three-point clustering measurements}}.
\bibinfo{journal}{{\em The Journal of Open Source Software}} \bibinfo{volume}{8} (\bibinfo{number}{91}), \bibinfo{eid}{5571}. \bibinfo{doi}{\doi{10.21105/joss.05571}}.
\eprint{2304.03643}.

\bibtype{Article}%
\bibitem[{Wang} et al.(2024)]{wang_window}
\bibinfo{author}{{Wang} MS}, \bibinfo{author}{{Beutler} F}, \bibinfo{author}{{Aguilar} J}, \bibinfo{author}{{Ahlen} S}, \bibinfo{author}{{Bianchi} D}, \bibinfo{author}{{Brooks} D}, \bibinfo{author}{{Claybaugh} T}, \bibinfo{author}{{de la Macorra} A}, \bibinfo{author}{{Doel} P}, \bibinfo{author}{{Font-Ribera} A}, \bibinfo{author}{{Gazta{\~n}aga} E}, \bibinfo{author}{{Gutierrez} G}, \bibinfo{author}{{Honscheid} K}, \bibinfo{author}{{Howlett} C}, \bibinfo{author}{{Kirkby} D}, \bibinfo{author}{{Lambert} A}, \bibinfo{author}{{Landriau} M}, \bibinfo{author}{{Miquel} R}, \bibinfo{author}{{Niz} G}, \bibinfo{author}{{Prada} F}, \bibinfo{author}{{P{\'e}rez-R{\`a}fols} I}, \bibinfo{author}{{Rossi} G}, \bibinfo{author}{{Sanchez} E}, \bibinfo{author}{{Schlegel} D}, \bibinfo{author}{{Schubnell} M}, \bibinfo{author}{{Sprayberry} D}, \bibinfo{author}{{Tarl{\'e}} G} and  \bibinfo{author}{{Weaver} BA} (\bibinfo{year}{2024}), \bibinfo{month}{Nov.}
\bibinfo{title}{{Window convolution of the galaxy clustering bispectrum}}.
\bibinfo{journal}{{\em arXiv e-prints}} , \bibinfo{eid}{arXiv:2411.14947}\bibinfo{doi}{\doi{10.48550/arXiv.2411.14947}}.
\eprint{2411.14947}.

\bibtype{Article}%
\bibitem[{Watkinson} et al.(2022)]{watkins_22}
\bibinfo{author}{{Watkinson} CA}, \bibinfo{author}{{Greig} B} and  \bibinfo{author}{{Mesinger} A} (\bibinfo{year}{2022}), \bibinfo{month}{Mar.}
\bibinfo{title}{{Epoch of reionization parameter estimation with the 21-cm bispectrum}}.
\bibinfo{journal}{{\em \mnras}} \bibinfo{volume}{510} (\bibinfo{number}{3}): \bibinfo{pages}{3838--3848}. \bibinfo{doi}{\doi{10.1093/mnras/stab3706}}.
\eprint{2102.02310}.

\bibtype{Article}%
\bibitem[{Weinberg} et al.(2013)]{weinberg}
\bibinfo{author}{{Weinberg} DH}, \bibinfo{author}{{Mortonson} MJ}, \bibinfo{author}{{Eisenstein} DJ}, \bibinfo{author}{{Hirata} C}, \bibinfo{author}{{Riess} AG} and  \bibinfo{author}{{Rozo} E} (\bibinfo{year}{2013}), \bibinfo{month}{Sep.}
\bibinfo{title}{{Observational probes of cosmic acceleration}}.
\bibinfo{journal}{{\em \physrep}} \bibinfo{volume}{530} (\bibinfo{number}{2}): \bibinfo{pages}{87--255}. \bibinfo{doi}{\doi{10.1016/j.physrep.2013.05.001}}.
\eprint{1201.2434}.

\bibtype{Article}%
\bibitem[{White}(1979)]{white_hier}
\bibinfo{author}{{White} SDM} (\bibinfo{year}{1979}), \bibinfo{month}{Jan.}
\bibinfo{title}{{The hierarchy of correlation functions and its relation to other measures of galaxy clustering.}}
\bibinfo{journal}{{\em \mnras}} \bibinfo{volume}{186}: \bibinfo{pages}{145--154}. \bibinfo{doi}{\doi{10.1093/mnras/186.2.145}}.

\bibtype{Article}%
\bibitem[{Williamson} et al.(2024)]{williamson_mhd}
\bibinfo{author}{{Williamson} V}, \bibinfo{author}{{Sunseri} J}, \bibinfo{author}{{Slepian} Z}, \bibinfo{author}{{Hou} J} and  \bibinfo{author}{{Greco} A} (\bibinfo{year}{2024}), \bibinfo{month}{Dec.}
\bibinfo{title}{{First Measurements of the 4-Point Correlation Function of Magnetohydrodynamic Turbulence as a Novel Probe of the Interstellar Medium}}.
\bibinfo{journal}{{\em arXiv e-prints}} , \bibinfo{eid}{arXiv:2412.03967}\bibinfo{doi}{\doi{10.48550/arXiv.2412.03967}}.
\eprint{2412.03967}.

\bibtype{Article}%
\bibitem[Wong(2008)]{wong2008higher}
\bibinfo{author}{Wong YY} (\bibinfo{year}{2008}).
\bibinfo{title}{Higher order corrections to the large scale matter power spectrum in the presence ofmassive neutrinos}.
\bibinfo{journal}{{\em Journal of Cosmology and Astroparticle Physics}} \bibinfo{volume}{2008} (\bibinfo{number}{10}): \bibinfo{pages}{035}.

\bibtype{Article}%
\bibitem[{Xu} et al.(2012)]{xu_2012}
\bibinfo{author}{{Xu} X}, \bibinfo{author}{{Padmanabhan} N}, \bibinfo{author}{{Eisenstein} DJ}, \bibinfo{author}{{Mehta} KT} and  \bibinfo{author}{{Cuesta} AJ} (\bibinfo{year}{2012}), \bibinfo{month}{Dec.}
\bibinfo{title}{{A 2 per cent distance to z = 0.35 by reconstructing baryon acoustic oscillations - II. Fitting techniques}}.
\bibinfo{journal}{{\em \mnras}} \bibinfo{volume}{427} (\bibinfo{number}{3}): \bibinfo{pages}{2146--2167}. \bibinfo{doi}{\doi{10.1111/j.1365-2966.2012.21573.x}}.
\eprint{1202.0091}.

\bibtype{Article}%
\bibitem[Yankelevich and Porciani(2019)]{yankelevich2019cosmological}
\bibinfo{author}{Yankelevich V} and  \bibinfo{author}{Porciani C} (\bibinfo{year}{2019}).
\bibinfo{title}{Cosmological information in the redshift-space bispectrum}.
\bibinfo{journal}{{\em Monthly Notices of the Royal Astronomical Society}} \bibinfo{volume}{483} (\bibinfo{number}{2}): \bibinfo{pages}{2078--2099}.

\bibtype{Article}%
\bibitem[{Yoo} et al.(2011)]{yoo_dalal}
\bibinfo{author}{{Yoo} J}, \bibinfo{author}{{Dalal} N} and  \bibinfo{author}{{Seljak} U} (\bibinfo{year}{2011}), \bibinfo{month}{Jul.}
\bibinfo{title}{{Supersonic relative velocity effect on the baryonic acoustic oscillation measurements}}.
\bibinfo{journal}{{\em \jcap}} \bibinfo{volume}{2011} (\bibinfo{number}{7}), \bibinfo{eid}{018}. \bibinfo{doi}{\doi{10.1088/1475-7516/2011/07/018}}.
\eprint{1105.3732}.

\bibtype{Article}%
\bibitem[{Zheng} et al.(2005)]{HODmodel}
\bibinfo{author}{{Zheng} Z}, \bibinfo{author}{{Berlind} AA}, \bibinfo{author}{{Weinberg} DH}, \bibinfo{author}{{Benson} AJ}, \bibinfo{author}{{Baugh} CM}, \bibinfo{author}{{Cole} S}, \bibinfo{author}{{Dav{\'e}} R}, \bibinfo{author}{{Frenk} CS}, \bibinfo{author}{{Katz} N} and  \bibinfo{author}{{Lacey} CG} (\bibinfo{year}{2005}), \bibinfo{month}{Nov.}
\bibinfo{title}{{Theoretical Models of the Halo Occupation Distribution: Separating Central and Satellite Galaxies}}.
\bibinfo{journal}{{\em \apj}} \bibinfo{volume}{633} (\bibinfo{number}{2}): \bibinfo{pages}{791--809}. \bibinfo{doi}{\doi{10.1086/466510}}.
\eprint{astro-ph/0408564}.

\end{thebibliography*}

\end{document}